\documentclass[aps,prd,twocolumn,preprintnumbers,amssymb,nofootinbib,floatfix,showpacs]{revtex4}
\usepackage{graphicx}% Include figure files
\usepackage{subfig}
\usepackage{dcolumn}% Align table columns on decimal point
\usepackage{bm}% bold math
\usepackage{amssymb}
\usepackage{amsmath}
\usepackage{amsfonts}
\usepackage{multirow}
\usepackage{hyperref}

\graphicspath{{PruningPaperFigures/}}

\newcommand{\be}[0]{\begin{equation}}
\newcommand{\ee}[0]{\end{equation}}

\newcommand{\kt}[0]{\text{k}_{\text{T}}}
\newcommand{\cut}[0]{\text{cut}}

\begin{document}

\title{Recombination Algorithms and Jet Substructure: Pruning as a Tool for Heavy Particle Searches}
\author{Stephen D. Ellis}
\author{Christopher K. Vermilion}
\author{Jonathan R. Walsh}
\affiliation{University of Washington, Seattle, WA 98195-1560}

\begin{abstract}
We discuss jet substructure in recombination algorithms for QCD jets and single jets from heavy particle decays.  We demonstrate that the jet algorithm can introduce significant systematic effects into the substructure.  By characterizing these systematic effects and the substructure from QCD, splash-in, and heavy particle decays, we identify a technique, pruning, to better identify heavy particle decays into single jets and distinguish them from QCD jets.  Pruning removes protojets typical of soft, wide angle radiation, improves the mass resolution of jets reconstructing a heavy particle decay, and decreases the QCD background.  We show that pruning provides significant improvements over unpruned jets in identifying top quarks and $W$ bosons and separating them from a QCD background, and may be useful in a search for heavy particles.

\end{abstract}

\maketitle

\section{Introduction}

The Large Hadron Collider (LHC) will present an exciting and challenging environment.  Efforts to tease out hints of Beyond the Standard Model (BSM) physics from complicated final states, typically dominated by Standard Model (SM) interactions, will almost surely require the use of new techniques applied to familiar quantities.  Of particular interest is the question of how we think about hadronic jets at the LHC \cite{Ellis:08.1}.  Historically jets have been employed as surrogates for \emph{individual} short distance energetic partons that evolve semi-independently into showers of energetic hadrons on their way from the interaction point through the detectors.  An accurate reconstruction of the jets in an event then provides an approximate description of the underlying short-distance, hard-scattering kinematics.  With this picture in mind, it is not surprising that the internal structure of jets, e.g., the fact that the experimentally detected jets exhibit nonzero masses, has rarely been used in analyses at the Tevatron.  However, we can anticipate that large-mass objects, which yield multijet decays at the Tevatron, e.g., $W/Z$'s (two jets) or top quarks (three jets), will often be produced with sufficient boosts to appear as single jets at the LHC.  Thus the masses of jets and further details of the internal structure of jets will be useful in identifying single jets not only as familiar objects like the aforementioned vector bosons and top quarks, but also as less familiar cascade decays of SUSY particles or the decays of V-particles \cite{Strassler:06.1}.  In fact, the idea of studying the subjet structure of jets has been around for some time, but initially this study took the form of discussing the number of jets as a function of the jet resolution scale, typically at $e^+e^-$ colliders, or the $p_T$ distribution within the cone of (cone) jets at the Tevatron. (See, for example, the analyses in \cite{Acosta:2005ix, Akers:1994wj, Akers:1994wc}.)  Recently a variety of studies \cite{Butterworth:02.1, Butterworth:07.1, Brooijmans:08.1, Butterworth:08.1, Thaler:08.1, Kaplan:08.1, Almeida:08.1, Krohn:09.1, Ellis:09.1, Butterworth:09.1, Plehn:09.1} have appeared suggesting a range of techniques for identifying jets with specific properties.  It is to this discussion that we intend to contribute.  Not surprisingly the current literature focuses on ``tagging'' the single jet decays of the particles mentioned above and the Higgs boson.  However, since we cannot be certain as to the full spectrum of new physics to be found at the LHC, it is important to keep in mind the underlying goal of separating QCD jets from \emph{any} other type of jet.  This will be challenging and the diversity of approaches currently being discussed in the literature is essential.  Successful searches for new physics at the LHC will likely employ a variety of techniques.  The analysis described below presents detailed properties of the ``pruning'' procedure outlined in \cite{Ellis:09.1}.

In the following discussion we will focus on jets defined by $\kt$-type jet algorithms.  The iterative recombination structure of these algorithms yields jets that, by definition, are assembled from a sequence of protojets, or subjets. It is natural to try to use this subjet structure (along with the $p_T$ and mass of the jet) to distinguish different types of jets. A combination of cuts and likelihood methods applied to this subjet structure can be used to identify jets, and thus events, likely to be enriched with vector or Higgs bosons, top quarks, or BSM physics.  Such jet-labeling techniques can then be used in conjunction with more familiar jet- and lepton-counting methods to isolate new physics at the LHC.

An essential aspect of high-$p_T$ jets at the LHC is that the jet algorithm ensures nonzero masses not only for the individual jets, but also for the subjets.  For recombination algorithms, we can analyze the $1 \to 2$ branching structure inherent in the substructure of the jet in terms of concepts familiar from usual two-body decays.  In fact, it is exactly such decays (say from $W$/$Z$ and top quark decays) that we want to compare in the current study to the structure of ``ordinary'' QCD (light quark and gluon) jets.  As we analyze the internal structure of jets we will attempt to keep in mind the various limitations of jets.  Jets are not intrinsically well-defined, but exhibit (often broad) distributions that are shaped by the very algorithms that define them.  Further, true experimental QCD jets are not identical to the leading-logarithm parton showers produced by Monte Carlos, but include also (perturbative) contributions from hard emissions, which may be important for precisely the properties of jets we want to discuss here, including masses.  Finally, the background particles from the underlying event, and from pile-up at higher luminosities, will influence the properties of the jets observed in the detector.

During the run-up to the LHC, jet substructure has increasingly drawn interest as an analysis tool \cite{Butterworth:02.1, Butterworth:07.1, Brooijmans:08.1, Butterworth:08.1, Thaler:08.1, Kaplan:08.1, Almeida:08.1, Krohn:09.1, Ellis:09.1, Butterworth:09.1, Plehn:09.1}.  The LHC will generate a deluge of multijet events that form a background to most interesting processes, and techniques to separate these signals will prove very useful.  To that end, various groups have shown that jet substructure can be used in $W$ \cite{Butterworth:02.1}, top \cite{Brooijmans:08.1, Thaler:08.1, Kaplan:08.1, Almeida:08.1, Plehn:09.1}, and Higgs \cite{Butterworth:08.1, Plehn:09.1} identification, as well as reconstruction of SUSY mass spectra \cite{Butterworth:07.1, Butterworth:09.1}.

We take a more general approach below.  Instead of describing a technique using jet substructure to find a particular signal, we study features of recombination algorithms.  We identify major systematic effects in jets found with the $\kt$ and CA algorithms, and discuss how they affect the found jet substructure.  To reduce these systematic effects we define a generic procedure, which we call pruning, that improves the jet substructure for the purposes of heavy particle identification.  We note that pruning is based on the same ideas as other jet substructure methods, \cite{Butterworth:08.1} and \cite{Kaplan:08.1}, in that these techniques also modify the jet substructure to improve heavy particle identification.  Pruning differs from these methods in that it is built as a broad jet substructure analysis tool, and one that can be used in a variety of searches.  To this end, the mechanics of the pruning procedure differ from other methods, allowing it to be generalized more easily.  Pruning can be performed using either the CA or $\kt$ algorithms to generate substructure for a jet, and the procedure can be implemented on jets coming from any algorithm, since the procedure is independent of the jet finder.  In the studies below, we will quantify several aspects of the performance of pruning to demonstrate its utility.

The following discussion includes a review of jet algorithms (Sec.~\ref{sec:recomb}) and a review of the expected properties of jets from QCD (Sec.~\ref{sec:QCDJets}) and those from heavy particles (Sec.~\ref{sec:reconHeavy}).  In studying QCD and heavy particle jets, we will discuss key systematic effects imposed on the jet substructure by the jet algorithm itself.  In Sec.~\ref{sec:algEffects} we then contrast the expected substructure for QCD and heavy particle jets and describe how the task of separating the two types of jets is complicated by systematic effects of the jet algorithm and the hadronic environment.  In Sec.~\ref{sec:pruning}, we show how these systematic effects can by reduced by a procedure we call ``pruning''.  Secs.~\ref{sec:MC} and \ref{sec:results} describe our Monte Carlo studies of pruning and their results.  Additional computational details are provided in Appendix \ref{sec:appendix}.  In Sec.~\ref{sec:conc} we summarize these results and provide concluding remarks.

%%%%%%%%%% END OF SECTION 1: INTRODUCTION %%%%%%%%%%

\section{Recombination Algorithms and Jet Substructure}
\label{sec:recomb}

Jet algorithms can be broadly divided into two categories, recombination algorithms and cone algorithms \cite{Ellis:08.1}.  Both types of algorithms form jets from protojets, which are initially generic objects such as calorimeter towers, topological clusters, or final state particles.  Cone algorithms fit protojets within a fixed geometric shape, the cone, and attempt to find stable configurations of those shapes to find jets.  In the cone-jet language, ``stable'' means that the direction of the total four-momentum of the protojets in the cone matches the direction of the axis of the cone. Recombination algorithms, on the other hand, give a prescription to \emph{pairwise} (re)combine protojets into new protojets, eventually yielding a jet.  For the recombination algorithms studied in this work, this prescription is based on an understanding of how the QCD shower operates, so that the recombination algorithm attempts to undo the effects of showering and approximately trace back to objects coming from the hard scattering.  The anti-$\kt$ algorithm \cite{Cacciari:08.2} functions more like the original cone algorithms, and its recombination scheme is not designed to backtrack through the QCD shower.  Cone algorithms have been the standard in collider experiments, but recombination algorithms are finding more frequent use.  Analyses at the Tevatron \cite{Aaltonen:2008eq} have shown that the most common cone and recombination algorithms agree in measurements of jet cross sections.

A general recombination algorithm uses a distance measure $\rho_{ij}$ between protojets to control how they are merged.  A ``beam distance'' $\rho_i$ determines when a protojet should be promoted to a jet.  The algorithm proceeds as follows:

\begin{itemize}
\item[0.] Form a list $L$ of all protojets to be merged.

\item[1.] Calculate the distance between all pairs of protojets in $L$ using the metric $\rho_{ij}$, and the beam distance for each protojet in $L$ using $\rho_i$.

\item[2.] Find the smallest overall distance in the set $\{\rho_i, \rho_{ij}\}$.

\item[3.] If this smallest distance is a $\rho_{ij}$, merge protojets $i$ and $j$ by adding their four vectors.  Replace the pair of protojets in $L$ with this new merged protojet.  If the smallest distance is a $\rho_i$, promote protojet $i$ to a jet and remove it from $L$.

\item[4.] Iterate this process until $L$ is empty, i.e., all protojets have been promoted to jets.\footnote{This defines an \emph{inclusive} algorithm.  For an \emph{exclusive} algorithm, there are no promotions, but instead of recombining until $L$ is empty, mergings proceed until all $\rho_{ij}$ exceed a fixed $\rho_\text{cut}$.}
\end{itemize}

For the $\kt$ \cite{Catani:92.1, Catani:93.1, Ellis:93.1} and Cambridge-Aachen (CA) \cite{Dokshitzer:97.1} recombination algorithms the metrics are
\be
\begin{split}
\kt: \rho_{ij} \equiv \min(p_{Ti},p_{Tj})\Delta R_{ij}/D, & \qquad \rho_i \equiv p_{Ti}; \\
\text{CA}: \rho_{ij} \equiv \Delta R_{ij}/D, & \qquad \rho_i \equiv 1.
\end{split}
\label{eq:algo}
\ee
Here $p_{Ti}$ is the transverse momentum of protojet $i$ and $\Delta R_{ij} \equiv \sqrt{(\phi_i - \phi_j)^2 + (y_i - y_j)^2}$ is a measure of the angle between two protojets that is invariant under boosts along and rotations around the beam direction.  $\phi$ is the azimuthal angle around the beam direction, $\phi = \tan^{-1} {p_y}/{p_x}$, and $y$ is the rapidity, $y = \tanh^{-1} {p_z}/{E}$, with the beam along the $z$-axis.  The angular parameter $D$ governs when protojets should be promoted to jets: it determines when a protojet's beam distance is less than the distance to other objects.  $D$ provides a rough measure of the typical angular size (in $y$--$\phi$) of the resulting jets.

The recombination metric $\rho_{ij}$ determines the \emph{order} in which protojets are merged in the jet, with recombinations that minimize the metric performed first.  From the definitions of the recombination metrics in Eq.~(\ref{eq:algo}), it is clear that the $\kt$ algorithm tends to merge low-$p_T$ protojets earlier, while the CA algorithm merges pairs in strict angular order. This distinction will be very important in our subsequent discussion.

\subsection{Jet Substructure}

A recombination algorithm naturally defines substructure for the jet.  The sequence of recombinations tells us how to construct the jet in step-by-step $2\to1$ mergings, and we can unfold the jet into two, three, or more subjets by undoing the last recombinations.  Because the jet algorithm begins and ends with physically meaningful information (starting at calorimeter cells, for example, and ending at jets), the intermediate (subjet) information generated by the $\kt$ and CA (but not the anti-$\kt$\footnote{The anti-$\kt$ algorithm has the metrics $\rho_{ij} \equiv \min (p_{Ti}^{-1}, p_{Tj}^{-1}) \Delta R_{ij}/D$, $\rho_i \equiv p_{Ti}^{-1}$, so it tends to cluster protojets with the hardest protojet, resulting in cone-like jets with uninteresting substructure.}) recombination algorithms is expected to have physical significance as well.   In particular, we expect the earliest recombinations to approximately reconstruct the QCD shower, while the last recombinations in the algorithm, those involving the largest-$p_T$ degrees of freedom, may indicate whether the jet was produced by QCD alone or a heavy particle decay plus QCD showering.  To discuss the details of jet substructure, we begin by defining relevant variables.

\subsection{Variables Describing Branchings and Their Kinematics}
\label{sec:recomb:variables}

In studying the substructure produced by jet algorithms, it will be useful to describe branchings using a set of kinematic variables.  Since we will consider the substructure of (massive) jets reconstructing kinematic decays and of QCD jets, there are two natural choices of variables.  Jet rest frame variables are useful to understand decays because the decay cross section takes a simple form.  Lab frame variables invariant under boosts along and rotations around the beam direction are useful because jet algorithms are formulated in terms of these variables, so algorithm systematics are most easily understood in terms of them.  The QCD soft/collinear singularity structure is also easy to express in lab frame variables.  We describe these two sets of variables and the relationship between them in this subsection.

Naively, there are twelve variables completely describing a $1 \to 2$ splitting.  Here we will focus on the top branching (the last merging) of the jet splitting into two daughter subjets, which we will label $J \to 1,2$.  Imposing the four constraints from momentum conservation to the branching leaves eight independent variables.  The invariance of the algorithm metrics under longitudinal boosts and azimuthal rotations removes two of these (they are irrelevant).  For simplicity we will use this invariance to set the jet's direction to be along the $x$-axis, defining the $z$-axis to be along the beam direction.  Therefore there are six relevant variables needed to describe a $1 \to 2$ branching.  Three of these variables are related to the three-momenta of the jet and subjets, and the other three are related to their masses.

The two sets of variables we will use to understand jet substructure share common elements.  Of the six variables, only one needs to be dimensionful, and we can describe all other scales in terms of this one.  The dimensionful variable we choose is the mass $m_J$ of the jet.  In addition, we use the masses of the two daughter subjets scaled by the jet mass:
\be
a_1 \equiv \frac{m_1}{m_J} \quad \text{and} \quad a_2 \equiv \frac{m_2}{m_J} .
\label{eq:a1def}
\ee
We choose the particle labeled by `1' to be the heavier particle, $a_1 > a_2$.  The three masses, $m_J$, $a_1$, and $a_2$, will be common to both sets of variables.  Additionally, we will typically want to fix the $p_{T}$ of the jet and determine how the kinematics of a system change as $p_{T_J}$ is varied.  For QCD, a useful dimensionless quantity is the ratio of the mass and $p_T$ of the jet, whose
square we call $x_J$:
\be
x_J \equiv \frac{m_J^2}{p_{T_J}^2} .
\label{eq:xJdef}
\ee
For decays, we will opt instead to use the familiar magnitude $\gamma$ of the boost of the heavy particle from its rest frame to the lab frame, which is related to $x_J$ by
\be
\gamma = \sqrt{\frac{1}{x_J} + 1},\quad x_J=\frac{1}{\gamma^2-1}.
\ee
The remaining two variables, which are related to the momenta of the subjets, will differ between the rest frame and lab frame descriptions of the splitting.

Unpolarized $1\to2$ decays are naturally described in their rest frame by two angles.  These angles are the polar and azimuthal angles of one particle (the heavier one, say) with respect to the direction of the boost to the lab frame, and we label them $\theta_0$ and $\phi_0$ respectively.  Since we are choosing that the final jet be in the $\hat{x}$ direction, $\theta_0$ is measured from the $\hat{x}$ direction while $\phi_0$ is the angle in the $y$--$z$ plane, which we choose to be measured from the $\hat{y}$ direction.  Putting these variables together, the set that most intuitively describes a heavy particle decay is the ``rest frame'' set
\be
\{m_J,\ a_1,\ a_2,\ \gamma,\ \cos\theta_0,\ \phi_0\} .
\label{eq:decayvars}
\ee
The requirement that the (last) recombination vertex being described actually ``fit'' in a single jet reconstructed in the lab frame yields the constraint $\Delta R_{12} < D$, where $\Delta R_{12}$ is treated as a function of the variables in Eq.~(\ref{eq:decayvars}).

Consider describing the same kinematics in the lab frame.  As noted above, we want to choose variables that are invariant under longitudinal boosts and azimuthal rotations, which can be mapped onto the recombination metrics of the jet algorithm.  The angle $\Delta R_{12}$ between the daughter particles is a natural choice, as is the ratio of the minimum daughter $p_T$ to the parent $p_T$, which is commonly called $z$:
\be
z \equiv \frac{\min(p_{T_1},p_{T_2})}{p_{T_J}} .
\label{eq:zdef}
\ee
These variables make the recombination metrics for the $\kt$ and CA algorithms simple:
\be
\rho_{12}(\kt) = p_{T_J}z\Delta R_{12} \quad \text{and} \quad \rho_{12}(\text{CA}) = \Delta R_{12} .
\label{eq:reconCAKT}
\ee
Note that for a generic recombination, the momentum factors in the denominator of Eq.~(\ref{eq:zdef}) and in the $\kt$ metric in Eq.~(\ref{eq:reconCAKT}) should be $p_{Tp}$, the momentum of the the parent or combined subjet of the $2\to1$ recombination.

From these considerations we choose to describe recombinations in the lab frame with the set of variables
\be
\{m_J,\ a_1,\ a_2,\ x_J,\ z,\ \Delta R_{12}\} .
\label{eq:labvars}
\ee

\begin{figure*}[htbp]
\subfloat[$a_1=a_2=0$] {\label{thetaphicontours1} \includegraphics[width = .235\textwidth] {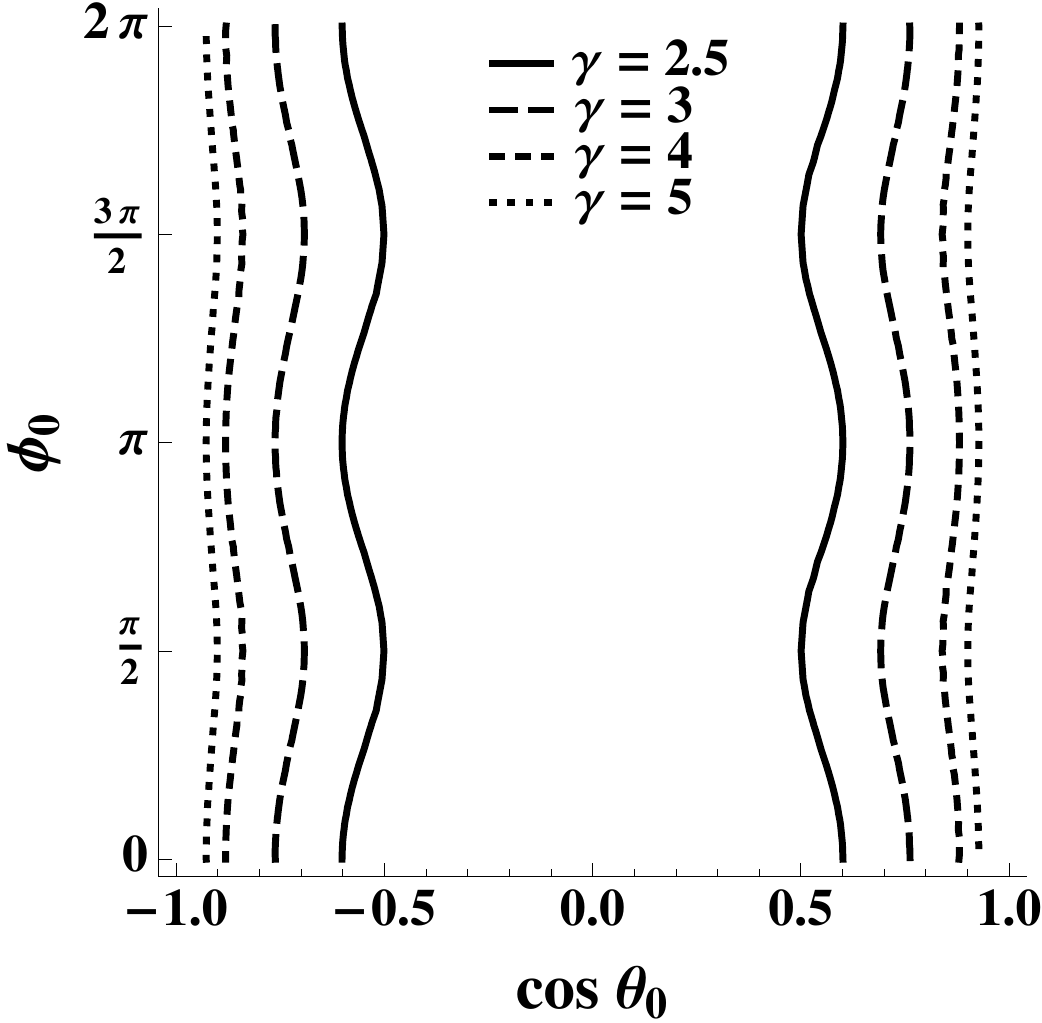}}
\subfloat[$a_1=0.46,\ a_2=0$] {\label{thetaphicontours2} \includegraphics[width = .235\textwidth] {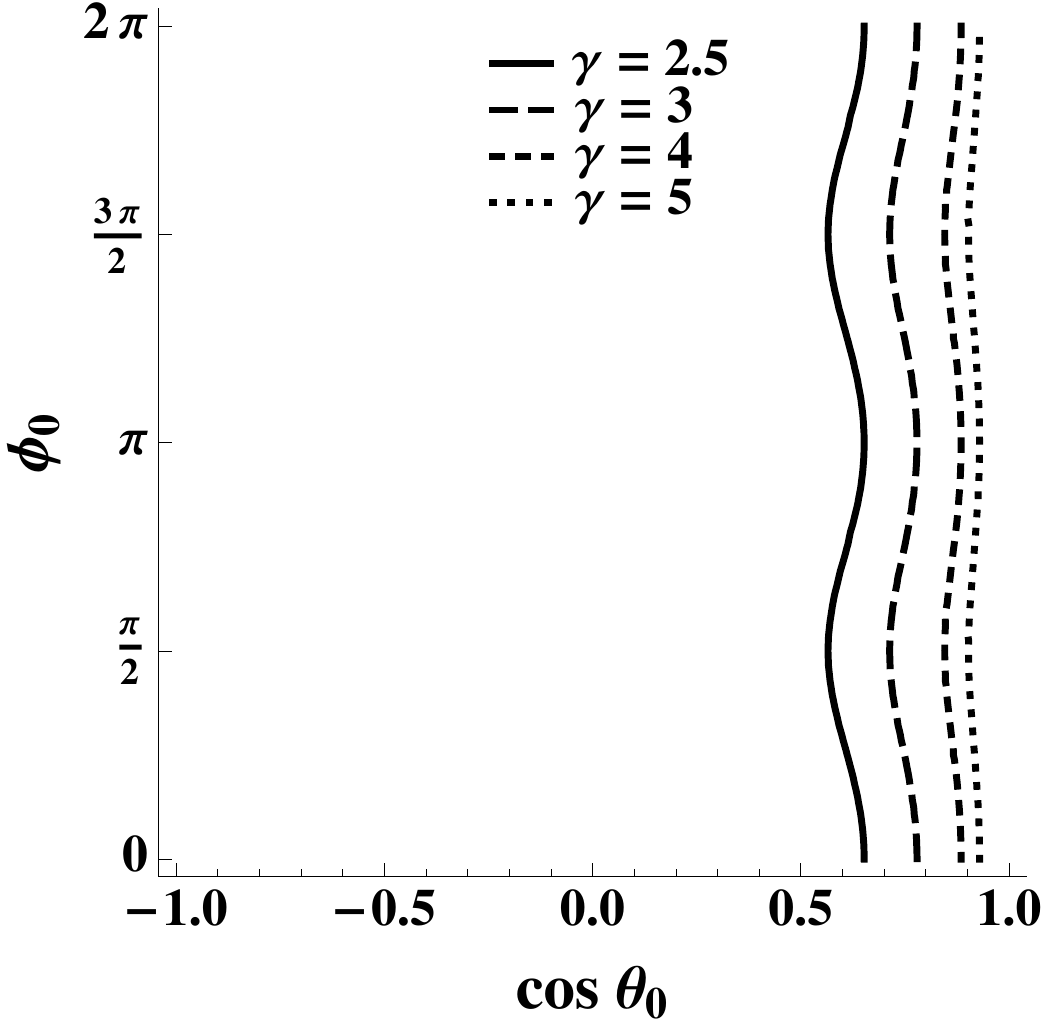}}
\subfloat[$a_1=0.9,\ a_2=0$] {\label{thetaphicontours3} \includegraphics[width = .235\textwidth]{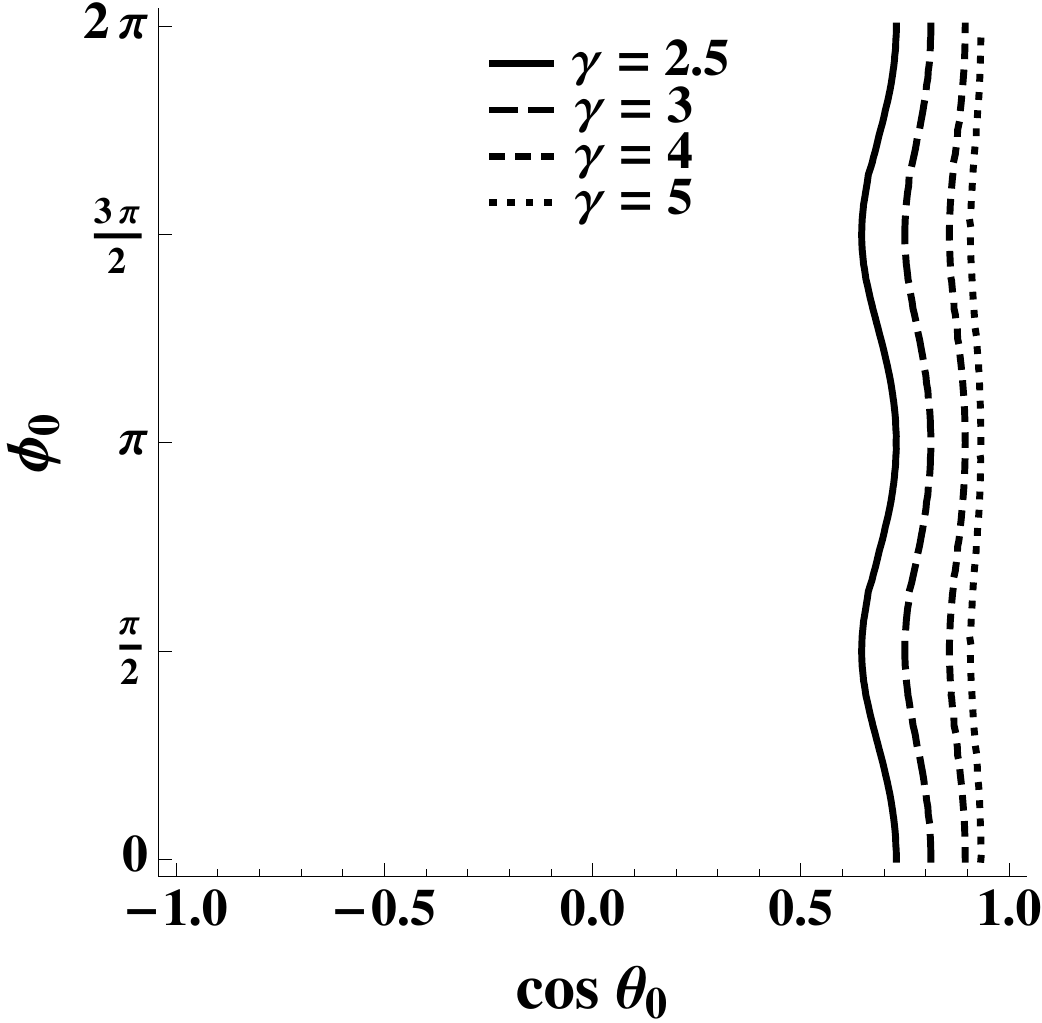}}
\subfloat[$a_1=0.3,\ a_2=0.1$] {\label{thetaphicontours4} \includegraphics[width = .235\textwidth] {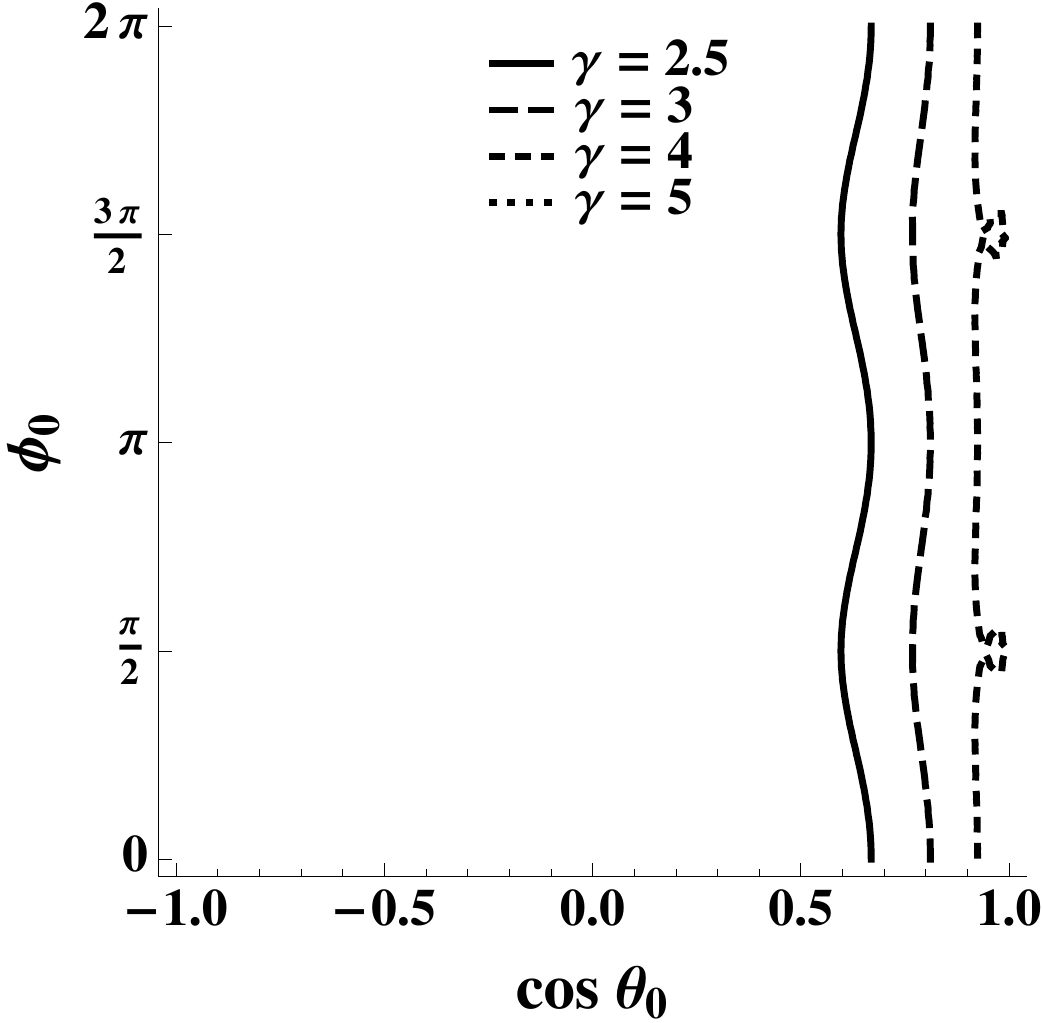}}
\caption{Boundaries in the $\cos\theta_0$--$\phi_0$ plane for a recombination step to fit in a jet of size $D=1.0$, for several values of the boost $\gamma$ and the subjet masses $\{a_1,\ a_2\}$.  The ``interior'' region has $\Delta R_{12} < D$.  }%\red{thetaphicontours}}
\label{thetaphicontours}
\end{figure*}

In using these variables it is essential to understand the structure of the corresponding phase space, especially for the last two variables in both sets.  Naively, for actual decays, we would expect that the phase space in $\cos\theta_0$ and $\phi_0$ of the rest frame variable set in  Eq.~(\ref{eq:decayvars}) is simple, with boundaries that are independent of the value of the other variables.  However, since we require that the decay ``fits'' in a jet (so that all the variables are defined), constraints and correlations appear.  The presence of these constraints and correlations is more apparent for the lab frame variables $\Delta R_{12}$ and $z$ since the recombination algorithm acts directly on the these variables.  As a first step in understanding these correlations we plot in Fig.~\ref{thetaphicontours}, the contour $\Delta R_{12} = D(=1.0)$ in the $(\cos\theta_0, \phi_0)$ phase space for different values of $\gamma$ and over different choices for $a_1$ and $a_2$.  These specific values of $a_1$ and $a_2$ correspond to a variety of interesting processes:  $a_1 = a_2 = 0$ gives the simplest kinematics and is therefore a useful starting point; $a_1 = 0.46, a_2 = 0$ gives the kinematics of the top quark decay; $a_1 = 0.9, a_2 = 0$ and $a_1 = 0.3, a_2 = 0.1$ are reasonable values for subjet masses from the CA and $\kt$ algorithms respectively.  The contour $\Delta R_{12} = D$ defines the boundary in phase space where a $1\to2$ process will no longer fit in a jet, with the interior region corresponding to splittings with $\Delta R_{12} < D$.  Note that the contour is nearly straight and vertical, increasingly so for larger $\gamma$.  This is a reflection of the fact that $\Delta R_{12}$ is nearly independent of $\phi_0$, up to terms suppressed by $\gamma^{-2}$.

\begin{figure*}[htbp]
\subfloat[$a_1=a_2=0$] {\label{zRcontours1} \includegraphics[width = .235\textwidth] {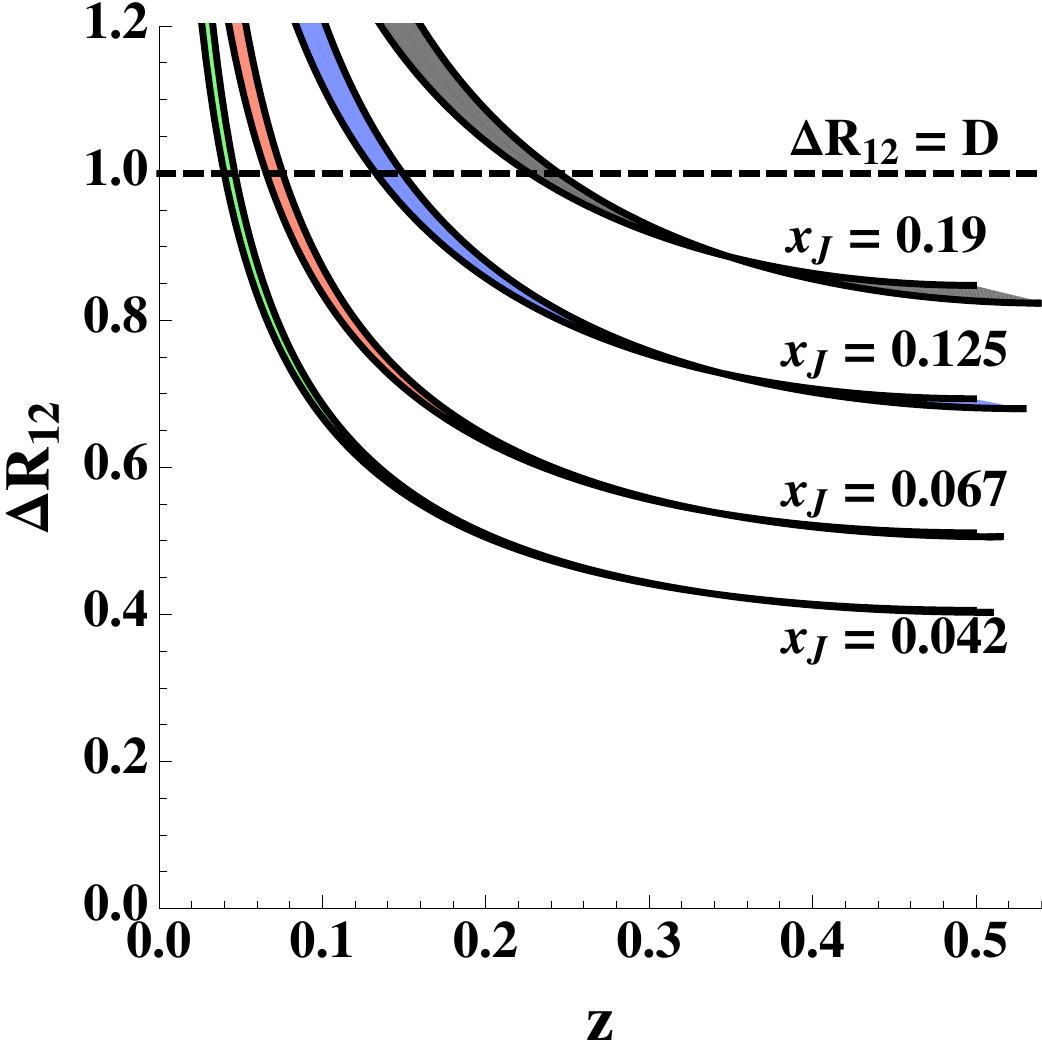}}
\subfloat[$a_1=0.46,\ a_2=0$] {\label{zRcontours2} \includegraphics[width = .235\textwidth] {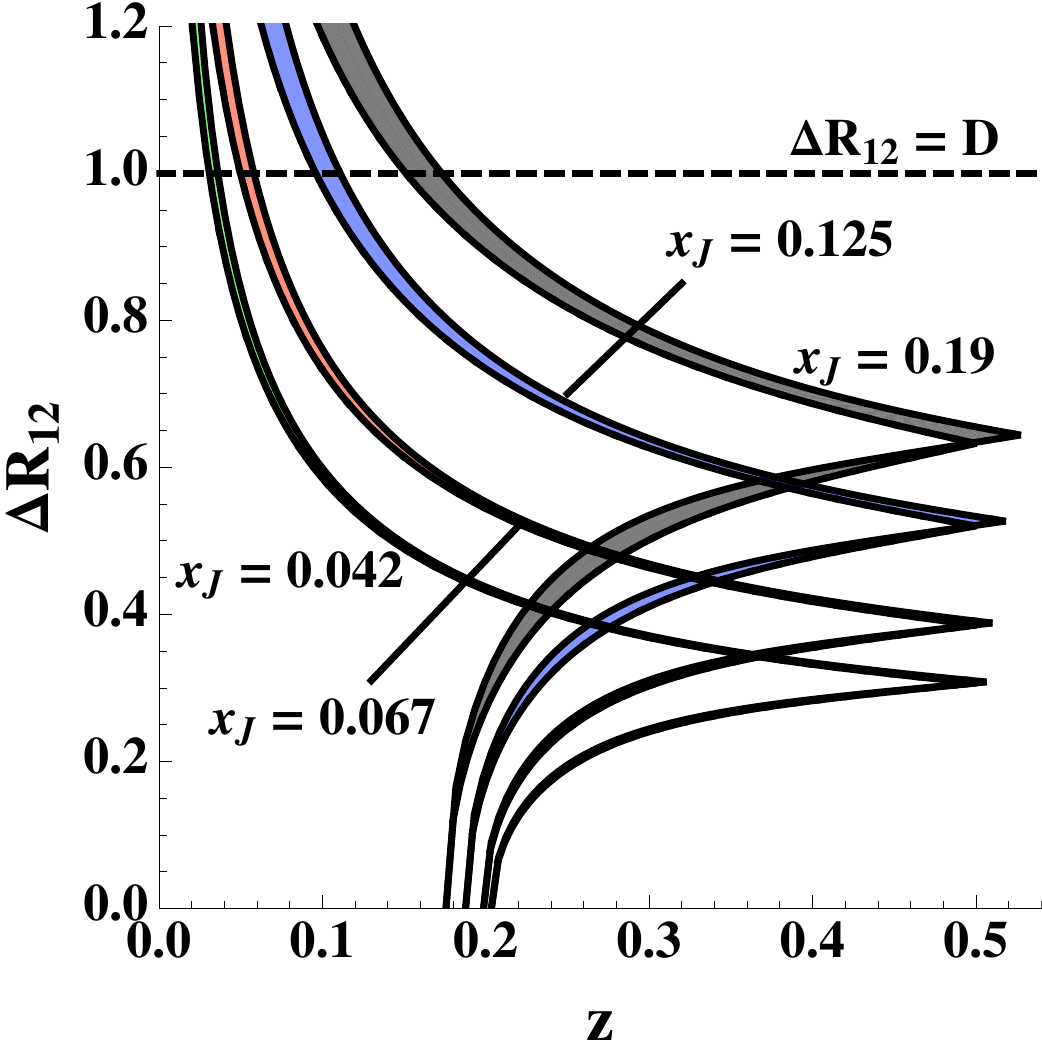}}
\subfloat[$a_1=0.9,\ a_2=0$] {\label{zRcontours3} \includegraphics[width = .235\textwidth] {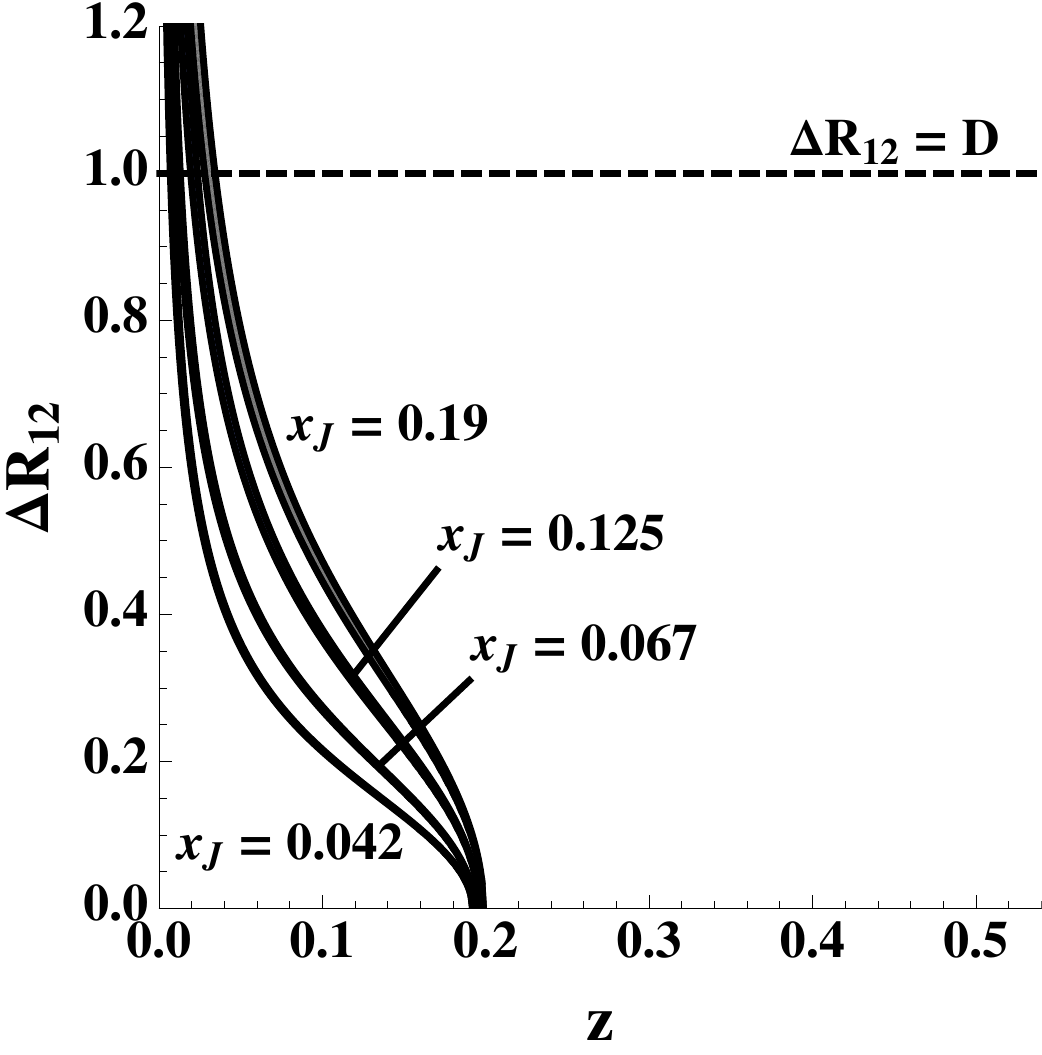}}
\subfloat[$a_1=0.3,\ a_2=0.1$] {\label{zRcontours4} \includegraphics[width = .235\textwidth]{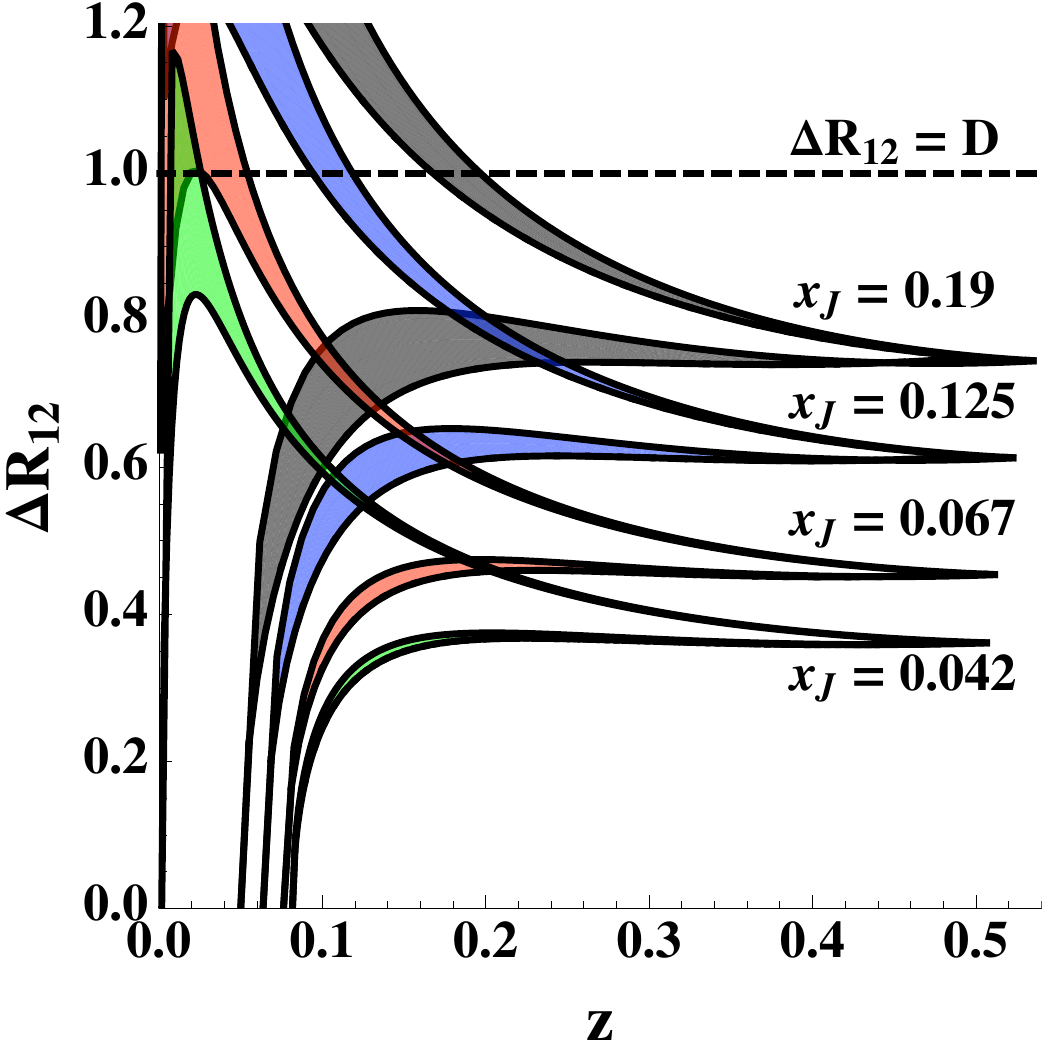}}
\caption{Boundaries in the $z$--$\Delta R_{12}$ plane for a recombination step of fixed $\{a_1,\ a_2,\ x_J\}$, for various values of $x_J$ and the subjet masses $\{a_1,\ a_2\}$.  Configurations with $\Delta R_{12} < D$ fit in a jet; $D = 1.0$ is shown for example.  }%\red{zRcontours}}
\label{zRcontours}
\end{figure*}

While the constraint $\Delta R_{12} < D$ for the $1\to2$ to fit in a jet becomes simpler in the $(z, \Delta R_{12})$ phase space, the boundaries of the phase space become more complex.  In Fig.~\ref{zRcontours}, we plot the available phase space in $(z, \Delta R_{12})$ for the same values of $x_J$, $a_1$, and $a_2$ as in Fig.~\ref{thetaphicontours}, translating the value of $\gamma$ into $x_J$.  The most striking feature is that for fixed $x_J$, $a_1$, and $a_2$, the phase space in ($z$, $\Delta R_{12}$) is nearly one-dimensional; this is again due to the fact that $\Delta R_{12}$ and also $z$ are nearly independent of $\phi_0$.  In particular, for $a_1 = a_2 = 0$ (as in Fig.~\ref{zRcontours1}), the phase space approximates the contour describing fixed $x_J$ for small $\Delta R_{12}$, which takes the simple form
\be
x_J \equiv \frac{m_{J}^{2}}{p_{T_J}^{2}} \approx z \left(  1-z \right) \Delta R_{12}^{2}.
\label{eq:simplejmass}
\ee
This approximation is accurate even for larger angles, $\Delta R_{12} \approx 1$, at the $10\%$ level. Note also that the width of the band about the contour described by Eq.~(\ref{eq:simplejmass}) is itself of order $x_J$.  As we decrease $x_J$ the band moves down and becomes narrower as indicated in Fig.~\ref{zRcontours1}).

As illustrated in Figs.~\ref{zRcontours2} and \ref{zRcontours4}, we can also see a double-band structure to the $(z, \Delta R_{12})$ phase space.  The upper band corresponds to the case where the lighter daughter is softer (smaller-$p_T$) than the heavier daughter (and determines $z$), while the lower band corresponds to the case where the heavier daughter is softer.  This does not occur in Fig.~\ref{zRcontours1} because $a_1 = a_2$ (the single band is double-covered), or in Fig.~\ref{zRcontours3} because the heavier particle is never the softer one for the chosen values of $x_J$.

Note that we have said nothing about the density of points in phase space for either pair of variables.  This is because the weighting of phase space is set by the dynamics of a process, while the boundaries are set by the kinematics.  Decays and QCD splittings weight the phase space differently, as we will show.

\subsection{Ordering in Recombination Algorithms}
\label{sec:AlgOrdering}

Having laid out variables useful to describe $1\to2$ processes, we can discuss how the jet algorithm orders recombinations in these variables.  Recombination algorithms merge objects according to the pairwise metric $\rho_{ij}$.  The sequence of recombinations is almost always monotonic in this metric: as the algorithm proceeds, the value typically increases.  Only certain kinematic configurations will decrease the metric from one recombination to the next, and the monotonicity violation is small and rare in practice.

This means it is rather straightforward to understand the typical recombinations that occur at different stages of the algorithm.  We can think in terms of a phase space boundary: the algorithm enforces a boundary in phase space at a constant value for the recombination metric which evolves to larger values as the recombination process proceeds.  If a recombination occurs at a certain value of the metric, $\rho_0$, then subsequent recombinations are very unlikely to have $\rho_{ij} < \rho_0$, meaning that region of phase space is unavailable for further recombinations.

\begin{figure}[htbp]
\subfloat[CA] {\label{algboundariesCA} \includegraphics[width=0.23\textwidth] {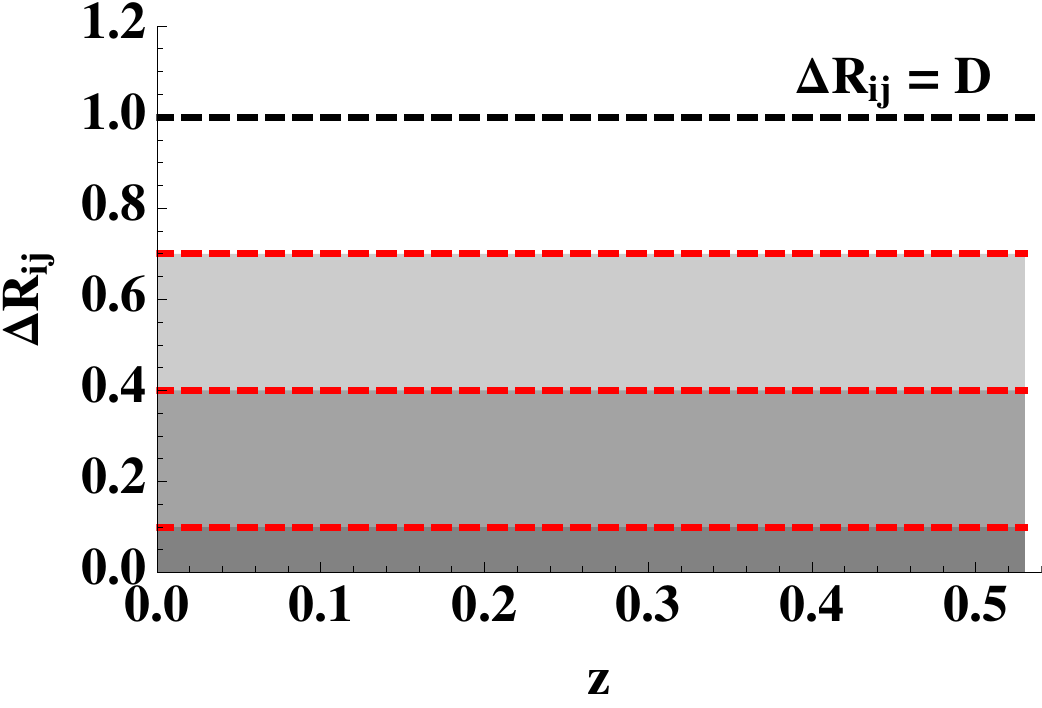}}
\subfloat[$\kt$] {\label{algboundariesKT} \includegraphics[width=0.23\textwidth] {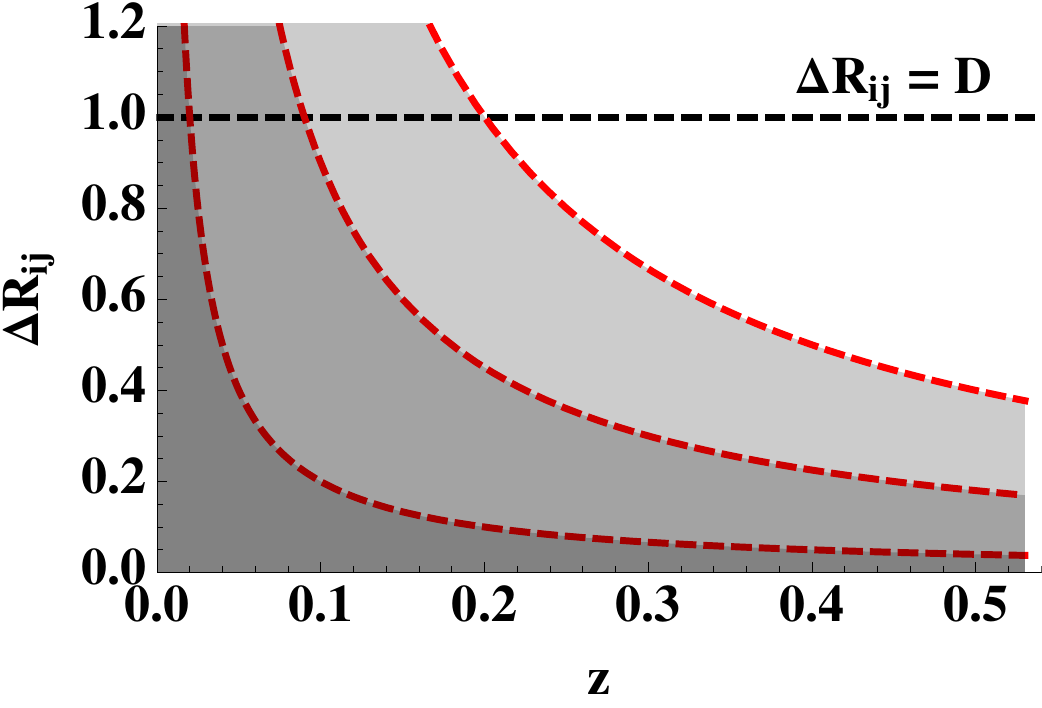}}
\caption{Typical boundaries (red, dashed lines) on phase space due to ordering in the CA and $\kt$ algorithms.  The shaded region below the boundaries is cut out, and the more heavily shaded regions correspond to earlier in the recombination sequence.  The cutoff $\Delta R_{ij} = D = 1.0$ is shown for reference (black, dashed line).   }%\red{algboundaries}}
\label{algboundaries}
\end{figure}

In Fig.~\ref{algboundaries}, we plot typical boundaries for the CA and $\kt$ algorithms in the $(z, \Delta R_{12})$ phase space.  For CA, these boundaries are simply lines of constant $\Delta R_{12}$, since the recombination metric is $\rho_{ij}(\textrm{CA}) = \Delta R_{ij}$.  For $\kt$, these boundaries are contours in $z\Delta R_{12}$, and implicitly depend on the $p_T$ of the parent particle in the splitting.  Because the $\kt$ recombination metric for $i,j\to p$ is $\rho_{ij}(\kt) = z\Delta R_{ij}p_{Tp}$, decreasing the value of $p_{Tp}$ will shift the boundary out to larger $z\Delta R_{ij}$.  These algorithm dependent ordering effects will be important in understanding the restrictions on the kinematics of the last recombinations in a jet.  For instance, we expect to observe no small-angle late recombinations in a jet defined by the CA algorithm.

\subsection{Studying the Substructure of Recombination Algorithms}

In the following sections we discuss various aspects of jet substructure, especially as applied toward identifying heavy particle decays within single jets and separating them from QCD jets.  To effectively discriminate between jets, we must have an understanding of the substructure expected from both QCD and decays.  To this end, we will study toy models of the underlying $1\to2$ processes with appropriate (but approximate) dynamics.  We will also study the substructure observed in jets found in simulated events, which include showering and hadronization, for both pure QCD and heavy particle decays.  In these more realistic jets, with many more degrees of freedom, we must understand the role of the jet algorithm in determining the features of the last recombinations in the jet.  This bias will impact how (and whether) we can interpret the last recombinations as relevant to the physics of the jet.

We will find that the differences in the metrics of $\kt$ and CA will introduce shaping effects on the recombinations.  We will observe these in the distributions of kinematic variables of interest, e.g., the jet and subjet masses, $z$, and $\Delta R_{12}$.  The major point of this work will be to motivate and develop a method to identify jet substructure most likely to come from the decay of a heavy particle and separate this substructure from recombinations likely to represent QCD.

In Sec.~\ref{sec:QCDJets}, we study QCD (only) jet masses and substructure in terms of the variables $x_J$, $z$ and $\Delta R_{12}$, starting with a leading-log approximation including only the soft and collinear singularities.  We find the distribution in $x_J$ in this approximation and discuss the implications for the substructure in a QCD jet, specifically the distributions in both $z$ and $\Delta R_{12}$ for fixed $x_J$.  Finally we look at the jet mass and substructure distributions found in jets from fully simulated events. Of particular interest is the algorithm dependence.

In Sec.~\ref{sec:reconHeavy}, we first study $1\to2$ decays with fixed boost and massless daughters (e.g., a $W$ decay into quarks) and a top quark decay into massless quarks.  The parton-level top quark decay into three quarks, which is made up of two $1\to2$ decays, is instructive because the jet algorithm matters: the CA and $\kt$ algorithms can reconstruct the jet in different ways.  For both kinds of decays, we consider both the full, unreconstructed decay distributions in $z$ and $\Delta R_{12}$, then proceed to study the shaping effects that reconstruction in a single jet has on the ``in-a-jet'' distributions of these variables.  We also look at the shaping in terms of the rest frame variable $\cos\theta_0$, which provides a good intuitive picture of which decays will be reconstructed in a single jet.  Understanding this shaping will be key to understanding the substructure we expect from decays and the effects of the jet algorithm.  We contrast this substructure with the expected substructure from QCD jets, pointing out key similarities and differences.  Finally we look at the distributions found in fully simulated events of both $W$ and top quark decays.

In Sec.~\ref{sec:algEffects}, we compare the results of Secs.~\ref{sec:QCDJets} and \ref{sec:reconHeavy}.  We also consider the impact of event effects such as the underlying event, which are common to all events.  In particular, we focus on understanding how these contributions manifest themselves in the substructure of the jet and the role that the algorithm plays in determining the substructure.  We will find that jet algorithms, acting on events that include these contributions, yield substructure that often obscures the recombinations reconstructing a heavy particle decay.  This is especially true of the CA algorithm, which we will show has a large systematic effect on its jet substructure.  We will use these lessons in later sections to construct the pruning procedure to modify the jet substructure, removing recombinations that are likely to obscure a heavy particle decay.

%%%%%%%%%% END OF SECTION 2: RECOMBINATION ALGORITHMS AND JET SUBSTRUCTURE%%%%%%%%%%

\section{QCD Jets}
\label{sec:QCDJets}

The LHC will be the first collider where jet masses play a serious role in analyses. The proton-proton center of mass energy at the LHC is sufficiently large that the mass spectrum of QCD jets will extend far into the regime of heavy particle production ($m_W$ and above). Because masses are such an important variable in jet substructure, masses of QCD jets will play an essential role in determining the effectiveness of jet substructure techniques at separating QCD jets from jets with new physics.  We expect that the jet mass distribution in QCD is smoothly falling due to the lack of any intrinsic mass scale above $\Lambda_{QCD}$, while jets containing heavy particles are expected to exhibit enhancements in a relatively narrow jet mass range (given by the particle's width, detector effects, and the systematics of the algorithm).

Understanding the more detailed substructure of QCD jets (beyond the mass of the jet) presents an interesting challenge.  QCD jets are typically characterized by the soft and collinear kinematic regimes that dominate their evolution, but QCD populates the entire phase space of allowed kinematics.  Due to its immense cross section relative to other processes, small effects in QCD can produce event rates that still dominate other signals, even after cuts. Furthermore, the full kinematic distributions in QCD jet substructure currently can only be approximately calculated, so we focus on understanding the key features of QCD jets and the systematic effects that arise from the algorithms that define them. Note that even when an on-shell heavy particle is present in a jet, the corresponding kinematic decay(s) will contribute to only a few of the branchings within the jet. QCD will still be responsible for bulk of the complexity in the jet substructure, which is produced as the colored partons shower and hadronize, leading to the high multiplicity of color singlet particles observed in the detector.

It is a complex question to ask whether the jet substructure is accurately reconstructing the parton shower, and somewhat misguided, as the parton shower represents colored particles while the experimental algorithm only deals with color singlets. A more sensible question, and an answerable one, is to ask whether the algorithm is faithful to the dynamics of the parton shower. This is the basis of the metrics of the $\kt$ and CA recombination algorithms --- the ordering of recombinations captures the dominant kinematic features of branchings within the shower. In particular, the cross section for an extra real emission in the parton shower contains both a soft ($z$) and a collinear ($\Delta R$) singularity:
\be
d\sigma_{n+1}\sim d\sigma_{n}\frac{dz}{z}\frac{d\Delta R}{\Delta R}.
\label{eq:socol}
\ee
While these singularities are regulated (in perturbation theory) by virtual corrections, the enhancement remains, and we expect emissions in the QCD parton shower to be dominantly soft and/or collinear.  Due to their different metrics, the $\kt$ and CA algorithms will recombine these emissions differently, producing distinct substructure.  We will discuss the interplay between the dynamics of QCD and the recombination algorithms in the next two subsections.  In the first, we will consider a simple leading-logarithm (LL) approximation to perturbative QCD jets with just a single branching and zero-mass subjets.  This will illustrate the simplest kinematics of Section \ref{sec:recomb:variables} coupled with soft/collinear dynamics.  In the second subsection we consider the properties of the more realistic QCD jets found in fully simulated events.

\subsection{Jets in a Toy QCD}
\label{sec:QCDJets:Toy}

\begin{figure}[htbp]
\includegraphics[width = \columnwidth]{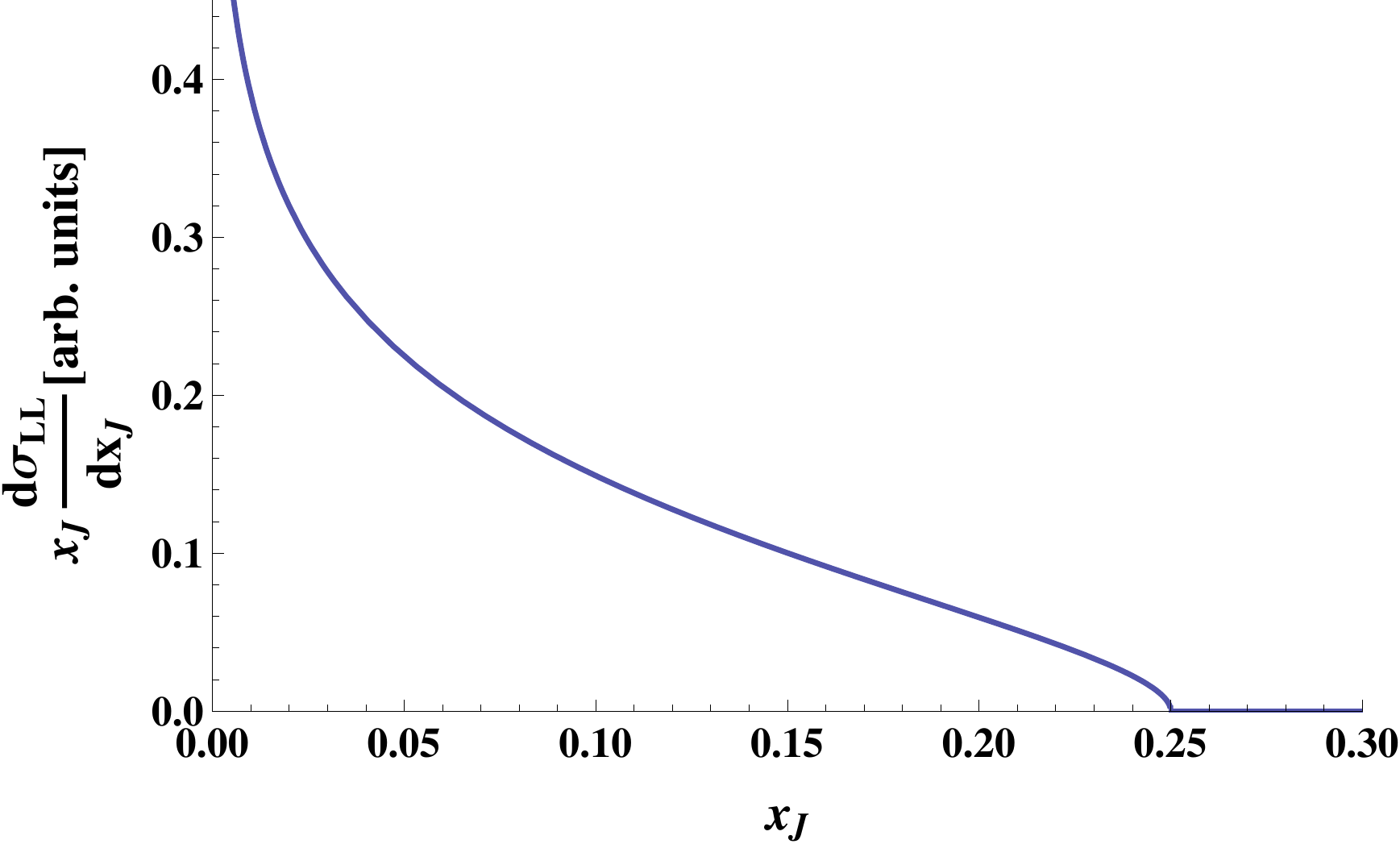}
\caption{Distribution in $x_J$ in simple LL toy model with D = 1.0.  }%\red{LLmjetplot}}
\label{LLmjetplot}
\end{figure}

To establish an intuitive level of understanding of jet substructure in QCD we consider a toy model description of jets in terms of a single branching and the variables $x_J$, $z$, and $\Delta R_{12}$.  We take the jet to have a fixed $p_{T_J}$.  We combine the leading-logarithmic dynamics of of Eq.~(\ref{eq:socol}) with the approximate expression for the jet mass in Eq.~(\ref{eq:simplejmass}), and we label this combined approximation as the ``LL'' approximation.  Recall that this approximation for the jet mass is useful for small subjet masses and small opening angles.  From Section \ref{sec:recomb:variables}, recall that fixing $x_J$ provides lower bounds on both $z$ and $\Delta R_{12}$ and ensures finite results for the LL approximation. This approach leads to the following simple form for the $x_J$ distribution,
\begin{eqnarray}
&&\frac{1}{\sigma}\frac{d\sigma_{LL}}{d(m_J^2/p_{T_J}^2)}\equiv\frac{1}{\sigma}
\frac{d\sigma_{LL}}{d x_J}\nonumber\\
&&\sim \int^{1/2}_0 \int^{D}_0 \frac{dz}{z}\frac{d\Delta R_{12}}{\Delta
R_{12}}\delta(x_J-z(1-z)\Delta R_{12}^2)\nonumber\\
&&= \frac{-\ln{\left(1-\sqrt{1-{4 x_J}/{D^2}}\right)}}{2x_J} \Theta\left[D^2/4 -
x_J\right] .
\label{eq:LL1}
\end{eqnarray}
Note we are integrating over the phase space of Fig.~\ref{zRcontours1}, treating it as one-dimensional.  The resulting distribution is exhibited in Fig.~\ref{LLmjetplot} for $D = 1.0$ where we have multiplied by a factor of $x_J$ to remove the explicit pole.  We observe both the cutoff at $x_J = D^2/4$ arising from the kinematics discussed in Section \ref{sec:recomb:variables} and the $-\ln(x_J)/x_J$ small-$x_J$ behavior arising from the singular soft/collinear dynamics.  Even if the infrared singularity is regulated by virtual emissions and the distribution is resummed, we still expect QCD jet mass distributions (with fixed $p_{T_{J}}$) to be peaked at small mass values and be rapidly cutoff for $m_J > p_{T_{J}} D/2$.

\begin{figure}[htbp]
\includegraphics[width = \columnwidth]{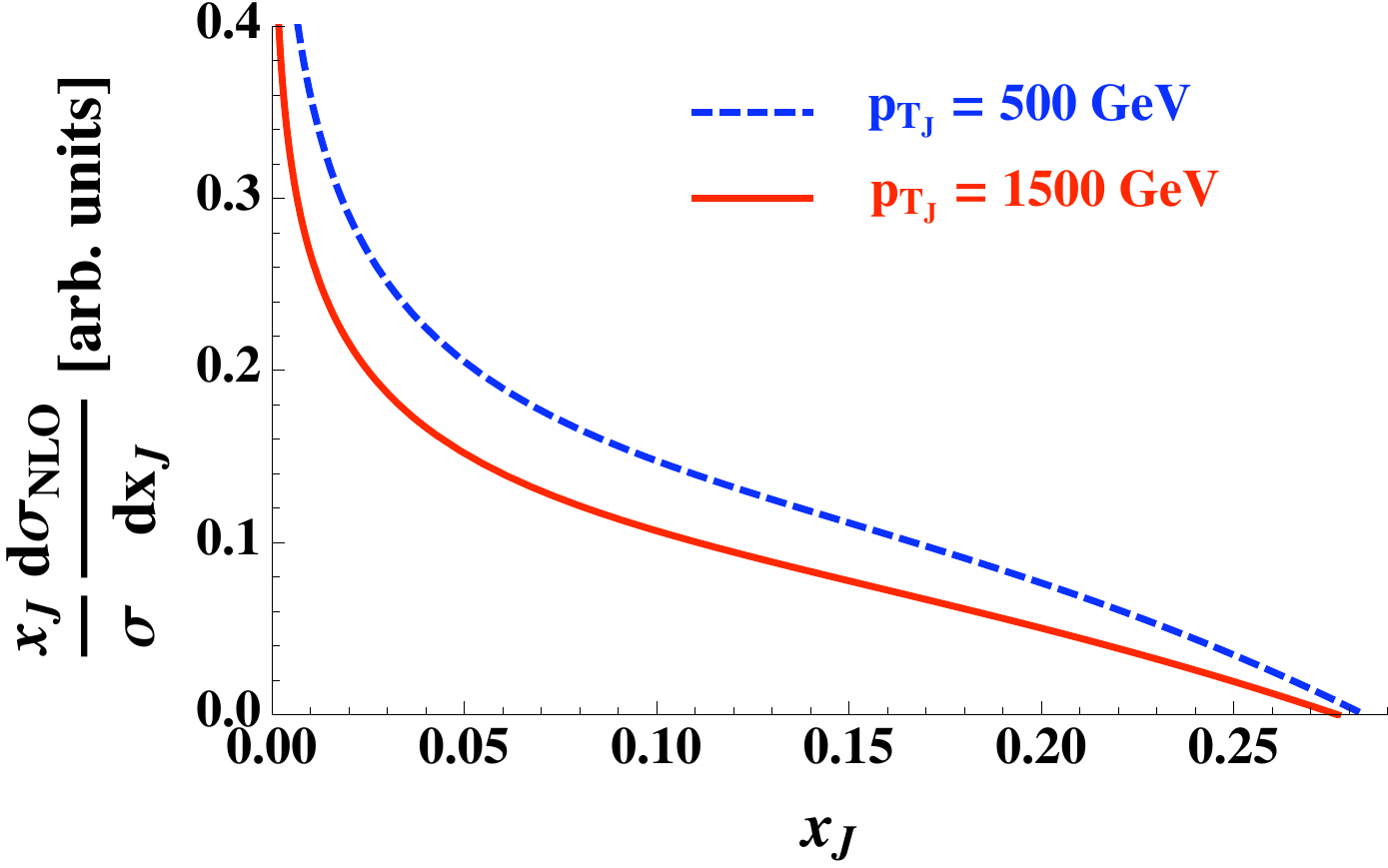}
\caption{NLO distribution in $x_J$ for $\kt$-style QCD jets with D = 1.0 and $\sqrt{s} = 14$ TeV and two values of $p_{T_{J}}$.  }%\red{QCDmassdist2}}
\label{QCDmassdist2}
\end{figure}

We can improve this approximation somewhat by using the more quantitative perturbative analysis described in \cite{Ellis:08.1}. In perturbation theory jet masses appear at next-to-leading order (NLO) in the overall jet process where two (massless) partons can be present in a single jet.  Strictly, the jet mass is then being evaluated at leading order (i.e., the jet mass vanishes with only one parton in a jet) and one would prefer a NNLO result to understand scale dependence (we take $\mu=p_{T_J}/2$).  Here we will simply use the available NLO tools \cite{Kunszt:92.1}. This approach leads to the very similar $x_J$ distribution displayed in Fig.~\ref{QCDmassdist2}, plotted for two values of $p_{T_J}$ (at the LHC, with $\sqrt{s} = 14$ TeV).
We are correctly including the full NLO matrix element (not simply the singular parts), the full kinematics of the jet mass (not just the small-angle approximation) and the effects of the parton distribution functions.  In this case the distribution is normalized by dividing by the Born jet cross section.  Again we see the dominant impact of the soft/collinear singularities for small jet masses.  Note also that there is little residual dependence on the value of the jet momentum (the distribution approximately scales with $p_{T_J}$) and that again the distribution essentially vanishes for $x_{J}>0.25$, $m_{J}/p_{T_J}>0.5\approx D/2$.\footnote{The fact that the $x_J$ distribution extends a little past $D^2/4$ arises from the fact that the true ($z, \Delta R_{12}$) phase space is really two-dimensional and there is still a small allowed phase space region below $\Delta R_{12} = D$ even when $x_J=D^2/4$.}  The average jet mass suggested by these results is $\langle m_{J}/p_{T_J} \rangle \approx 0.2 D$.  However, because the jet only contains two partons at NLO, we are still ignoring the effects of the nonzero subjet masses and the effects of the ordering of mergings imposed by the algorithm itself.  For example, at this order there is no difference between the CA and $\kt$ algorithms.

\begin{figure}[htbp]
\includegraphics[width = \columnwidth]{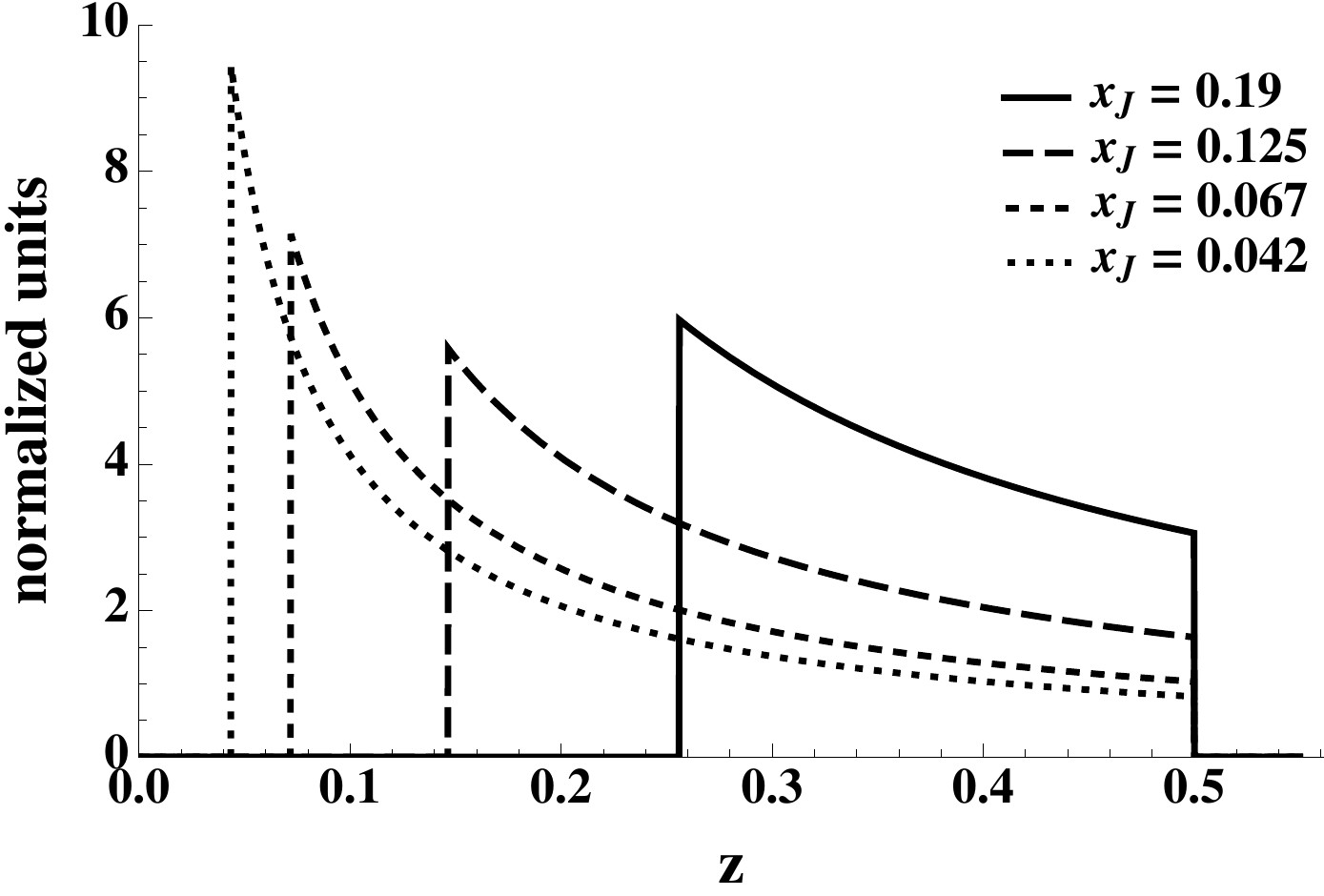}
\caption{Distribution in $z$ for LL QCD jets for $D = 1.0$ and various values of $x_J$.  The curves are normalized to have unit area. }%\red{NLOzplot}}
\label{NLOzplot}
\end{figure}

Next we consider the $z$ and $\Delta R_{12}$ distributions for the LL approximation where a single recombination of two (massless) partons is required to reconstruct as a jet of definite $p_{T_J}$ and mass (fixed $x_J$).  To that end we can ``undo'' one of the integrals in Eq.~(\ref{eq:LL1}) and consider the distributions for $z$ and $\Delta R_{12}$ .  We find for the $z$ distribution the form
\be
\frac{1}{\sigma}\frac{d\sigma_{LL}}{dx_J dz} \sim \frac{1}{2 z x_J}\Theta{\left[z-\frac{1-\sqrt{1- 4 x_J/D^2}}{2}\right]} \Theta \left[ \frac{1}{2}-z \right].
\label{eq:LL2}
\ee
As expected, we see the poles in $z$ and $x_J$ from the soft/collinear dynamics, but, as in Section \ref{sec:recomb:variables}, the constraint of fixed $x_J$ yields a lower limit for $z$.  Recall that the upper limit for $z$ arises from its definition, again applied in the small-angle limit.  Thus the LL QCD distribution in $z$ is peaked at the lower limit but the characteristic turn-on point is fixed by the kinematics, requiring the branching at fixed $x_J$ to be in a jet of size $D$.  This behavior is illustrated in Fig.~\ref{NLOzplot} for various values of $x_J = 1/(\gamma^2-1)$ corresponding to those used in Section \ref{sec:recomb:variables}.

\begin{figure}[htbp]
\includegraphics[width = \columnwidth]{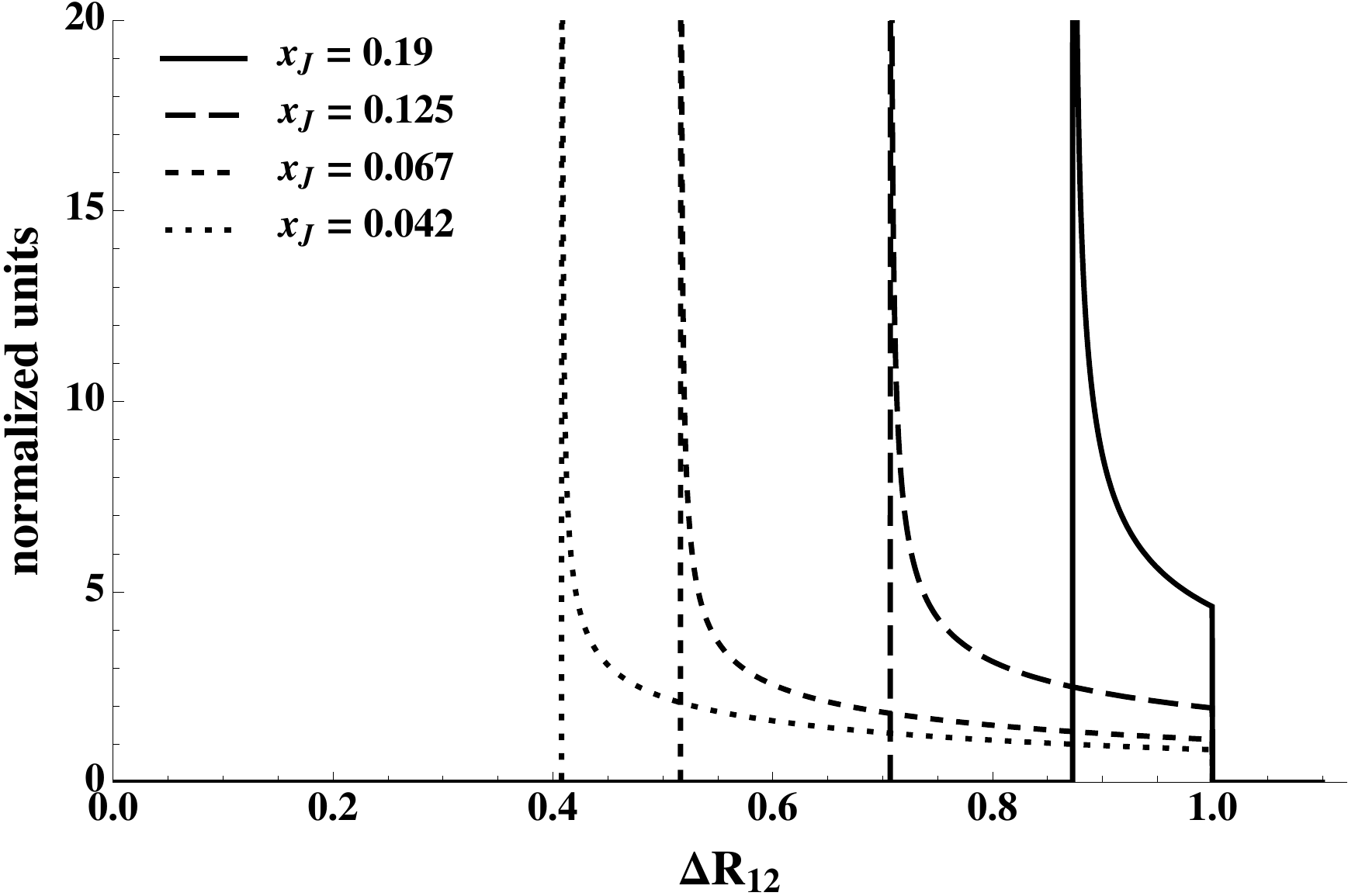}
\caption{Distribution in $\Delta R_{12}$ for LL QCD jets for $D = 1.0$ and various values of $x_J$.  The curves are normalized to have unit area.  }%\red{NLOthetaplot}}
\label{NLOthetaplot}
\end{figure}

The expression for the $\Delta R_{12}$ dependence in the LL approximation is
\begin{eqnarray}
&&\frac{1}{\sigma}\frac{d\sigma_{LL}}{dx_J d\Delta R_{12}}\\
&&\sim \frac{2}{\Delta R_{12}^2}\frac{\Theta{\left[\Delta R_{12}-2 \sqrt{x_J}\right]}\Theta{ \left[ D - \Delta R_{12} \right]} }{\sqrt{\Delta R_{12}^2-{4 x_J}}\left(1-\sqrt{1-{4 x_J}/{\Delta R_{12}^2 }}\right)}. \nonumber
\label{eq:LL3}
\end{eqnarray}
This distribution is illustrated in Fig.~\ref{NLOthetaplot} for the same values of $x_J$ as in Fig.~\ref{NLOzplot}.  As with the $z$ distribution the kinematic constraint of being a jet with a definite $x_J$ yields a lower limit, $\Delta R_{12} \gtrsim 2\sqrt{x_J}$, along with the expected upper limit, $\Delta R_{12} \leq D$. However, for $\Delta R_{12}$ the change of variables also introduces an (integrable) square root singularity at the lower limit. This square root factor tends to be numerically more important than the $1/\Delta R_{12}^2$ factor.  (One factor of $\Delta R_{12}$ arises from the collinear QCD dynamics while the other comes from change of variables.  The soft QCD singularity is contained in the denominator factor $\left(1-\sqrt{1-{4 x_J}/{\Delta R_{12}^2 }}\right) \to 2z$ for $x_J\ll \Delta R^2$ (equivalently, $z \ll 1$).)  Since this square root singularity arises from the choice of variable (a kinematic effect), we will see that it is also present for heavy particle decays, suggesting that the $\Delta R_{12}$ variable will not be as useful as $z$ in distinguishing QCD jets from heavy particle decay jets.

Thus, in our toy QCD model with a single recombination, leading-logarithm dynamics and the small-angle jet mass definition, the constraints due to fixing $x_J$ tend to dominate the behavior of the $z$ and $\Delta R_{12}$ distributions, with limited dependence on the QCD dynamics and no distinction between the CA and $\kt$ algorithms.  However, this situation changes dramatically when we consider more realistic jets with full showering, a subject to which we now turn.

\subsection{Jet Substructure in Simulated QCD events}
\label{sec:QCDJets:simulated}

To obtain a more realistic understanding of the properties of QCD jet masses we now consider jet substructure that arises in more fully simulated events.  In particular, we focus on Monte Carlo simulated QCD jets with transverse momenta in the range $p_{T_J}=$ 500--700 GeV ($c$ = 1 throughout this paper) found in matched QCD multijet samples created as described in Appendix \ref{sec:appendix}. The matching process means that we are including, to a good approximation, the full NLO perturbative probability for energetic, large-angle emissions in the simulated showers, and not just the soft and collinear terms.  As suggested earlier, we anticipate two important changes from the previous discussion.  First, the showering ensures that the daughter subjets at the last recombination have nonzero masses.  More importantly and as noted in Section \ref{sec:AlgOrdering}, the sequence of recombinations generated by the jet algorithm tends to force the final recombination into a particular region of phase space that depends on the recombination metric of the algorithm.  For the CA algorithm this means that the final recombination will tend to have a value of $\Delta R_{12}$ near the limit $D$, while the $\kt$ algorithm will have a large value of $z \Delta R_{12} p_{T{_J}}$.  This issue will play an important role in explaining the observed $z$ and $\Delta R_{12}$ distributions.

\begin{figure}[htbp]
\includegraphics[width = \columnwidth]{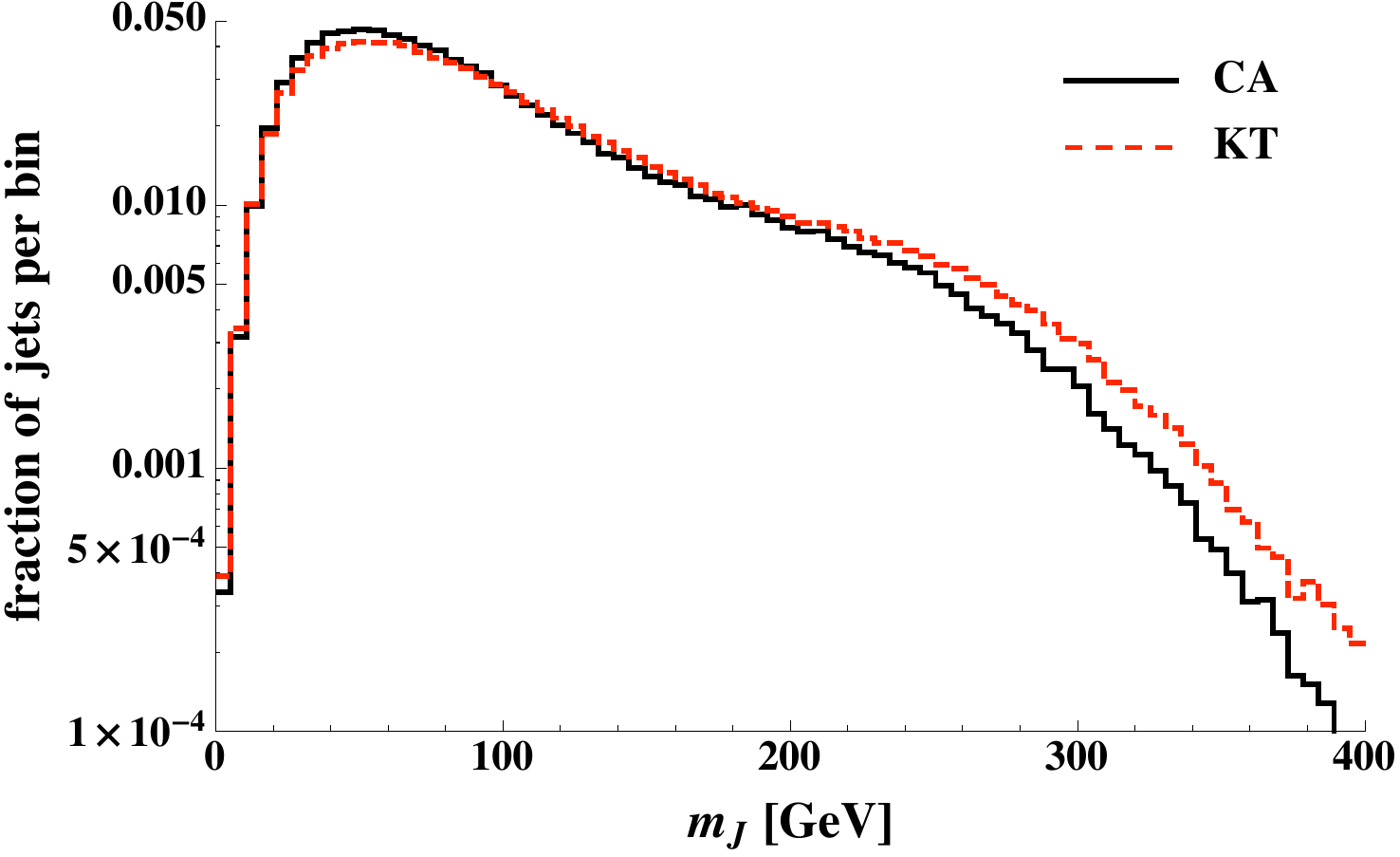}
\caption{Distribution in $m_J$ for QCD jets with $p_T$ between 500 and 700 GeV with D = 1.0.  }%\red{mjetplot}}
\label{mjetplot}
\end{figure}

First, consider the jet mass distributions from the simulated event samples.  In Fig.~\ref{mjetplot}, we plot the jet mass distributions for the $\kt$ and CA algorithms for all jets in the stated $p_T$ bin (500--700 GeV).
As expected, for both algorithms the QCD jet mass distribution smoothly falls from a peak only slightly displaced from zero (the remnant of the perturbative $-\ln(m^2)/m^2$ behavior).  There is a more rapid cutoff for $m_J > p_{T_J}D/2$, which corresponds to the expected kinematic cutoff of $m_J = p_{T_J}D/2$ from the LL approximation, but smeared by the nonzero width of the $p_T$ bin, the nonzero subjet masses and the other small corrections to the LL approximation. The average jet mass, $\left\langle m_{J} \right\rangle \approx 100$ GeV, is in crude agreement with the perturbative expectation $\left\langle m_{J}/p_{T_J}\right\rangle \approx 0.2$.  Note that the two algorithms now differ somewhat in that the $\kt$ algorithm displays a slightly larger tail at high masses.  As we will see in more detail below, this distinction arises from the difference in the metrics leading to recombining protojets over a slightly larger angular range in the $\kt$ algorithm.  On the other hand, the two curves are remarkably similar.  Note that we have used a logarithmic scale to ensure that the difference is apparent.  Without the enhanced number of energetic, large-angle emissions characteristic of this \emph{matched} sample, the distinction between the two algorithms is much smaller, i.e., a typical dijet, LO Monte Carlo sample yields more similar distributions for the two algorithms.

Other details of the QCD jet substructure are substantially more sensitive to the specific algorithm than the jet mass distribution.  To illustrate this point we will discuss the distributions of $z$, $\Delta R_{12}$, and the subjet masses for the last recombination in the jet.  We can understand the observed behavior by combining a simple picture of the geometry of the jet with the constraints induced on the phase space for a recombination from the jet algorithm.  In particular, recall that the ordering of recombinations defined by the jet algorithm imposes relevant boundaries on the phase space available to the late recombinations (see Fig.~\ref{algboundaries}).

While the details of how the $\kt$ and CA algorithms recombine protojets within a jet are different, the overall structure of a large-$p_T$ jet is set by the shower dynamics of QCD, i.e., the dominance of soft/collinear emissions.  Typically the jet has one (or a few) hard core(s), where a hard core is a localized region in $y$--$\phi$ with large energy deposition.  The core is surrounded by regions with substantially smaller energy depositions arising from the radiation emitted by the energetic particles in the core (i.e., the shower), which tend to dominate the area of the jet.  In particular, the periphery of the jet is occupied primarily by the particles from soft radiation, since even a wide-angle hard parton will radiate soft gluons in its vicinity.  This simple picture leads to very different recombinations with the $\kt$ and CA algorithms, especially the last recombinations.

The CA algorithm orders recombinations only by angle and ignores the $p_T$ of the protojets.  This implies that the protojets still available for the last recombination steps are those at large angle with respect to the core of the jet.  Because the core of the jet carries large $p_T$, as the recombinations proceed the directions of the protojets in the core do not change significantly.  Until the final steps, the recombinations involving the soft, peripheral protojets tend to occur only locally in $y$--$\phi$ and do not involve the large-$p_T$ protojets in the core of the jet. Therefore, the last recombinations defined by the CA algorithm are expected to involve two very different protojets.  Typically one has large $p_T$, carrying most of the four-momentum of the jet, while the other has small $p_T$ and is located at the periphery of the jet.  As we illustrate below, the last recombination will tend to exhibit large $\Delta R_{12}$, small $z$, large $a_1$ (near 1), and small $a_2$, where the last two points follow from the small $z$ and correspond to the $(z, \Delta R_{12})$ phase space of Fig.~\ref{zRcontours3}.

In contrast, the $\kt$ algorithm orders recombinations according to both $p_T$ and angle.  Thus the $\kt$ algorithm tends to recombine the soft protojets on the periphery of the jet earlier than with the CA algorithm.  At the same time, the reduced dependence on the angle in the recombination metric implies the angle between protojets for the final recombinations will be lower for $\kt$ than CA.  While there is still a tendency for the last recombination in the $\kt$ algorithm to involve a soft protojet with the core protojet, the soft protojet tends to be not as soft as with the CA algorithm (i.e., the $z$ value is larger), while the angular separation is smaller.  Since this final soft protojet in the $\kt$ algorithm has participated in more previous recombinations than in the CA case, we expect the average $a_2$ value to be farther from zero and the $a_1$ value to be farther from 1.  Generally the $(z, \Delta R_{12})$ phase space for the final $\kt$ recombination is expected to be more like that illustrated in Figs.~\ref{zRcontours2} and \ref{zRcontours4} (coupled with the boundary in Fig.~\ref{algboundariesKT}).

\begin{figure}[htbp]
\subfloat[$z$, CA] {\includegraphics[width = .23\textwidth] {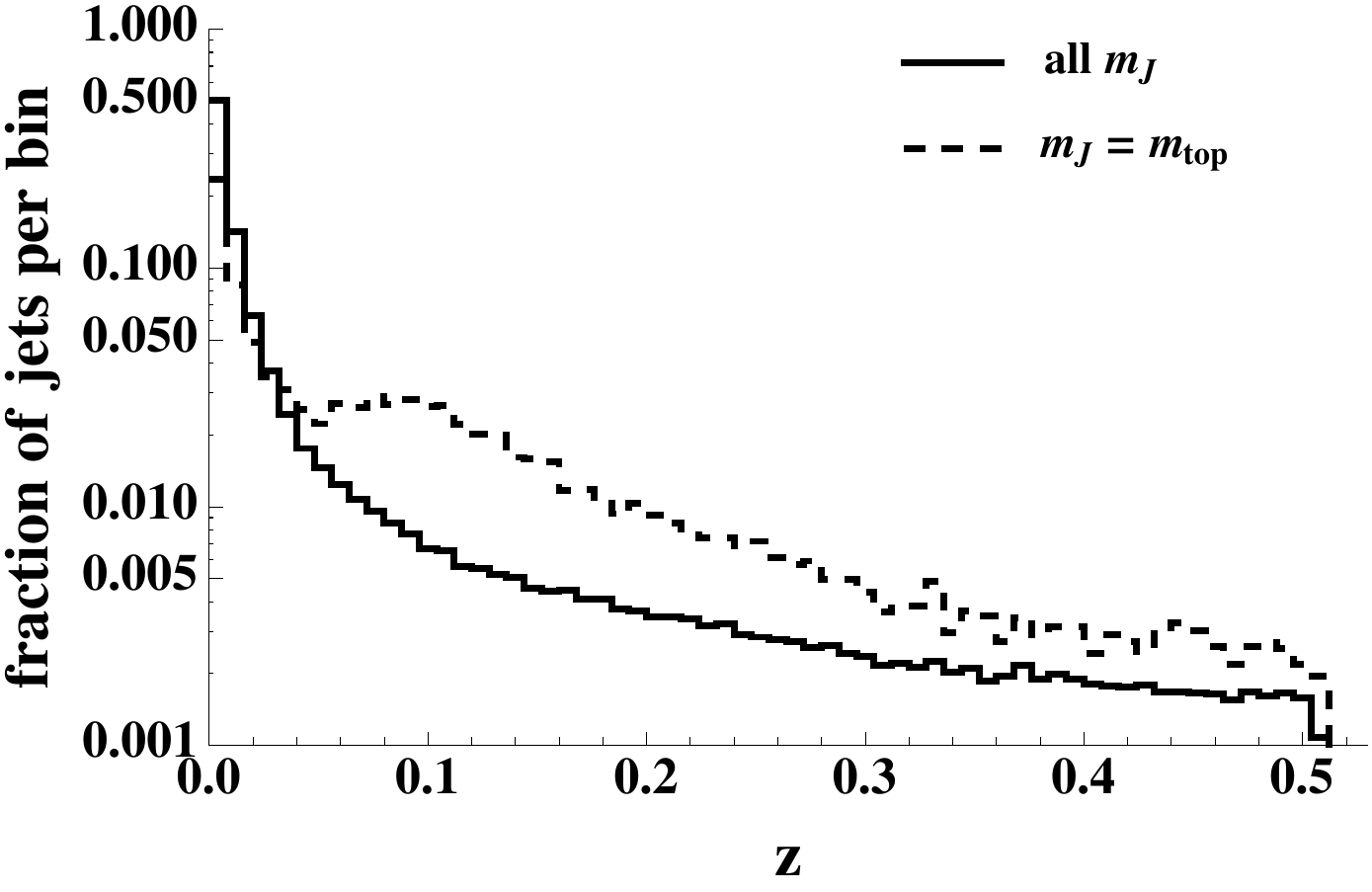} \label{zplotCA}}
\subfloat[$z$, $\kt$]{\includegraphics[width = .23\textwidth]{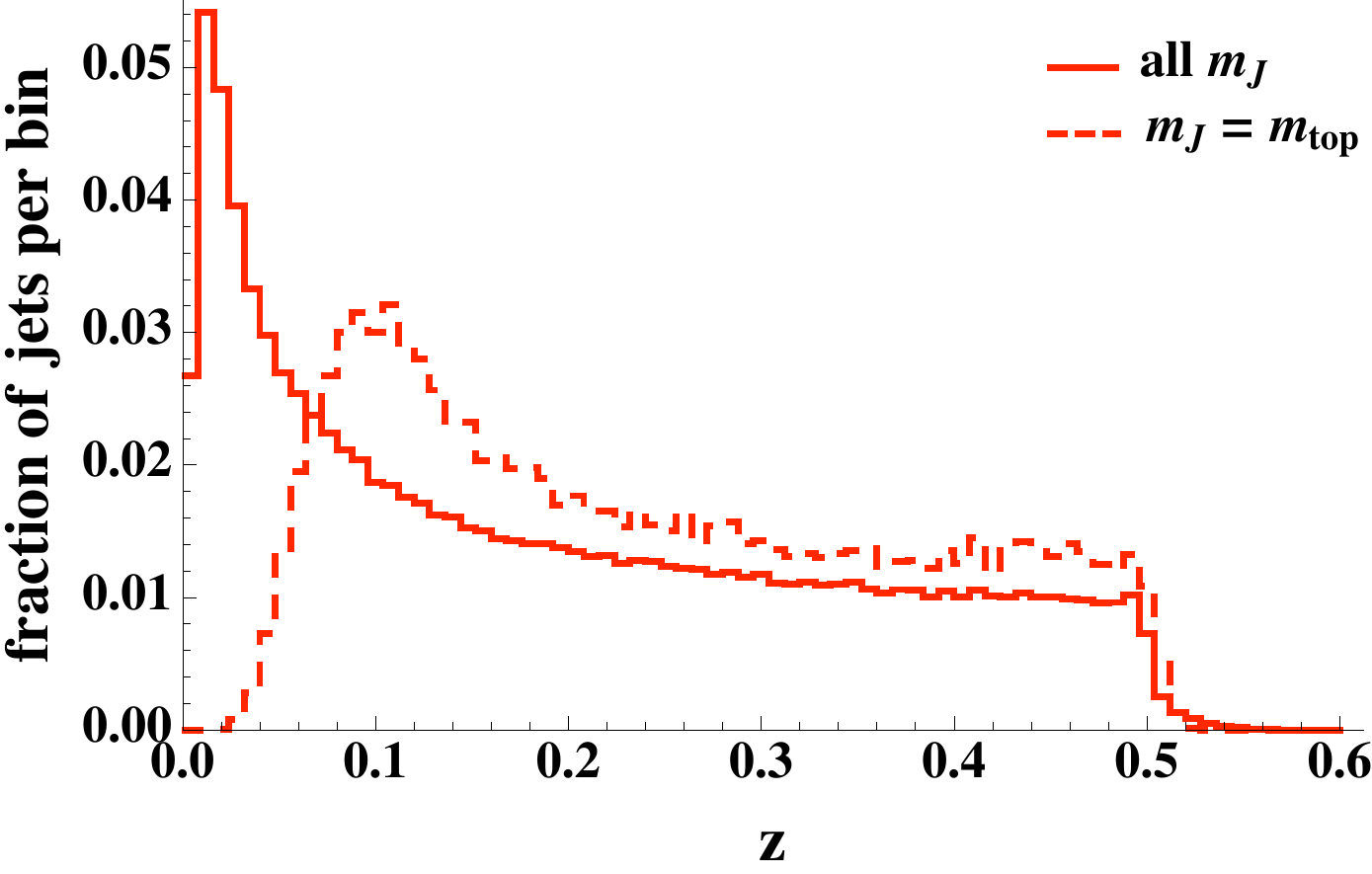} \label{zplotKT}}

\subfloat[$\Delta R_{12}$, CA]{\includegraphics[width = .23\textwidth] {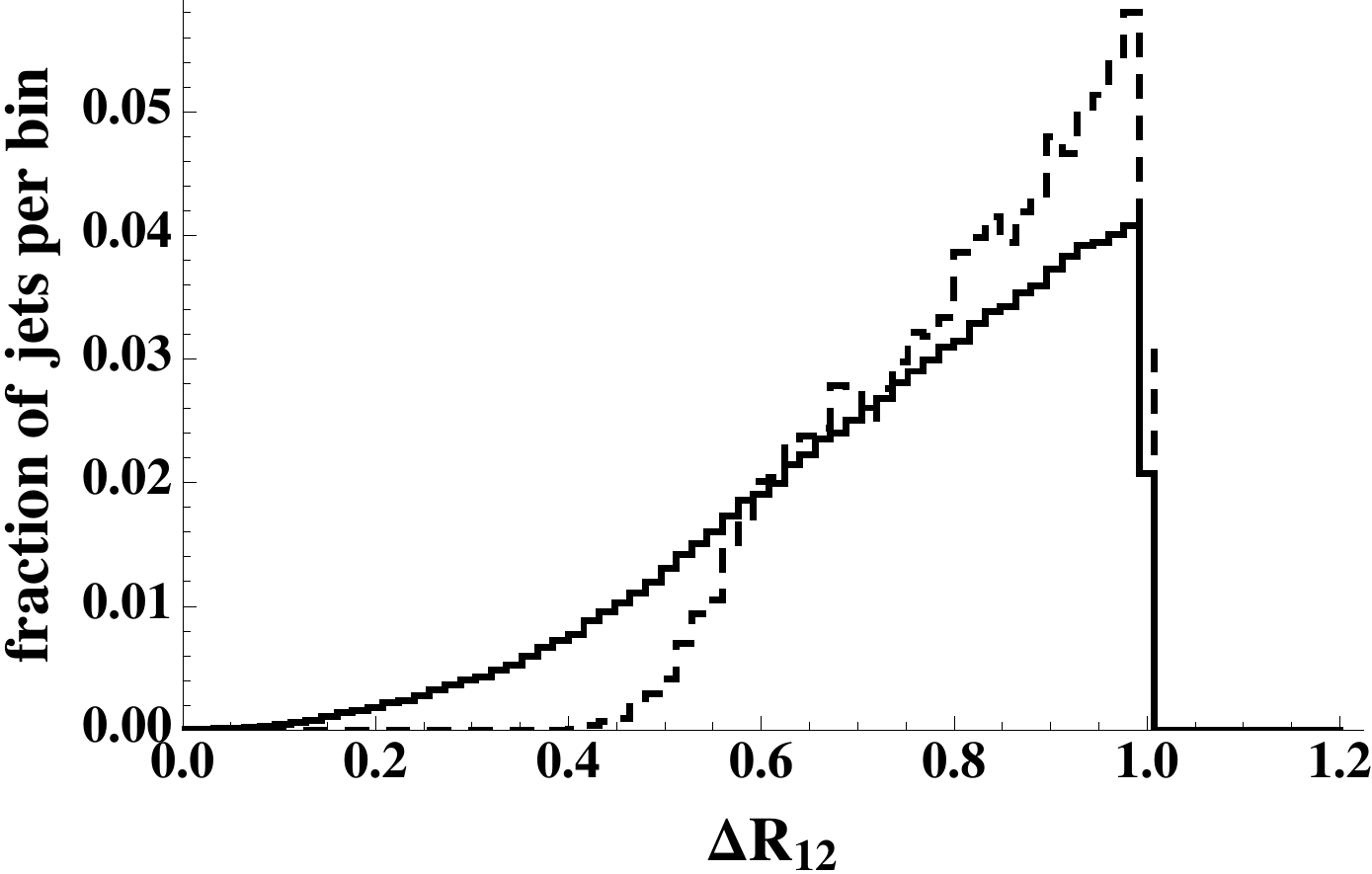}
\label{thetaplotCA}}
\subfloat[$\Delta R_{12}$, $\kt$]{\includegraphics[width = .23\textwidth] {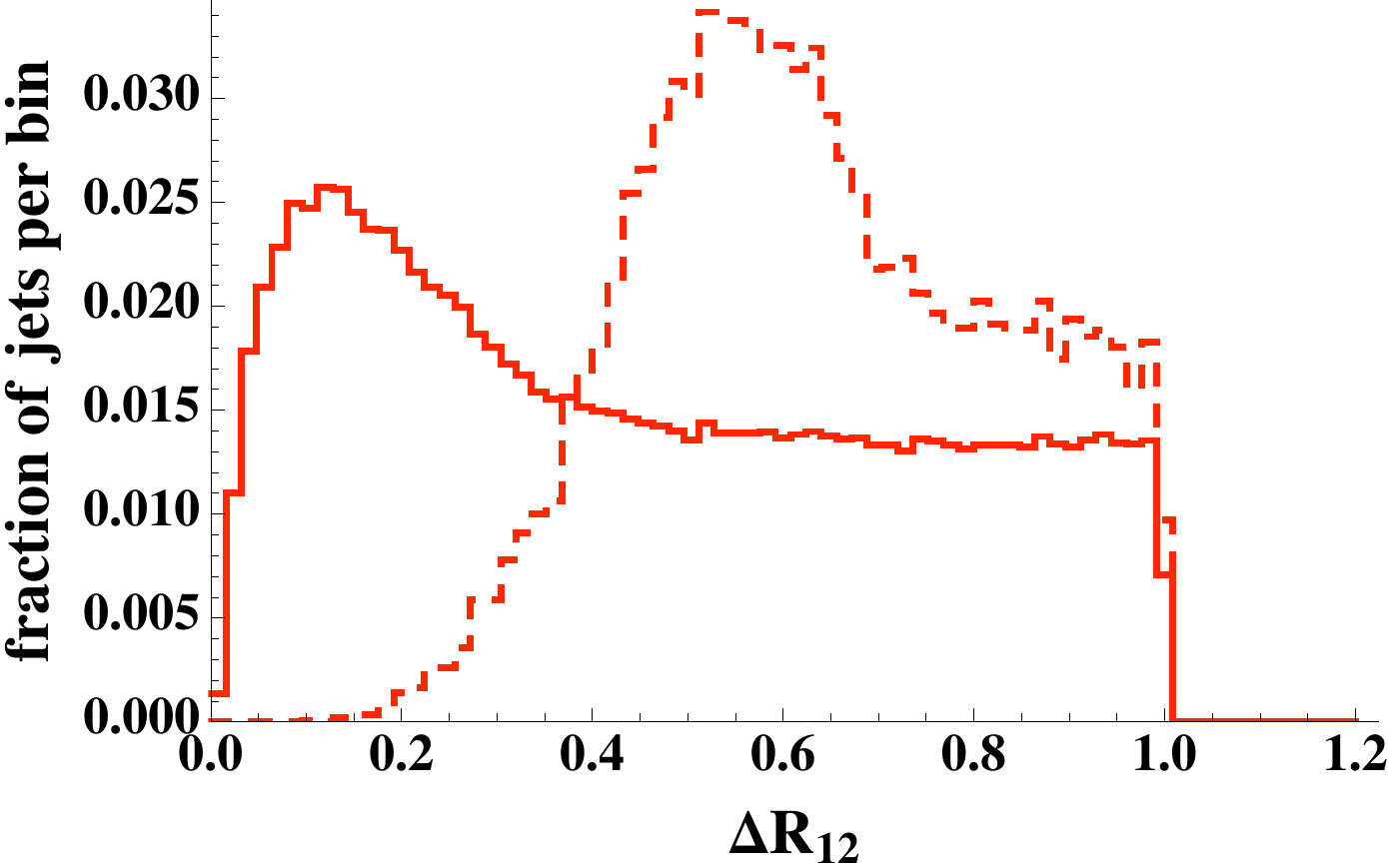} \label{thetaplotKT}}

\subfloat[$a_1$, CA] {\includegraphics[width = .23\textwidth]{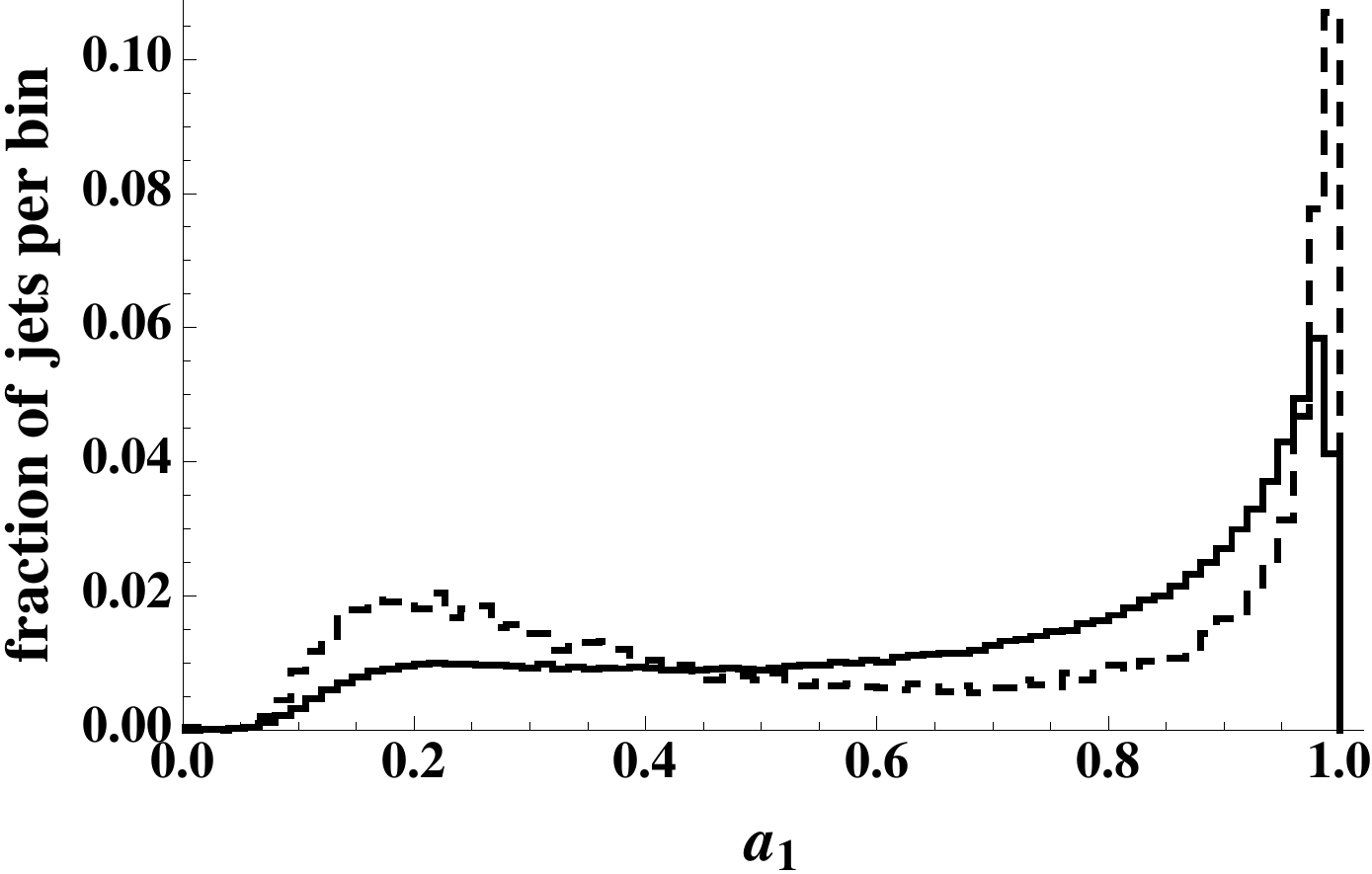} \label{a1plotCA}}
\subfloat[$a_1$, $\kt$]{\includegraphics[width = .23\textwidth]{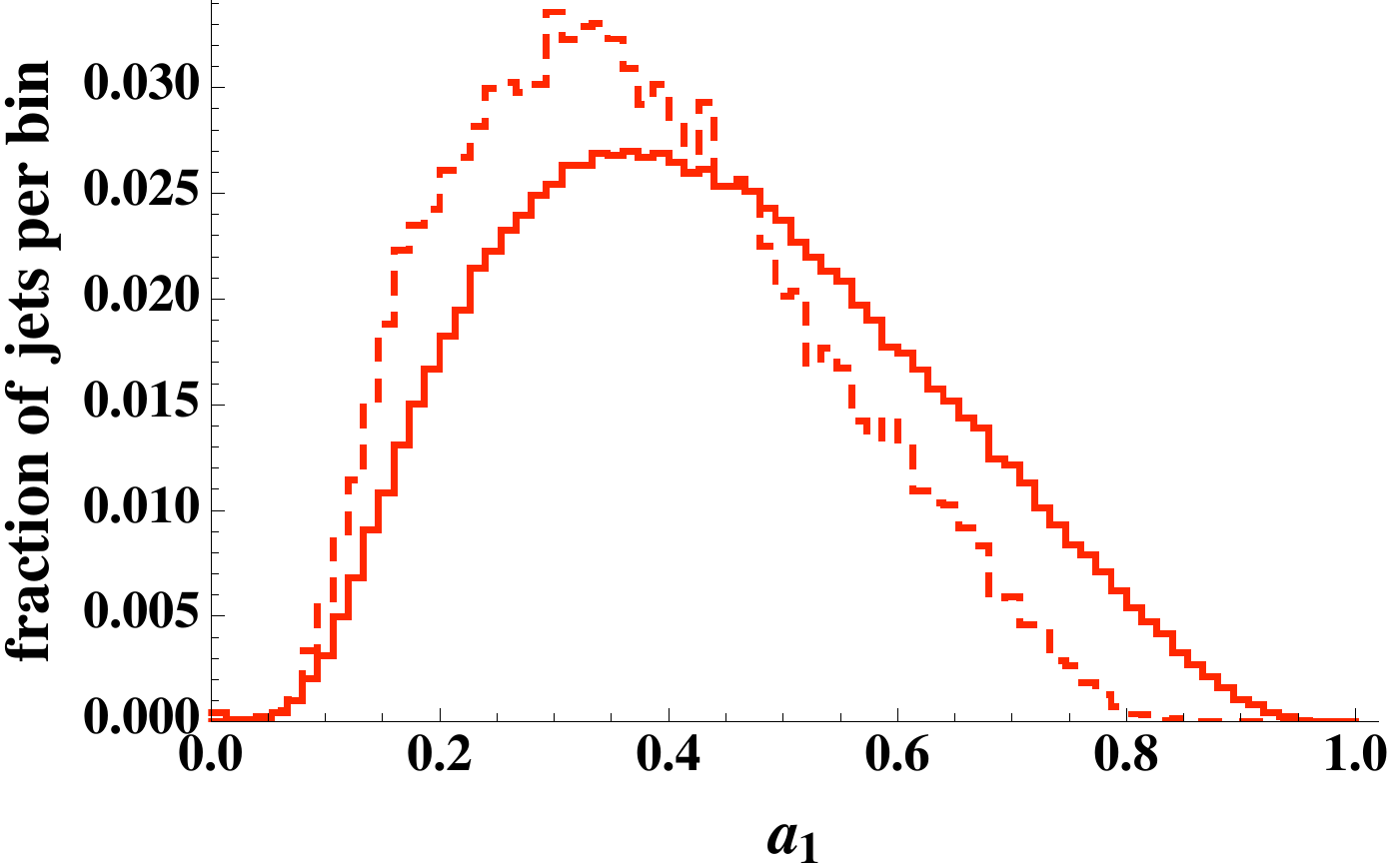}
\label{a1plotKT}}
\caption{Distribution in $z$, $\Delta R_{12}$, and the scaled (heavier) daughter mass $a_1$ for QCD jets, using the CA and $\kt$ algorithms, with (dashed) and without (solid) a cut around the top quark mass.  The jets have $p_T$ between 500 and 700 GeV with D = 1.0.  Note the log scale for the $z$ distribution of CA jets. }%\red{\{z,theta,a1\}$\times$plot$\times$\{CA,KT\}, QCDkinematicplots}}
\label{QCDkinematicplots}
\end{figure}

To summarize and illustrate this discussion, we have plotted distributions of $z$, $\Delta R_{12}$, and $a_1$ for the last recombination in a jet for the $\kt$ and CA algorithms in Figs.~\ref{QCDkinematicplots}(a-f) for the matched QCD sample described previously.  We plot distributions with and without a cut on the jet mass, where the cut is a narrow window ($\approx$ 15 GeV) around the top quark mass.  This cut selects heavy QCD jets, and for the $p_T$ window of 500--700 GeV it corresponds to a cut on $x_J$ of 0.06--0.12.
These distributions reflect the combined influence of the QCD shower dynamics, the restricted kinematics from being in a jet, and the algorithm-dependent ordering effects discussed above.  Most importantly, note the very strong enhancement at the smallest values of $z$ for the CA algorithm in Fig.~\ref{zplotCA}, which persists even after the heavy jet mass cut.  Note there is a log scale in Fig.~\ref{zplotCA} to make the differences between the distributions clearer and better show the dynamic range.  While the $\kt$ result in Fig.~\ref{zplotKT} is still peaked near zero when summed over all jet masses, the enhancement is not nearly as strong.  After the heavy jet mass cut is applied, the distribution shifts to larger values of $z$, with an enhancement remaining at small values.  Only in this last plot is there evidence of the lower limit on $z$ of order 0.1 expected from the earlier LL approximation results.  Note also that the $z$ distributions all extend slightly past $z = 0.5$, indicating another small correction to the LL approximation arising the the true two-dimensional nature of the ($z,\Delta R_{12}$) phase space.

Fig.~\ref{thetaplotCA} illustrates the expected enhancement near $\Delta R_{12} = D = 1.0$ for CA.  Fig.~\ref{thetaplotKT} shows that $\kt$ exhibits a much broader distribution than CA with an enhancement for small $\Delta R_{12}$ values.  Once the heavy jet mass cut is applied, both algorithms exhibit the lower kinematic cutoff on $\Delta R_{12}$ suggested in the LL approximation results, as both distributions shift to larger values of the angle.  This shift serves to enhance the CA peak at the upper limit and moves the the lower end enhancement in $\kt$ to substantially larger values of $\Delta R_{12}$.

\begin{figure}[htbp]
\includegraphics[width = \columnwidth]{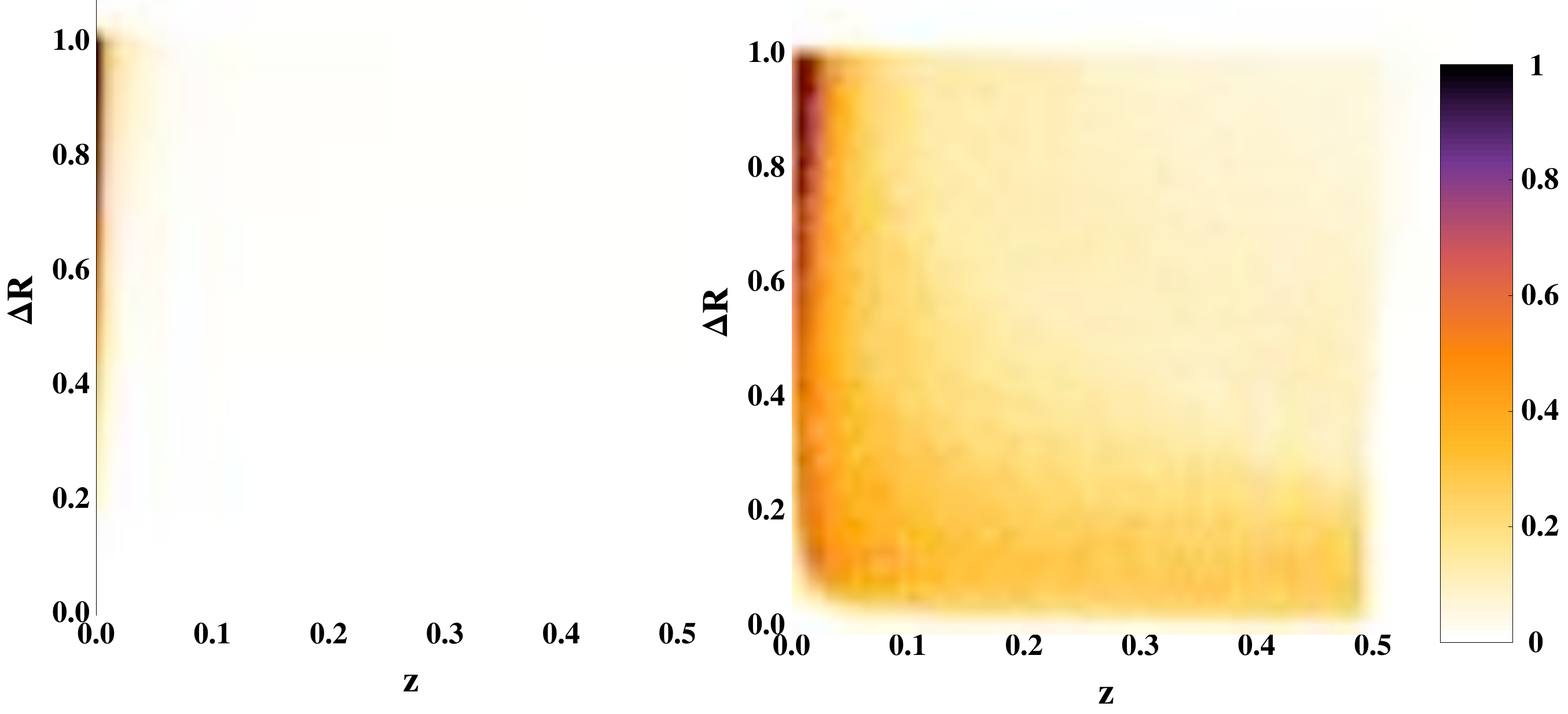}
\caption{Combined distribution in $z$ and $\Delta R_{12}$ for QCD jets, using the CA (left) and $\kt$ (right) algorithms, for jets with $p_T$ between 500 and 700 GeV with D = 1.0.  Each bin represent a \emph{relative} density in each bin, normalized to 1 for the largest bin.  }%\red{zthetaplot}}
\label{zthetaplot}
\end{figure}

The CA algorithm bias toward large $a_1$ is demonstrated in Fig.~\ref{a1plotCA}.  We can see that requiring a heavy jet enhances the large-$a_1$ peak and also results in a much smaller enhancement around $a_1 \approx 0.2$.  The $\kt$ distribution in $a_1$, shown in Fig.~\ref{a1plotKT}, exhibits a broad enhancement around $a_1 \approx 0.4$.  This distribution is relatively unchanged after the jet mass cut.  To give some insight into the correlations between $z$ and $\Delta R_{12}$, in Fig.~\ref{zthetaplot} we plot the distribution of both variables simultaneously for both algorithms, with no jet mass cut applied.
The very strong enhancement at small $z$ and large $\Delta R_{12}$ for CA is evident in this plot.  For $\kt$, there is still an enhancement at small $z$ and large $\Delta R_{12}$, but there is support over the whole range in $z$ and $\Delta R_{12}$ with the impact of the shaping due to the $z \times \Delta R_{12}$ dependence in the metric clearly evident.  Note that the $\kt$ distribution is closer to what one would expect from QCD alone, with enhancements at \emph{both} small $z$ and small $\Delta R_{12}$, while the CA distribution is asymmetrically shaped away from the QCD-like result.  Finally we should recall, as indicated by Fig.~\ref{mjetplot}, that the jets found by the two algorithms tend to be slightly different, with the $\kt$ algorithm recombining slightly more of the original (typically soft) protojets at the periphery and leading to slightly larger jet masses.

Because the QCD shower is present in all jets, and is responsible for the complexity in the jet substructure, the systematic effects discussed above will be present in all jets.  While the kinematics of a heavy particle decay is distinct from QCD in certain respects, we will find that these effects still present themselves in jets containing the decay of a heavy particle.  This reduces our ability to identify jets containing a heavy particle, and will lead us to propose a technique to reduce them.  In the following section, we study the kinematics of heavy particle decays and discuss where these systematic effects arise.

%%%%%%%%%% END OF SECTION 3: QCD JETS %%%%%%%%%%

\section{Reconstructing Heavy Particles}
\label{sec:reconHeavy}
Recombination algorithms have the potential to reconstruct the decay of a heavy particle.  Ideally, the substructure of a jet may be used to identify jets coming from a decay and reject the QCD background to those jets.  In this section, we investigate a pair of unpolarized parton-level decays, a heavy particle decaying into two massless quarks (a $1 \to 2$ decay) and a top quark decay into three massless quarks (a two-step decay).  For each decay, we study the available phase space in terms of the lab frame variables $\Delta R_{12}$ and $z$ and the shaping of kinematic distributions imposed by the requirement that the decay be reconstructed in a single jet.  We will determine the kinematic regime where decays are reconstructed, and contrast this with the kinematics for a $1\to2$ splitting in QCD.

\subsection{\texorpdfstring{$1\to2$ Decays}{1 -> 2 Decays}}
\begin{figure}[htbp]
\includegraphics[width = \columnwidth]{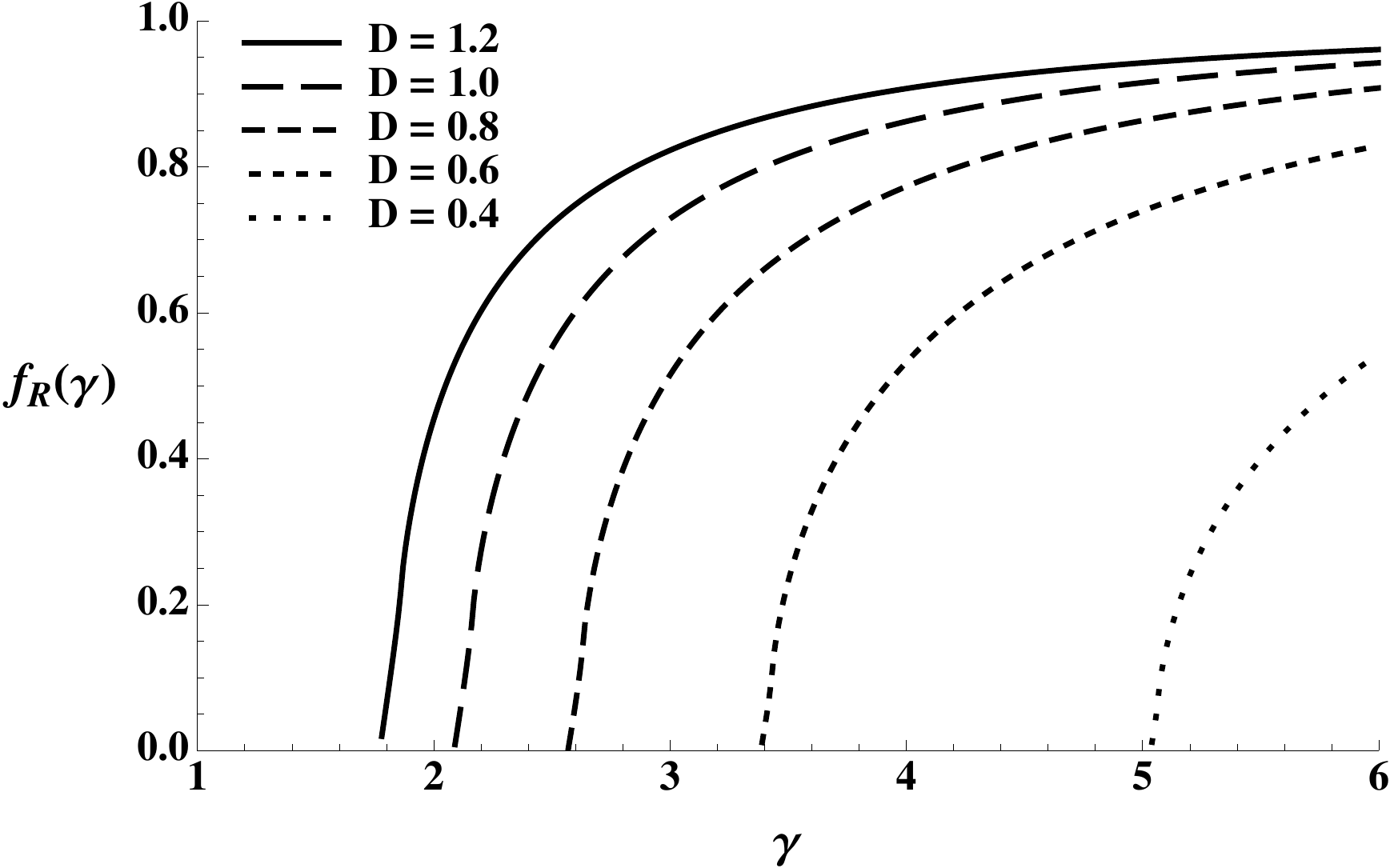}
\caption{Reconstruction fractions $f_R(\gamma)$ as a function of $\gamma$ for various $D$.  }%\red{fRdist}}
\label{fRdist}
\end{figure}

We begin by considering a $1\to2$ decay with massless daughters.  An unpolarized decay has a simple phase space in terms of the rest frame variables $\cos\theta_0$ and $\phi_0$:
\be
\frac{d^2N_0}{d\cos\theta_0 d\phi_0} = \frac{1}{4\pi} .
\label{eq:restframedist}
\ee
Recall from Sec.~\ref{sec:recomb:variables} that $\cos\theta_0$ and $\phi_0$ are the polar and azimuthal angles of the heavier daughter particle (when the daughters are identical, we can take these to be the angles for a randomly selected daughter of the pair) in the parent particle rest frame relative to the direction of the boost to the lab frame.  In general, we will use $N_0$ to label the distribution of \emph{all} decays, while $N$ will label the distribution of decays \emph{reconstructed} inside a single jet. $N_0$ is normalized to unity, so that for any variable set $\Phi$,
\be
\int d\Phi\frac{dN_0}{d\Phi} = 1 .
\label{eq:normN0}
\ee
The distribution $N$ is defined from $N_0$ by selecting those decays that fit in a single jet, so that generically
\be
\frac{dN}{d\Phi} \equiv \int d\Phi' \frac{dN_0}{d\Phi'}\delta(\Phi' - \Phi)\Theta(\text{single jet reconstruction}) .
\label{eq:dNdef}
\ee
$N$ is naturally normalized to the total fraction of reconstructed decays.  The constraints of single jet reconstruction will depend on the decay and on the jet algorithm used, and abstractly take the form of a set of $\Theta$ functions specifying the ordering and limits on recombinations.  For a $1\to2$ decay and a recombination-type algorithm, the only constraint is that the daughters must be separated by an angle less than $D$:
\be
\Delta R_{12} < D .
\label{eq:recon1to2}
\ee
Since the kinematic limits imposed by reconstruction are sensitive to the boost $\gamma$ of the parent particle, we will want to consider the quantities of interest at a variety $\gamma$ values.  To illustrate this $\gamma$ dependence, we first find the total fraction of all decays that are reconstructed in a single jet for a given value of the boost.  We call this fraction $f_R(\gamma)$:
\be
f_R(\gamma) \equiv \int d\cos\theta_0 d\phi_0 \frac{d^2N_0}{d\cos\theta_0d\phi_0}\Theta\left(D - \Delta R_{12}\right) .
\label{eq:fRgammadef}
\ee
In Fig.~\ref{fRdist}, we plot $f_R(\gamma)$ vs. $\gamma$ for several values of $D$.
The reconstruction fraction rapidly rises from no reconstruction to nearly complete reconstruction in a very narrow range in $\gamma$.  This indicates that $\Delta R_{12}$ is highly dependent on $\gamma$ for fixed $\cos\theta_0$ and $\phi_0$, which we will see below.  Furthermore, the cutoff where $f_R(\gamma) = 0$ is very sensitive to the value of $D$, with very large boosts required to reconstruct a particle in a single jet except for larger values of $D$.  This turn-on for increasing $\gamma$ is the same effect as the ($z,\Delta R_{12}$) phase space moving into the allowed region below $\Delta R_{12}=D$ in Fig.~\ref{zRcontours1} as $x_J$ is reduced.

To better understand the effect that reconstruction has on the phase space for decays, we would like to find the distribution of $1\to2$ decays in terms of lab frame variables,
\be
\frac{d^2N_0}{dzd\Delta R_{12}} .
\label{eq:decaylabvars}
\ee
With two massless daughters, $\Delta R_{12}$ is given in terms of rest frame variables by
\begin{eqnarray}
\Delta R_{12}^2  =  \left[\tanh^{-1}\left(\frac{2\gamma\sin\theta_0\sin\phi_0}{ \sin^2\theta_0(\beta^2\gamma^2 + \sin^2\phi_0) + 1}\right)\right]^2 \nonumber\\
+ \left[\tan^{-1}\left(\frac{2\beta\gamma\sin\theta_0\cos\phi_0}{ \sin^2\theta_0(\beta^2\gamma^2 + \sin^2\phi_0) - 1}\right)\right]^2 .
\label{eq:dR}
\end{eqnarray}
with $\beta \equiv \sqrt{1 - \gamma^{-2}}$.  This relation is analytically
non-invertible, meaning we cannot write the Jacobian for the transformation
\be
\frac{d^2N_0}{d\cos\theta_0d\phi_0} \to \frac{d^2N_0}{dzd\Delta R_{12}}
\label{eq:disttransform}
\ee
in closed form.  However, $\Delta R_{12}$ has some simple limits.  In particular, when the boost $\gamma$ is large, to leading order in $\gamma^{-1}$,
\be
\Delta R_{12} = \frac{2}{\gamma\sin\theta_0} + \mathcal{O}\left(\gamma^{-3}\right) .
\label{eq:dRlimit}
\ee
This limit is only valid for $\sin\theta_0 \gtrsim \gamma^{-1}$, but as we will see this is the region of phase space where the decay will be reconstructed in a single jet.  The large-boost approximation describes the key features of the kinematics and is useful for a simple picture of kinematic distributions when particles are reconstructed in a single jet.

Since $\gamma = \sqrt{1+ 1/x_J}$, this limit is equivalent to the small-angle limit we took in Sec.~\ref{sec:QCDJets:Toy}.  (For $\Delta R^2 \ll 1$, $x_J \approx z(1-z)\Delta R^2 \ll 1$.)  We can see this in Eq.~(\ref{eq:dR}), where $\Delta R \approx 1/\gamma$.

The value of $z$ is also simple in the large-boost approximation.  In this limit,
\be
z = \frac{1 - \left|\cos\theta_0\right|}{2} +
\mathcal{O}\left(\gamma^{-2}\right) .
\label{eq:zlimit}
\ee
With the large-boost approximation, $z$ and $\Delta R_{12}$ are both independent of $\phi_0$.  As noted earlier both $\Delta R_{12}$ and $z$ depend on $\phi_0$ only through terms that are suppressed by inverse powers of $\gamma$ (cf. Figs.~\ref{thetaphicontours} and \ref{zRcontours}), and taking the large-boost limit eliminates this dependence.  Therefore, in this limit we can integrate out $\phi_0$ and find the distributions in $z$ and $\Delta R_{12}$ for all decays.  For $z$ the distribution is simply flat:
\be
\frac{dN_0}{dz} \approx 2\Theta\left(\frac12 - z\right)\Theta(z) .
\label{eq:dN0dz}
\ee
We have included the limits for clarity.  For $\Delta R_{12}$, the distribution is
\be
\frac{dN_0}{d\Delta R_{12}} \approx \frac{4}{\gamma^2\Delta R_{12}^2}\frac{\Theta\left(\Delta R_{12} - 2\gamma^{-1}\right)}{\sqrt{\Delta R_{12}^2 - 4\gamma^{-2}}} .
\label{eq:dN0ddR}
\ee
This distribution has a lower cutoff requiring $\Delta R_{12} \ge 2\gamma^{-1}$.  This is close to the true lower limit on $\Delta R_{12}$, which comes from setting $\phi_0 = 0$ in the exact formula for $\Delta R_{12}$ and simplifying.  The exact lower limit is
\be
\Delta R_{12} \ge 2\csc^{-1}\gamma ,
\label{eq:dRlowerlimit}
\ee
which is within 5\% of $2\gamma^{-1}$ for values of $\gamma$ for which $f_R(\gamma) > 0$.  Note that in Eq.~(\ref{eq:dN0ddR}), there is a enhancement at the lower cutoff in $\Delta R_{12}$ due to the square root singularity arising from the change of variables, just as there was in the QCD result in Eq.~(\ref{eq:LL3}).  Thus the distribution in $\Delta R_{12}$ is highly localized at the cutoff, which is a function of $\gamma$.

\begin{figure}[htbp]
\includegraphics[width=\columnwidth]{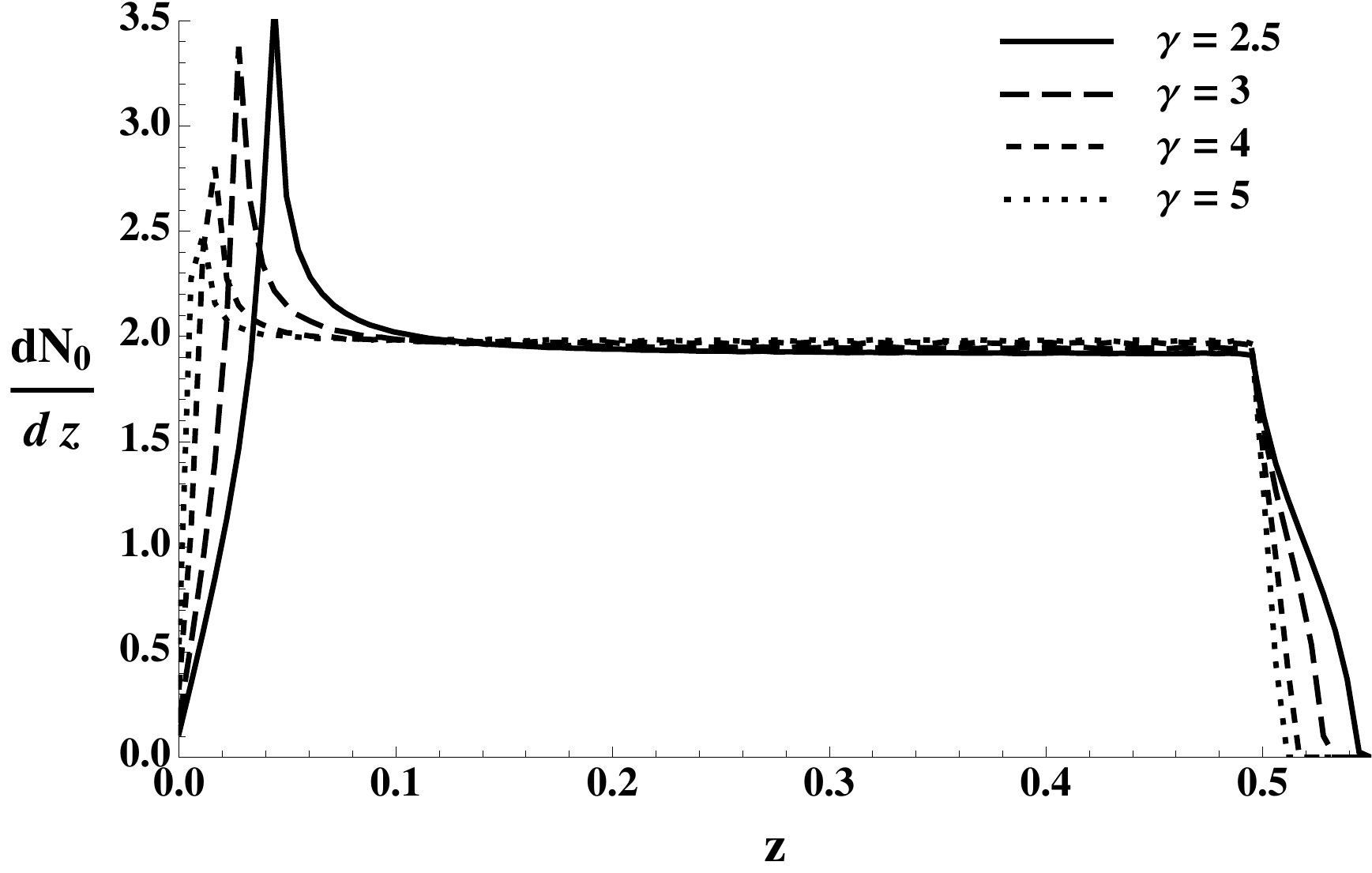}
\caption{The distribution of all decays in $z$ for several values of $\gamma$.  }%\red{zdistall}}
\label{zdistall}
\end{figure}

\begin{figure}[htbp]
\includegraphics[width=\columnwidth]{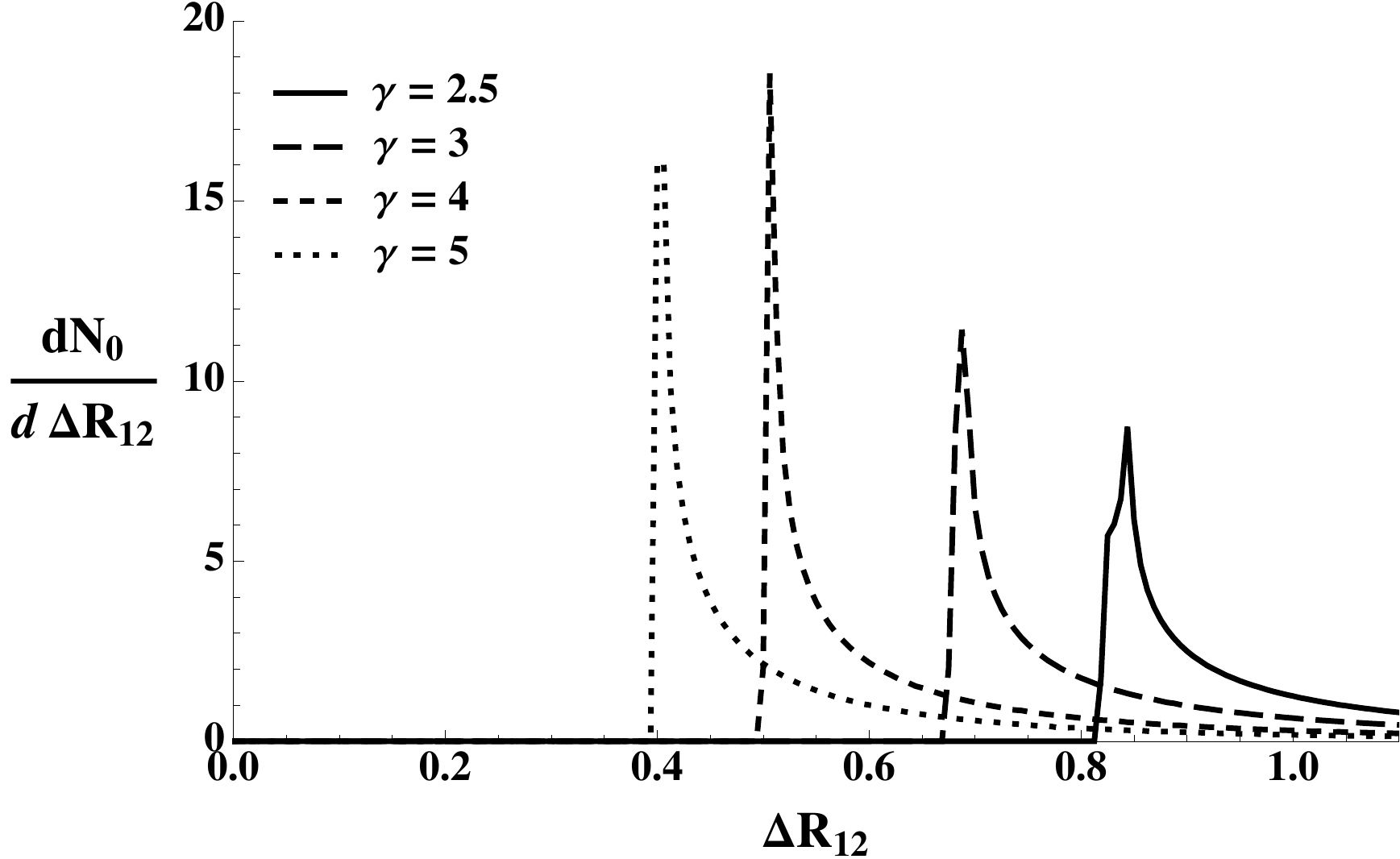}
\caption{The distribution of all decays in $\Delta R_{12}$ for several values of $\gamma$.  }%\red{dRdistall}}
\label{dRdistall}
\end{figure}

In Fig.~\ref{zdistall}, we plot the true distribution $dN_0/dz$, found numerically using no large-boost approximation, for several values of $\gamma$.
Qualitatively, the true distribution is very similar to the approximate one in Eq.~(\ref{eq:dN0dz}), which is flat.  The peak in the distribution at small $z$ values comes from the reduced phase space as $z\to0$, and the peak is lower for larger boosts.  Likewise, the exact distribution $dN_0/d\Delta R_{12}$ is very similar to the large-boost result; in Fig.~\ref{dRdistall}, we plot $dN_0/d\Delta R_{12}$ with no approximation.
The distribution in $\Delta R_{12}$ is localized at the lower limit, especially for larger boosts.  This provides a useful rule: the opening angle of a decay is highly correlated with the transverse boost of the parent particle.  Note that the relevant boost is the transverse one because the angular measure $\Delta R$ is invariant under longitudinal boosts (recall that in the example here, we have set the parent particle to be transverse).

\begin{figure}[htbp]
\includegraphics[width = \columnwidth]{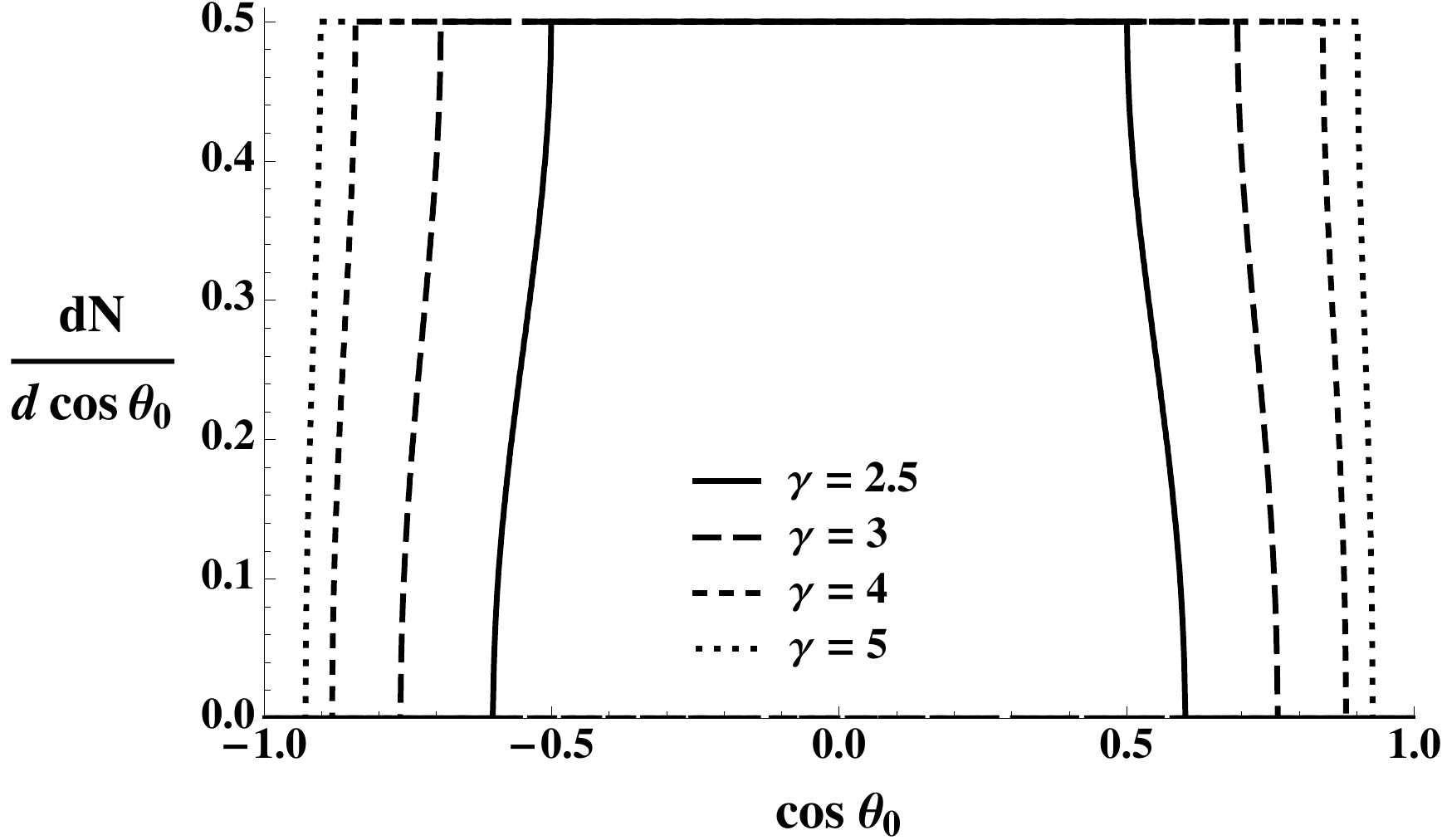}
\caption{The reconstructed distribution $dN/d\cos\theta_0$ with $D = 1.0$ for various values of $\gamma$.  }%\red{costhdist}}
\label{costhdist}
\end{figure}

The constraint imposed by reconstruction is simple to interpret in the large-boost approximation.  In terms of $\sin\theta_0$, the constraint $\Delta R_{12} < D$ requires $\sin\theta_0 > 2/\gamma D$, which excludes the region where the approximation breaks down.  Therefore the large-boost approximation is apt for describing the kinematics of a reconstructed decay.  In Fig.~\ref{costhdist}, we plot the distribution, $dN/d\cos\theta_0$, where the implied sharp cutoff is apparent (and should be compared to what we observed in Fig.~\ref{thetaphicontours1}).
This distribution is easy to understand in the rest frame of the decay.  When $|\cos\theta_0|$ is close to 1, one of the daughters is nearly collinear with the direction of the boost to the lab frame, and the other is nearly anti-collinear.  The anti-collinear daughter is not sufficiently boosted to have $\Delta R_{12} < D$ with the collinear daughter, and the parent particle is not reconstructed.  As $|\cos\theta_0|$ decreases, the two daughters can be recombined in the same jet; this transition is rapid because the $\phi_0$ dependence of the kinematics is small.  We now look at the distributions of $z$ and $\Delta R_{12}$ when we require reconstruction.

\begin{figure}[htbp]
\includegraphics[width=\columnwidth]{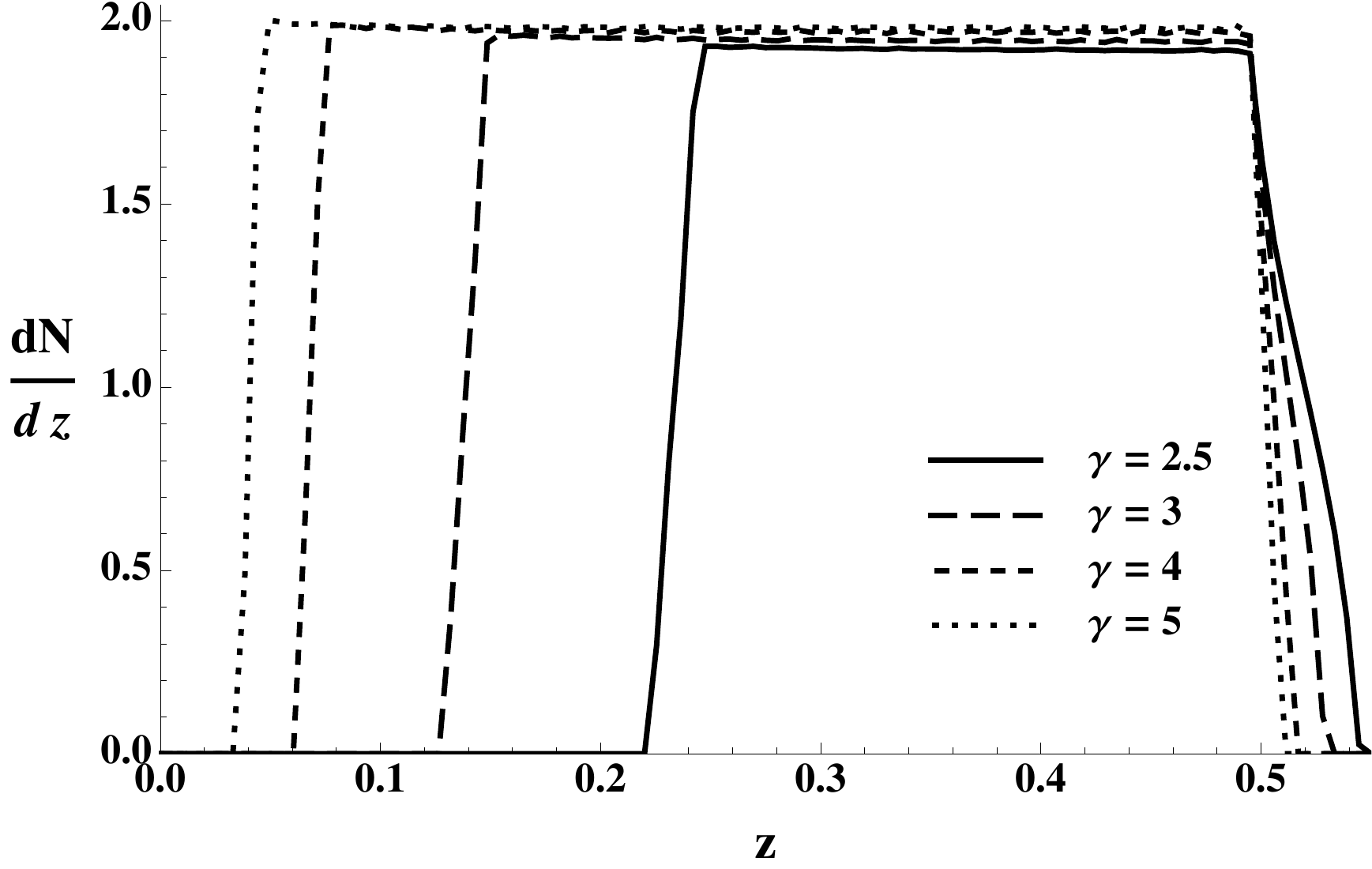}
\caption{The distribution of reconstructed decays in $z$ for several values of $\gamma$.  }%\red{zdistrecon}}
\label{zdistrecon}
\end{figure}

Because $z$ is linearly related to $\cos\theta_0$ at large boosts, the distribution in $z$ has a simple form:
\be
\frac{dN}{dz} \approx 2 \Theta\left(z - \frac{1 - \sqrt{1 - 4/(\gamma^2D^2)}}{2}\right) \Theta\left(\frac12 - z\right) .
\label{eq:dNdz}
\ee
Comparing to Eq.~(\ref{eq:dN0dz}), we see that requiring reconstruction simply cuts out the region of phase space at small $z$.  This is confirmed in the exact distribution $dN/dz$, shown in Fig.~\ref{zdistrecon}.
The small-$z$ decays that are not reconstructed come from the regions of phase space with $|\cos\theta_0|$ near 1, just as in the previous discussion.  In these decays, the backwards-going (anti-collinear) daughter in the parent rest frame is boosted to have small $p_T$ in the lab frame.  Comparing to Fig.~\ref{NLOzplot}, the distribution in $z$ for QCD splittings, we see first that the cutoffs on the distributions are similar (they are not identical because of the LL approximation used in Fig.~\ref{NLOzplot}).  However, the QCD distribution has an enhancement at small $z$ values, due to the QCD soft singularity, that the distribution for reconstructed decays does not exhibit.

The distribution of reconstructed particles in the variable $\Delta R_{12}$ is related simply to the distribution of all decays in the same variable:
\be
\frac{dN}{d\Delta R_{12}} = \frac{dN_0}{d\Delta R_{12}}\Theta\left(D - \Delta R_{12}\right) ,
\label{eq:dNddR}
\ee
which means that the distribution $dN/d\Delta R_{12}$ is given by Fig.~\ref{dRdistall} with a cutoff at $\Delta R_{12} = D$.  Note that this distribution is very close in shape to the distribution of QCD branchings versus $\Delta R_{12}$ displayed in Eq.~(\ref{eq:LL3}) and Fig.~\ref{NLOthetaplot}.  This similarity arises from that the fact that the most important factor in the shape is the square root singularity, which arises from the change of variables in both cases and is not indicative of the underlying differences in dynamics.

In this subsection, we have considered $1\to2$ decays with massless daughters and a fixed boost and the shaping effects that arise from requiring that the decay be reconstructed in a jet.  We have found that decays share many kinematic features with QCD branchings into two massless partons at fixed $x_J$.  In particular, the cutoffs on distributions are set by the kinematics, and do not depend on the process.  Comparing Eqs.~(\ref{eq:LL2}, \ref{eq:dNdz}) and Eqs.~(\ref{eq:LL3}, \ref{eq:dNddR}), we see that the upper and lower cutoffs are the same within the approximations used.  The dominant feature in the $\Delta R_{12}$ distribution, the square root singularity at the lower bound, is also a kinematic effect shared by both decays and QCD branchings.  On the other hand, the $z$ distributions are distinct.  While QCD branchings are enhanced at small $z$, for decays the distribution in $z$ is flat over the allowed range.

%%%%%%%%%%%%%%%%

\subsection{Two-step Decays}
\label{sec:reconHeavy:2step}

We now turn our attention to two-step decays, which exhibit a more complex substructure than a single $1\to2$ decay.  Compared to one-step decays, two-step decays offer new insights into the ordering effects of the $\kt$ and CA algorithms, highlight the shaping effects from the algorithm on the jet substructure and offer a surrogate for the cascade decays that are often featured in new physics scenarios.  The top quark is a good example of such a decay, and we focus on it in this section.  Unlike a $1\rightarrow2$ decay, in reconstructing a multi-step decay at the parton level the choice of jet algorithm matters; different algorithms can give different substructure.  We take the same approach as for the $1\to2$ decay, studying the kinematics of the parton-level top quark decay in terms of the lab frame variables $\Delta R_{12}$ and $z$.

We will label the top quark decay $t\to Wb$, with $W\to qq'$.  In this discussion requiring that the top quark be reconstructed means that the $W$ must be recombined from $q$ and $q'$ first, followed by the $b$.  This recombination ordering reproduces the decay of the top, and the $W$ is a daughter subjet of the top quark.  For the $\kt$ algorithm, reconstructing the top quark in a single jet imposes the following constraints on the partons:
\be
\begin{split}
\min(p_{Tq},p_{T{q'}})\Delta R_{qq'} &< \min(p_{Tq},p_{Tb})\Delta R_{bq}, \\
\min(p_{Tq},p_{T{q'}})\Delta R_{qq'} &< \min(p_{T{q'}},p_{Tb})\Delta R_{bq'}, \\
\Delta R_{qq'} &< D,~\text{and}\\
\Delta R_{bW} &< D.
\end{split}
\ee
For the CA algorithm the relations are strictly in terms of the angle:
\be
\begin{split}
\Delta R_{qq'} &< \Delta R_{bq}, \\
\Delta R_{qq'} &< \Delta R_{bq'}, \\
\Delta R_{qq'} &< D,~\text{and}\\
\Delta R_{bW} &< D.
\end{split}
\ee
The kinematic limits requiring the decay to be reconstructed in a single jet are the same for the two algorithms, but fixing the ordering of the two recombinations requires a different restriction for each algorithm, which in turn biases the distributions of kinematic variables.

The common requirement that the top quark be reconstructed in a single jet, $\Delta R_{qq'} < D$ and $\Delta R_{Wb} < D$, is straightforward to understand in terms of the rest frame variable $\cos\theta_0$, which here is the polar angle in the top quark rest frame between the $W$ and the boost direction to the lab frame.  For $\cos\theta_0 \approx 1$, the $W$ has a large transverse boost in the lab frame, so $\Delta R_{qq'} < D$, but the angle between the $W$ and $b$ will be large (as was the case for the corresponding $1\to2$ decay in the previous section).  For $\cos\theta_0 \approx -1$, the $W$ transverse boost is small, and $\Delta R_{qq'}$ will be large.  Therefore, we only expect to reconstruct top quarks in a single jet when $|\cos\theta_0|$ is not near $1$.  Specifically which decays will be reconstructed, though, depends on the algorithm.

If the CA algorithm correctly reconstructs the top quark, the two quarks from the $W$ decay must be the closest pair (in $\Delta R$) of the three final state particles.  This requirement highly selects for decays where the $W$ opening angle, $\Delta R_{qq'}$, is smaller than the top quark opening angle, $\Delta R_{Wb}$.  Therefore, only decays with a large (transverse) $W$ boost will be reconstructed by the CA algorithm.  In terms of $\cos\theta_0$, the fraction of decays that are reconstructed will increase as we increase $\cos\theta_0$ towards the upper limit where $\Delta R_{Wb} \ge D$, and the reconstruction fraction will be small for lower values of $\cos\theta_0$.

The $\kt$ algorithm orders recombinations by $p_T$ as well as angle, and the set of reconstructed decays is understood most easily by contrasting with CA.  As the transverse boost of the $W$ decreases, on average the $p_T$ of the $q$ and $q'$ decrease while the $p_T$ of the $b$ increases.  Therefore, while $\Delta R_{qq'}$ is increasing, $\min(p_{Tq}, p_{Tq'})$ is decreasing, and these competing effects suggest that $\kt$ reconstructs decays with smaller values of $\cos\theta_0$ than CA, and that the dependence on $\cos\theta_0$ is not as strong.

\begin{figure}[htbp]
\includegraphics[width = \columnwidth]{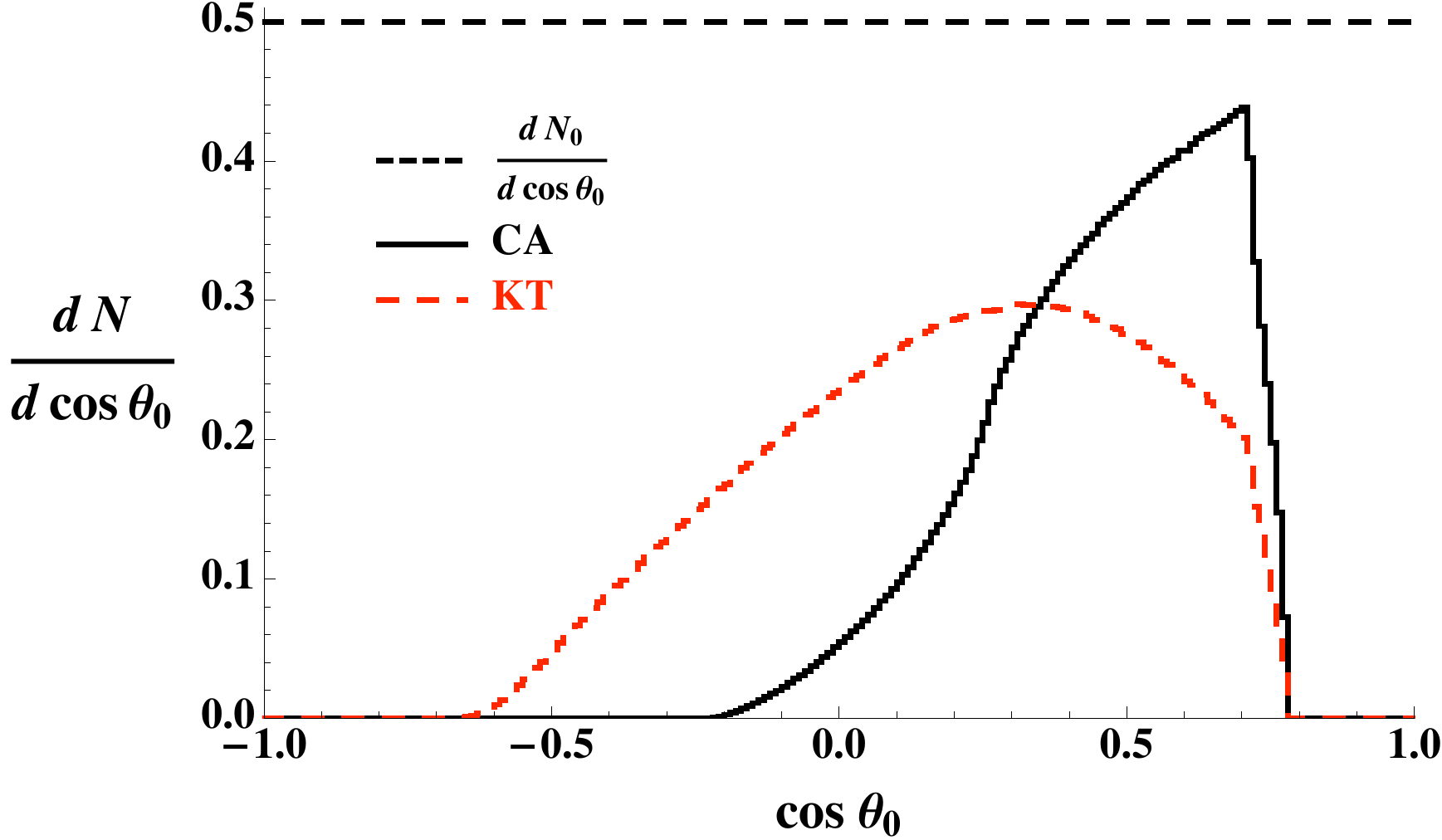}
\caption{$dN/d\cos\theta_0$ vs. $\cos\theta_0$, with $\gamma = 3$, for both the $\kt$ and CA algorithms.  The underlying distribution $dN_0/d\cos\theta_0 = 1/2$ is plotted as the dotted line for reference.  }%\red{topcosthdist}}
\label{topcosthdist}
\end{figure}

The effect of the CA and $\kt$ algorithms on the observed distribution in $\cos\theta_0$ is shown in Fig.~\ref{topcosthdist}, where we plot the distribution of $\cos\theta_0$ for reconstructed top quarks for both algorithms.  The top boost is fixed to $\gamma = 3$.
We observe the kinematic limit near $\cos\theta_0\approx 0.8$ is common between algorithms, and that $\cos\theta_0\approx-1$ is not accessed by either algorithm.  As expected, the distribution for the CA algorithm falls off more sharply than for $\kt$ at lower values of $\cos\theta_0$.

\begin{figure}[htbp]
\includegraphics[width = \columnwidth]{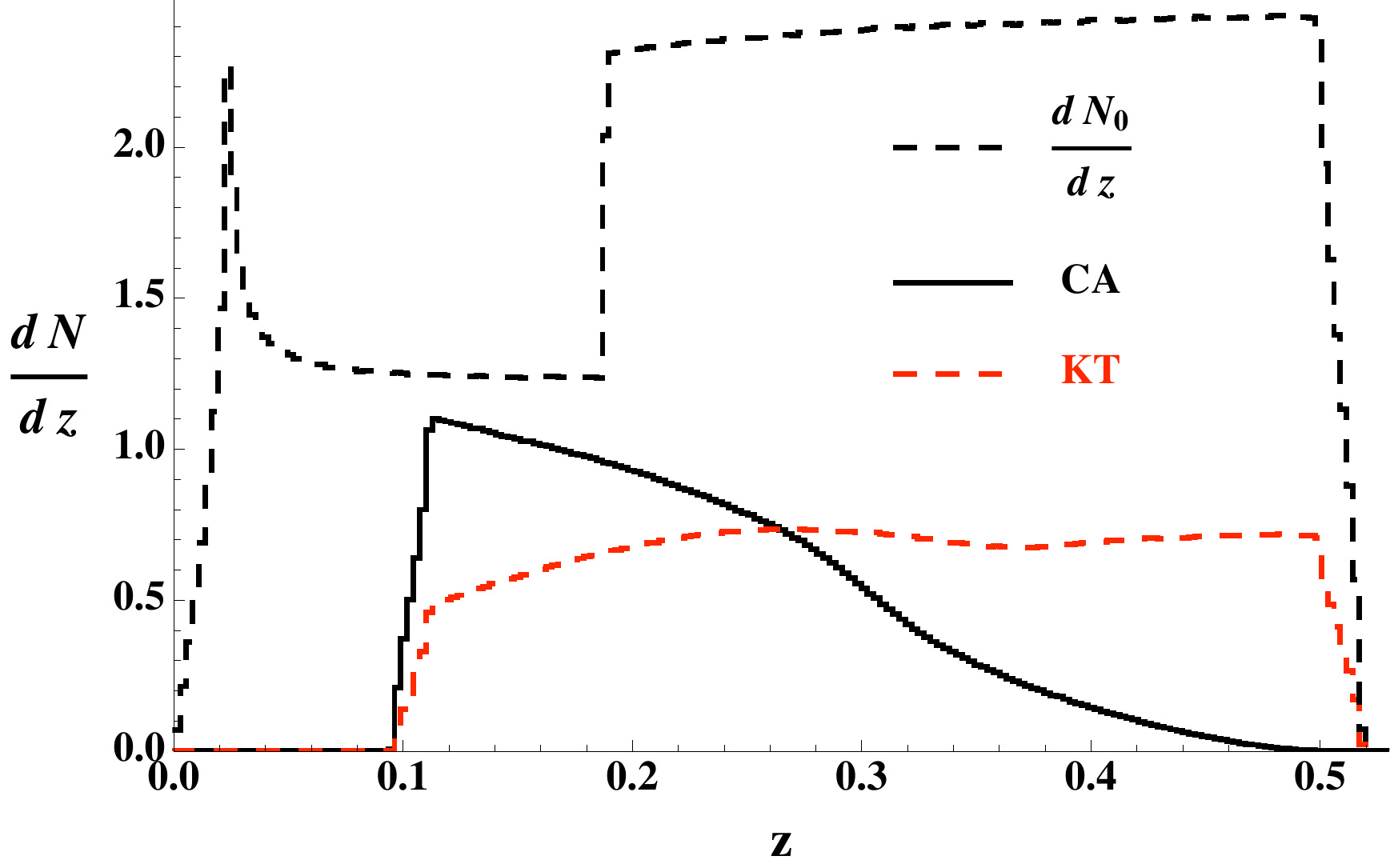}
\caption{$dN_0/dz$ (all decays) and $dN/dz$ (reconstructed decays), with $\gamma = 3$.  }%\red{topzdist}}
\label{topzdist}
\end{figure}

Next, we look at distributions in $z$ and $\Delta R_{Wb}$.  Just as in the $1\rightarrow2$ decay, we expect decays with small $z$ not to be correctly reconstructed.  Small values of $z$ will come when the $W$ or $b$ is soft, and therefore produced very backwards-going in the top rest frame.  This corresponds to $\cos\theta_0 \approx \pm 1$, and from Fig.~\ref{topcosthdist} these decays are not reconstructed.  In Fig.~\ref{topzdist}, we plot the distribution in $z$ for all decays, $dN_0/dz$, and the distribution for reconstructed decays, $dN/dz$, for a boost of $\gamma = 3$.

In $dN_0/dz$, the discontinuity at $z\approx0.2$ arises from the fact that the $W$ is sometimes softer than the $b$, but has a minimum $p_T$.  The extra weight in $dN_0/dz$ for $z$ above this value comes from the decays where the $W$ is softer than the $b$.  Note that these decays are rarely reconstructed, especially for CA: the distribution $dN/dz$ is smooth, and has little additional support in the region where the $W$ is softer.  This correlates with the fact that decays with negative $\cos\theta_0$ values are rarely reconstructed with CA, but more frequently with $\kt$.  The distribution $dN/dz$ has a lower cutoff that corresponds to the upper cutoff in Fig.~\ref{topcosthdist}.  As the boost $\gamma$ of the top increases, the cutoff at small $z$ decreases, since the limit in $\cos\theta_0$ for which $\Delta R_{Wb} > D$ will increase towards 1.

\begin{figure}[htbp]
\includegraphics[width = \columnwidth]{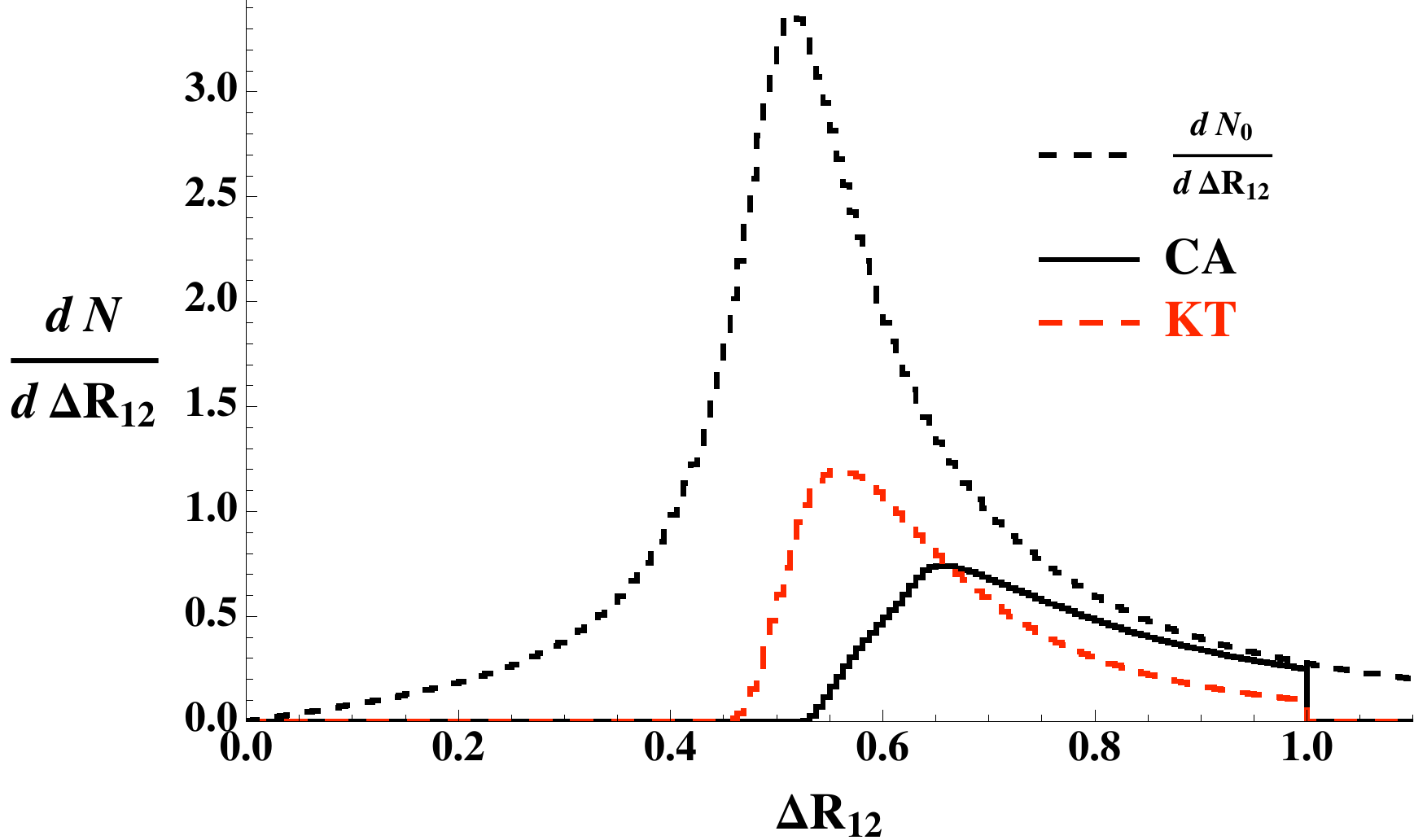}
\caption{$dN_0/d\Delta R_{Wb}$ (all decays) and $dN/d\Delta R_{12}$ (reconstructed decays), with $\gamma = 3$.  }%\red{topdRdist}}
\label{topdRdist}
\end{figure}

The opening angle $\Delta R_{Wb}$ of the top quark decay also illustrates how strongly the kinematics are shaped by the jet algorithm.  When $\cos\theta_0 \approx -1$, for sufficient boosts $\Delta R_{Wb}$ is small because the $W$ is boosted forward in the lab frame, but these decays are not reconstructed because the ordering of recombinations will typically be incorrect and the $W$ decay may not within $\Delta R_{qq'} < D$.  For $\cos\theta_0 \approx 1$, $\Delta R_{Wb}$ will exceed $D$ and the top will not be reconstructed.  In Fig.~\ref{topdRdist}, we plot the distribution $dN_0/d\Delta R_{Wb}$ of the angle between the $W$ and $b$ in all top decays for a top boost of $\gamma = 3$, as well as the distribution $dN/d\Delta R_{12}$ of the angle of the last recombination for reconstructed top quarks with the $\kt$ and CA algorithms.  Note that when the top quark is reconstructed at the parton level, $\Delta R_{12} = \Delta R_{Wb}$.
The difference in $dN/d\Delta R_{12}$ between the $\kt$ and CA algorithms reflects their different recombination orderings.  Because CA orders strictly by angle, the angle $\Delta R_{12}$ tends to be larger than for $\kt$ because CA requires $\Delta R_{12} = \Delta R_{Wb} > \Delta R_{qq'}$.  The $\kt$ algorithm prefers smaller angles for $\Delta R_{Wb}$, because in these cases the $W$ is softer so that the value of the $\kt$ metric to recombine the $q$ and $q'$, $\min(p_{Tq},p_{Tq'})\Delta R_{qq'}$, is smaller.

\subsection{Hadron-level Decays}

To this point, we have looked at parton-level kinematics of the top decay.  However, we cannot expect the jet algorithm to faithfully represent the kinematics of the parton-level top decay in jets which include the physics of showering and hadronization.  That is, the systematic effects of the jet algorithm, similar to those seen in QCD jets in Section \ref{sec:QCDJets}, can be expected to appear in distributions of kinematic variables for jets reconstructing the top quark mass.  The substructure of a jet that reconstructs the top quark mass may not match onto the kinematics of that decay due to systematic effects of the jet algorithm.  For instance, in the CA algorithm we expect that soft recombinations will occur at the last recombination step, even for jets that contain the decay products of a top quark.  This can make the substructure look more like a heavy QCD jet than a top quark decay, and subsequently the jet may not be properly identified.

\begin{figure}[htbp]
\includegraphics[width = \columnwidth]{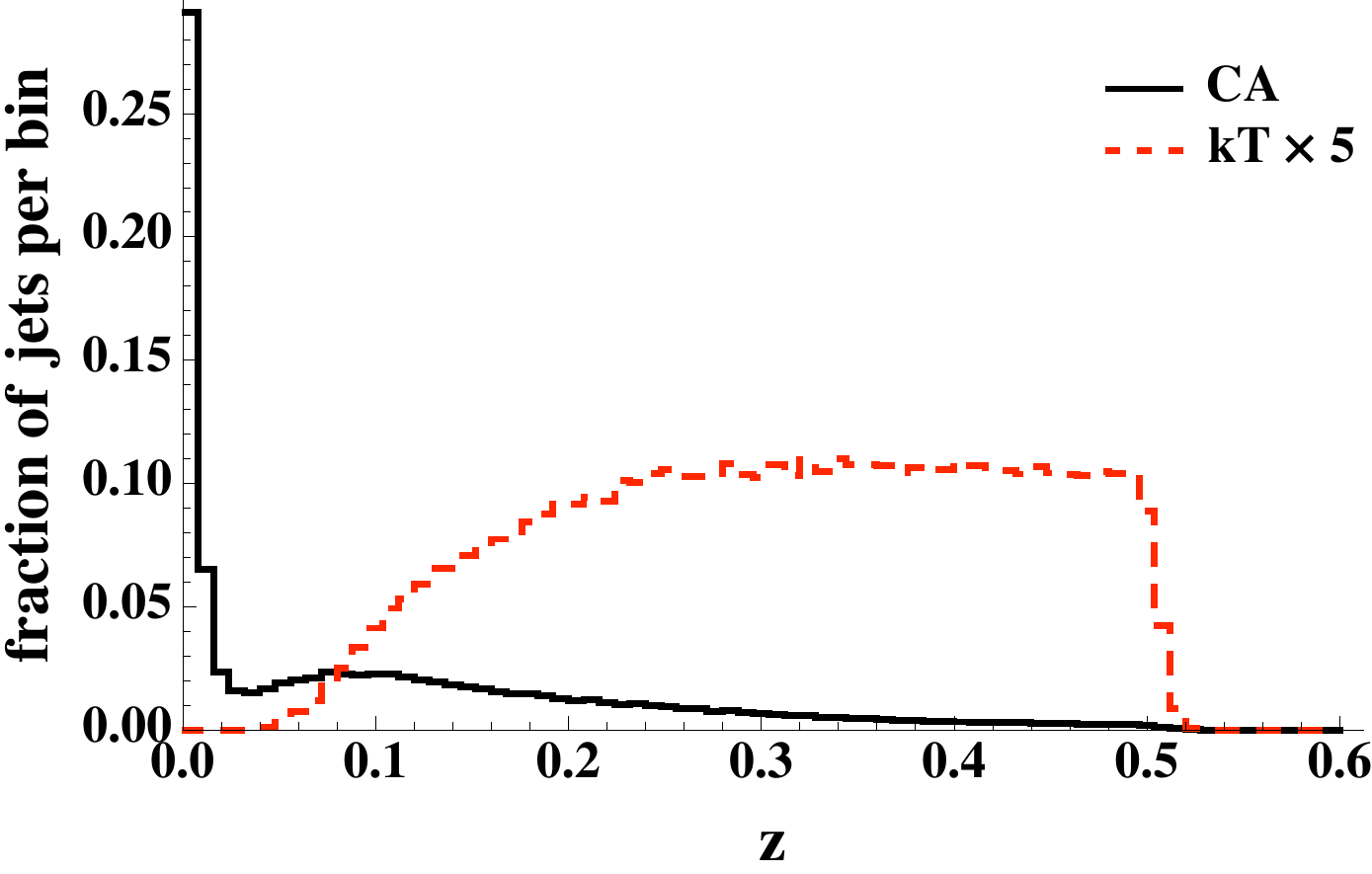}
\caption{Distribution in $z$ for jets with the top mass in the $t\bar{t}$ sample.  The jets have $p_T$ between 500 and 700 GeV, and D = 1.0.  Note the $\kt$ distribution is scaled up by a factor of 5 to make the scales comparable.  }%\red{topquarkjetz}}
\label{topquarkjetz}
\end{figure}

To demonstrate this point, in Fig.~\ref{topquarkjetz} we plot the distribution in $z$ for jets with mass within a window around the top quark mass.  The data represent simulated $t\bar t$ events as described in Appendix \ref{sec:appendix}.  In this sample, the top quarks have a $p_T$ between 500--700 GeV, so that many are expected to be reconstructed in a single jet.

\begin{figure}[htbp]
\includegraphics[width = \columnwidth]{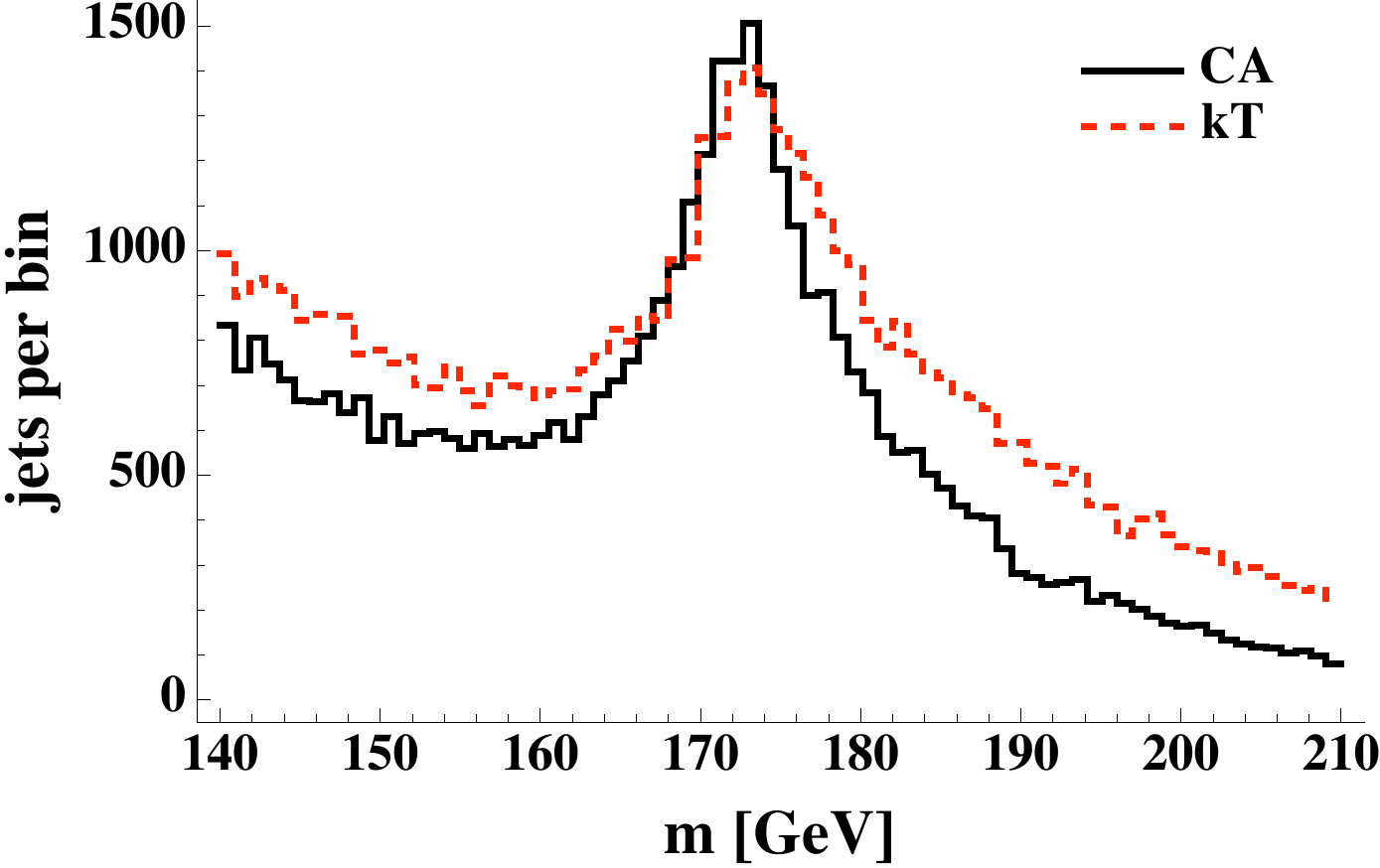}
\caption{Distribution in jet mass for jets in the neighborhood of the top mass in $t\bar{t}$ events for the CA (black) and $\kt$ (red) algorithms.  }%\red{topmassdist}}
\label{topmassdist}
\end{figure}

The distribution for CA jets is very different from the parton-level distribution, plotted in Fig.~\ref{topzdist}.  The excess at small values of $z$ arises from soft recombinations in the CA algorithm, which make the distribution similar to the distribution in $z$ from QCD jets shown in Figs.~\ref{zplotCA} and \ref{zplotKT}.  For the $\kt$ algorithm, there are rarely soft recombinations late in the algorithm, because the metric orders according to $z$ as well as $\Delta R$.  However, the $\kt$ algorithm tends to have a much broader mass distribution for reconstructed tops than the CA algorithm, since soft particles that dominate the periphery of the jet are recombined early in the algorithm.  This means that soft energy depositions in the calorimeter near the decay products of a top quark have a higher probability of being included in the jet and broadening the reconstructed top mass distribution.  In Fig.~\ref{topmassdist}, we plot the jet mass distribution in the neighborhood of the top mass for jets in the same $t\bar{t}$ sample as in Fig.~\ref{topquarkjetz} for both algorithms.

The top mass peak is broadened for the $\kt$ algorithm relative to CA.  From the point of view of jet substructure, we cannot identify vertex-specific variables (such as $z$ and $\Delta R$) that characterize this broadening, because it is due to recombinations early in the algorithm.  However, we will find that techniques used to remove the systematic effects of the algorithm from the substructure of jets are effective in narrowing mass distributions.

\section{Identifying Reconstructed Heavy Particles with Jet Substructure}
\label{sec:algEffects}

In the previous two sections we examined several kinematic distributions for QCD splittings and for heavy particle decays.  These studies fall into two categories: parton-level, dealing with the fundamental $1\to2$ processes, and hadron-level, including the physics of showering and hadronization.  While the parton-level studies are important to understand the kinematics of reconstructed decays and the differences from QCD, the hadron-level studies encompass the effects of the QCD shower and the jet algorithm.  We will explore these effects more in this section, and give a more complete picture of jet substructure.  Since our focus is on reconstructing heavy particles, we will discuss the difficulties that arise in interpreting jet substructure.

Our parton-level studies can be briefly summarized.  In Sec.~\ref{sec:QCDJets}, we used a toy model for QCD splittings in jets that contained the dominant soft and collinear physics of QCD, and studied the kinematics for fixed $m/p_T$ of the parent parton in the splitting.  In Sec.~\ref{sec:reconHeavy}, we looked at $1\to2$ and $1\to3$, two-step decays with fixed boost, requiring that the decay be reconstructed in a jet.  For the two-step top quark decay, requiring full reconstruction of the top (including the $W$ as a subjet) from the three final state quarks imposed kinematic restrictions that depended on the algorithm used.  These studies led to the $z$ and $\Delta R_{12}$ distributions seen in Figs.~\ref{NLOzplot} and \ref{NLOthetaplot} (QCD), \ref{zdistrecon} and \ref{dRdistall} (one-step decays), and \ref{topzdist} and \ref{topdRdist} (two-step decays).  We can see that the distributions in $\Delta R_{12}$ are quite similar, but that QCD splittings tend to have smaller $z$ values than heavy particle decays.  However, the kinematics of a heavy particle decay are not always simple to detect in a jet that includes showering, as our hadron-level studies have demonstrated.

The QCD shower and the jet algorithm both play a significant role in shaping the jet substructure.  The ordering of recombinations for the $\kt$ and CA algorithms imposes significant kinematic constraints on the phase space for the last recombinations in a jet.  This leads to kinematic distributions for the last recombination in a jet that depend as much on the algorithm as the underlying physics of the jet.  For instance, in Figs.~\ref{zplotCA}--\ref{a1plotKT}, we find that the kinematics of the last recombination in QCD jets is very different between the $\kt$ and CA algorithms.  In particular, we can compare Figs.~\ref{zplotCA} and \ref{zplotKT}, the distribution in $z$ of the last recombination for QCD jets, with Fig.~\ref{topquarkjetz}, the distribution in $z$ of the last recombination for jets in a $t\bar{t}$ sample that reconstruct the top quark mass.  For the $\kt$ algorithm, the differences reflect the different physics of QCD splittings and decays.  However, the CA algorithm has shaped the distributions to have a large enhancement at small $z$ for both processes.  This implies that it is difficult to discern the physics of the jet simply from the value of $z$ in the last recombination for CA.  For the $\kt$ algorithm, because of the ordering of recombinations, the final recombinations better discriminate between decays and QCD, but the mass resolution is poorer than for CA.  In Fig.~\ref{topmassdist}, we see that the mass distribution of a reconstructed top quark is degraded for the $\kt$ algorithm relative to CA.

There is one more important contribution to jet substructure common to QCD jets and heavy particle decays that we have not yet discussed.  This is the combined effect of splash-in from several sources: soft radiation from other parts of the hard scattering, from the underlying event (UE), i.e., from the rest of the individual $pp$ scattering, and from pile-up, i.e., from other $pp$ collisions that occur in the same time bin.  All of these sources add particles to jets that are typically soft and approximately uncorrelated.  Splash-in particles will mostly be located at large angle to the jet core, simply because there is more area there.  How these particles affect jet substructure depends on the algorithm used, and we expect them to contribute similarly to soft radiation from the QCD shower, discussed at the ends of Secs.~\ref{sec:QCDJets} and \ref{sec:reconHeavy}.  For concreteness, we now examine briefly the effect of adding an UE to our Monte Carlo events.  We expect other splash-in effects to be similar.

\begin{figure}[htbp]
\subfloat[QCD jets, CA] {\includegraphics[width=0.23\textwidth] {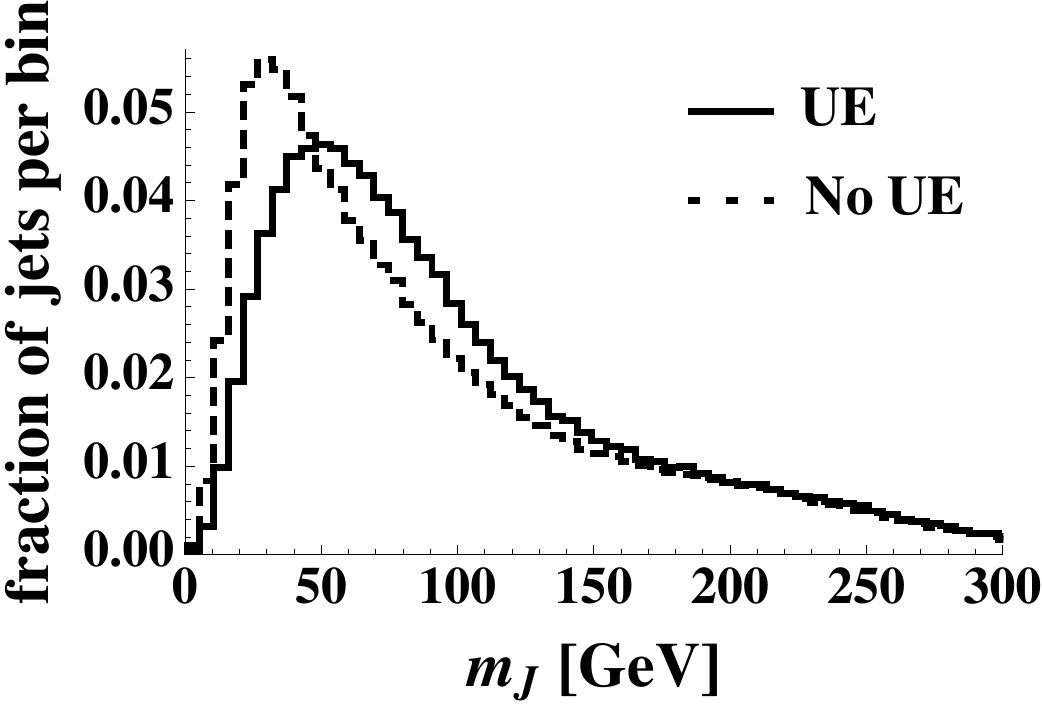}}
\subfloat[QCD jets, $\kt$] {\includegraphics[width=0.23\textwidth] {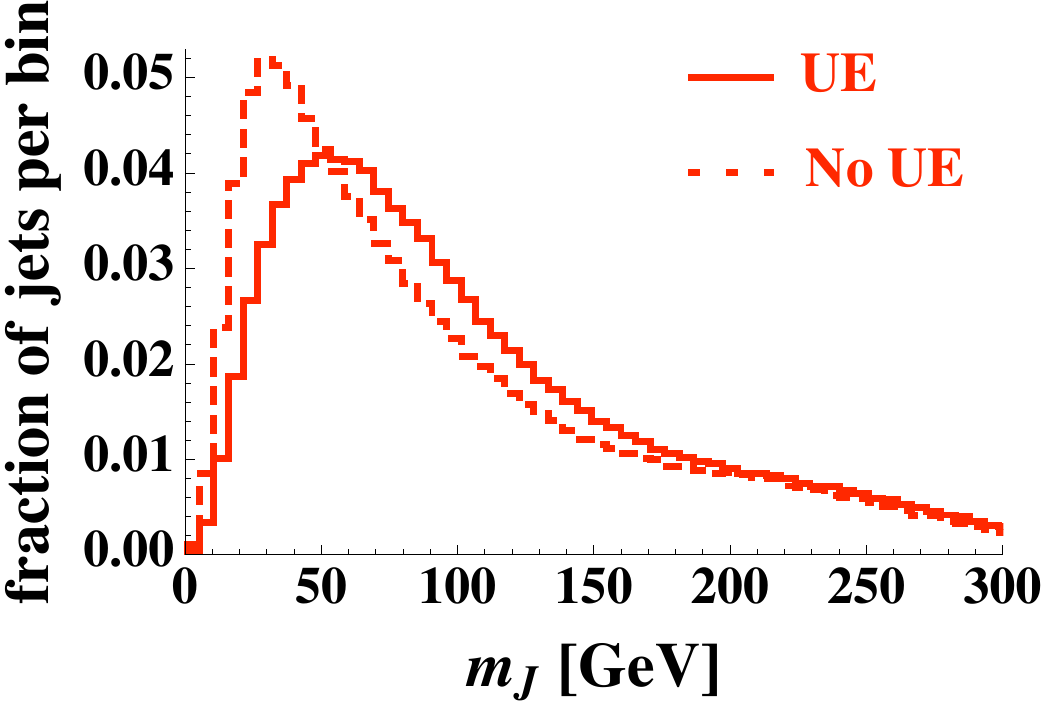}}

\subfloat[top jets, CA] {\includegraphics[width=0.23\textwidth] {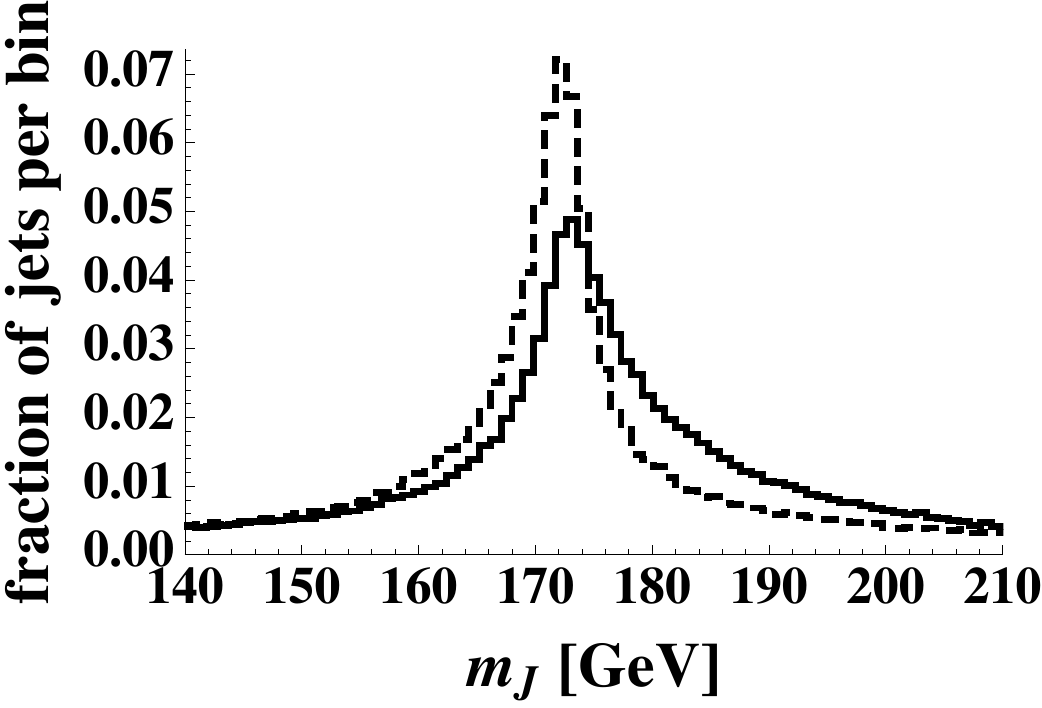}}
\subfloat[top jets, $\kt$] {\includegraphics[width=0.23\textwidth] {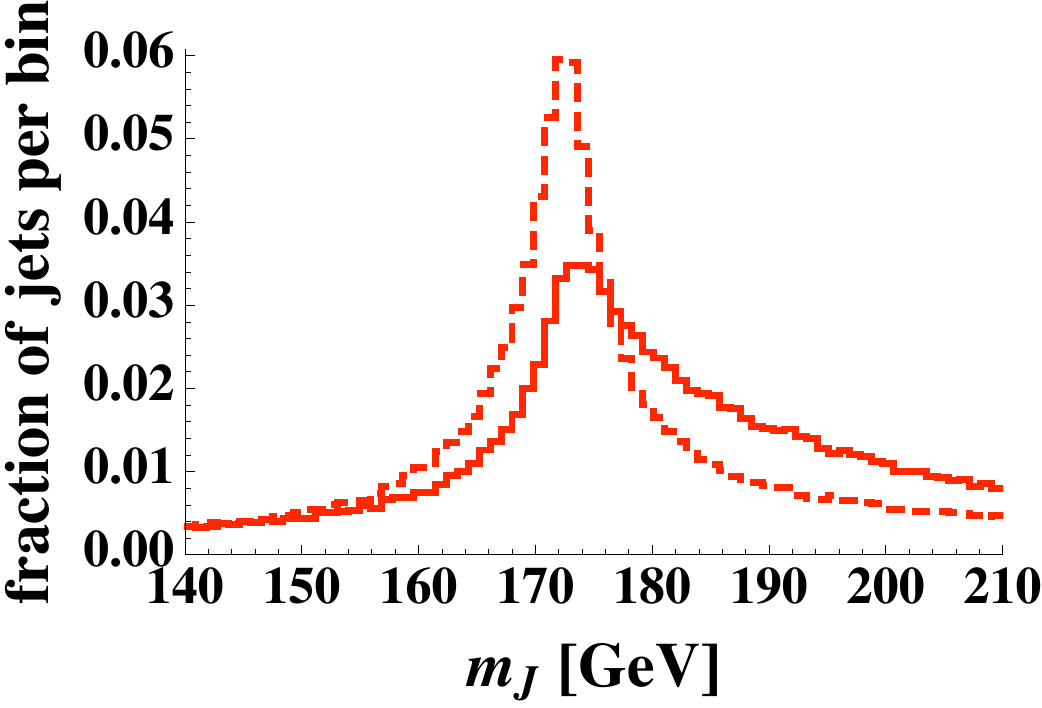}}
\caption{Distribution in $m_J$ with and without underlying event for QCD and top jets, using the CA and $\kt$ algorithms.  The jets have $p_T$ between 500 and 700 GeV, and D = 1.0.  The samples are described further in Appendix~\ref{sec:appendix}. }%\red{UEcompMasses}}
\label{UEcompMasses}
\end{figure}

\begin{figure}[htbp]
\subfloat[QCD jets, z] {\includegraphics[width=0.23\textwidth] {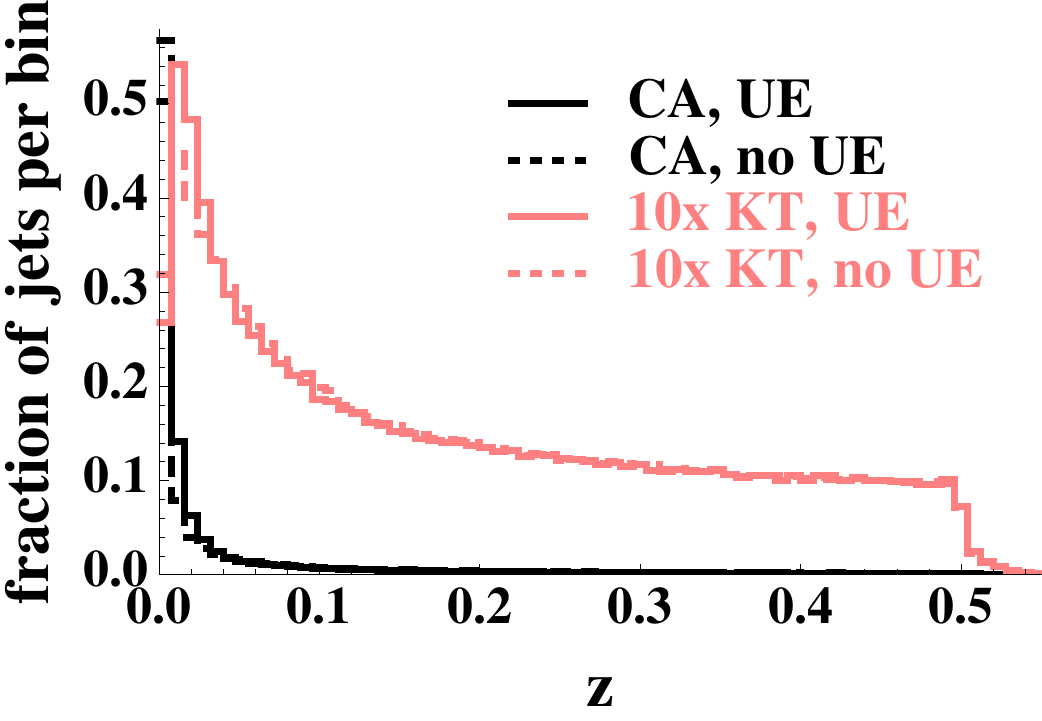}}
\subfloat[QCD jets, $\Delta R_{12}$] {\includegraphics[width=0.23\textwidth] {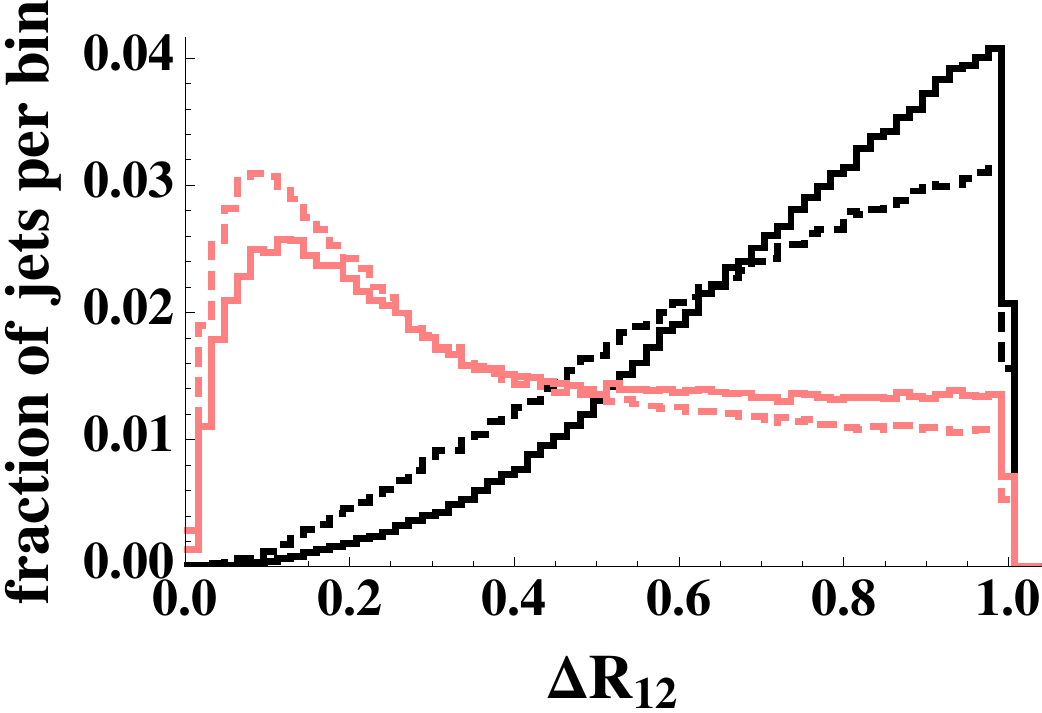}}

\subfloat[top jets, z] {\includegraphics[width=0.23\textwidth] {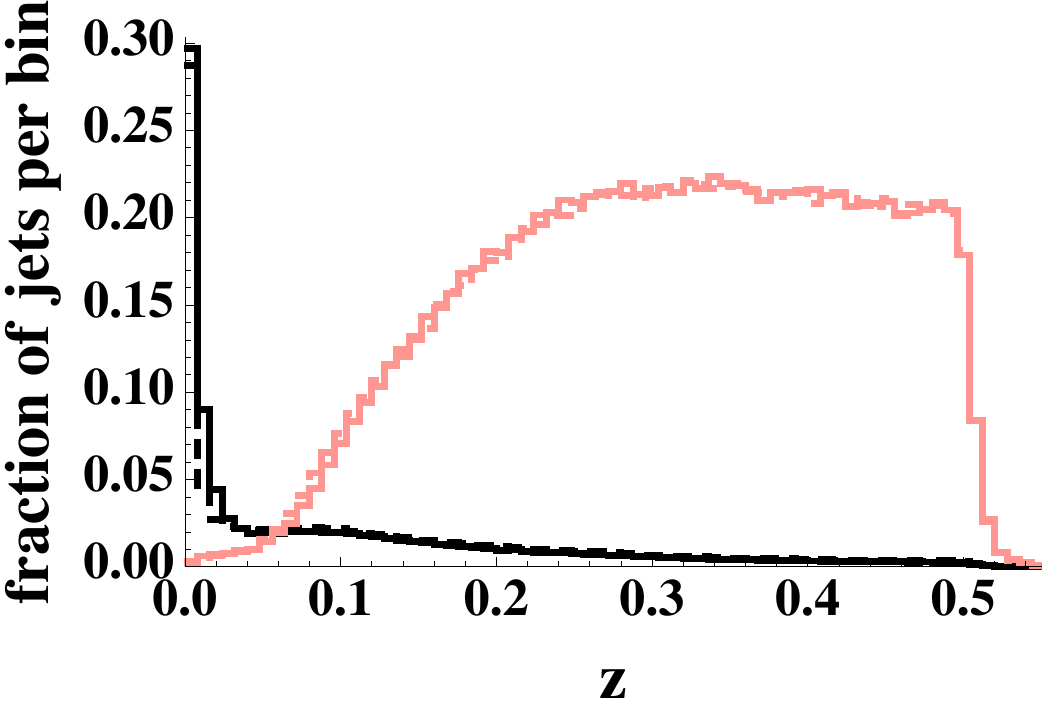}}
\subfloat[top jets, $\Delta R_{12}$] {\includegraphics[width=0.23\textwidth] {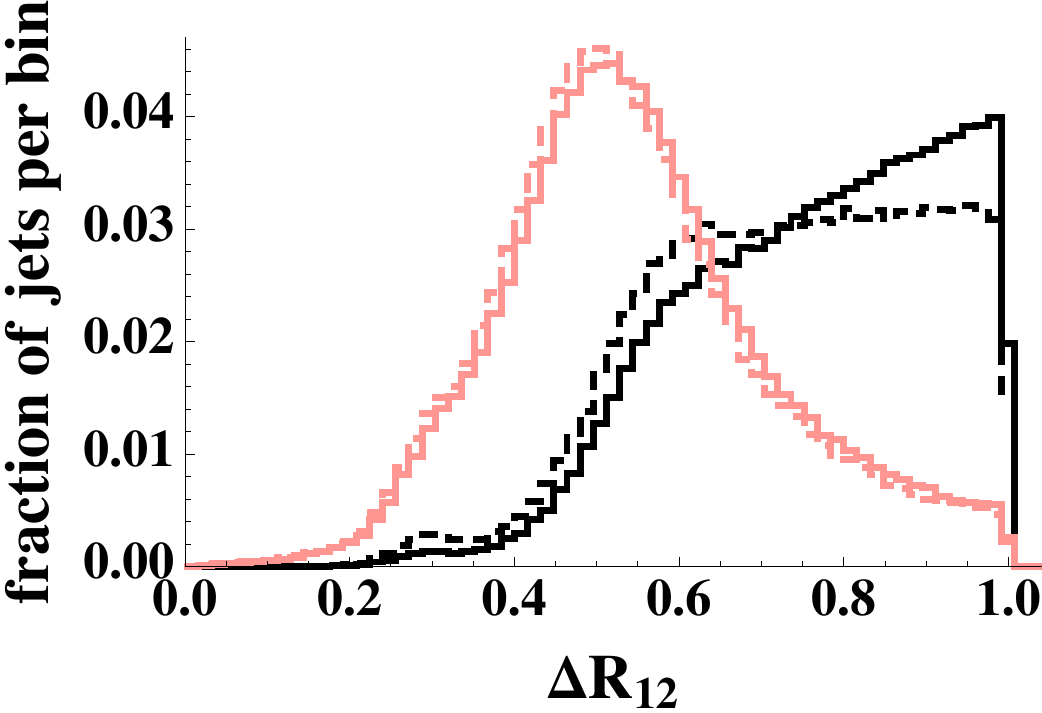}}
\caption{Distributions in $z$ and $\Delta R_{12}$ with and without underlying event for QCD and top jets, using the CA and $\kt$ algorithms.  The jets have $p_T$ between 500 and 700 GeV, and D = 1.0.  The samples are described further in Appendix~\ref{sec:appendix}. }%\red{UEcompZDR}}
\label{UEcompZDR}
\end{figure}

In Fig.~\ref{UEcompMasses}, we show the effect of adding an UE on jet masses.  The effect here is simple: adding extra energy to jets pushes the mass distribution higher.  Note that for top jets, the mass peak has also broadened, making it harder to find the signal mass bump over the background distribution.  In Fig.~\ref{UEcompZDR}, we show how distributions in $z$ and $\Delta R_{12}$ are affected by the UE.  Due to the extra radiation at large angles from the UE, the distribution in the angle of the last recombination, $\Delta R_{12}$, is systematically shifted to larger values.  The UE populates the same region in the jet as soft radiation from the hard partons, meaning the distribution in $z$ is not significantly altered by the UE.


We have seen numerous examples that the kinematics of the jet substructure in the last recombination for CA is a poor indicator for the physics of the jet.  However, we can characterize the aberrant substructure very simply.  For the CA algorithm, late recombinations (necessarily at large $\Delta R$) with small $z$ are more likely to arise from systematics effects of the algorithm than from the dynamics of the underlying physics in the jet.  For the $\kt$ algorithm, the poor mass resolution of the jet arises from earlier recombinations of soft protojets.  The last recombination for $\kt$ is representative of the physics of the jet, but the degraded mass resolution makes it difficult to efficiently discriminate between jets reconstructing heavy particle decays and QCD.  While small-$z$, large-$\Delta R$ recombinations are not as frequent late in the $\kt$ algorithm as in CA, they do contribute the most to the poor mass resolution of $\kt$.

As a simple example of the sensitivity of the mass to small-$z$, large-$\Delta R$ recombinations, consider the recombination $i,j\to p$ of two massless objects in the small-angle approximation.  The mass of the parent $p$ is given by $m_p^2 = p_{T_p}^2z(1-z)\Delta R_{ij}^2$, as in Eq.~(\ref{eq:simplejmass}).  Suppose the value of the $\kt$ recombination metric, $\rho_{ij}(\kt) = p_{T_p}z\Delta R_{12}$ is bounded below by a value $\rho_0$ (say by previous recombinations), and the recombination $i,j\to p$ occurs at $\rho_{ij}(\kt) = \rho_0$.  Then the mass of the parent is $m_p^2 = \rho_0^2(1-z)/z$, which is maximized for small $z$.  Therefore, at a given stage of the algorithm, small-$z$ recombinations have a large effect on the mass of the jet.

\begin{figure}[htbp]
\subfloat[$m_J$ cut, $z$]{\includegraphics[width = 0.23\textwidth]{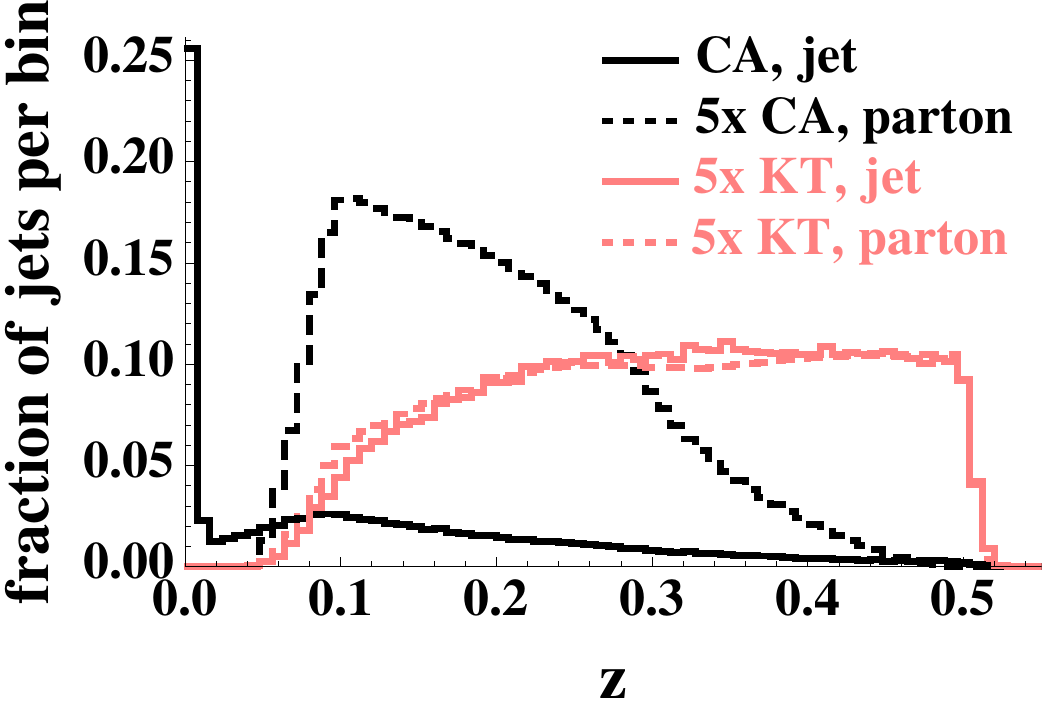}}
\subfloat[$m_J$ cut, $\Delta R_{12}$]{\includegraphics[width = 0.23\textwidth]{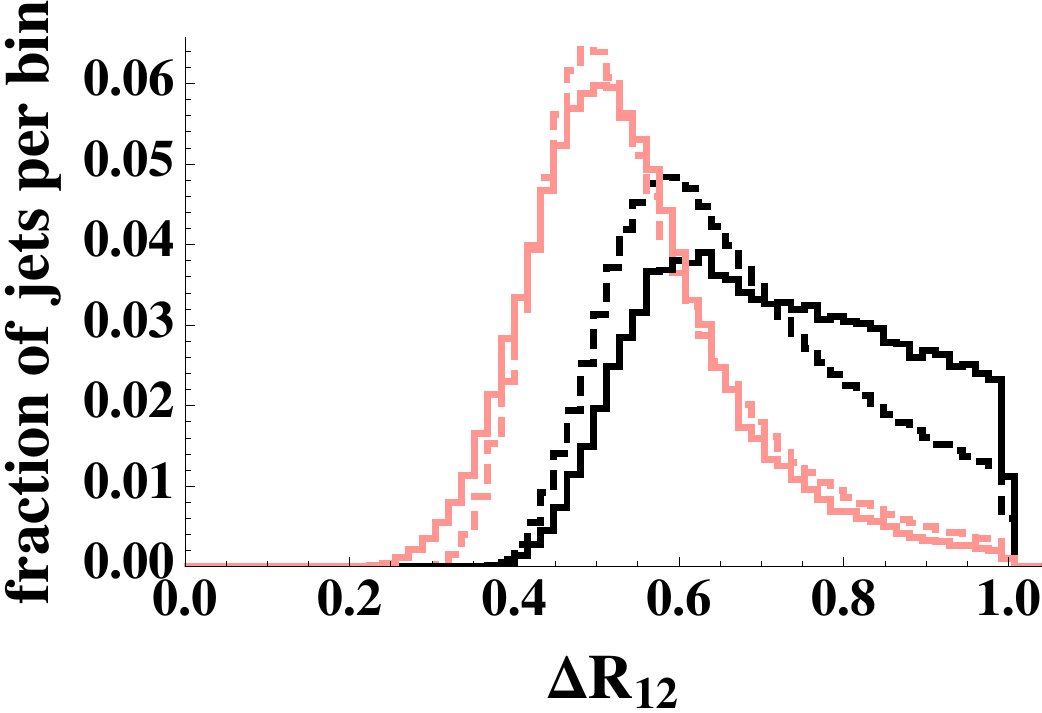}}

\subfloat[$m_J$ and $m_{\text{Sub}J}$ cuts] {\includegraphics[width = 0.23\textwidth] {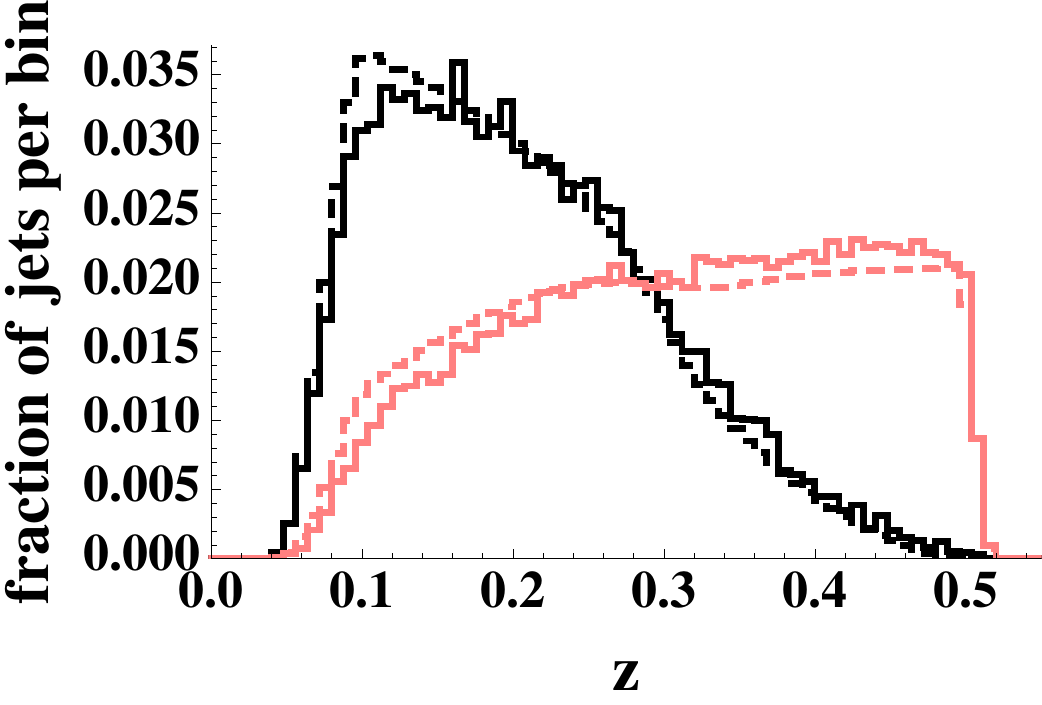}}
\subfloat[$m_J$ and $m_{\text{Sub}J}$ cuts] {\includegraphics[width =
0.23\textwidth] {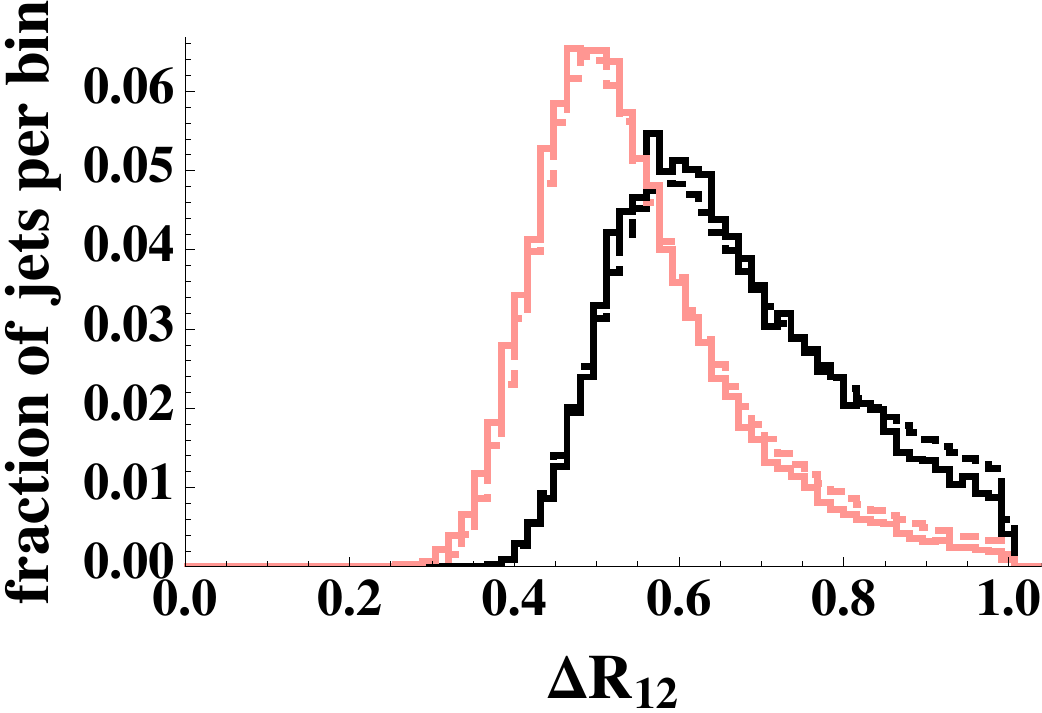}}

\caption{Distributions in $z$ and $\Delta R_{12}$ comparing for top quark decays at the parton-level and from Monte Carlo events.  The jets have $p_T$ between 500 and 700 GeV, and have D = 1.0.  The parton-level top decays have the same distribution of boosts as the Monte Carlo top jets.  Jets in the upper plots have a mass cut on the jet; the lower plots includes a subjet mass cut.  The details of these cuts are described in Sec.~\ref{sec:MC}. }%\red{topPartonVsReconZDR}}
\label{topPartonVsReconZDR}
\end{figure}

When we can resolve the mass scales of a decay in a jet, the distribution of kinematic variables matches closely what we expect from the parton-level kinematics of the decay.  For the example of the top quark decay, if we select jets with the top mass that have a daughter subjet with the $W$ mass, the kinematic distributions of $z$ and $\Delta R_{12}$ closely match the distributions from the parton-level decay of the top quark.  We show this in Fig.~\ref{topPartonVsReconZDR}, where we make a top quark ``hadron-parton'' comparison for $z$ and $\Delta R_{12}$.  In the hadron-level events, we take jets from $t\bar{t}$ production and either make a cut on the jet mass, requiring a mass near the top mass, or both the jet mass and the subjet mass, requiring proximity to the top and W masses.  The specifics of the mass cuts are described in Sec.~\ref{sec:MC}.  In the parton-level events, we simply require that the top quark decay to three partons be fully reconstructed by the algorithm in a single jet, namely that the $W$ is correctly recombined first from its decay products before recombination with the $b$ quark to make the top.  The parton-level events have the same distribution of top quark boosts as the top jets in the hadron-level events.
It is clear that simply requiring the hadron-level jet have the top mass, which makes no cut on the substructure, leads to kinematic distributions in $z$ and $\Delta R_{12}$ for CA that do not match the parton-level distributions, although the distributions do match quite well for the $\kt$ algorithm.  The excess of small-$z$ recombinations for CA in the hadron-level jet with only a jet mass cut arises from jet algorithm effects discussed previously.  After the subjet mass cut, these are removed and the distribution of $z$ in the jet matches the reconstructed parton-level decay very well.

Therefore, when we can accurately reconstruct the mass scales of a decay in a jet, the kinematics of the jet substructure tend to reproduce the parton-level kinematics of the decay.  This suggests that if we can reduce systematic effects that generate misleading substructure, we can improve heavy particle identification and separation from background.  Reducing these systematic effects can also improve the mass resolution of the jet, which will aid in identifying a heavy particle decay reconstructed in a jet and in rejecting the QCD background.  We now discuss a technique that aims to accomplish this goal.

%%%%%%%%%% END OF SECTION 5: ALGORITHM EFFECTS %%%%%%%%%%

\section{The Pruning Procedure}
\label{sec:pruning}

In this section we define a technique that modifies the jet substructure to reduce the systematic effects that obscure heavy particle reconstruction.  In general, we will think of a \emph{pruning procedure} as using a criterion on kinematic variables to determine whether or not a branching is likely to represent accurate reconstruction of a heavy particle decay.  This takes the form of a cut: if a branching does not pass a set of cuts on kinematic variables, that recombination is vetoed.  This means that one of the two branches to be combined (determined by some test on the kinematics) is discarded and the recombination does not occur.

In Sec.~\ref{sec:algEffects}, we identified recombinations that are unlikely to represent the reconstruction of a heavy particle.  These can be characterized in terms of the variables $z$ and $\Delta R$: recombinations with large $\Delta R$ and small $z$ are much more likely to arise from systematic effects of the jet algorithm and in QCD jets rather than heavy particle reconstruction.  We expect that removing (\emph{pruning}) these recombinations will tend to improve our ability to measure the mass of a jet reconstructing a heavy particle.  We also expect that this procedure will systematically shift the QCD mass distribution lower, reducing the background in the signal mass window.  Finally this procedure is expected to reduce the impact of uncorrelated soft radiation from the underlying event and pile-up. We therefore define the following pruning procedure:

\begin{itemize}
\item[0.]  Start with a jet found by any jet algorithm, and collect the objects (such as calorimeter towers) in the jet into a list $L$.  Define parameters $D_{\cut}$ and $z_{\cut}$ for the pruning procedure.
\item[1.]  Rerun a jet algorithm on the list $L$, checking for the following condition in each recombination $i,j\to p$:
\[
z = \frac{\min(p_{Ti},p_{Tj})}{p_{Tp}} < z_{\cut} \quad \text{and} \quad \Delta R_{ij} > D_{\cut}.
\]
This algorithm must be a recombination algorithm such as the CA or $\kt$ algorithms, and should give a ``useful'' jet substructure (one where we can meaningfully interpret recombinations in terms of the physics of the jet).
\item[2.]  If the conditions in 1. are met, do not merge the two branches $1$ and $2$ into $p$.  Instead, discard the softer branch, i.e., veto on the merging.  Proceed with the algorithm.
\item[3.]  The resulting jet is the \emph{pruned jet}, and can be compared with the jet found in Step 0.
\end{itemize}

This technique is intended to be generically applicable in heavy particle searches.  It generalizes analysis techniques suggested by other authors \cite{Butterworth:08.1, Kaplan:08.1}, in that these methods also modify the jet substructure to assist separate a particular signal from backgrounds.  We emphasize that pruning can be broadly applied.  We have endeavored to justify this claim with the discussions in Secs.~\ref{sec:QCDJets}-\ref{sec:algEffects}, which demonstrate that the interpretation of jet substructure is subject to systematic effects that can be well characterized.  Pruning is not the only option, but offers some advantages which we explore in further studies below.

In the analysis of pruning, we will explore the dependence of the pruned jets on the value of $D$ from the jet algorithm.  When reconstructing a boosted heavy particle in a single jet, without pruning the reconstruction is optimized if the value of $D$ is fit to the expected opening angle of the decay.  However, this angle depends on the mass of the particle (which is not known in a search) and its $p_T$.  We will show that pruning reduces the sensitivity to $D$ and allows one to use large $D$ jets over a broad range in $p_T$ to search for heavy particles.  This makes a search much more straightforward to carry out by using pruning.

Values for the two parameters of the pruning procedure, $z_{\cut}$ and $D_{\cut}$, can be well motivated.  In the following studies, we will show that the results of pruning are rather insensitive to the parameters, and that the optimal parameters are similar for different searches.  That is, it is not necessary to tune the pruning procedure for individual searches.

The parameter $z_{\cut}$ can be chosen based on the analysis of single-step and multi-step decays in Sec.~\ref{sec:reconHeavy}.  Near the limit in boost where decays are reconstructed in a single jet, the value of $z$ is typically large.  It is only at large boosts, where the production rate of heavy particles is much smaller, that small values of $z$ are allowed for reconstructed decays.  Therefore, we can choose a value of $z_{\cut}$ that will keep all reconstructed parton-level decays at small boost, and only remove a small fraction of decays at larger boosts.  For both the $\kt$ and CA algorithms, we set $z_{\cut} = 0.10$ initially, and will study the performance of pruning as $z_{\cut}$ is varied for different searches.

The parameter $D_{\cut}$ can be determined on a jet-by-jet basis, allowing pruning to be more adaptive than a fixed parameter procedure.  $D_{\cut}$ essentially determines how much of the jet substructure can be pruned, with smaller values allowing for more pruning.  $D_\cut$ should be sufficiently small so that if a decay is ``hidden'' inside the jet substructure by late recombinations of, say, UE particles, the substructure can be pruned and the decay can be found.  A value that is too small, however, will result in over-pruning.  A natural scale for $D_{\cut}$ is the opening angle of the jet.  However, this is an infrared unsafe quantity, as soft radiation can change the opening angle.  Instead, the dimensionless ratio $m_J/p_{T_J}$ for the jet is related to the opening angle: typically, $\Delta R_{12} \approx 2m_J/p_{T_J}$.  Therefore, we choose $D_{\cut}$ to scale with $2m_J/p_{T_J}$, and a value $D_{\cut} = m_J/p_{T_J}$ is a reasonable starting value.  We will study the performance of pruning as a function of the scaling of $D_{\cut}$ with $2m_J/p_{T_J}$.

\subsection{Effects of Pruning}

\begin{figure}[htbp]
\subfloat[$m_J$ cut, $z$]{\includegraphics[width = 0.23\textwidth]{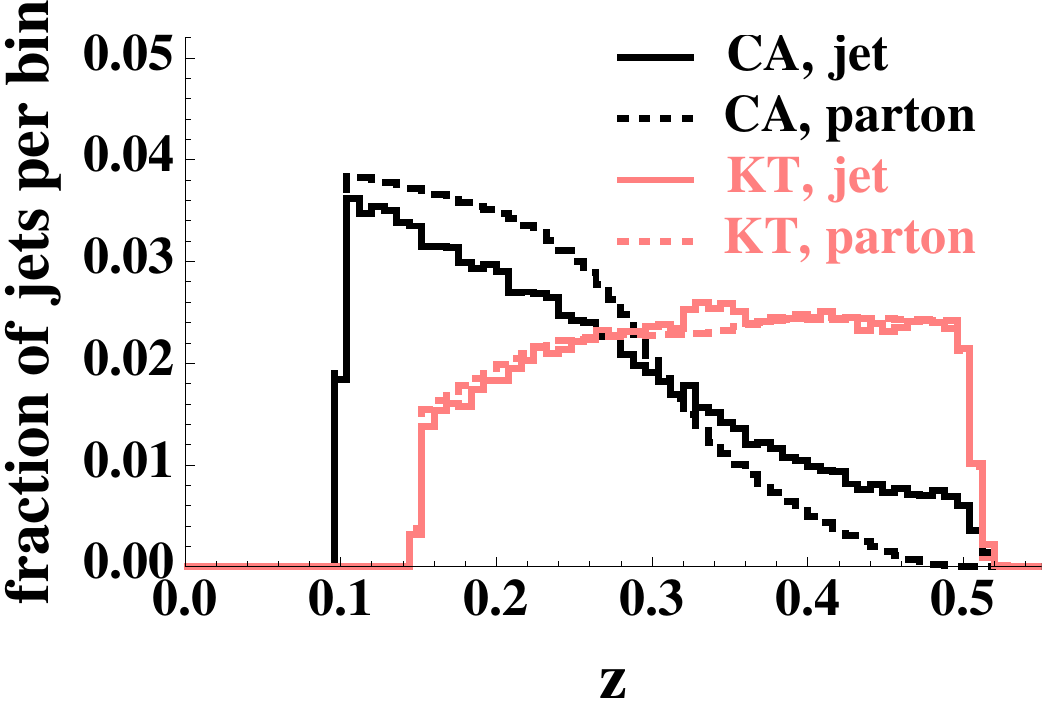}}
\subfloat[$m_J$ cut, $\Delta R_{12}$]{\includegraphics[width = 0.23\textwidth]{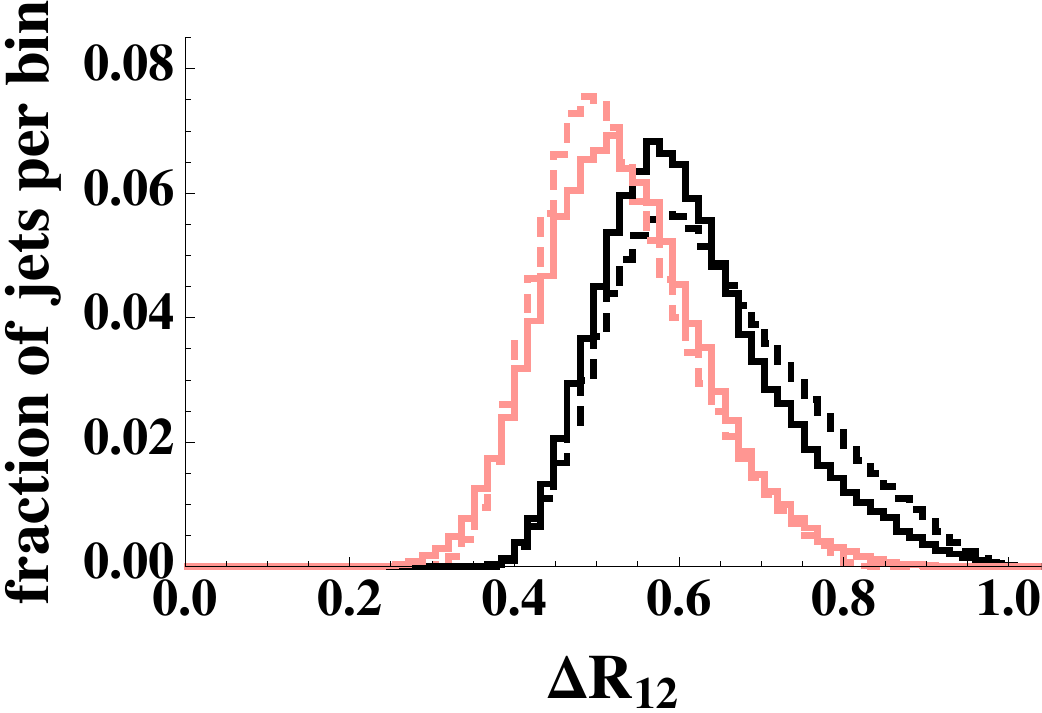}}

\subfloat[$m_J$ and $m_{\text{Sub}J}$ cuts, $z$] {\includegraphics[width = 0.23\textwidth] {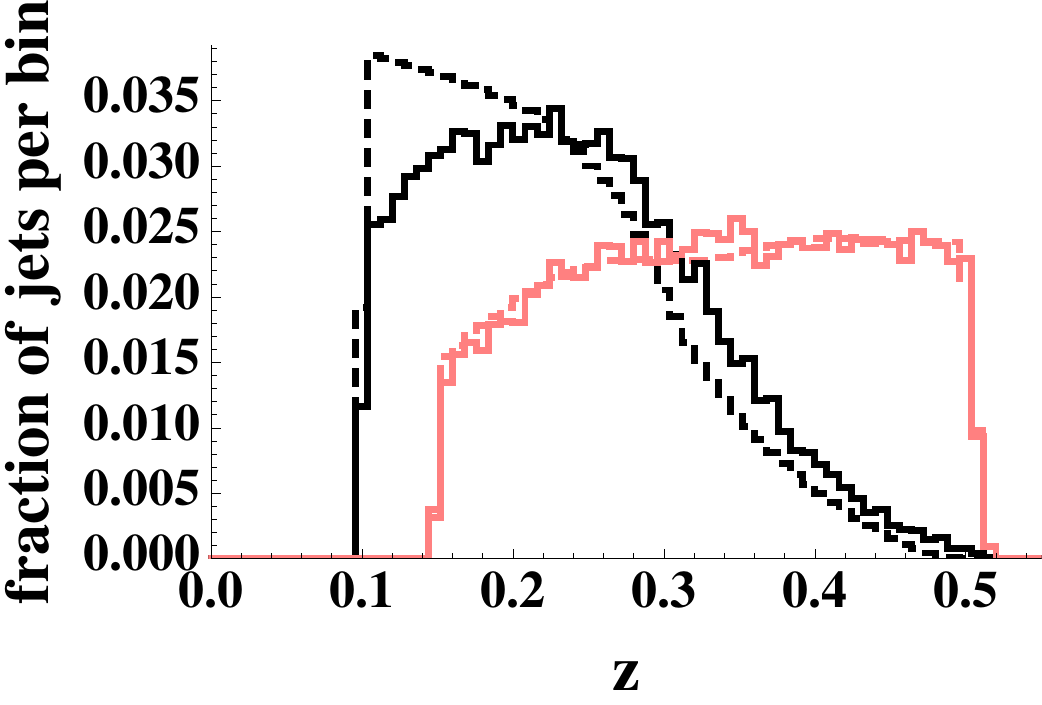}}
\subfloat[$m_J$ and $m_{\text{Sub}J}$ cuts, $\Delta R_{12}$] {\includegraphics[width =
0.23\textwidth] {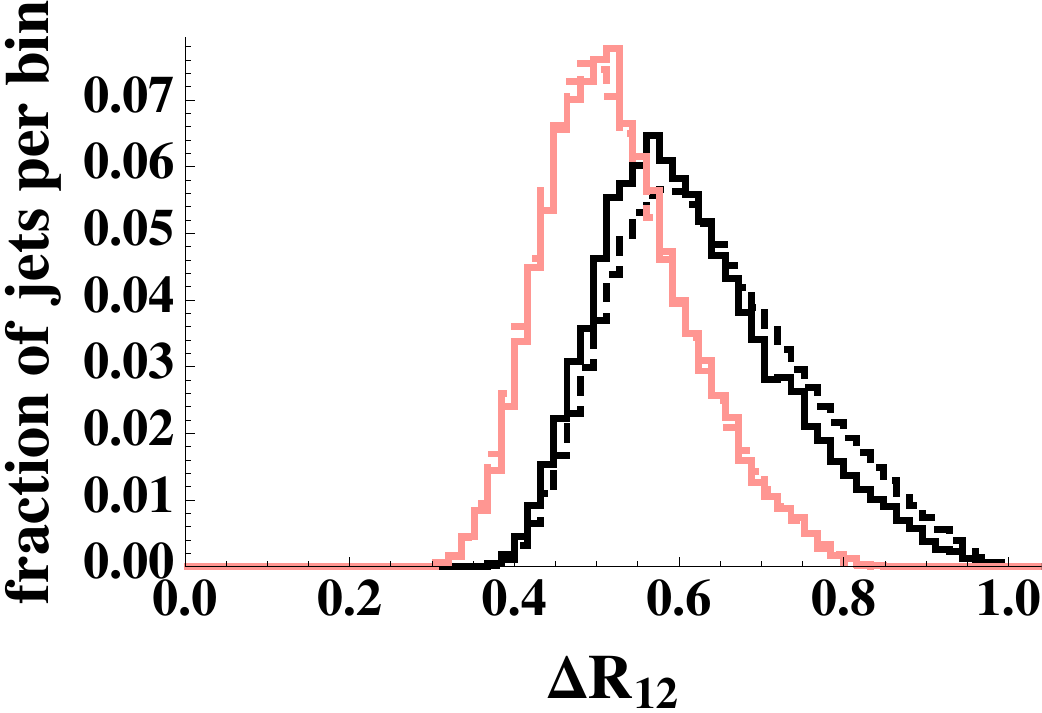}}

\caption{Distributions in $z$ and $\Delta R_{12}$ comparing for top quark decays at the parton-level and from Monte Carlo events after implementing pruning.  This figure uses the same samples and cuts as Fig.~\ref{topPartonVsReconZDR}. }%\red{topPartonVsReconZDRpruned}}
\label{topPartonVsReconZDRpruned}
\end{figure}

Having defined the pruning procedure, we can demonstrate how effective it is in reducing systematic effects and improving the mass resolution of jets.  In this study, we use the parameters $D_{\cut} = m_J/p_{T_J}$ for both algorithms, and $z_{\cut} = 0.10$ for the CA algorithm and 0.15 for the $\kt$ algorithm.  We will motivate these parameters with the study in Sec.~\ref{sec:results:params}.  First, in Fig.~\ref{topPartonVsReconZDRpruned}, we reproduce the ``hadron-parton'' comparison in Fig.~\ref{topPartonVsReconZDR} from Sec.~\ref{sec:algEffects}, using pruning at both the hadron level and the parton level.  The parton-level pruning is implemented in the same way as defined above, treating the three partons of the reconstructed top quark as the jet.

It is clear by comparing Figs.~\ref{topPartonVsReconZDR} and \ref{topPartonVsReconZDRpruned} that pruning has removed much of the systematic effects in the CA algorithm; when only a jet mass cut is made, the distribution in $z$ and $\Delta R_{12}$ for pruned jets match the parton-level distribution much better than unpruned jets.  When both mass and subjet mass cuts are made, pruning shows a slightly poorer agreement to the parton-level kinematics than the unpruned case.  This arises from the fact that the value of $z_{\cut}$ is fixed, while the distribution in $z$ is dependent on the kinematics of the decay.

\begin{figure}[htbp]
\subfloat[unpruned QCD jets] {\includegraphics[width=0.23\textwidth] {mQCDUEcompare_mjet_CA.pdf}}
\subfloat[pruned QCD jets] {\includegraphics[width=0.23\textwidth] {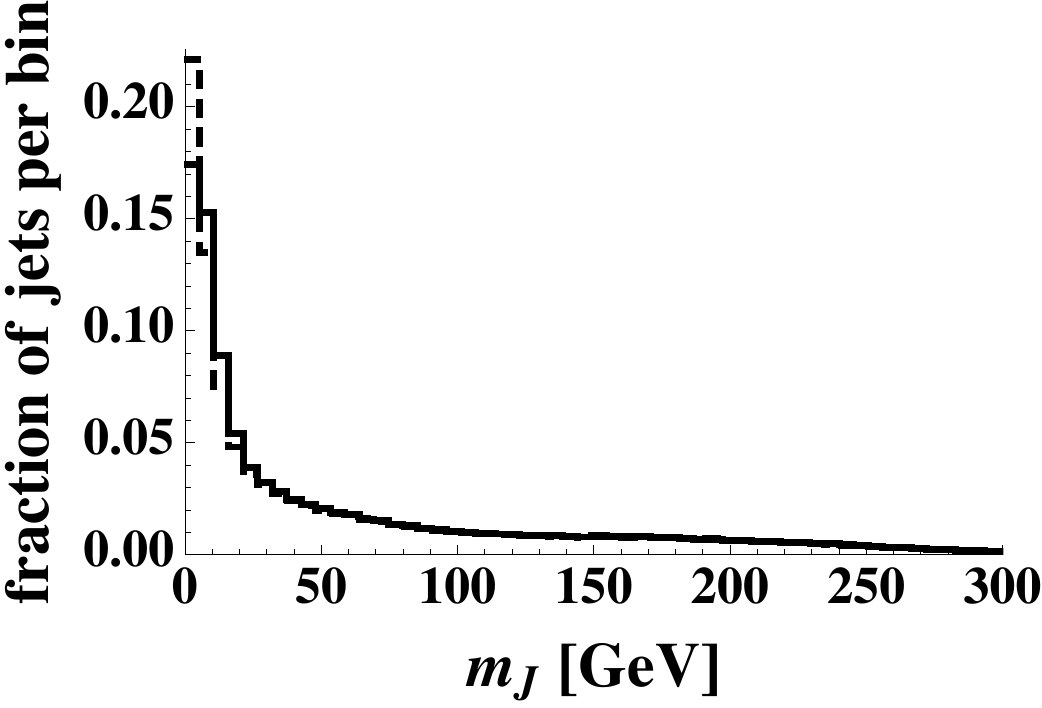}}

\subfloat[unpruned top jets] {\includegraphics[width=0.23\textwidth] {ttbarUEcompare_mjet_CA.pdf}}
\subfloat[pruned top jets] {\includegraphics[width=0.23\textwidth] {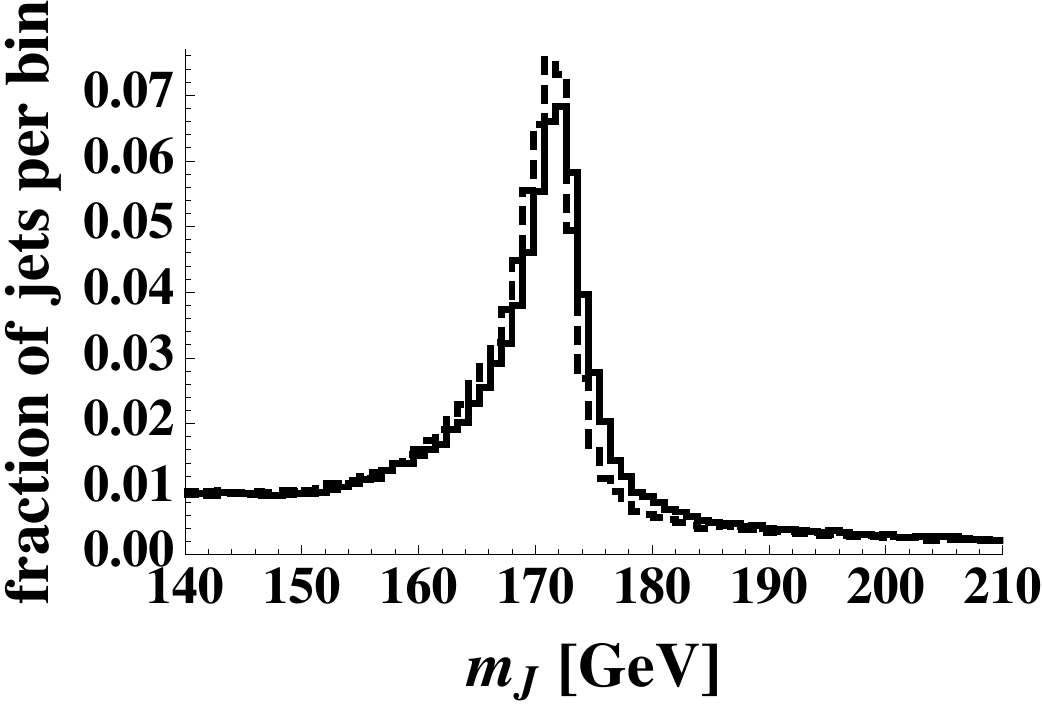}}
\caption{Distributions in $m_J$ with and without underlying event, for QCD and top jets, using the CA algorithm, with and without pruning.  The jets have $p_T$ between 500 and 700 GeV, and D = 1.0.  }%\red{PrunedUEcompMCA}}
\label{PrunedUEcompMCA}
\end{figure}

\begin{figure}[htbp]
\subfloat[unpruned QCD jets] {\includegraphics[width=0.23\textwidth] {mQCDUEcompare_mjet_KT.pdf}}
\subfloat[pruned QCD jets] {\includegraphics[width=0.23\textwidth] {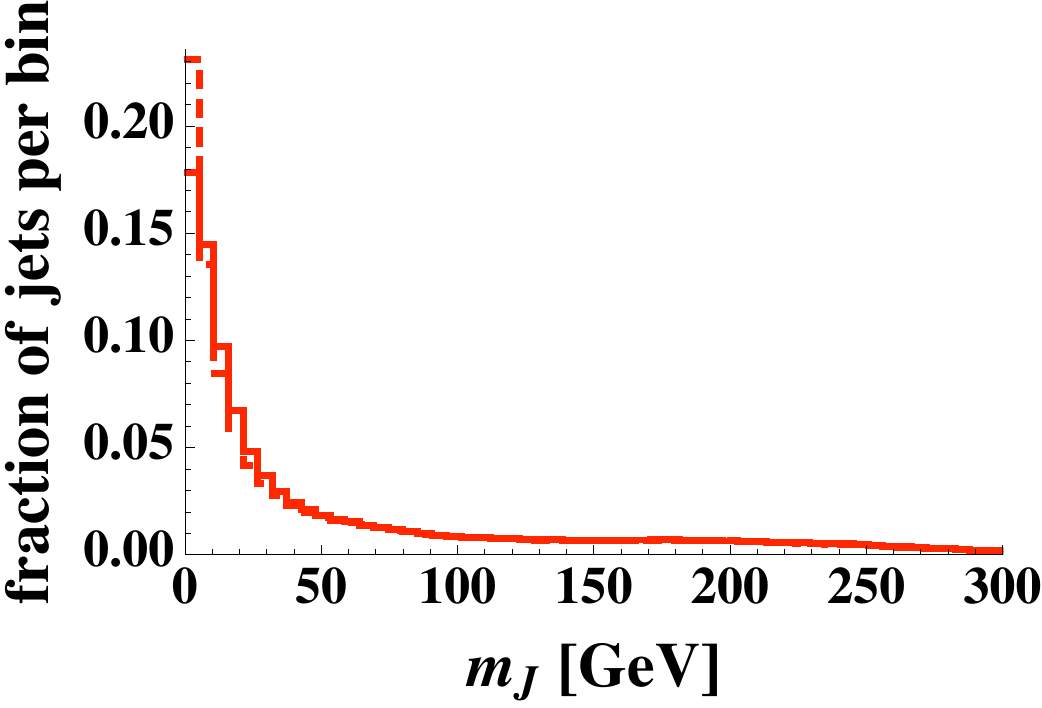}}

\subfloat[unpruned top jets] {\includegraphics[width=0.23\textwidth] {ttbarUEcompare_mjet_KT.pdf}}
\subfloat[pruned top jets] {\includegraphics[width=0.23\textwidth] {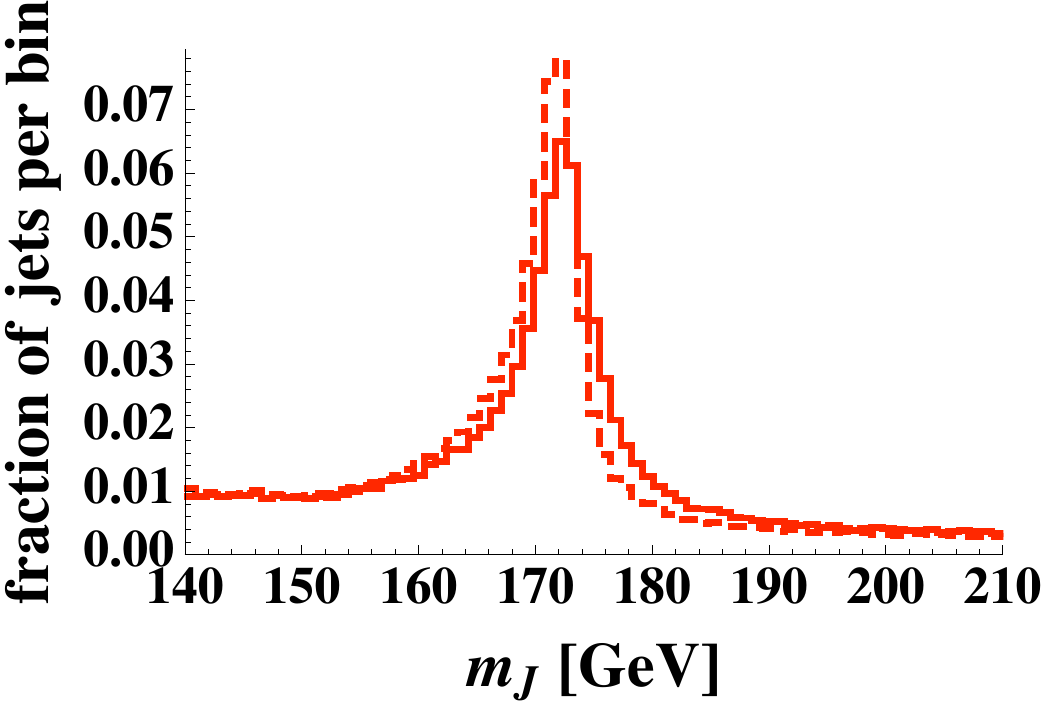}}
\caption{Distributions in $m_J$ with and without underlying event, for QCD and top jets, using the $\kt$ algorithm, with and without pruning.  The jets have $p_T$ between 500 and 700 GeV, and D = 1.0.  }%\red{PrunedUEcompMKT}}
\label{PrunedUEcompMKT}
\end{figure}

\begin{figure}[htbp]
\subfloat[unpruned QCD jets] {\includegraphics[width=0.23\textwidth] {mQCDUEcompare_z_CAKT.pdf}}
\subfloat[pruned QCD jets] {\includegraphics[width=0.23\textwidth] {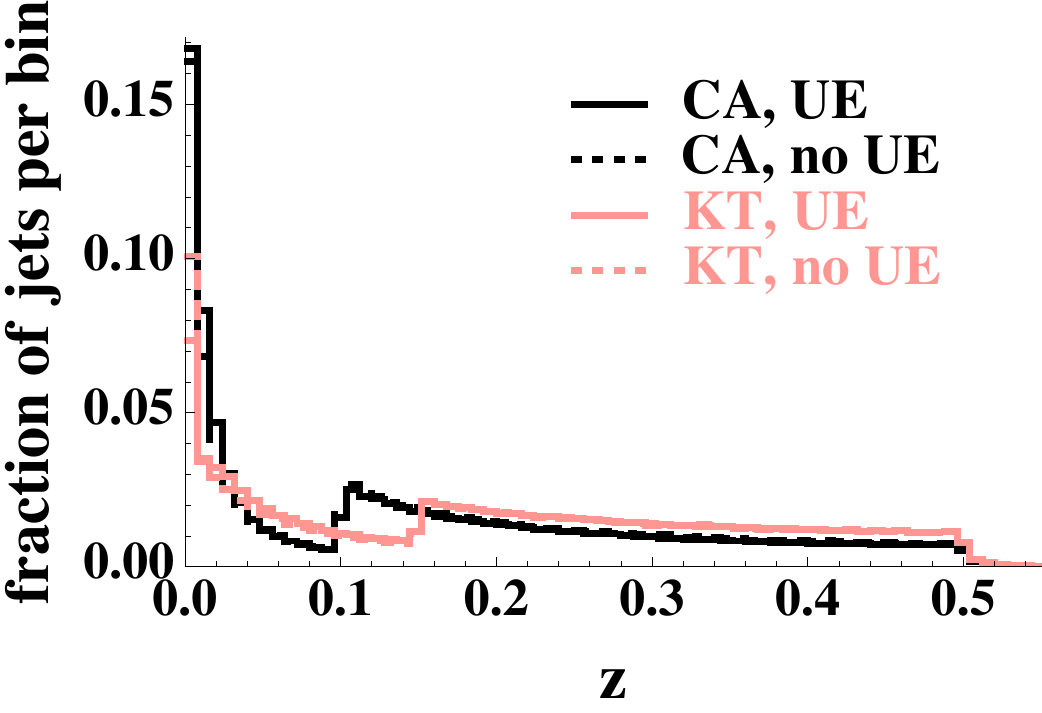}}

\subfloat[unpruned top jets] {\includegraphics[width=0.23\textwidth] {ttbarUEcompare_z_CAKT.pdf}}
\subfloat[pruned top jets] {\includegraphics[width=0.23\textwidth] {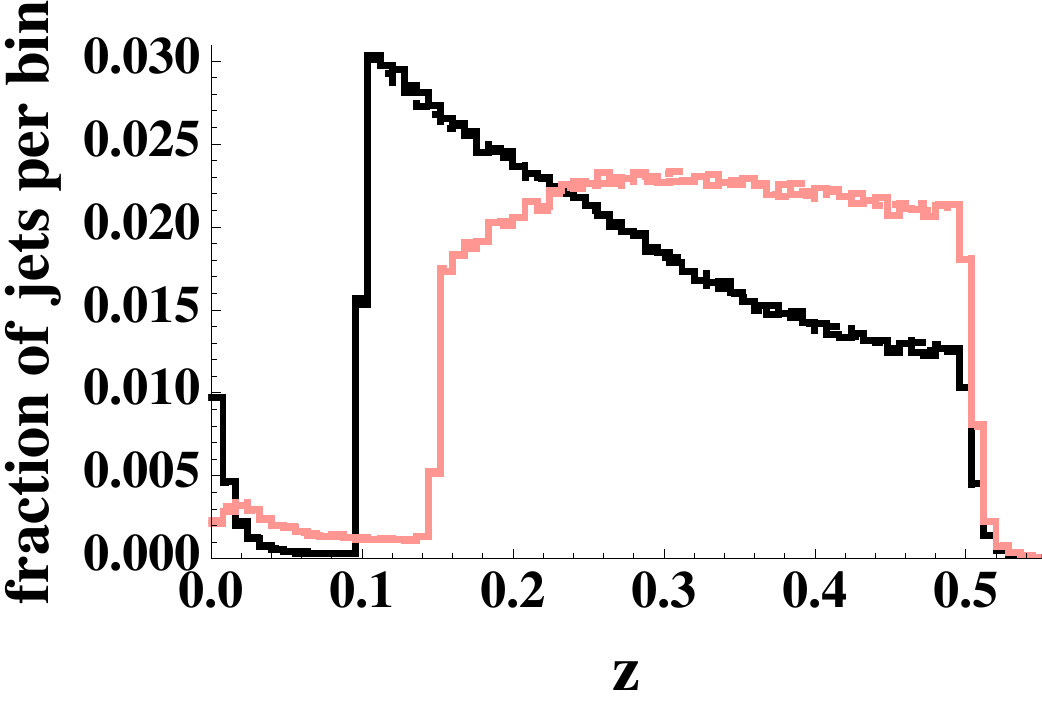}}
\caption{Distribution in $z$ with and without underlying event, for QCD and top jets, using the CA and $\kt$ algorithms, with and without pruning.  The legends for plots (c) and (d) correspond to (a) and (b), respectively.  The jets have $p_T$ between 500 and 700 GeV, and D = 1.0.  }%\red{PrunedUEcompZ}}
\label{PrunedUEcompZ}
\end{figure}

\begin{figure}[htbp]
\subfloat[unpruned QCD jets] {\includegraphics[width=0.23\textwidth] {mQCDUEcompare_theta_CAKT.pdf}}
\subfloat[pruned QCD jets] {\includegraphics[width=0.23\textwidth] {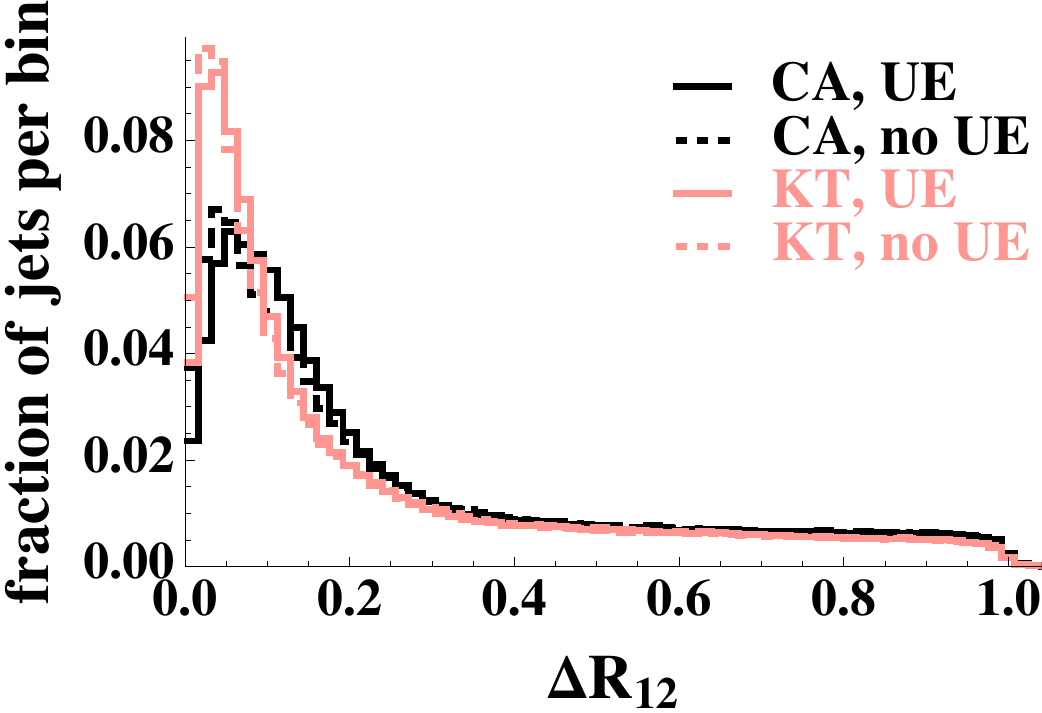}}

\subfloat[unpruned top jets] {\includegraphics[width=0.23\textwidth] {ttbarUEcompare_theta_CAKT.pdf}}
\subfloat[pruned top jets] {\includegraphics[width=0.23\textwidth] {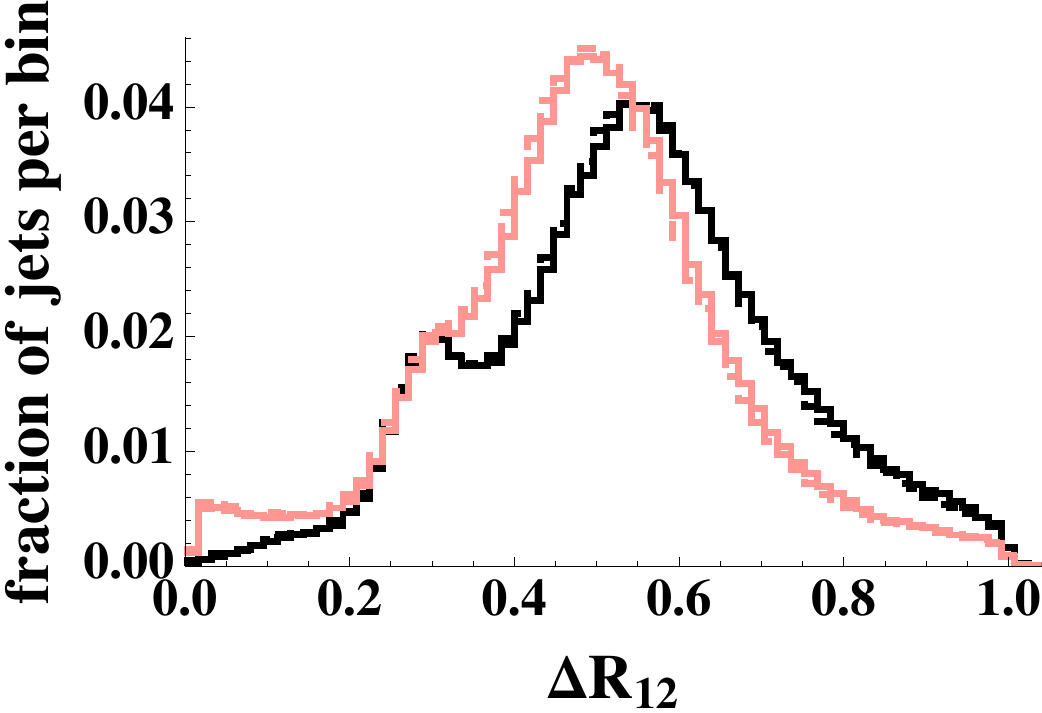}}
\caption{Distribution in $\Delta R_{12}$ with and without underlying event, for QCD and top jets, using the CA and $\kt$ algorithms, with and without pruning.  The jets have $p_T$ between 500 and 700 GeV, and D = 1.0.  }%\red{PrunedUEcompDR}}
\label{PrunedUEcompDR}
\end{figure}

In addition to improving the kinematics of the jet substructure, pruning reduces the contribution of the underlying event and improves the mass resolution of reconstructed decays.  In Figs.~\ref{PrunedUEcompMCA} and \ref{PrunedUEcompMKT} we give the mass distribution of jets with and without the UE in both the QCD and $t\bar{t}$ samples for the CA and $\kt$ algorithms, but now with and without pruning.  In Figs.~\ref{PrunedUEcompZ} and \ref{PrunedUEcompDR} we show how the effect of UE on distributions in $z$ and $\Delta R_{12}$, also with and without pruning.


Three distinctions between pruned and unpruned jets are clear.  First, the distributions with and without the UE are very similar for pruned jets, while they noticeably differ for unpruned jets.  This shows that pruning has drastically reduced the contribution of the underlying event.  Second, the mass peak of jets near the top quark mass in the $t\bar{t}$ sample is significantly narrowed by the introduction of pruning (especially when the UE is included).  This is evidence of the improved mass resolution of pruning, and will contribute to the improvement in heavy particle identification with pruning.  And finally, the mass distribution of QCD jets is pushed significantly downward by pruning.  The QCD jet mass is dominantly built from the soft, large-angle recombinations --- most recombinations are soft, and for fixed $p_T$, larger-angle recombinations contribute more to the jet mass.  Removing these by pruning the jets reduces the QCD mass distribution in the large mass range and will contribute to the reduction of the QCD background.

We move on to examine pruning through a set of studies using Monte Carlo simulated events.  We will investigate the parameter dependence of pruning, motivating the parameters used above.  We will extensively study both top and $W$ reconstruction with pruning, and quantify the improvements of pruning in terms of basic statistical measures.  These studies will provide evidence of the insensitivity of pruning to the value of $D$ in the jet algorithm.

%%%%%%%%%% END OF SECTION 6: PRUNING ALGORITHMS %%%%%%%%%%

\section{Monte Carlo Studies}
\label{sec:MC}

\subsection{Study Layout}

The parameter space for questions about pruning procedures is very large.  In this work, we want to ask whether pruning is a viable data analysis tool, and how effective it can be.  We use Monte Carlo samples to study $W$ reconstruction and the rejection of $W$ + jets backgrounds, as well as top quark reconstruction and the rejection of QCD multijet backgrounds.  To test the usefulness of pruning across a range of jet $m/p_T$, and hence the heavy particle boost, we study both signals in four $p_T$ bins.  We will also be able to compare a signal with a single mass scale (the $W$) to one with two (the top).  The details of the Monte Carlo samples and their generation are described in Appendix~\ref{sec:appendix}.

In the following sections, we define a particular method to identify the heavy particles using jet substructure, and examine pruning in this context.  In this work, we are more concerned with the \emph{improvements} provided by pruning than its absolute performance.  Therefore, we compare pruning to an analysis procedure where the jets are left unpruned.  This comparison removes dependence on quantities that have large uncertainties, such as signal and background cross sections, or are not specified, such as the integrated luminosity.  Instead, the performance of pruning is quantified in terms of key measures --- how much \emph{better} pruning resolves the physically relevant substructure of the jet and separates signal and background processes than using the substructure from unpruned jets.

Additionally, we test the performance of pruning as parameters of the jet algorithm and the pruning procedure are varied.  The performance will change with the parameter $D$, since it controls how boosted the decay must be to be reconstructed in a single jet.  We expect the $D$ dependence to be closely correlated with the jet $p_T$, as it is a direct measure of the boost of the heavy particle.  We also test the sensitivity of the pruning procedure to the parameters $z_{\cut}$ and $D_{\cut}$.  We aim to draw some basic conclusions about how pruning should be applied in a search.

\subsection{Measures used to quantify pruning}
\label{sec:MC:Metrics}

Mass variables are by far the strongest discriminator between QCD jets and jets reconstructing heavy particle decays.  QCD jets have a smooth mass distribution set by the jet $p_T$ (see Sec.~\ref{sec:QCDJets}), while a decaying particle can have multiple intrinsic mass scales.  This allows us to define simple criteria to identify a jet as coming from a top quark: if the jet mass is in the top mass window and one of the two subjets has a mass in the $W$ mass window, then we tag the jet as a \emph{top jet}.  The top and $W$ mass windows are defined by fitting the relevant mass peaks of the signal sample, which we describe in detail below.  The $W$ study proceeds analogously with only a jet mass cut.  In a real search for a particle of unknown mass, one obviously cannot fit a ``signal sample''.  However, we employ this method to demonstrate two effects of pruning: sharpening the signal mass peak and reducing the QCD background in this region.  These two effects will determine how well pruning improves our ability to find bumps in jet mass distributions.

Jet algorithms can be compared using a variety of measures depending on how the algorithm is used.  Our focus is on heavy particle identification and separation from background.  In particular, we compare analyses performed with and without pruning to quantify the improvement that pruning provides.  We use a common set of variables to measure the relative difference between a jet algorithm and its pruned version.  Let $N_{\text{\tiny{S}}}(A)$ be the number of jets in the signal sample identified as a reconstructed heavy particle for algorithm $A$, and $N_{\text{\tiny{B}}}(A)$ the analogous number of jets in the background sample.  Use $pA$ to denote the pruning procedure run on jets found with algorithm $A$.  Then the variables we use are:
\be
\begin{split}
\epsilon &= \frac{N_{\text{\tiny{S}}}(pA)}{N_{\text{\tiny{S}}}(A)} , \\
R &= \frac{N_{\text{\tiny{S}}}(pA) / N_{\text{\tiny{B}}}(pA)}{N_{\text{\tiny{S}}}(A) / N_{\text{\tiny{B}}}(A)}, ~\text{and}\\
S &= \frac{N_{\text{\tiny{S}}}(pA) / \sqrt{N_{\text{\tiny{B}}}(pA)}}{N_{\text{\tiny{S}}}(A) / \sqrt{N_{\text{\tiny{B}}}(A)}} .
\end{split}
\ee
$\epsilon$ is the relative efficiency of pruning in identifying heavy particles in the signal sample, while $R$ and $S$ are the relative signal-to-background and signal-to-noise ratios for the pruned and unpruned algorithms.  We also evaluate the relative mass window widths, which we label $w_\text{rel}$.  For the $W$ study, this is the ratio of the $W$ mass window width for pruning relative to not pruning; for the top study it is the ratio in the top mass window width.  Note that in the top study, a $W$ subjet mass cut is also used.  A value of $w_\text{rel} < 1$ means pruning has improved the mass resolution of the jets.  These ratios are independent of the integrated luminosity and the total cross sections, and are representative of the improvements that pruning would provide in an analysis.

\begin{figure}[htbp]
\includegraphics[width=\columnwidth]{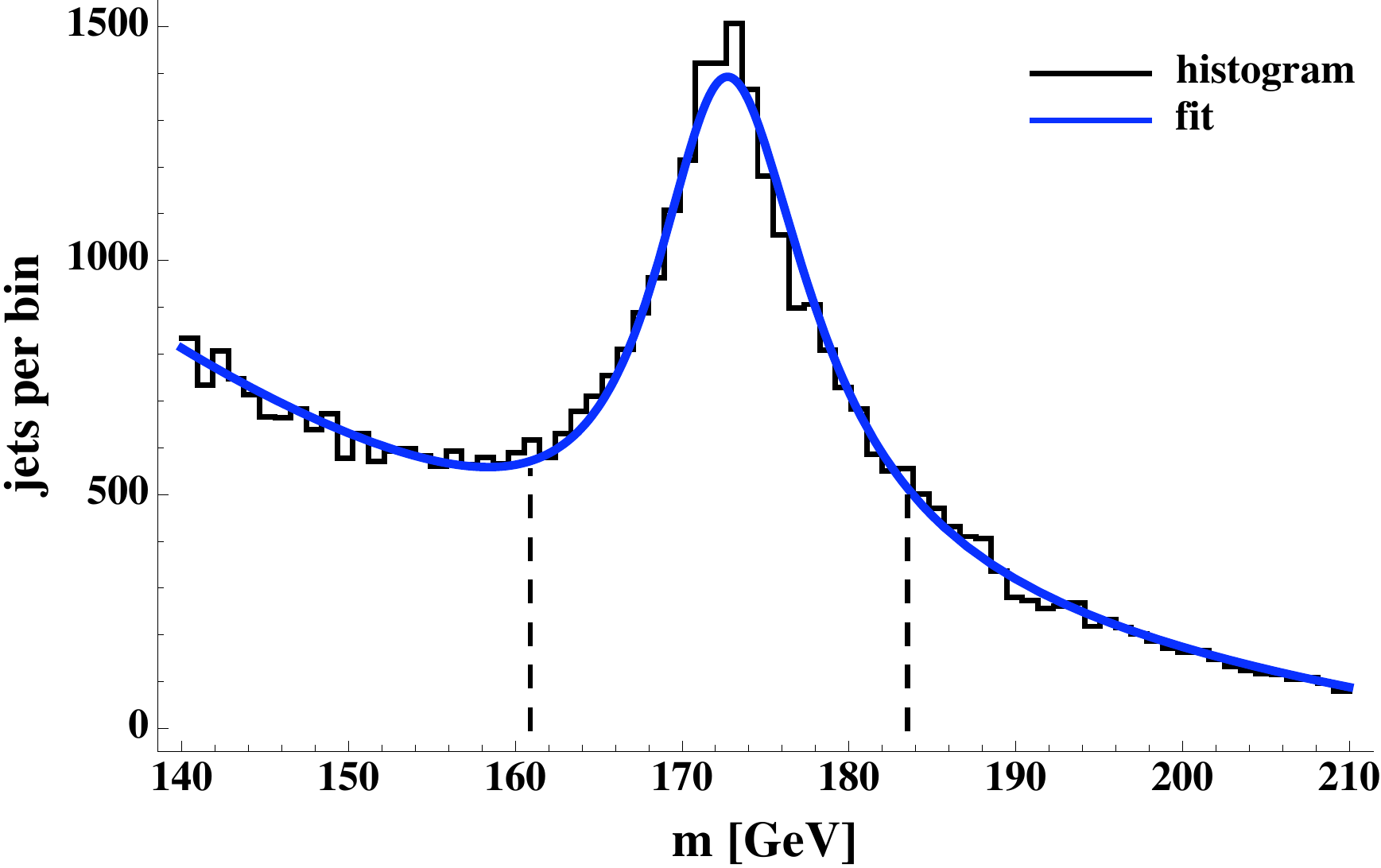}
\caption{A sample fit showing the jet mass distribution (black histogram) and sample fit (blue curve) for CA jets from $t\bar{t}$ events. }%\red{SampleBWfit}}
\label{SampleBWfit}
\end{figure}

To determine the mass window for a particular signal sample, we fit the mass peak to determine the window width.  In these studies, a skewed Breit-Wigner is sufficient to fit the peak, with a power law continuum background.  These functions used to fit mass peaks are:
\be
\begin{split}
\text{peak: } f(m) &= \frac{M^2\Gamma^2}{(m^2-M^2)^2 + M^2\Gamma^2}\left(a + b(m-M)\right) ; \\
\text{continuum: } g(m) &= \frac{c}{m} + \frac{d}{m^2} .
\end{split}
\ee
$M$ is the location of the mass peak; $\Gamma$ is the width of the peak.  A sample fit it shown in Fig.~\ref{SampleBWfit}.

  The mass window $[M-\Gamma,M+\Gamma]$ is found to be nearly optimal, given this functional form, in measures similar to $\epsilon$, $R$, and $S$: the area in the window ($\sim\epsilon$), the ratio of area to the window width ($\sim R$), and the ratio of area to the square root of the width ($\sim S$).  In the next section, we will use these statistics to quantify the improvements gained by pruning in identifying heavy particles and separating them from backgrounds, and explore the advantages that can be achieved by pruning.

%%%%%%%%%% END OF SECTION 7: MONTE CARLO STUDIES %%%%%%%%%%

\section{Study Results}
\label{sec:results}

In this section we present results comparing analyses with pruned jets to unpruned jets.  We demonstrate two main points: first, pruning is useful and broadly applicable, and second, its parameters do not need fine tuning for it to provide significant improvement.

The natural starting point is to investigate the parameters particular to the pruning procedure, $D_{\cut}$ and $z_{\cut}$.  The most important question is whether these need to be tuned to the signal.  To answer this, in Sec.~\ref{sec:results:params} we study the performance of pruning as we vary its parameters for two different signals across the full $p_T$ range for the samples.  We find that optimal choices of $z_\cut$ and $D_\cut$ vary slowly with $m/p_T$, but that our choice of parameters is not far from optimal in all cases.

After fixing $z_\cut$ and $D_\cut$, we consider the effect of varying $D$ in the jet algorithm.  In Sec.~\ref{sec:results:fixedD} we study pruning with $D$ fixed at 1.0 over all $p_T$ bins.  This type of analysis is like a search where the mass (and hence $m/p_T$) of the new heavy particle is not known.  For comparison, in Sec.~\ref{sec:results:varD} we redo the analysis, but with $D$ adjusted for each bin to fit the expected angular size of the decay in that bin.  In this case, the unpruned jet algorithm performs better than with a constant $D$, as expected, but pruning still shows improvements in finding $W$'s and tops.  In all cases, pruned jets are a better way to identify heavy particles than unpruned.  In Sec.~\ref{sec:results:compD} we compare the results of Secs.~\ref{sec:results:fixedD} and \ref{sec:results:varD}.  Significantly, if jets are pruned, we find that it does not make much difference what the initial $D$ value was, indicating that searches with large fixed $D$ do not suffer in power compared to searches with $D$ tuned to known or suspected $m/p_T$.

In Sec.~\ref{sec:results:absolute} we give some absolute measures of top-finding with pruned jets for comparison to other methods.  In Sec.~\ref{sec:results:algComparison} we directly compare the CA and $\kt$ algorithms, before and after pruning.  Finally, in Sec.~\ref{sec:results:smearing} we consider the effect of a crude detector model where we smear the energies of all particles in the calorimeter.  We find that the performance of the pruned and unpruned algorithms are degraded, but that pruning still provides significant improvement.

\subsection{Dependence on Pruning Parameters}
\label{sec:results:params}

The pruning procedure we have defined has two free parameters (in addition to those of the jet algorithms themselves).  In introducing the procedure, we argued that $z_{\cut} = 0.10$ and $D_{\cut} = m_J/p_{T_J}$ were sensible choices.  In this subsection we will investigate how pruning performs when each of these parameters is varied while the other is held fixed, for both ($W$ and top) signals and across the four $p_T$ bins for each signal.

We will look at the values of the metrics $w_\text{rel}$, $\epsilon$, $R$, and $S$ defined in Sec.~\ref{sec:sec:MC:Metrics}.  The priority in choosing particular values for $z_{\cut}$ and $D_{\cut}$ should be in optimizing $S$, as it is the criterion for discovery.  That being said, $\epsilon$ and $R$ are still important measures as they determine the total size of the signal and remaining fraction relative to the background.  As we will see, the dependence of $\epsilon$ and $R$ on the parameters is not strong.  We also evaluate $w_\text{rel}$ because the mass window width drives the other three metrics.  As the relative width decreases, in general the measures $R$ and $S$ will increase because the heavy particle is better resolved and more of the background is rejected, but $\epsilon$ will decrease simply because the narrower width selects fewer signal jets.

\begin{figure*}[htbp]
\subfloat[$W$'s, CA jets]{\label{VaryZcut:WCA}\includegraphics[width=0.20\textwidth]{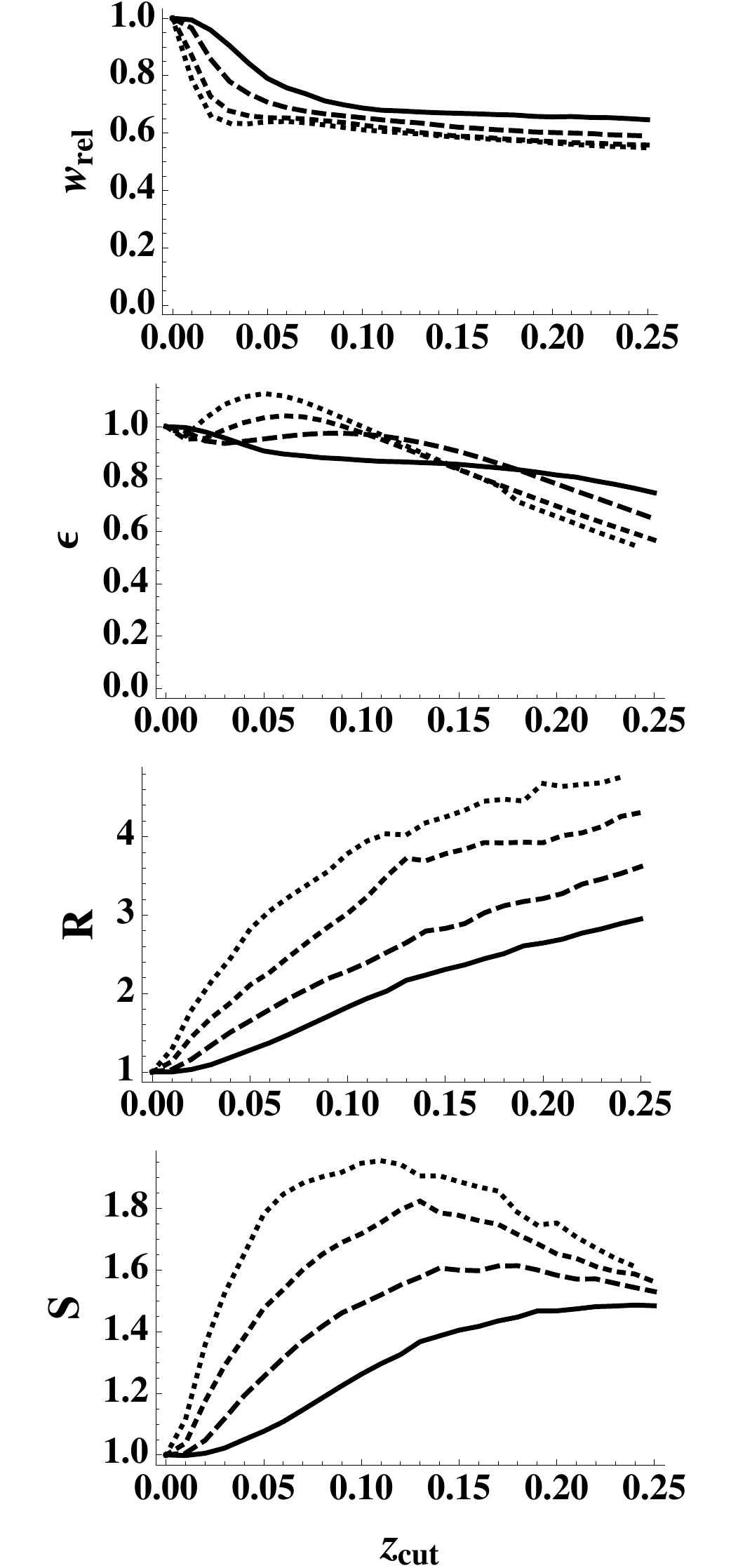}}
\subfloat[tops, CA jets]{\label{VaryZcut:tCA}\includegraphics[width=0.20\textwidth]{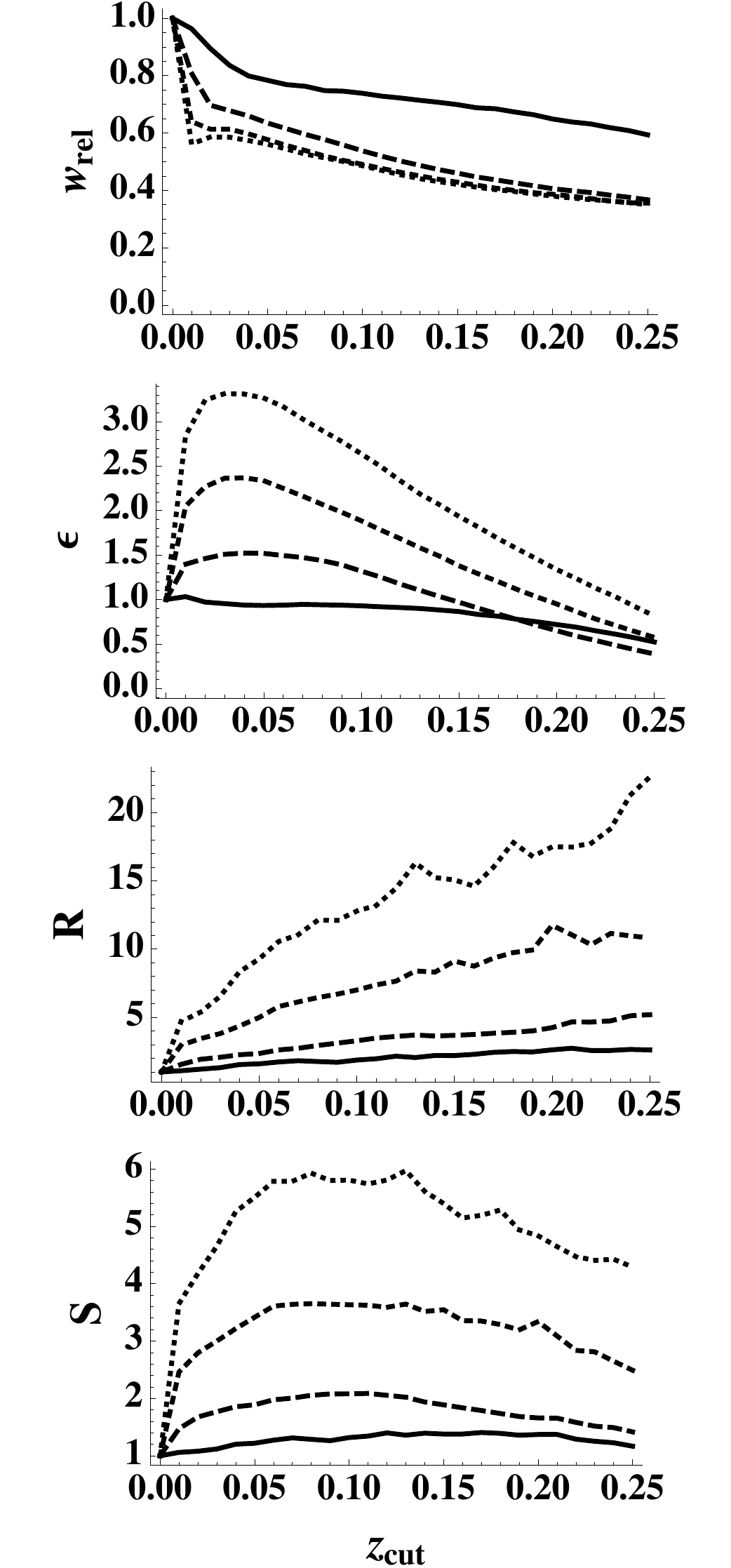}}
\subfloat[$W$'s, $\kt$ jets]{\label{VaryZcut:WkT}\includegraphics[width=0.20\textwidth]{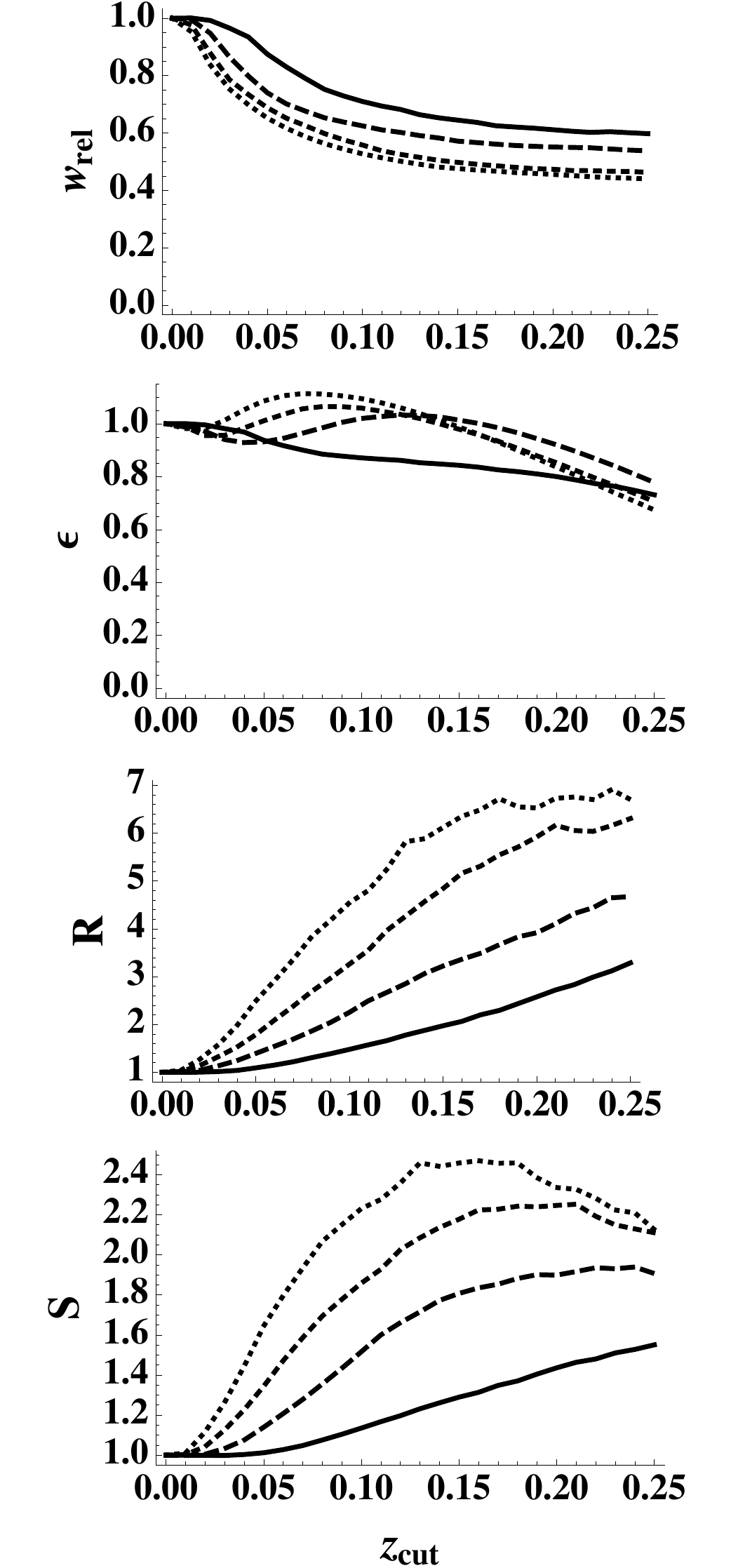}}
\subfloat[tops, $\kt$ jets]{\label{VaryZcut:tkT}\includegraphics[width=0.20\textwidth]{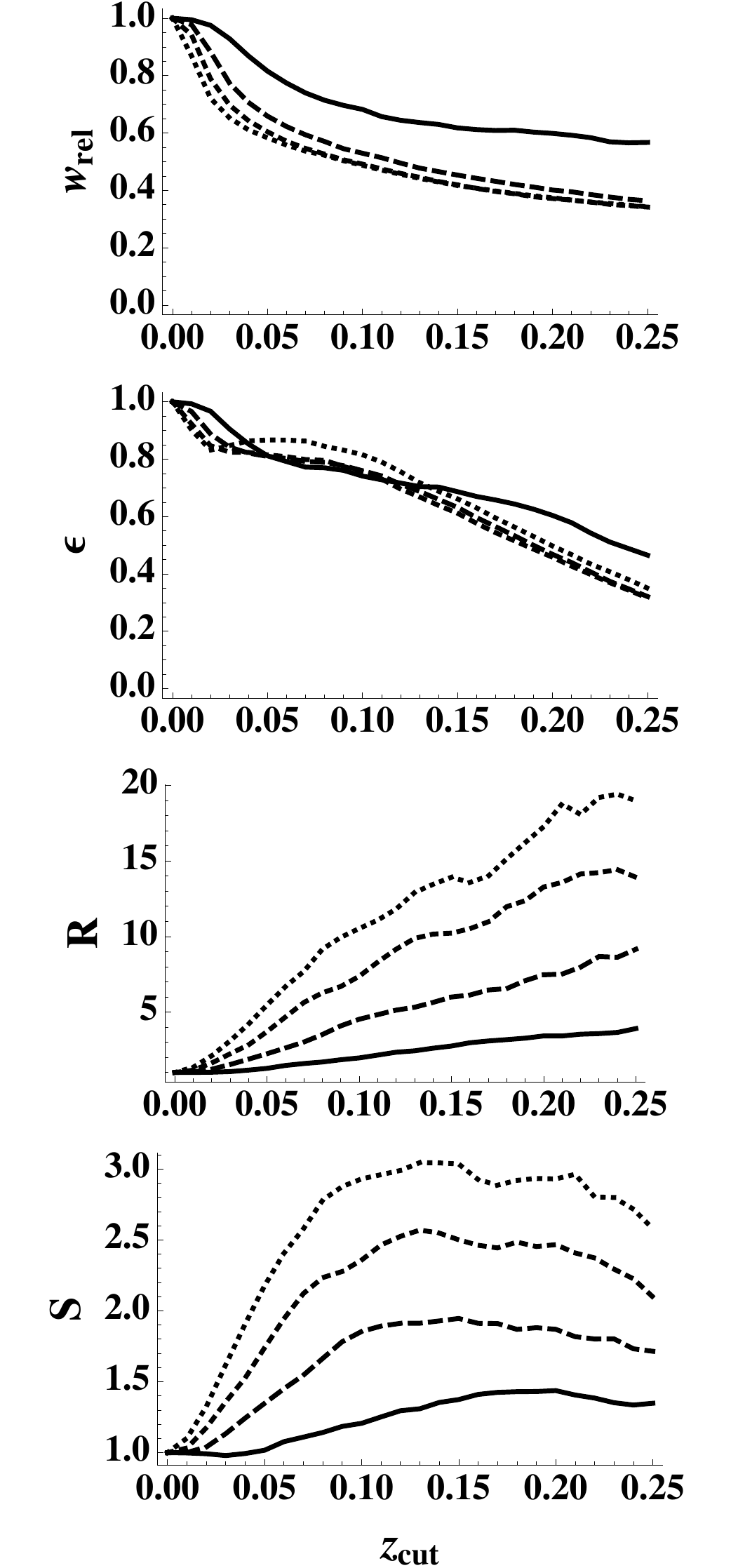}}
\includegraphics[width=0.10\textwidth]{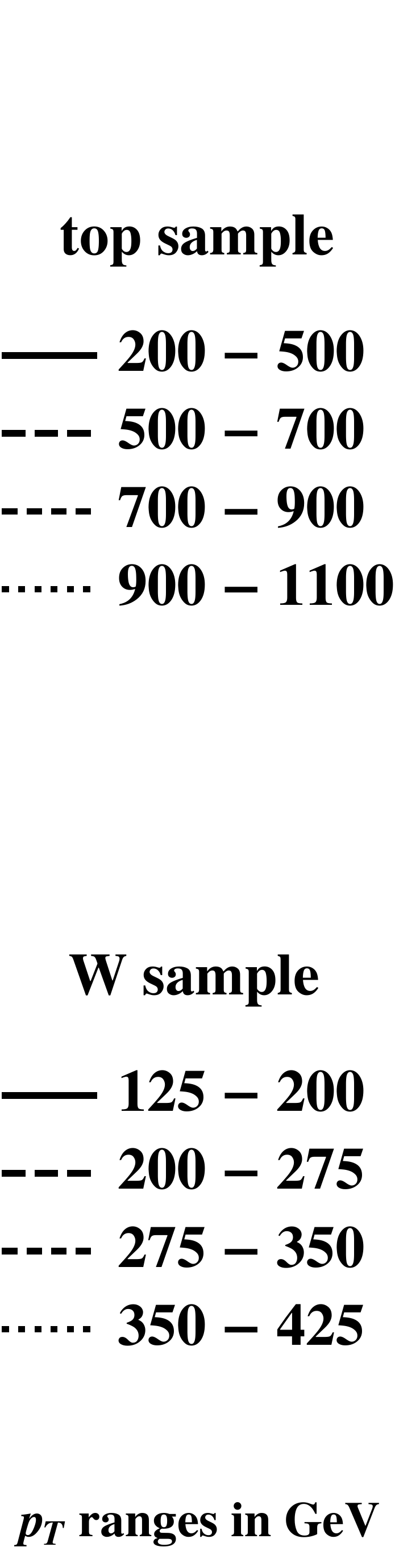}
\caption{Relative statistical measures $w_\text{rel}$, $\epsilon$, $R$, and $S$ vs. $z_{\cut}$ for $W$'s and tops, using CA and $\kt$ jets.  Four $p_T$ bins are shown for each sample.  Statistical errors (not shown) are $\mathcal{O}(1\%)$ for $w_\text{rel}$ and $\epsilon$, and $\mathcal{O}(10\%)$ for $R$ and $S$.  }%\red{VaryZcut}}
\label{VaryZcut}
\end{figure*}

In Fig.~\ref{VaryZcut}, we show all four metrics for top and $W$ jets, for both CA and $\kt$ jets.  $D_\cut$ is set to $m_J/p_{T_J}$ throughout, and $z_\cut$ is varied in [0, 0.25].  $z_\cut = 0$ represents no pruning and we can see that all metrics are 1 here.  With increasing pruning, the mass window width initially decreases rapidly, then levels out.  In all but the smallest $p_T$ bin, the relative signal efficiency $\epsilon$ increases as the width narrows, suggesting that signal jets that had ``vacuumed up'' too much UE or soft radiation are being pruned back into the mass window.  Note that for the top quark sample with the $\kt$ algorithm, $\epsilon$ merely flattens out for a range in $z_\cut$, and does not increase as it does for the other samples.  Once the window stops shrinking significantly (around $z_\cut = 0.05$), the relative signal efficiency starts decreasing; now the dominant effect is over-pruning signal jets out of the mass window.  Note, however, that even though the relative signal efficiency is \emph{decreasing}, the relative signal-to-background ratio $R$ is \emph{increasing} over the full range.  So even as signal jets are being removed from the mass window, background jets are being removed even faster.  If we look at signal-to-noise, $S$, there appears to be a broad optimal range in $z_{\cut}$ that depends somewhat on the signal, on the $p_T$ bin and on the jet algorithm.

There are two important lessons to be learned from these plots.  First, more pruning is required for $\kt$ jets than for CA to achieve similar results.  The right two columns ($\kt$) are similar to the left two (CA) except that features are shifted out in $z_\cut$.  Second, the peak in $S$ does not depend strongly on the signal or the $p_T$, in the three largest $p_T$ bins.  The dependence on $S$ in the smallest $p_T$ bin, however, is different from the others due to threshold effects of the heavy particle being reconstructed in a single jet.  In the smallest $p_T$ bin, the boosts of the $W$'s or tops are small enough that many decays are just at the threshold for being reconstructed.  Decays at the reconstruction threshold typically have poor mass resolution, and cutting more aggressively on $z$ reduces these threshold effects and significantly decreases the background, leading to an increase in $S$ over the whole range in $z_{\cut}$.  For CA, our ``reasonable choice'' of $z_\cut$ of 0.1 looks close to optimal for the upper three bins, and not far off for the smallest.  For $\kt$, a larger $z_\cut$ is needed; 0.15 is close to optimal.

\begin{figure}[htbp]
\subfloat[$p_T$ bin 1, 200--500 GeV]{\includegraphics[width = 0.23\textwidth]{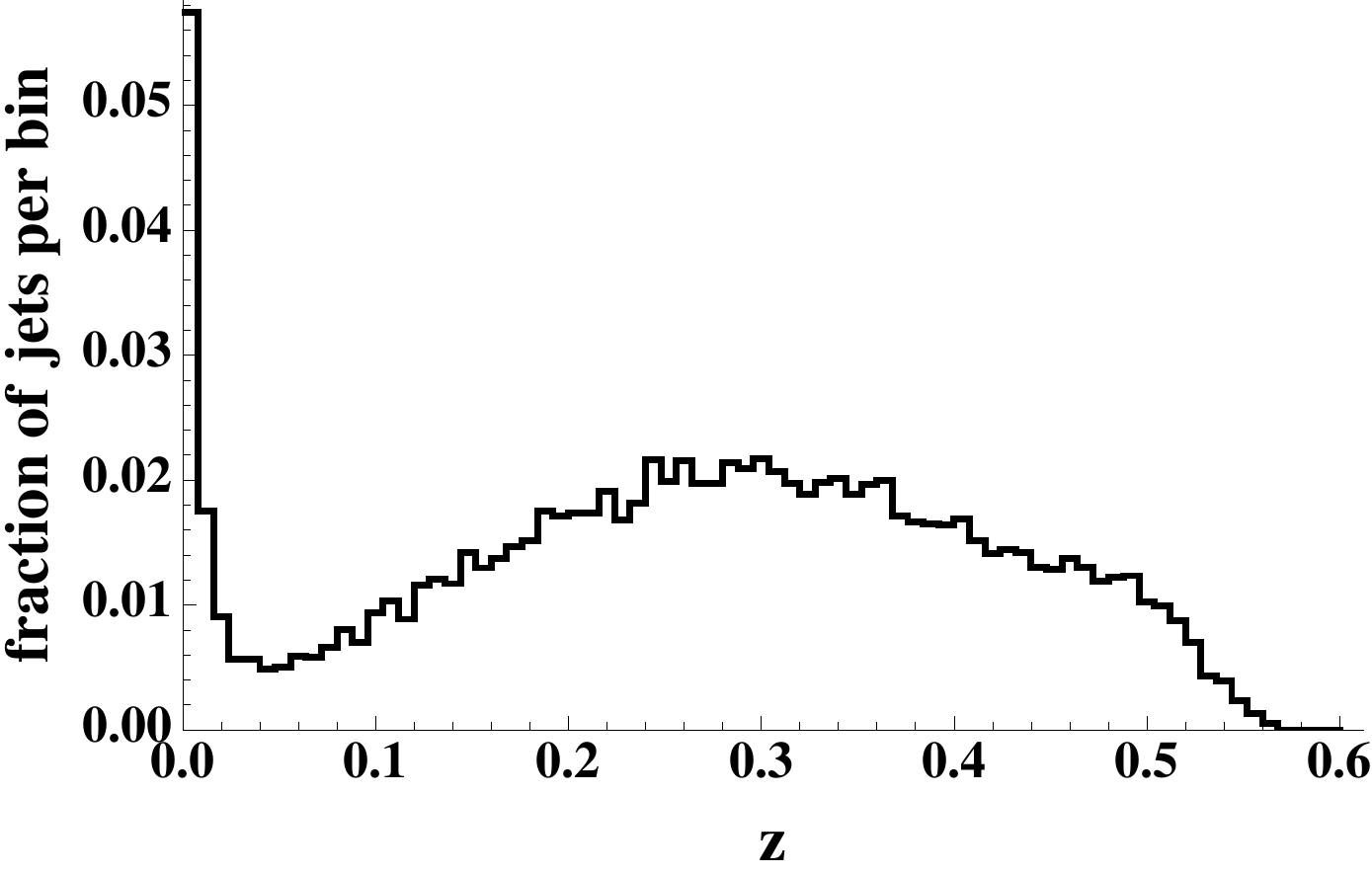}}
\subfloat[$p_T$ bin 4, 900--1100 GeV]{\includegraphics[width = 0.23\textwidth]{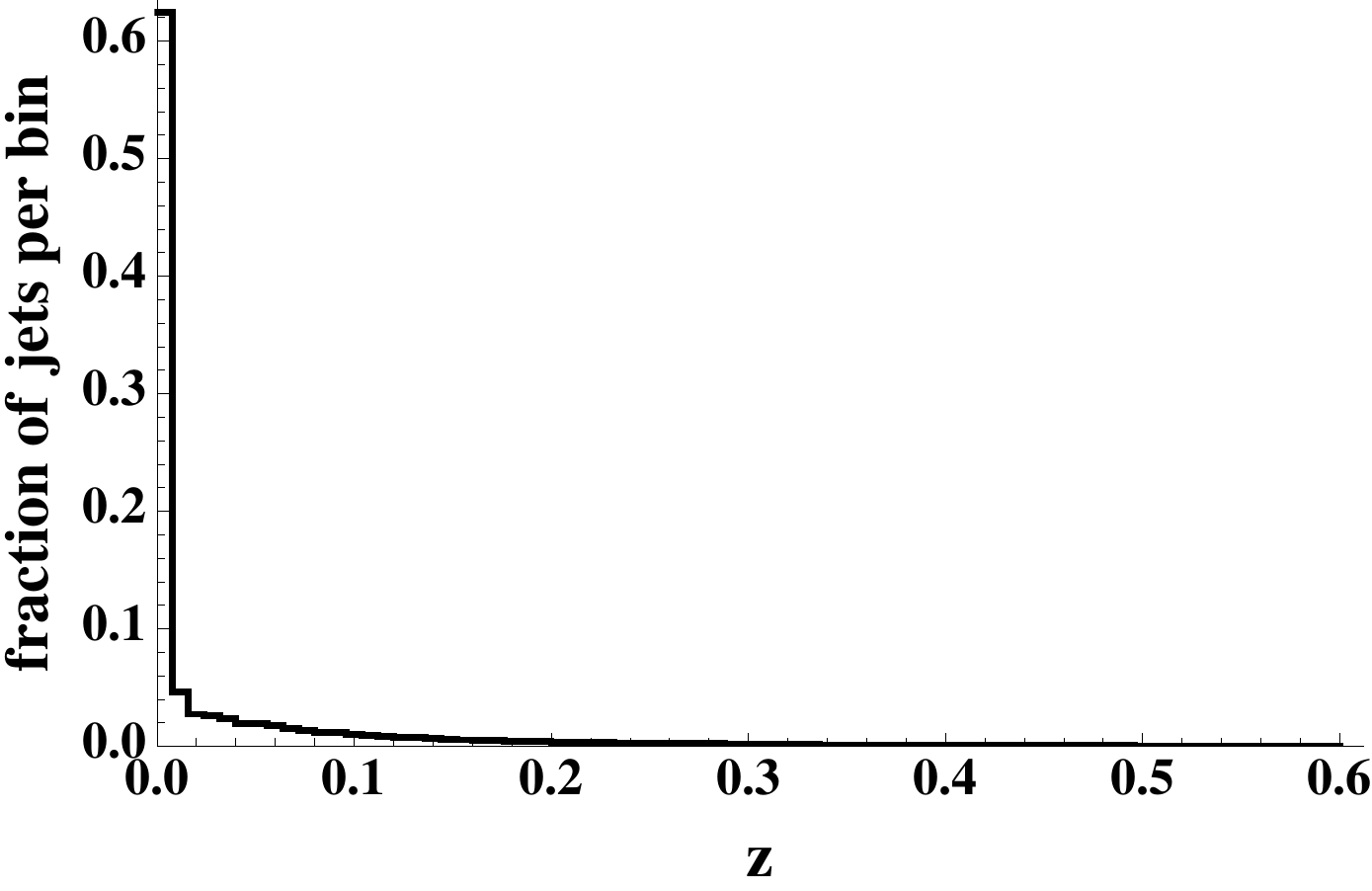}}

\caption{Distribution in $z$ for unpruned CA jets in the top mass window for two $p_T$ bins.  The small $p_T$ bin distribution (left) has only a small enhancement of entries at small $z$, while the large $p_T$ bin distribution (right) is dominated by small $z$.  }%\red{zCAmtopPT1\&4}}
\label{zCAmtopPT1&4}
\end{figure}

\begin{figure}[htbp]
\includegraphics[width=\columnwidth]{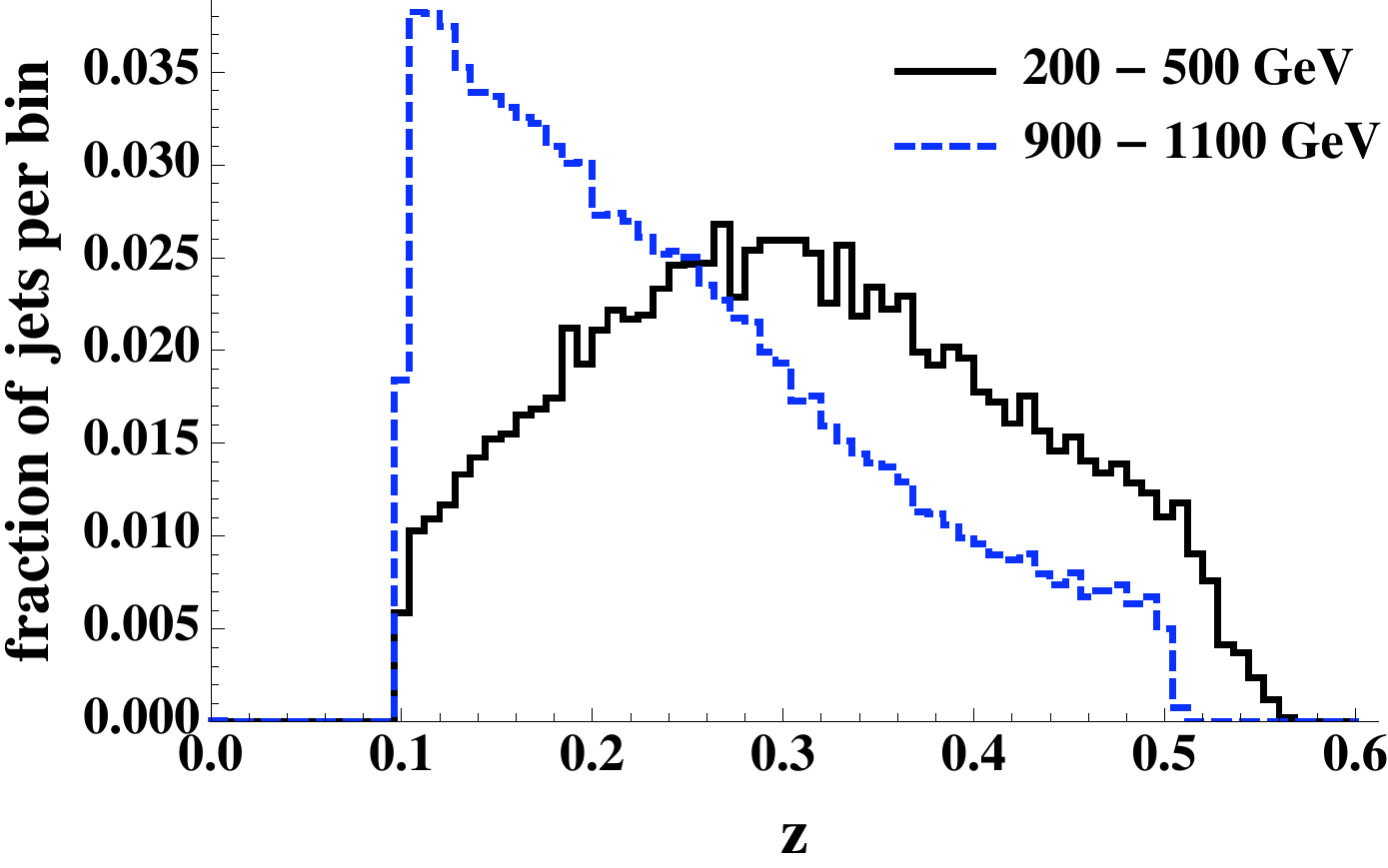}
\caption{Distribution in $z$ for pruned CA jets in the top mass window for two $p_T$ bins, using $z_\text{cut}$ = 0.10.  }%\red{zpCAmtopPT1\&4}}
\label{zpCAmtopPT1&4}
\end{figure}

Additionally, these plots offer an interesting perspective on the substructure dependence on $z$ for both algorithms.  The $t\bar{t}$ sample for the CA algorithm is the most instructive.  In this case, small values of $z_{\cut}$ lead to dramatically increased efficiency for finding top jets in the larger $p_T$ bins.  This is due to the improved ability after pruning to find the $W$ as a subjet of the top.  At large $p_T$ with a fixed $D = 1.0$, the opening angle of the top quark decay is much smaller than $D$.  This means that the top quark decay is very localized in the jet, and much of the jet area includes soft radiation.  For the CA algorithm, which recombines solely by the angle between protojets, this tends to delay recombining the soft peripheral radiation until the end of the algorithm.
The result is substructure with small $z$ at the last recombination that is not representative of the top quark decay --- neither daughter protojet of the top has the $W$ mass.  As an illustration of this point, in Fig.~\ref{zCAmtopPT1&4} we plot the distribution of $z$ for unpruned jets in the top mass range for the CA algorithm in the largest and smallest $p_T$ bins.  Note that in the largest $p_T$ bin, where the top quark decay is highly localized in the jet and the decay angle is much less than $D$, there is a substantially increased fraction of jets with a small value of $z$.  This does not occur in the smallest $p_T$ bin, where most of the reconstructed tops are at threshold for being just inside the jet.
When pruning is implemented, however, much of this soft radiation is removed.  In Fig.~\ref{zpCAmtopPT1&4}, we plot the same distributions as in Fig.~\ref{zCAmtopPT1&4}, but for pruned jets.  In this case, no jets with the top mass have small $z$, since pruning has removed those recombinations.  This leads to a highly enhanced efficiency to resolve the $W$ subjet and identify the jet and a top jet.  In Sec.~\ref{sec:results:fixedD}, we will study pruning when the value of $D$ is matched to the average angle of the heavy particle decay, and we will see that the performance of the unpruned CA algorithm improves.

By contrast, this situation does not occur for the $\kt$ algorithm.  Even when the value of $D$ is mismatched with the top quark decay angle, the soft radiation on the periphery of the jet is recombined early in the $\kt$ algorithm because of the $p_T$ weighting in the recombination metric.  Therefore, there is no increase in efficiency with increasing $z_\cut$ for large $p_T$, and the decrease in $\epsilon$ comes from the narrower width of the top and $W$ mass distributions.  The small variation in the measures $R$ and $S$ for the $\kt$ algorithm at small $z_{\cut}$ is evidence of the fact that $\kt$ tends to have many fewer small-$z$ recombinations at the end of the algorithm, and supports the larger value of $z_{\cut} = 0.15$ for the $\kt$ algorithm that we will use in the remainder of the study.

\begin{figure*}[htbp]
\subfloat[$W$'s, CA jets]{\label{VaryDcut:WCA}\includegraphics[width=0.20\textwidth]{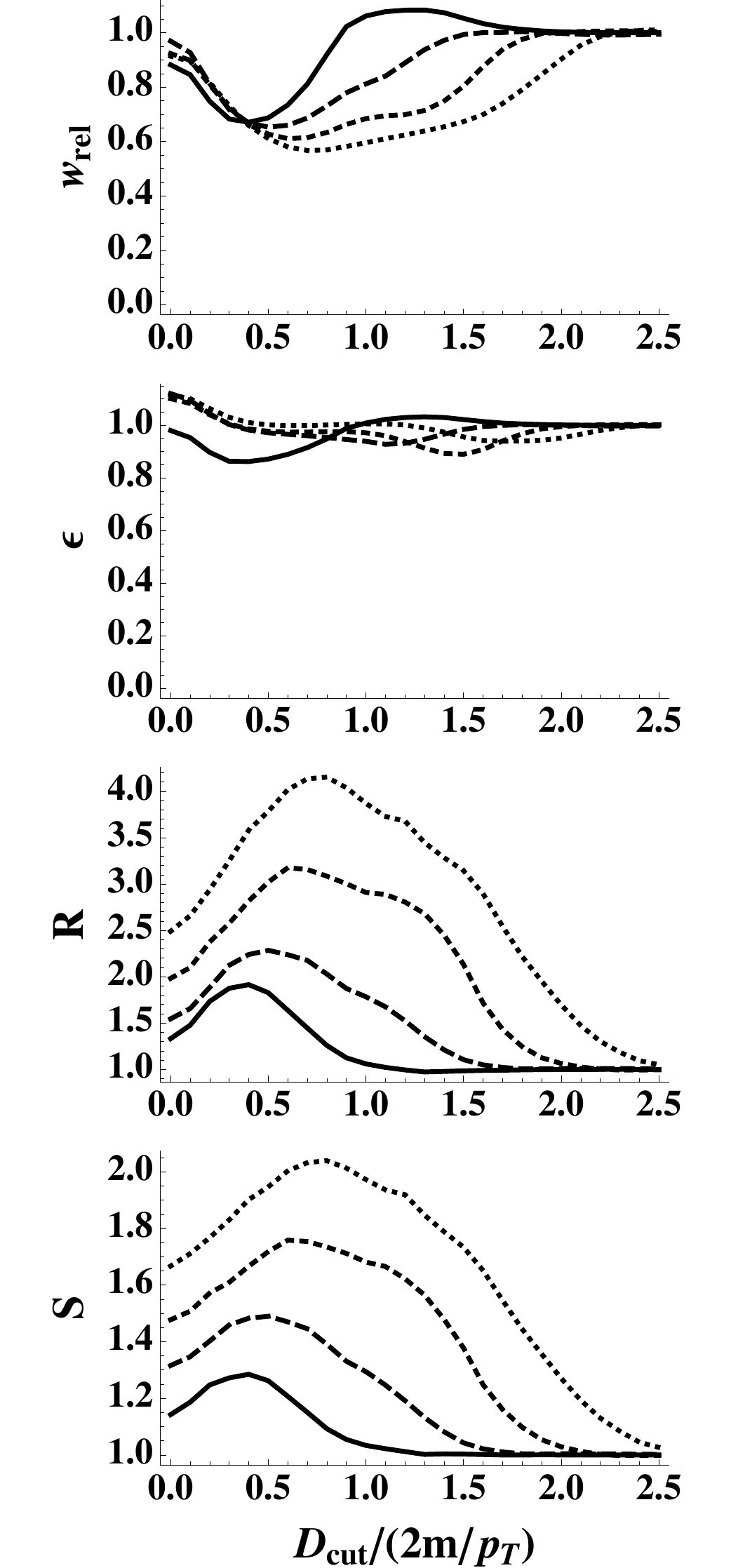}}
\subfloat[tops, CA jets]{\label{VaryDcut:tCA}\includegraphics[width=0.20\textwidth]{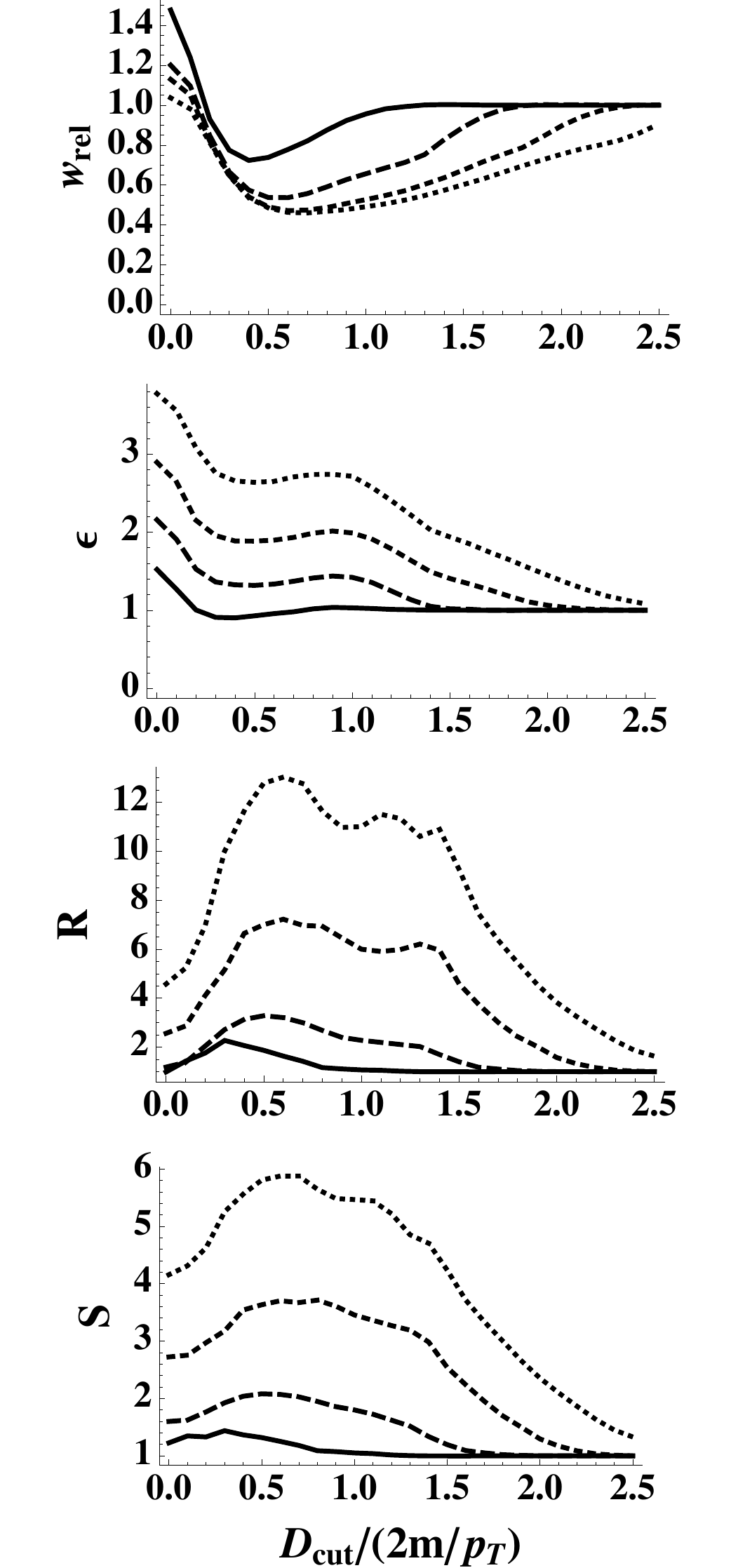}}
\subfloat[$W$'s, $\kt$ jets]{\label{VaryDcut:WkT}\includegraphics[width=0.20\textwidth]{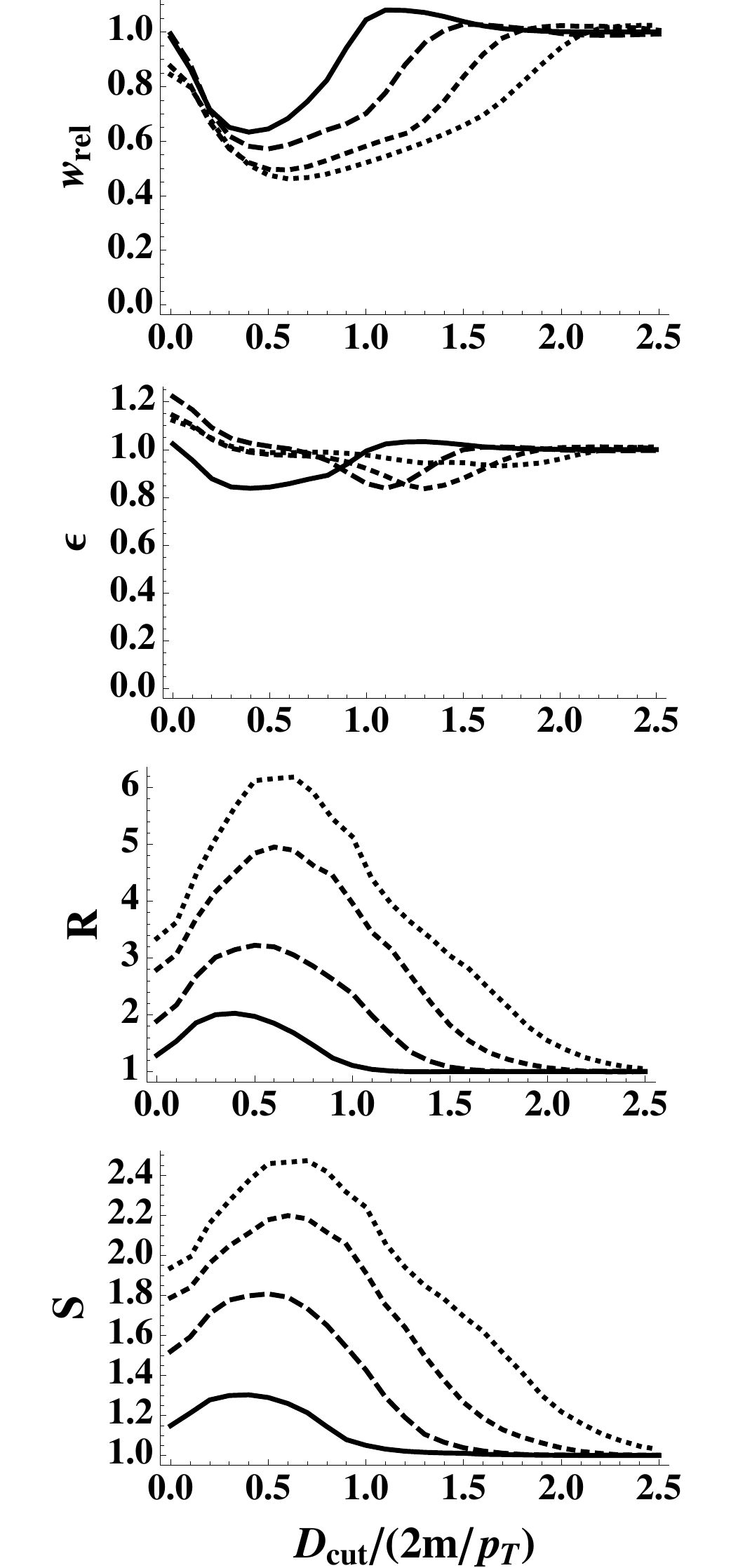}}
\subfloat[tops, $\kt$ jets]{\label{VaryDcut:tkT}\includegraphics[width=0.20\textwidth]{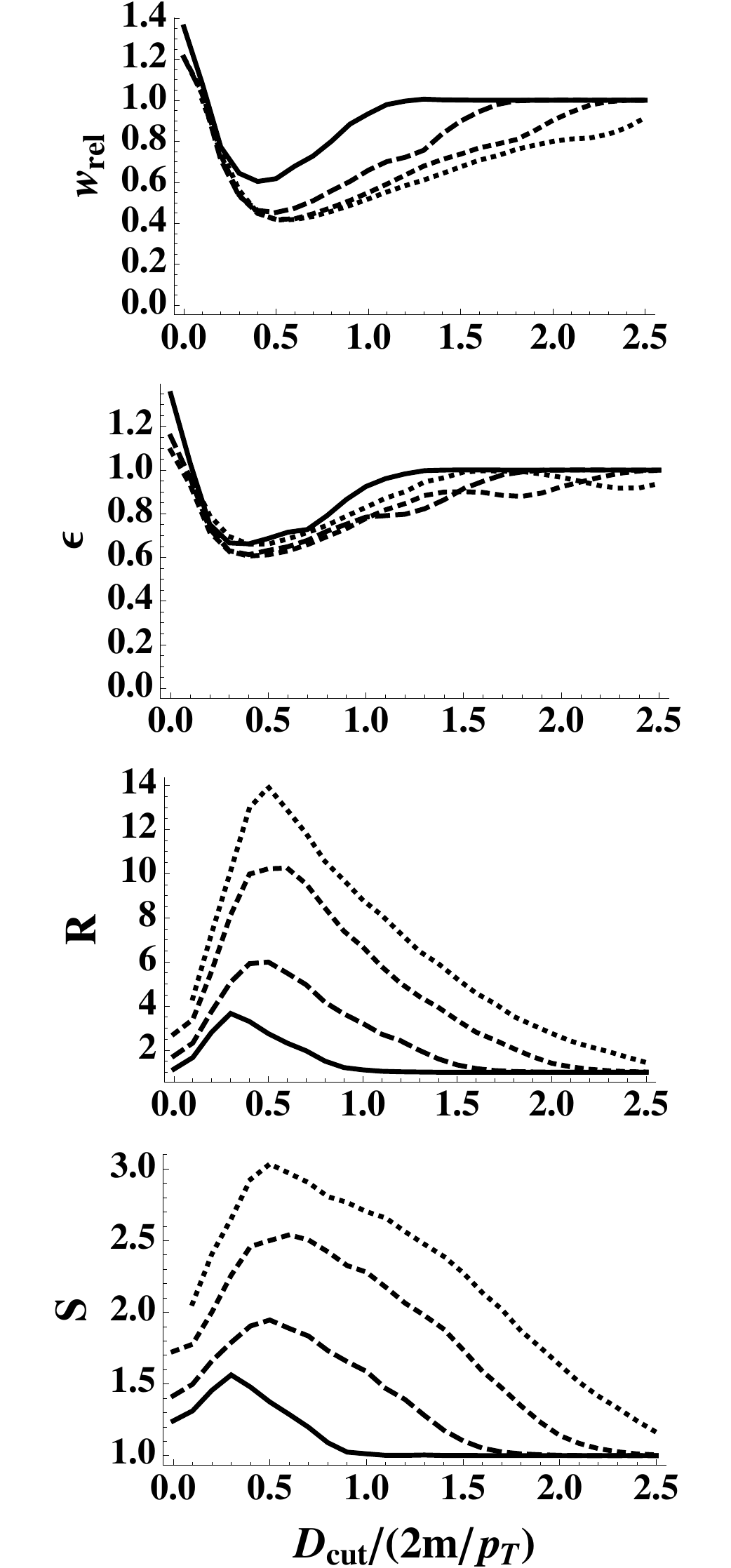}}
\includegraphics[width=0.10\textwidth]{pTbinlegend.pdf}
\caption{Relative statistical measures $w_\text{rel}$, $\epsilon$, $R$, and $S$ vs. $D_{\cut}/\frac{2 m_J}{p_{T_J}}$ for $W$'s and tops, using CA and $\kt$ jets.  Four $p_T$ bins are shown for each sample.  Statistical errors (not shown) are $\mathcal{O}(1\%)$ for $w_\text{rel}$ and $\epsilon$, and $\mathcal{O}(10\%)$ for $R$ and $S$.  }%\red{VaryDcut}}
\label{VaryDcut}
\end{figure*}

We now fix $z_{\cut}$ to study the dependence on $D_{\cut}$.  For the CA algorithm we choose $z_{\cut} = 0.1$, and for $\kt$ we choose 0.15.  In Fig.~\ref{VaryDcut}, we plot $w_\text{rel}$, $\epsilon$, $R$, and $S$ as $D_{\cut}$ is varied in [0, $5 m_J/p_{T_J}$].  While $z_\cut$ sets the minimum $p_T$ asymmetry that recombinations can have, $D_\cut$ sets the minimum opening angle for recombinations that can be pruned.  We can think of $D_\cut$ as determining which recombinations can be pruned, and $z_\cut$ as determining whether or not that pruning takes place.  This difference is clearer when we consider two limiting values of $D_\cut$ and their impact on the pruned jet substructure.

As $D_\cut$ grows past $2m_J/p_{T_J}$, any recombination must have a large opening angle between the daughters to be pruned.  Note that the limit $D_\cut \to \infty$ is the limit of no pruning.  For both the CA and $\kt$ algorithms, in this limit only very late recombinations in the algorithm can be pruned (if the jet can be pruned at all).  In this limit, we expect the statistical measures to tend to one as the amount of pruning decreases.

The second limit is $D_\cut \to 0$.  In this limit any recombination can be pruned, since the minimum opening angle needed is very small.  As $D_\cut$ decreases towards zero, more of the jet substructure can be pruned.  In particular, earlier recombinations --- those with smaller opening angle on average --- can be pruned as $D_\cut$ decreases.  In general, these early recombinations are associated with the QCD shower, and pruning them can degrade the mass resolution of the jet because too much radiation is being removed.  Therefore, we expect the performance of pruning to be poor in this region.

Both of these limits are present in Fig.~\ref{VaryDcut}, and our expectations about these limits are correct.  It is in the intermediate region, where $D_\cut \approx m_J/p_{T_J}$, that the performance of pruning is optimal, with a maximum in $S$ that is not very sensitive to the $p_T$ bin, sample, or algorithm.  This value of $D_\cut = m_J/p_{T_J}$ is sensible when we recognize that the average opening angle of the jet is approximately $2m_J/p_{T_J}$, and half this value allows for pruning of late recombinations but not the soft, small-angle recombinations associated with the QCD shower.

For the remainder of the study, we fix the pruning parameters $z_{\cut} = 0.1$ for the CA algorithm and $z_{\cut} = 0.15$ for the $\kt$ algorithm, as well as $D_{\cut} = m_J/p_{T_J}$ for both algorithms.  With these parameters fixed, we move on to discuss more interesting tests of the pruning procedure, namely the improvements conferred by pruning over a range in heavy particle boost and the $D$ dependence of the pruning procedure.

\subsection{Top and \texorpdfstring{$W$}{W} Identification with Constant \texorpdfstring{$D$}{D}}
\label{sec:results:fixedD}

\begin{figure*}[htbp!]
\subfloat[$W$'s, CA jets]{\label{VarypT:WCA}\includegraphics[width=0.24\textwidth]{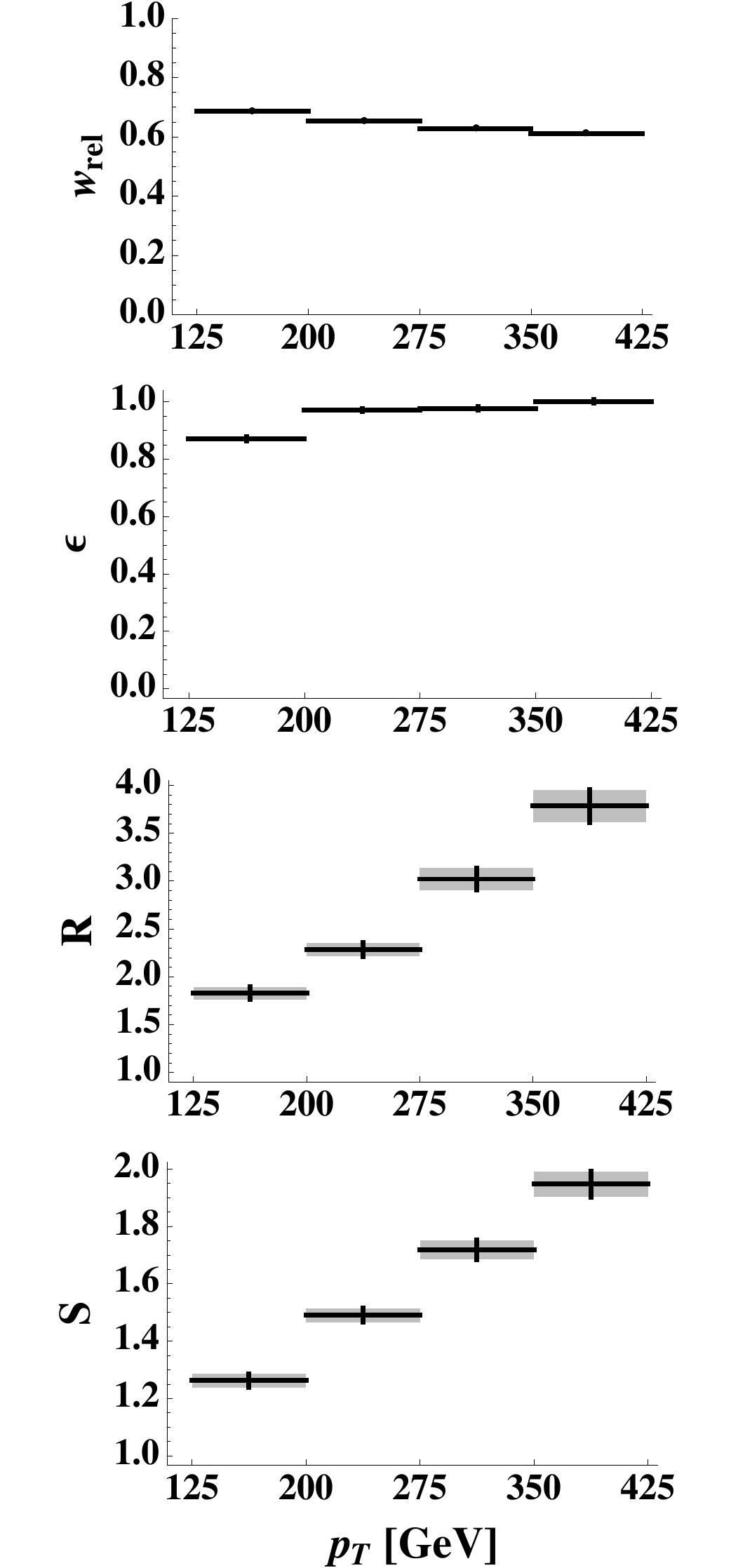}}
\subfloat[tops, CA jets]{\label{VarypT:tCA}\includegraphics[width=0.24\textwidth]{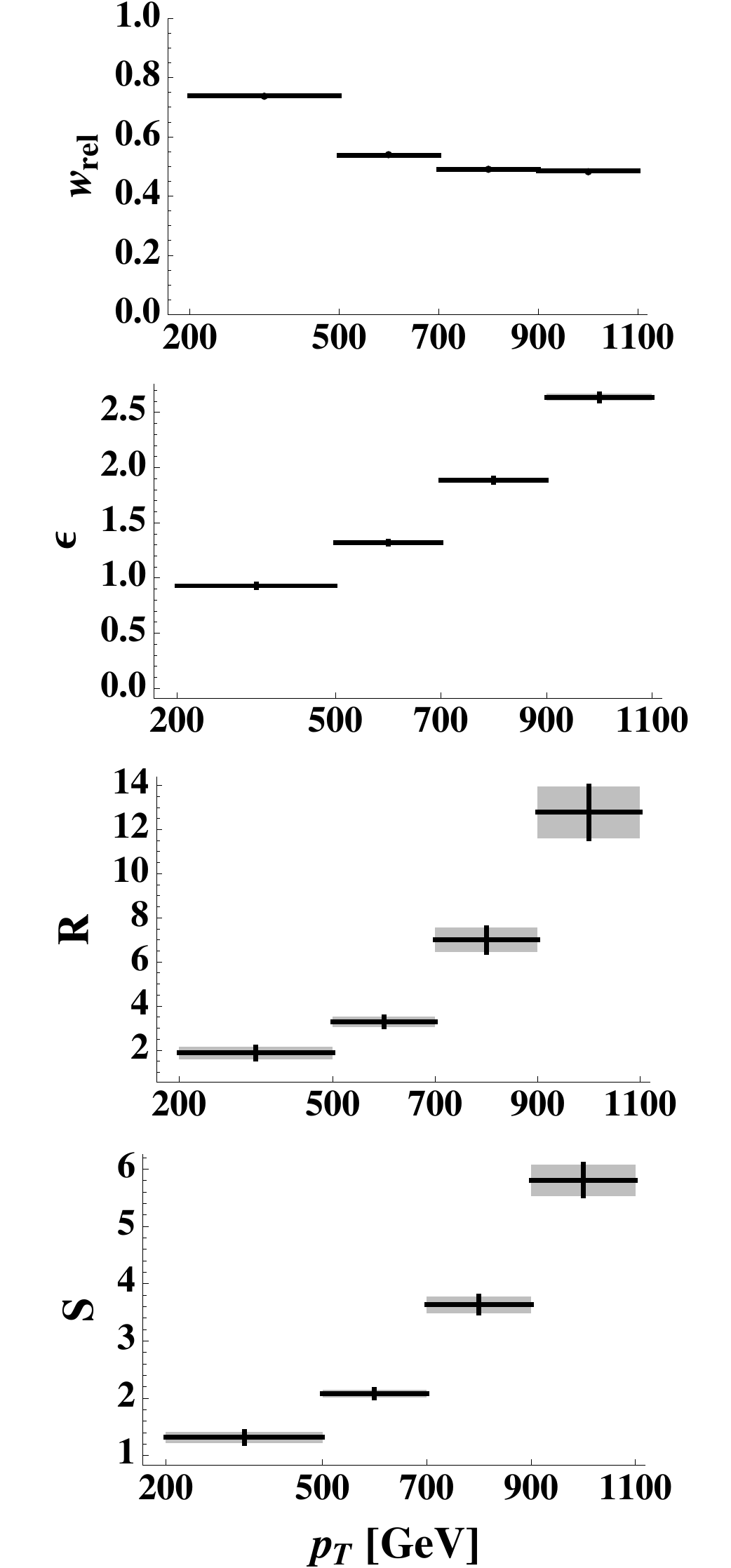}}
\subfloat[$W$'s, $\kt$ jets]{\label{VarypT:WkT}\includegraphics[width=0.24\textwidth]{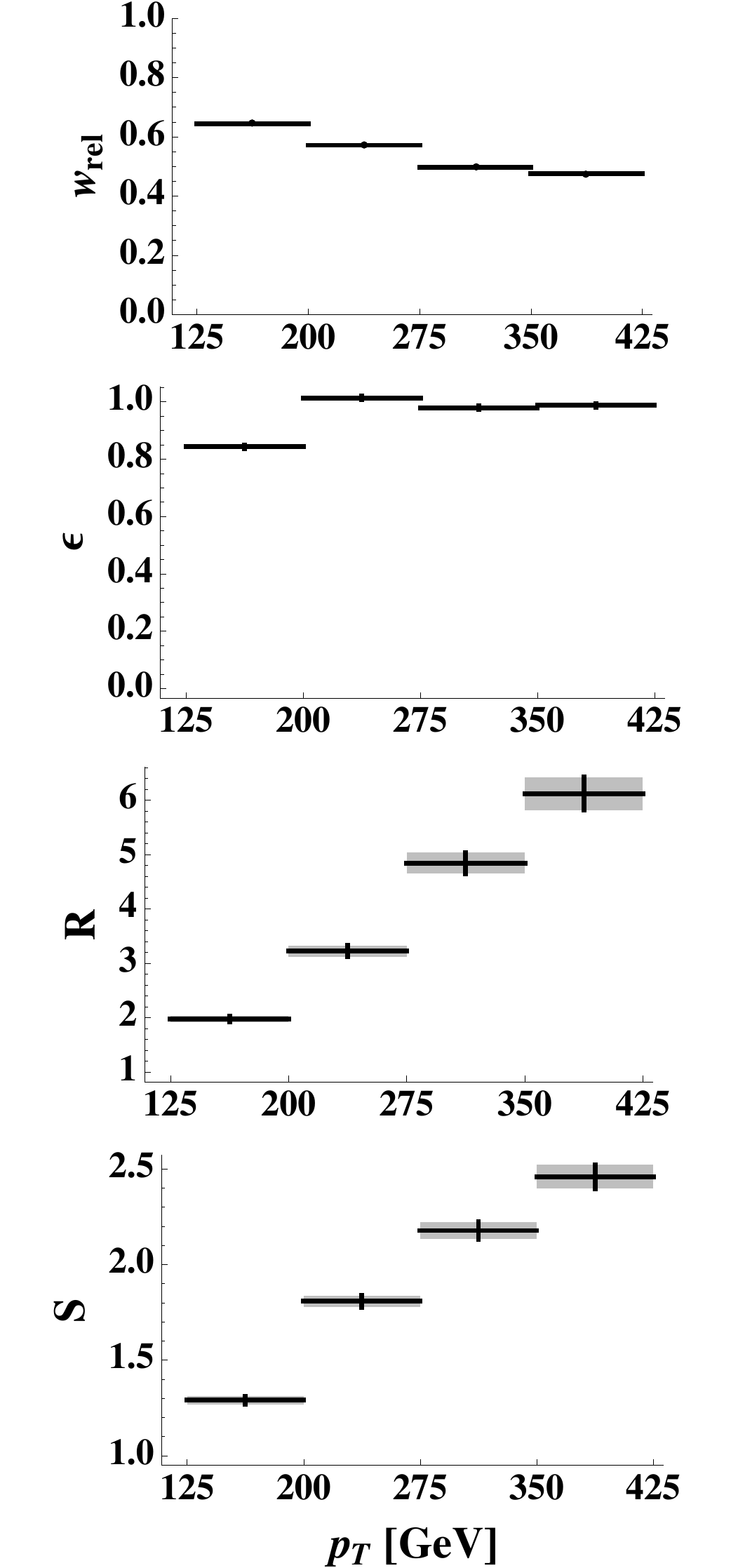}}
\subfloat[tops, $\kt$ jets]{\label{VarypT:tkT}\includegraphics[width=0.24\textwidth]{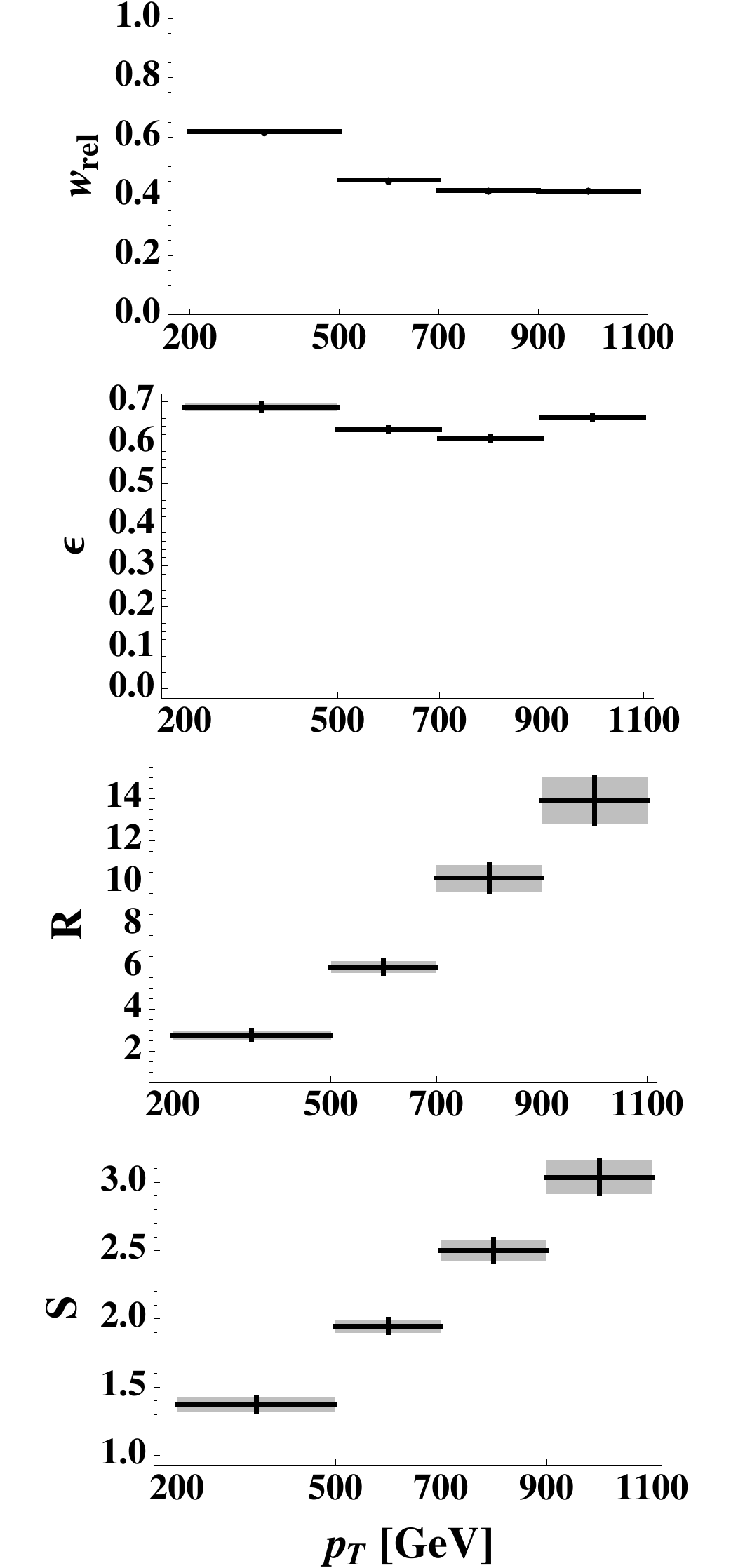}}
\caption{Relative statistical measures $w_\text{rel}$, $\epsilon$, $R$, and $S$ vs. $p_T$ for $W$'s and tops, using CA and $\kt$ jets with D = 1.0.  Statistical errors are shown.  }%\red{VarypT}}
\label{VarypT}
\end{figure*}

In a search for heavy particles decaying into jets, it may be unfeasible to divide a sample into $p_T$ bins and use a tailored jet algorithm to look for local excesses in the jet mass distribution in each $p_T$ bin.  (A ``variable-$R$'' method for avoiding $p_T$-binning, which we do not consider here, has recently been suggested \cite{Krohn:09.1}.  This still requires knowing or guessing the mass of the new particle, since it is $m/p_T$ that determines the relevant angular size.)  For instance, the appropriate angular scale may be unknown because the mass of the heavy particle is not known or the production mechanism is not well understood (so that the spectrum of heavy particle boosts is not known).  In this case, a large-$D$ jet algorithm may be used to search for heavy particles reconstructed in single jets.  To mimic such an analysis, and provide a reference point for further tests of pruning, we find our statistical measures for $W$ and top quark jets, over a range of jet $p_T$ bins and with a fixed $D$ of 1.0.

In Fig.~\ref{VarypT} we plot the values for $w_\text{rel}$, $\epsilon$, $R$, and $S$ versus $p_T$ bin for $W$'s and tops, using the CA and $\kt$ algorithms.  For both algorithms, pruning improves $W$ and top finding, with substantial improvements for large $p_T$.  The measure $S$ in the smallest $p_T$ bins ranges from 30--40\%, growing to values between 100--600\% in the largest $p_T$ bins.  At large $p_T$ in the top quark study, the improvement in signal-to-noise for the CA algorithm is larger than for the $\kt$ algorithm, as is the relative efficiency to identify tops.  This arises because the CA algorithm is poor at reconstructing the $W$ as a subjet of the top jet at large $p_T$ when the value of $D$ is not matched to the opening angle of the decay.  We will investigate this case further in the rest of the analysis.

\subsection{Top Identification with Variable \texorpdfstring{$D$}{D}}
\label{sec:results:varD}

\begin{figure*}[htbp]
\subfloat[$W$'s, CA jets]{\label{VarypToptD:WCA}\includegraphics[width=0.24\textwidth]{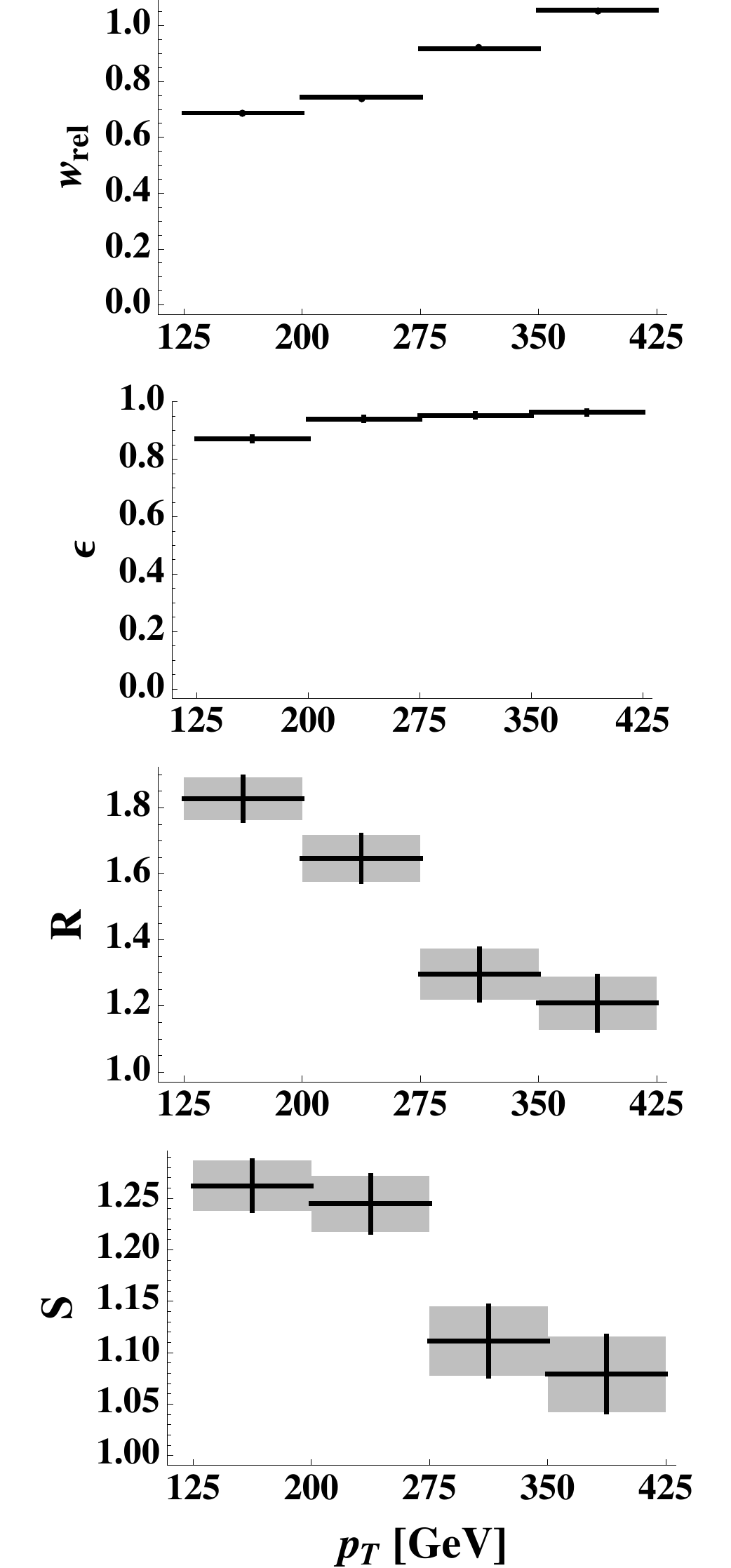}}
\subfloat[tops, CA jets]{\label{VarypToptD:tCA}\includegraphics[width=0.24\textwidth]{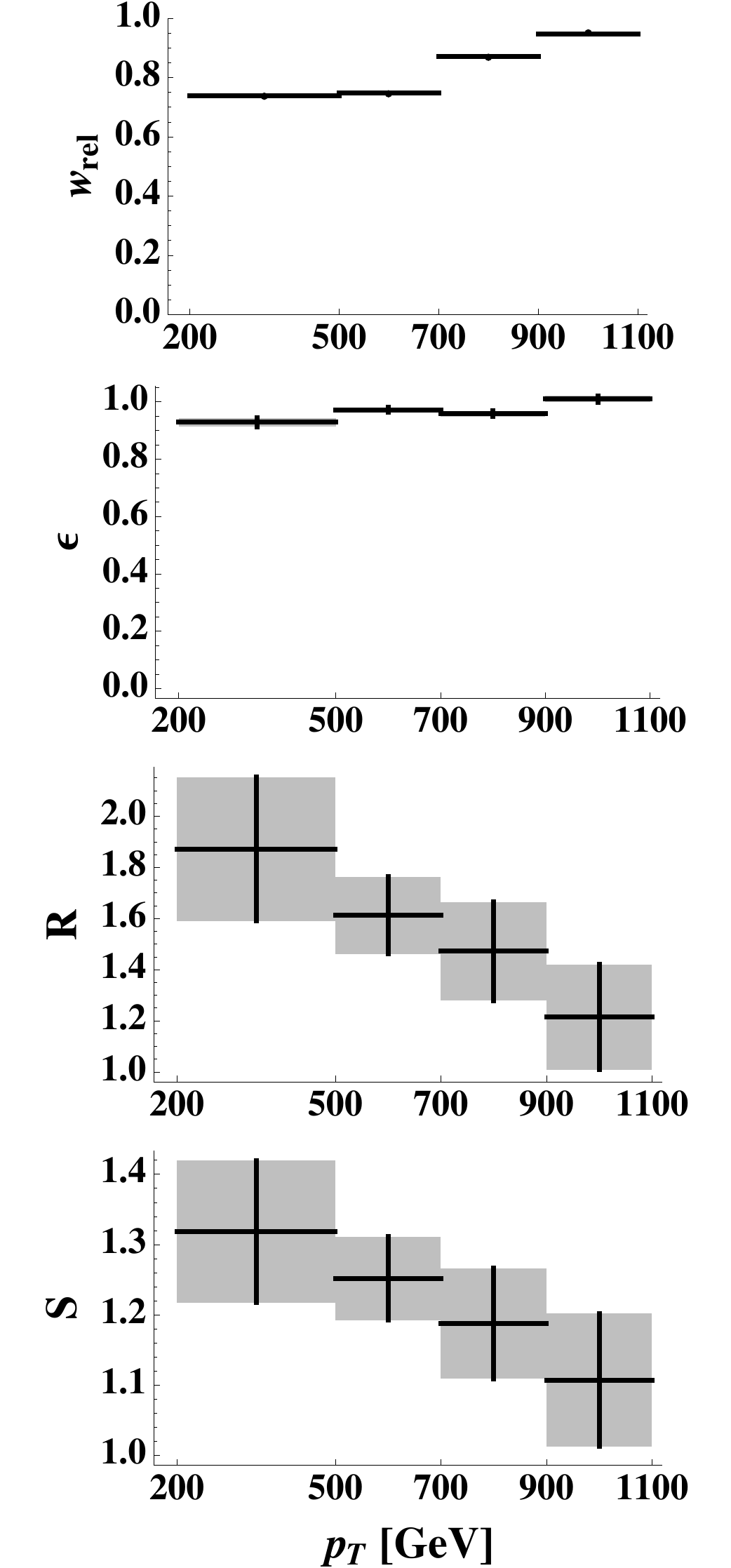}}
\subfloat[$W$'s, $\kt$ jets]{\label{VarypToptD:WkT}\includegraphics[width=0.24\textwidth]{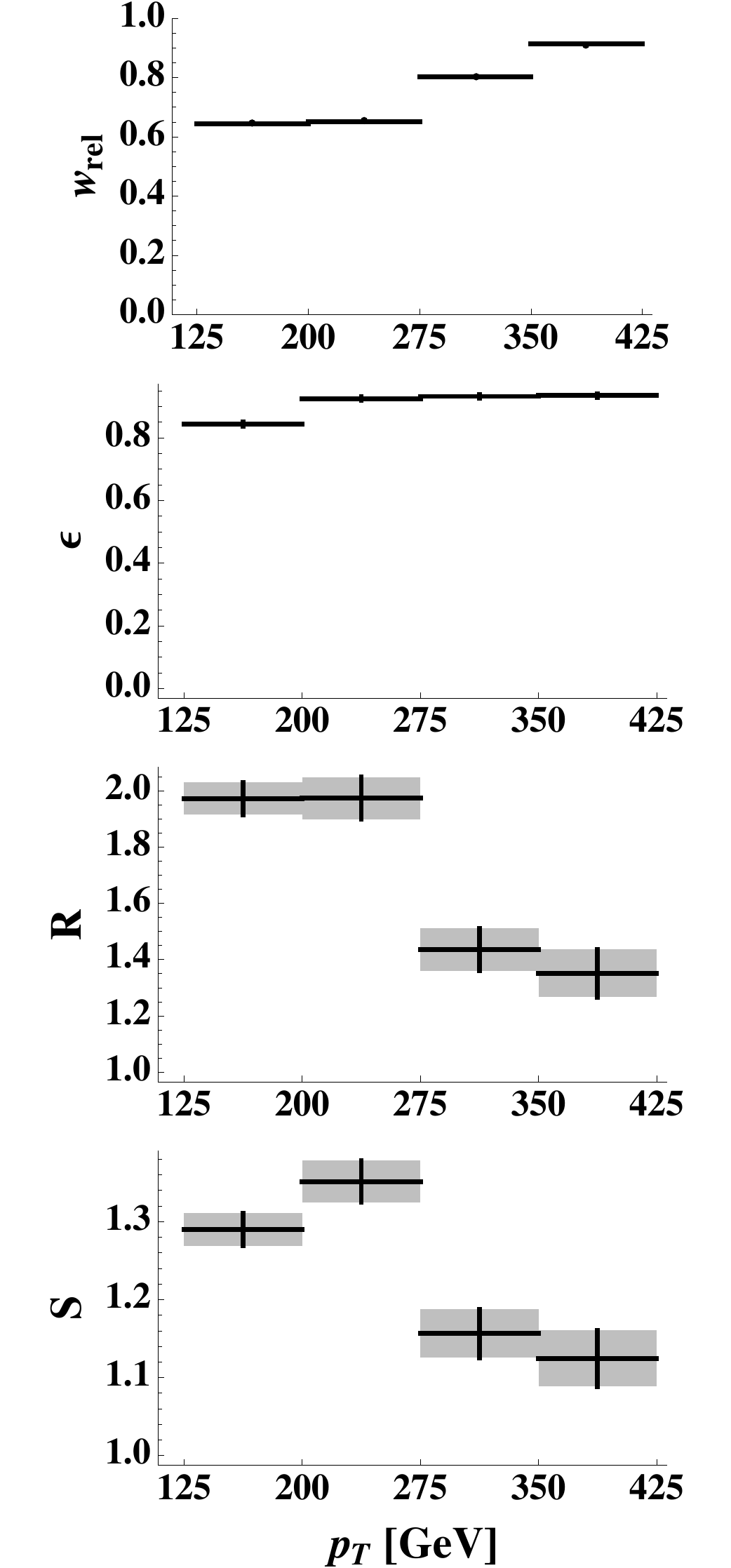}}
\subfloat[tops, $\kt$ jets]{\label{VarypToptD:tkT}\includegraphics[width=0.24\textwidth]{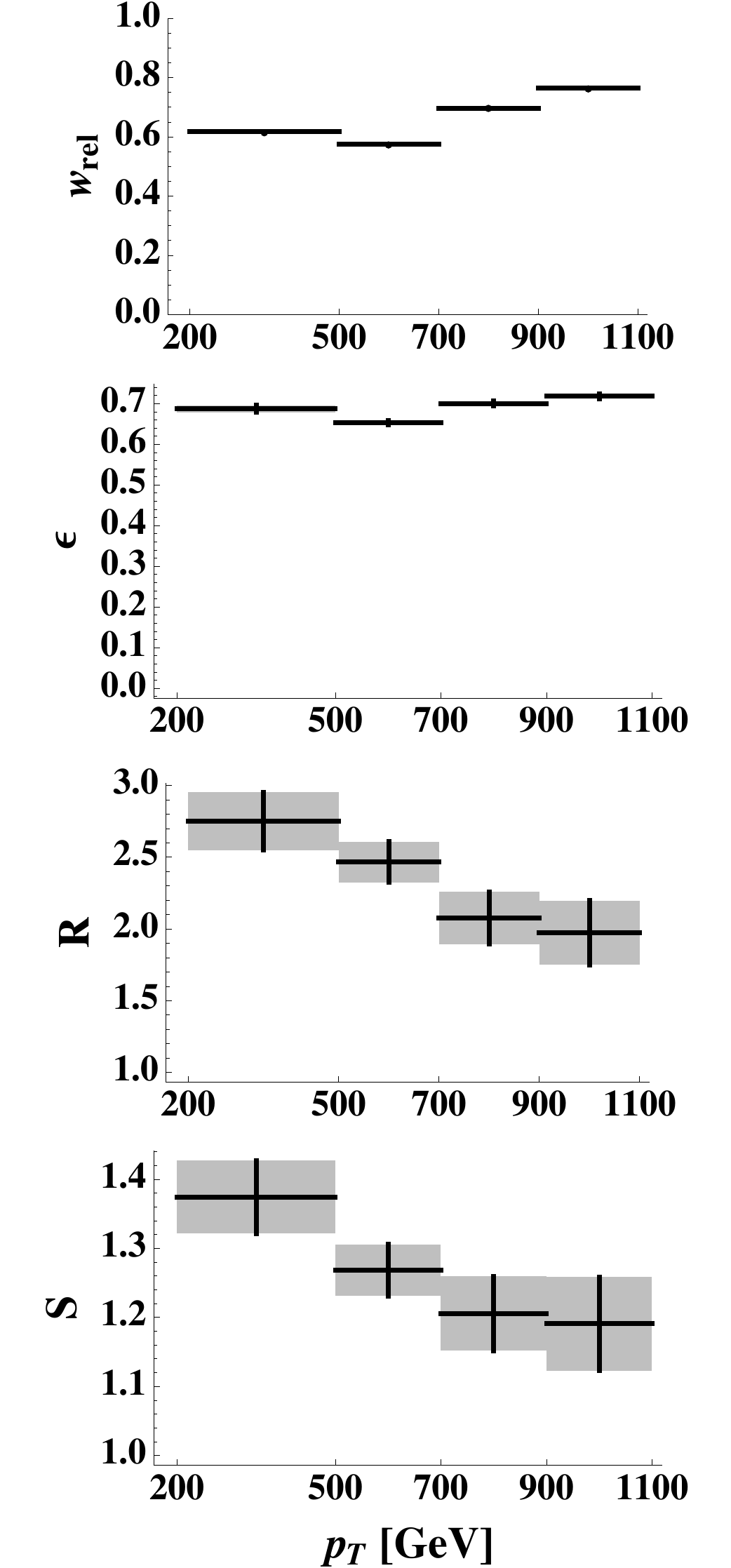}}
\caption{Relative statistical measures $w_\text{rel}$, $\epsilon$, $R$, and $S$ vs. $p_T$ for $W$'s and tops, using CA and $\kt$ jets.  Instead of a fixed $D = 1.0$, a tuned $D$ is used for each $p_T$ bin (see Table \ref{table:D}).  Statistical errors are shown.  }%\red{VarypToptD}}
\label{VarypToptD}
\end{figure*}

\begin{table}[htbp]
\begin{tabular}{|r||c|c|c|c|}
\hline
 & \multicolumn{4}{c|}{$W$} \\
 \hline
$p_T$ (GeV) & 125--200 & 200--275 & 275--350 & 350-425 \\
\hline
``tuned'' $D$ & 1.0 & 0.8 & 0.6 & 0.4 \\
\hline
\hline
 & \multicolumn{4}{c|}{top} \\
\hline
$p_T$ (GeV) &  200--500 & 500--700 & 700--900 & 900--1100 \\
\hline
``tuned'' $D$ & 1.0 & 0.7 & 0.5 & 0.4 \\
\hline
\end{tabular}
\caption{``Tuned'' D values for $W$ and top $p_T$ bins.  The fixed-$D$ analysis used $D = 1.0$, so the smallest bin does not change.  }%\red{table:D}}
\label{table:D}
\end{table}

For an analysis where the heavy particle mass is known, the jet algorithm can be tailored to the jet $p_T$ when searching for the heavy particle reconstructed in a single jet.  In this case, the $D$ value can be chosen using the relation
\be
D = \min\left(1.0, 2\frac{m}{p_T}\right) .
\label{varD}
\ee
where $m$ is the heavy particle mass and $p_T$ is the transverse momentum of the jet.  We take 1.0 to be the maximum allowed value of $D$.  The $D$ values we use are given in Table~\ref{table:D}.  In Fig.~\ref{VarypToptD}, we plot $w_\text{rel}$, $\epsilon$, $R$, and $S$ for jets with these $D$ values used for each $p_T$ bin.  Note that Eq.~(\ref{varD}) neglects the differences between algorithms, which depend on the particular decay.  As an example of the fidelity of this relation for $D$, recall Fig.~\ref{topdRdist}, which plotted the distribution in $\Delta R$ for reconstructed parton-level top quark decays with a top boost of $\gamma = 3$.  Eq.~(\ref{varD}) suggests the value $D = 0.7$, while the means of the CA and $\kt$ distributions for the reconstructed parton-level decay are 0.75 and 0.65 respectively.  Because the distribution in opening angles of the reconstructed decay is broad, by using a smaller, fixed $D$ some decays will not be reconstructed by the jet algorithm.

The difference between the case of constant $D = 1.0$ and variable $D$ is readily apparent.  When the $D$ value is matched to the expected opening angle of the decay, the improvements in pruning are flatter over the whole range in $p_T$, and generally decreasing towards high $p_T$.  The decreased efficiency for pruning, especially for the $\kt$ algorithm, is outweighed by the increases in $R$ and $S$ over the whole range in $p_T$.

\begin{figure*}[htbp!]
\subfloat[$W$'s, CA jets]{\label{VarypTcompareD:WCA}\includegraphics[width=0.24\textwidth]{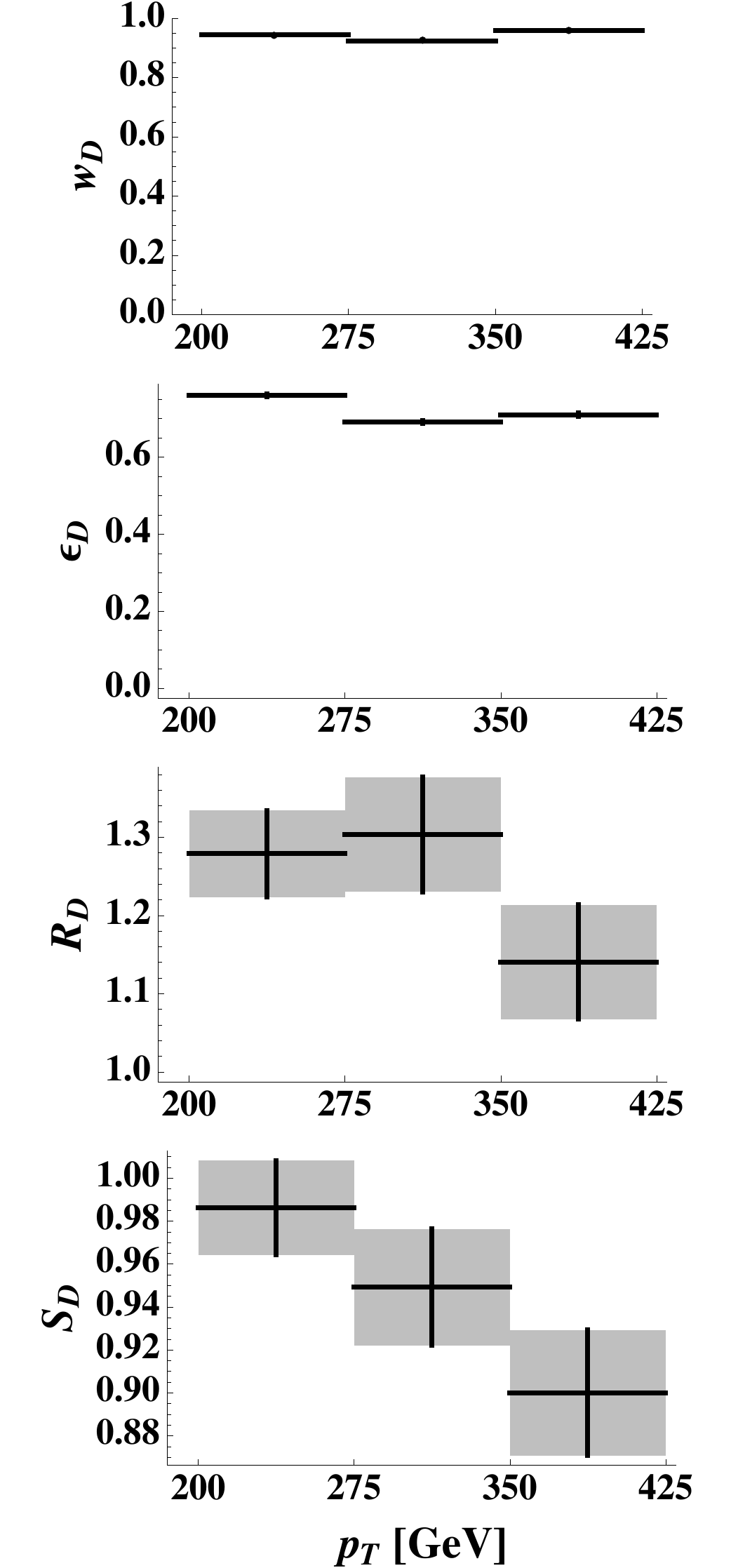}}
\subfloat[tops, CA jets]{\label{VarypTcompareD:tCA}\includegraphics[width=0.24\textwidth]{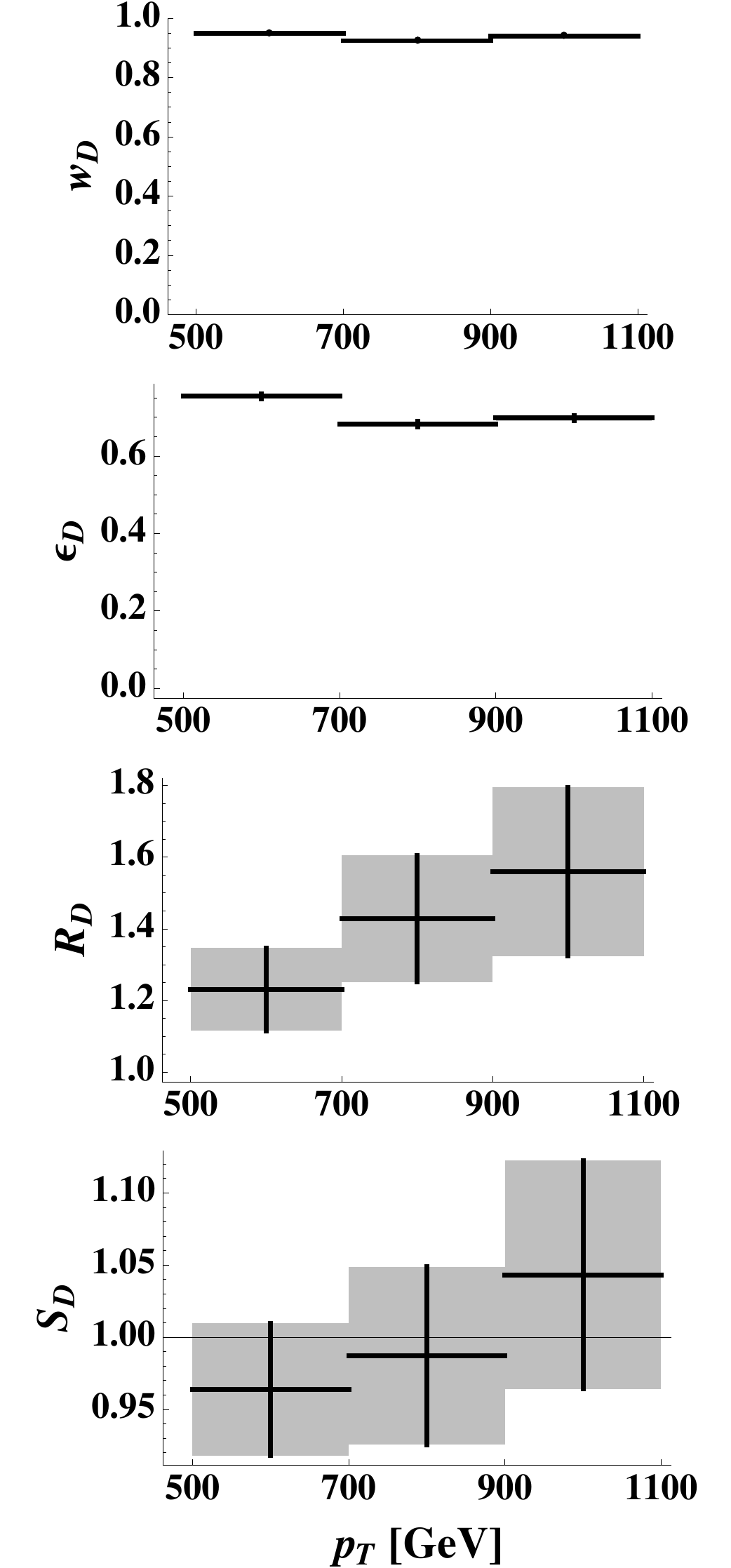}}
\subfloat[$W$'s, $\kt$ jets]{\label{VarypTcompareD:WkT}\includegraphics[width=0.24\textwidth]{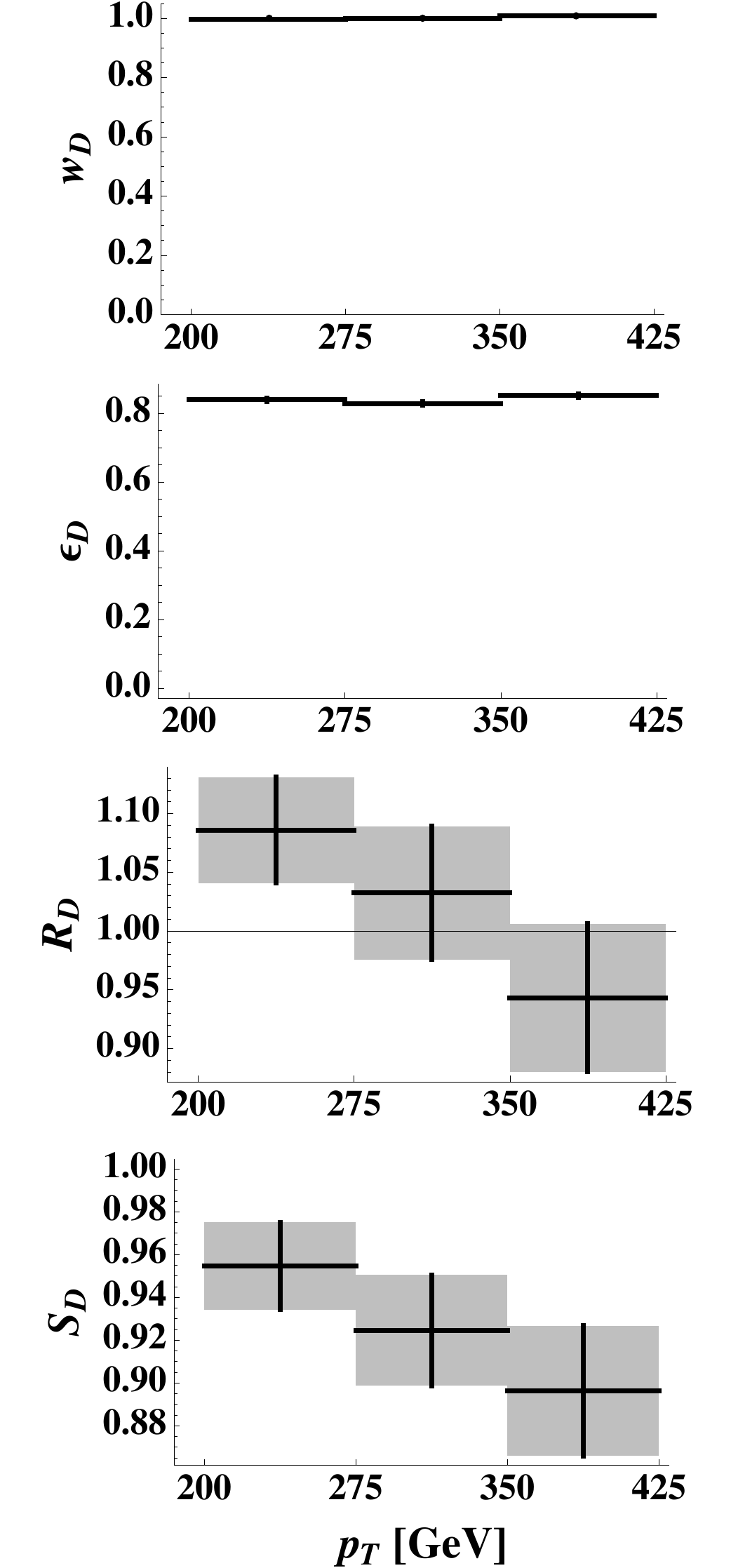}}
\subfloat[tops, $\kt$ jets]{\label{VarypTcompareD:tkT}\includegraphics[width=0.24\textwidth]{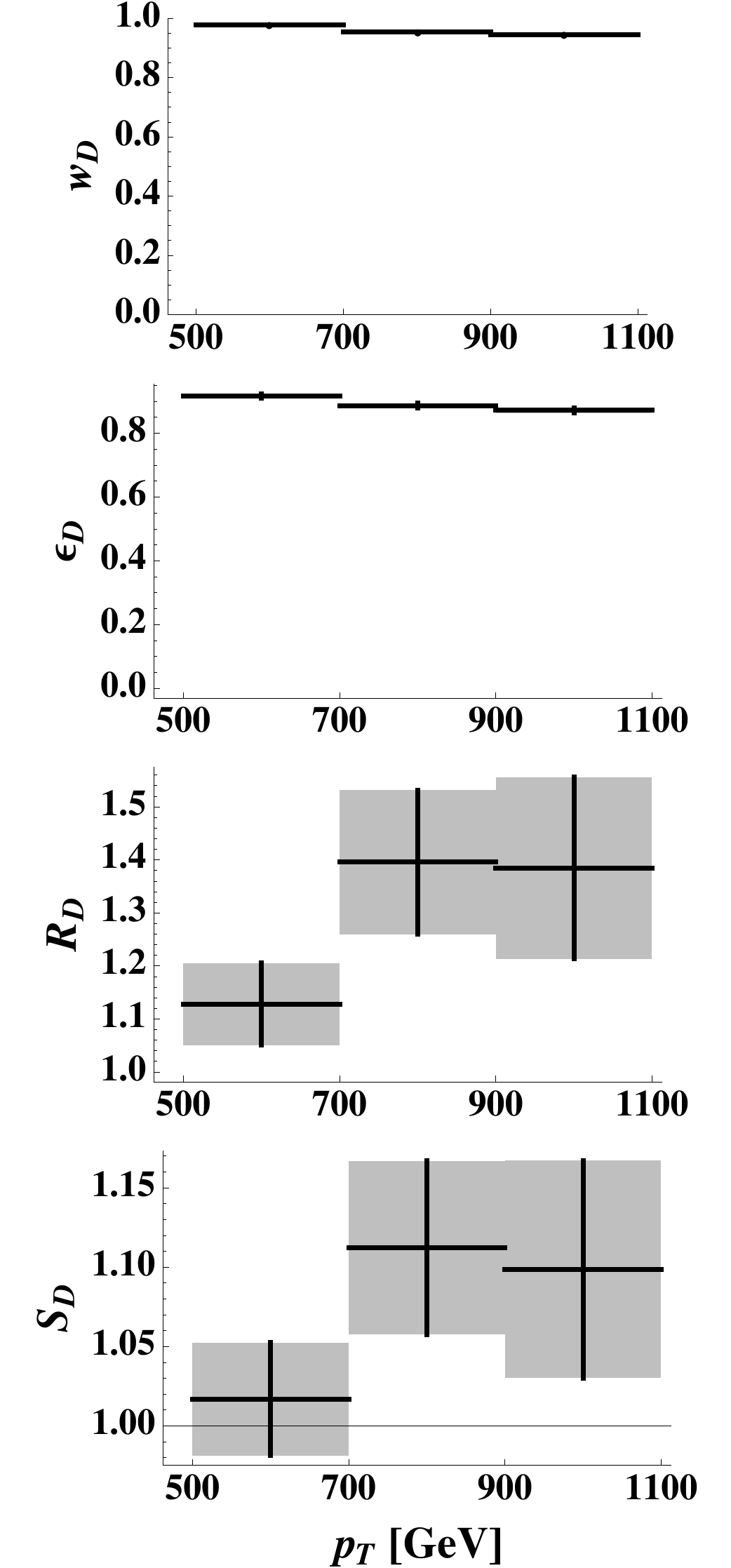}}
\caption{Relative statistical measures $w_D$, $\epsilon_D$, $R_D$, and $S_D$ vs. $p_T$ for $W$'s and tops, using CA and $\kt$ jets.  The measures now compare pruning with a tuned $D$ value in each $p_T$ bin to pruning with a fixed $D$.  Statistical errors are shown.  }%\red{VarypTcompareD}}
\label{VarypTcompareD}
\end{figure*}

Although pruning shows improvements over a broad range in $p_T$ for both constant and variable $D$, we want to compare results for both approaches.  This serves as an indicator of how sensitive the final pruned jet is to the value of $D$ from the jet algorithm.

\subsection{Comparing Pruning with Different \texorpdfstring{$D$}{D} Values}
\label{sec:results:compD}

In the previous two subsections we have seen that an unpruned analysis performs much better when $D$ is tuned to the $m/p_T$ of the signal.  We now consider whether this is true of a pruned analysis.

In each $p_T$ bin, we can compare the results of pruned jets with $D = 1.0$ with pruned jets using value of $D$ fit to the expected size of the decay.  Because the naive expectation is that the tuned value of $D$ will yield better separation from background, we find the improvements in pruning when $D$ is tuned, relative to pruning with a fixed $D$ of 1.0.  Analogous metrics, $w_D$, $\epsilon_D$, $R_D$, and $S_D$, are used, but now they compare the results from pruning with the tuned $D$ value to the results from pruning with $D = 1.0$.  For instance,
\be
R_D \equiv \frac{S/B\text{ from pruning with tuned }D}{S/B\text{ from pruning with } D = 1.0}.
\ee
Note that $x_D > 1$ indicates that tuning $D$ yields an improvement.  The values of these four measures are shown in Fig.~\ref{VarypTcompareD} over the range of $p_T$.  Note that since the tuned value of $D$ in the smallest $p_T$ bin is 1.0, the comparison there is trivial and so is not shown.

These results show only small improvements in $S_D$, with the statistical error bars at most data points including the value $S_D = 1$.  They indicate that the improvements after pruning are roughly independent of the value of $D$ used in the jet algorithm, as long as that $D$ is large enough to fit the expected size of the decay in a single jet.  From the point of view of heavy particle searches, we can conclude that pruning removes much of the $D$ dependence of the jet algorithm in the search.

\subsection{Absolute Measures of Pruning}
\label{sec:results:absolute}

So far, we have only considered measures of pruning relative to a similar analysis without pruning, because this factors out much of the dependence on details of the samples.  However, several recent studies report absolute performance metrics for heavy particle identification, so we examine similar measures here for completeness.  In addition, we directly compare the CA and $\kt$ algorithms, with and without pruning.

\begin{figure}[htbp!]
\subfloat[top mass window width]{\label{MassWidths:top}\includegraphics[width=0.23\textwidth]{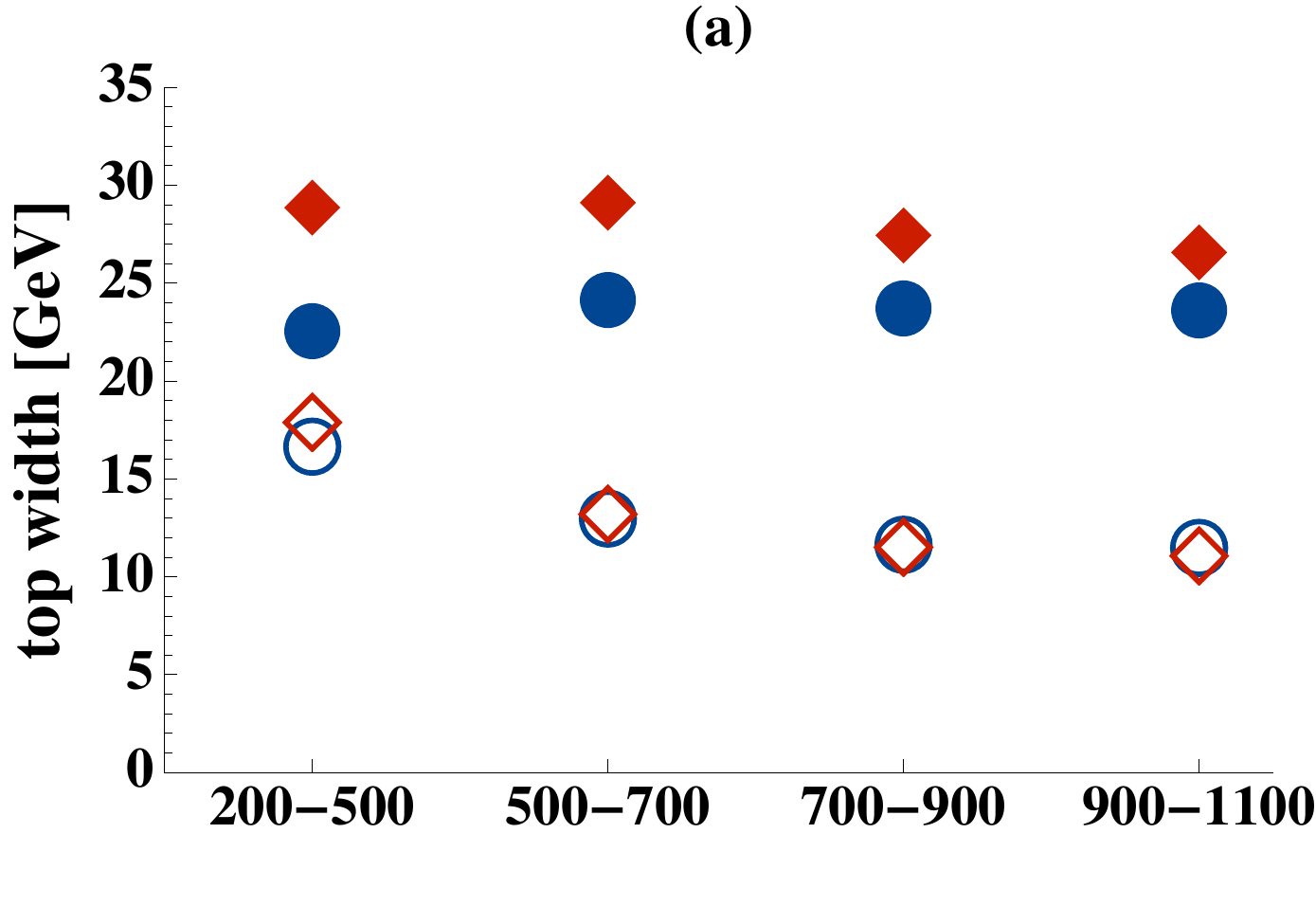}}
{\label{MassWidths:legend}\includegraphics[width=0.23\textwidth]{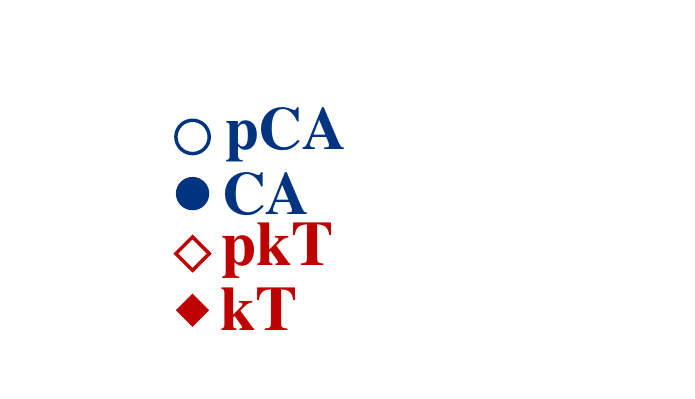}}

\subfloat[$W$ mass window width, top sample]{\label{MassWidths:topW}\includegraphics[width=0.23\textwidth]{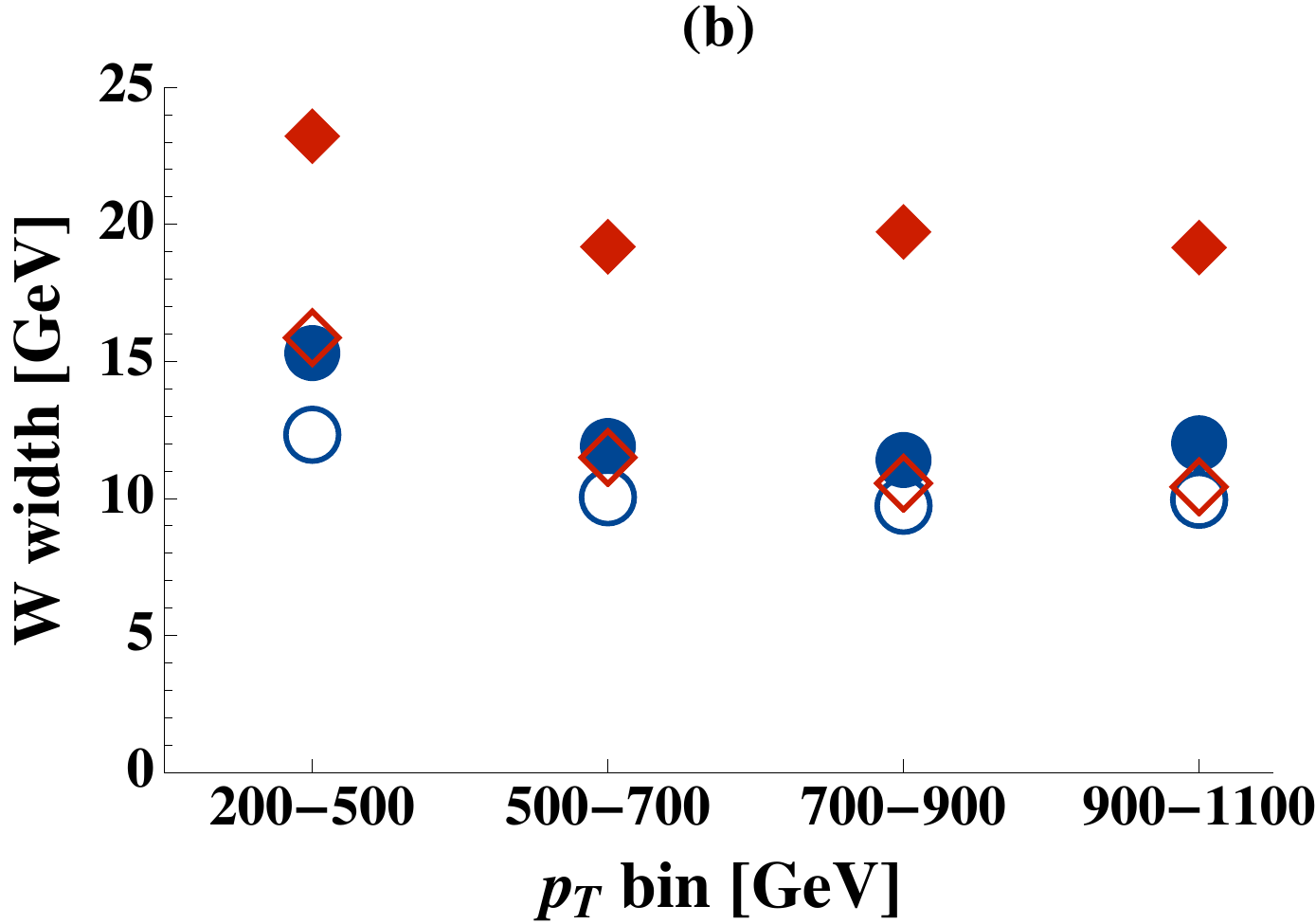}}
\subfloat[$W$ mass window width, $W$ sample]{\label{MassWidths:W}\includegraphics[width=0.23\textwidth]{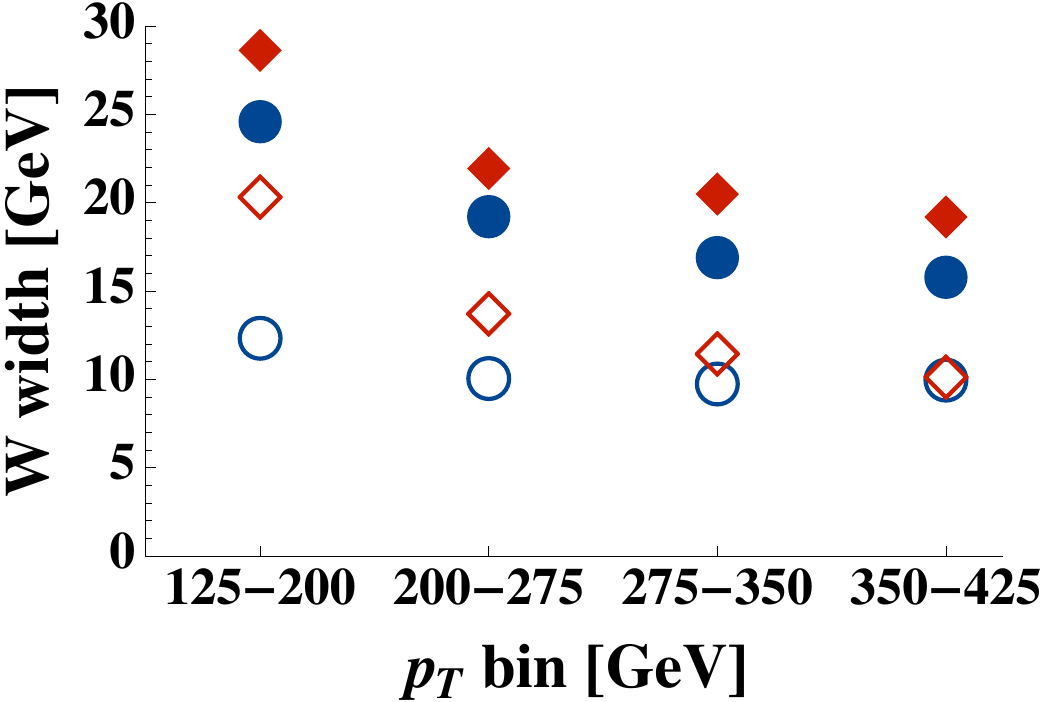}}
\caption{Widths of the top jet (a), $W$ subjet of the top jet (b), and $W$ jet (c) mass windows for the top and $W$ signal samples.}
\label{MassWidths}
\end{figure}

As can be seen from the plots of $w_\text{rel}$ in previous sections, pruning reduces the width of the mass distribution for heavy particles.  In Figs.~\ref{MassWidths:top}, \ref{MassWidths:topW}, and \ref{MassWidths:W}, we plot the absolute widths of the fitted mass distributions for both the top and $W$ in the $t\bar{t}$ sample and the $W$ in the $WW$ sample, over all $p_T$ bins.  We plot this width for the pruned and unpruned version of the CA and $\kt$ algorithms.

%\begin{figure}[htbp]
%\includegraphics[width=\columnwidth]{ttbar_WindowWidths.pdf}
%\caption{Top (a, upper) and W (b, lower) window widths vs. $p_T$ bin in the $t\bar{t}$ sample, for the CA and $\kt$ algorithms with and without pruning, using $D = 1.0$.  A ``p'' before the algorithm name denotes the pruned version.  }%\red{TopWindowWidths}}
%\label{TopWindowWidths}
%\end{figure}

%\begin{figure}[htbp]
%\includegraphics[width=\columnwidth]{W_WindowWidths.pdf}
%\caption{W window widths vs. $p_T$ bin in the $WW$ sample, for the CA and $\kt$ algorithms with and without pruning, using $D = 1.0$.  A ``p'' before the algorithm name denotes the pruned version.  }%\red{WWindowWidths}}
%\label{WWindowWidths}
%\end{figure}

Note that the heavy particle identification method we use in this work selects jets within a range of width $2\Gamma$, with $\Gamma$ coming from a fit to the signal sample.  This gives rise to a mass range cut that is typically much narrower than fixed width ranges used in other studies, and hence the absolute efficiency to identify heavy particles is lower.

\begin{figure}[htbp!]
\subfloat[$\epsilon_{\text{abs}}$, tops]{\label{AbsoluteEfficiencies:topabs}\includegraphics[width=0.23\textwidth]{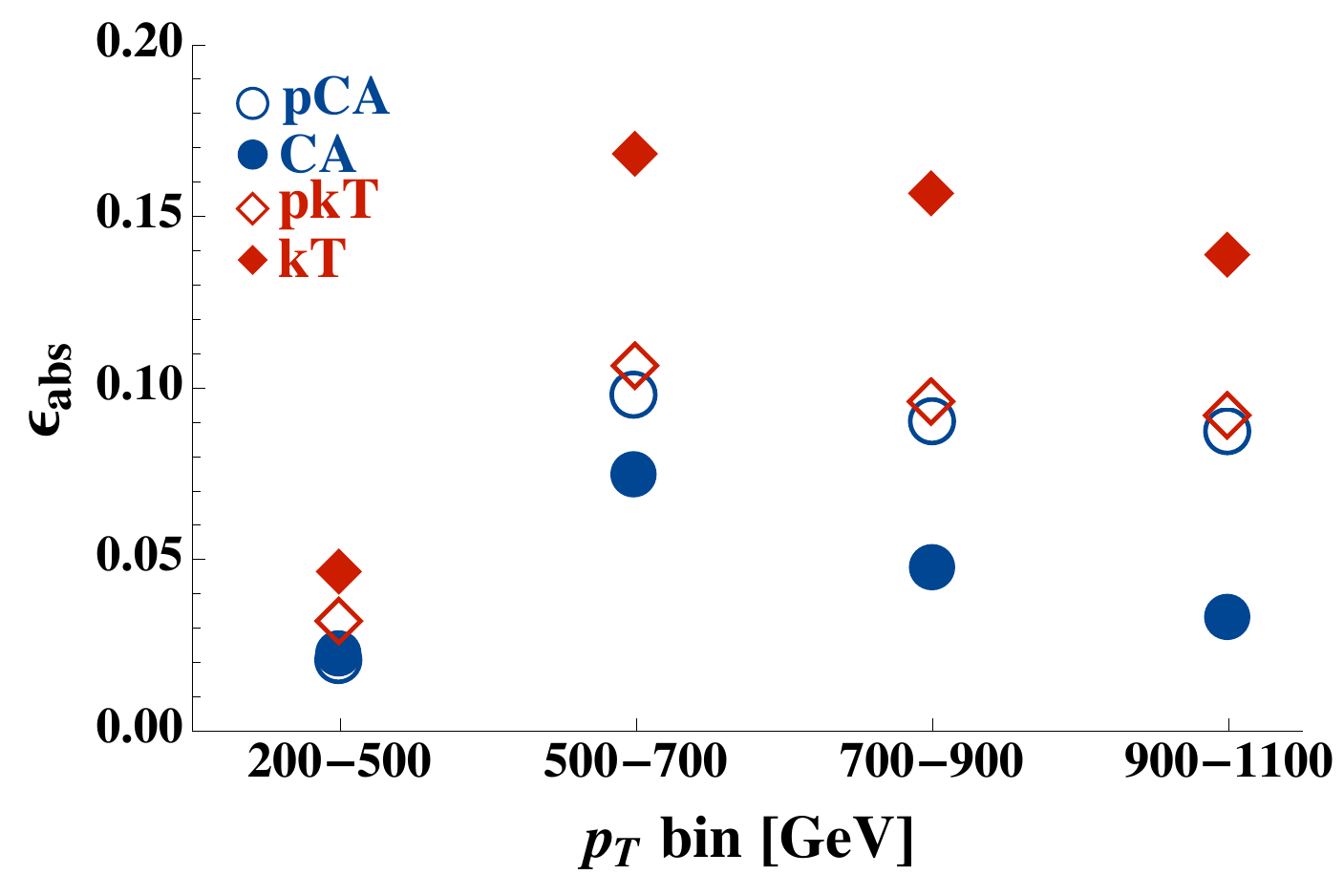}}
\subfloat[$\epsilon_{\text{abs}}$, $W$s]{\label{AbsoluteEfficiencies:Wabs}\includegraphics[width=0.23\textwidth]{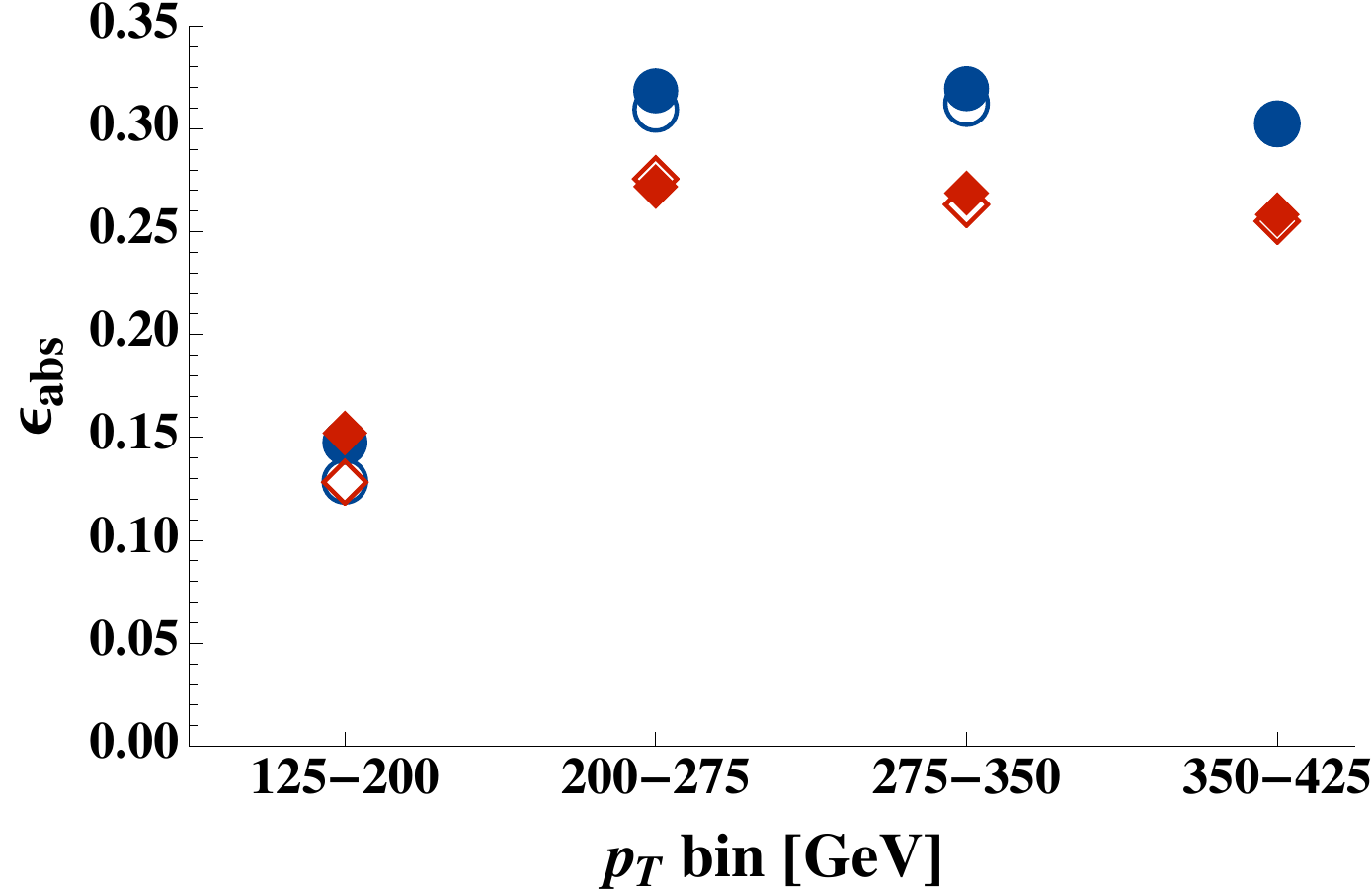}}

\subfloat[$\epsilon_{\text{fake}}$, tops]{\label{AbsoluteEfficiencies:topfake}\includegraphics[width=0.23\textwidth]{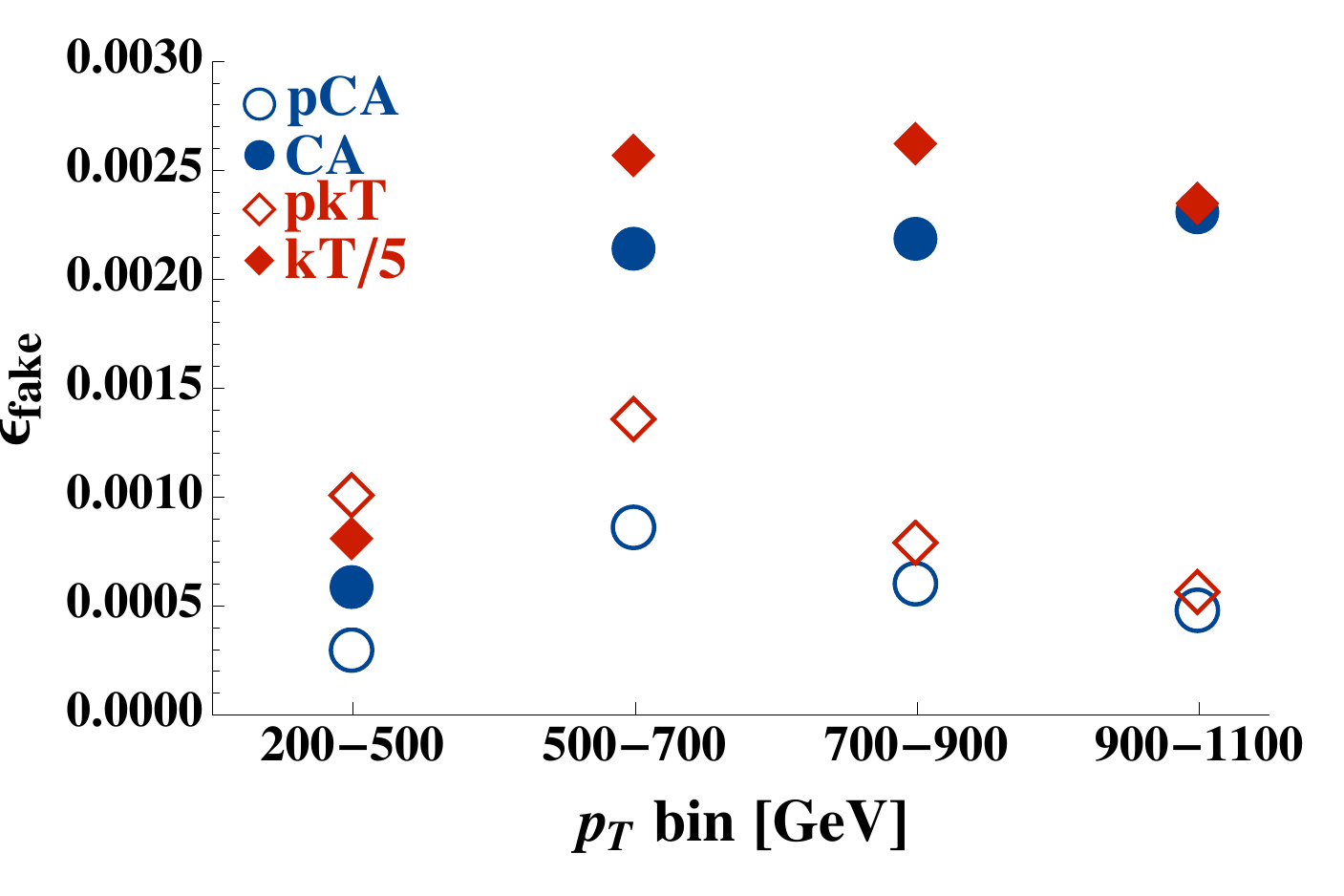}}
\subfloat[$\epsilon_{\text{fake}}$, $W$s]{\label{AbsoluteEfficiencies:Wfake}\includegraphics[width=0.23\textwidth]{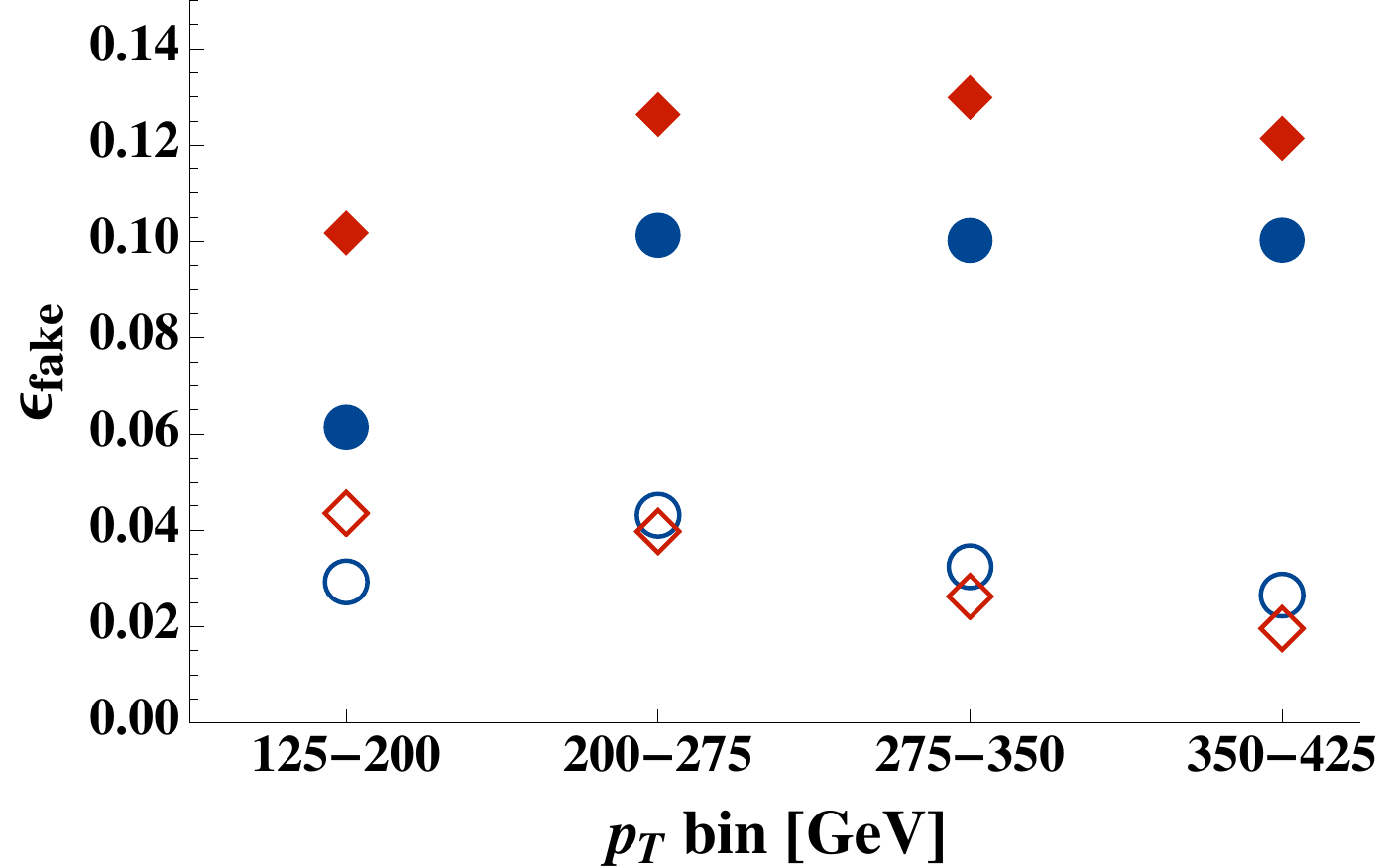}}
\caption{$\epsilon_\text{abs}$ and $\epsilon_\text{fake}$ vs. $p_T$ bin, for the CA and $\kt$ algorithms with and without pruning, using $D = 1.0$.  A ``p'' before the algorithm name denotes the pruned version.  The legend for figure (a) applies to figures (b) and (d).}
\label{AbsoluteEfficiences}
\end{figure}

In Figs.~\ref{AbsoluteEfficiencies:topabs} and \ref{AbsoluteEfficiencies:Wabs}, we plot the absolute efficiency to identify tops and $W$s in the two signal samples for both algorithms, with and without pruning.  For the top sample, this efficiency $\epsilon_\text{abs}$ is the ratio
\be
\epsilon_\text{abs} \equiv \frac{\text{\# of top jets in the signal sample}}{\text{\# of parton-level tops in the $p_T$ range}}
\ee
for each $p_T$ bin, with $\epsilon_\text{abs}$ defined analogously for the $W$ sample.  Because the substructure of the $W$ decay is much simpler than the top decay, with no secondary mass cut, the absolute identification efficiencies are similar between all algorithms.

%\begin{figure}[htbp]
%\includegraphics[width=\columnwidth]{ttbar_AbsPartonEff_legend.pdf}
%\caption{$\epsilon_\text{abs}$ vs. $p_T$ bin in the $t\bar{t}$ sample, for the CA and $\kt$ algorithms with and without pruning, using $D = 1.0$.  A ``p'' before the algorithm name denotes the pruned version.  }%\red{TopVarypTpartoneff}}
%\label{TopVarypTpartoneff}
%\end{figure}

The efficiency to find top quarks is only meaningful when compared to the fake rate for QCD jets to be misidentified as a top quark.  We define this fake rate as
\be
\epsilon_\text{fake} \equiv \frac{\text{\# of fake top jets in the background sample}}{\text{\# of unpruned jets in the $p_T$ range}}
\ee
for each $p_T$ bin, and analogously for the $W$ sample.  In Figs.~\ref{AbsoluteEfficiencies:topfake} and \ref{AbsoluteEfficiencies:Wfake}, we plot $\epsilon_\text{fake}$ for tops and $W$s in the two background samples for both algorithms, with and without pruning.  The fake rate is significantly reduced for pruned jets compared to unpruned jets, for both the top and $W$ studies.  The decrease in absolute efficiency arising from using a narrow mass window is compensated by a correspondingly small fake rate for QCD jets.

\begin{figure*}[htbp!]
\subfloat[$W$'s, CA vs. $\kt$]{\label{CompareCAKT:W}\includegraphics[width=0.24\textwidth]{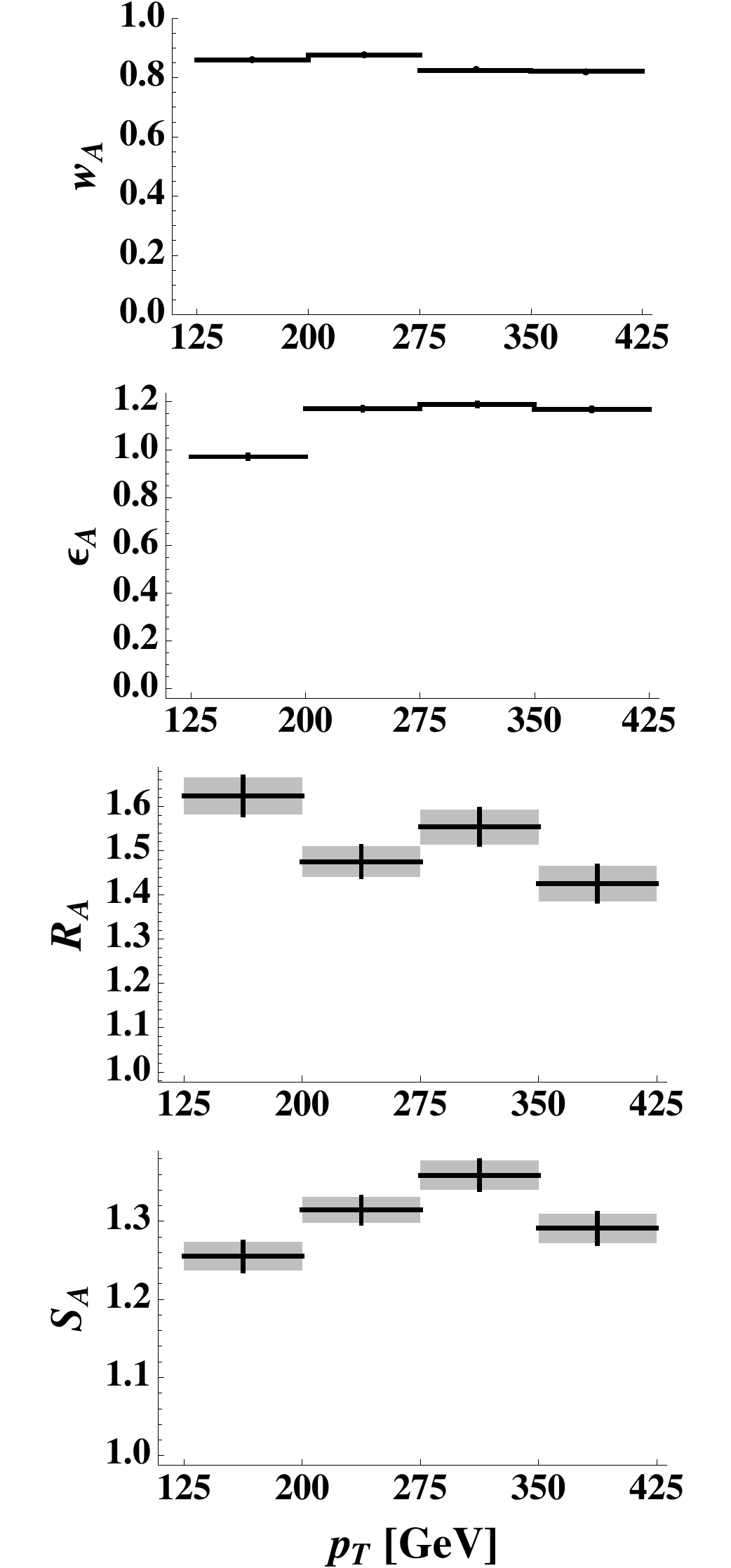}}
\subfloat[tops, CA vs. $\kt$]{\label{CompareCAKT:t}\includegraphics[width=0.24\textwidth]{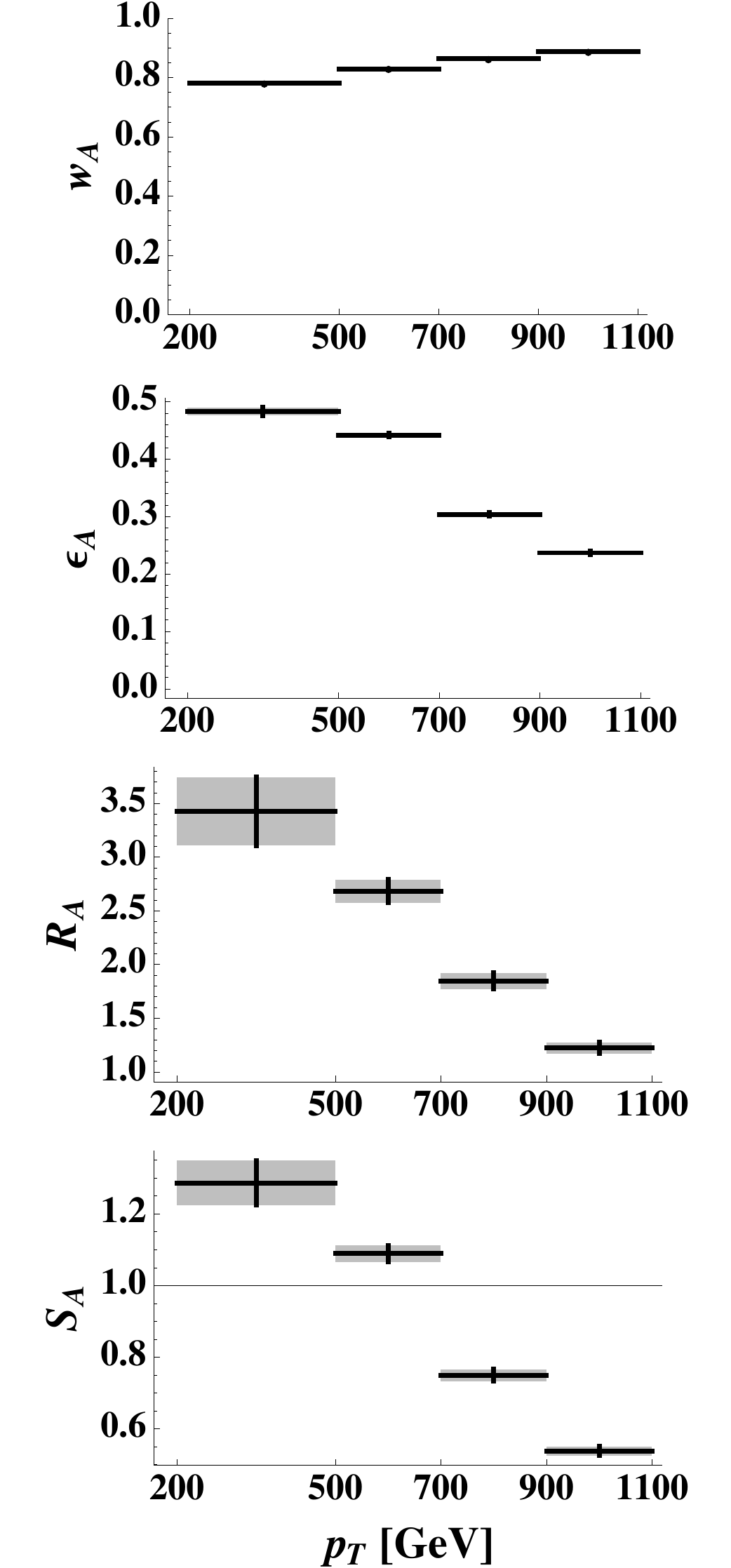}}
\subfloat[$W$'s, pCA vs. p$\kt$]{\label{ComparepCApKT:W}\includegraphics[width=0.24\textwidth]{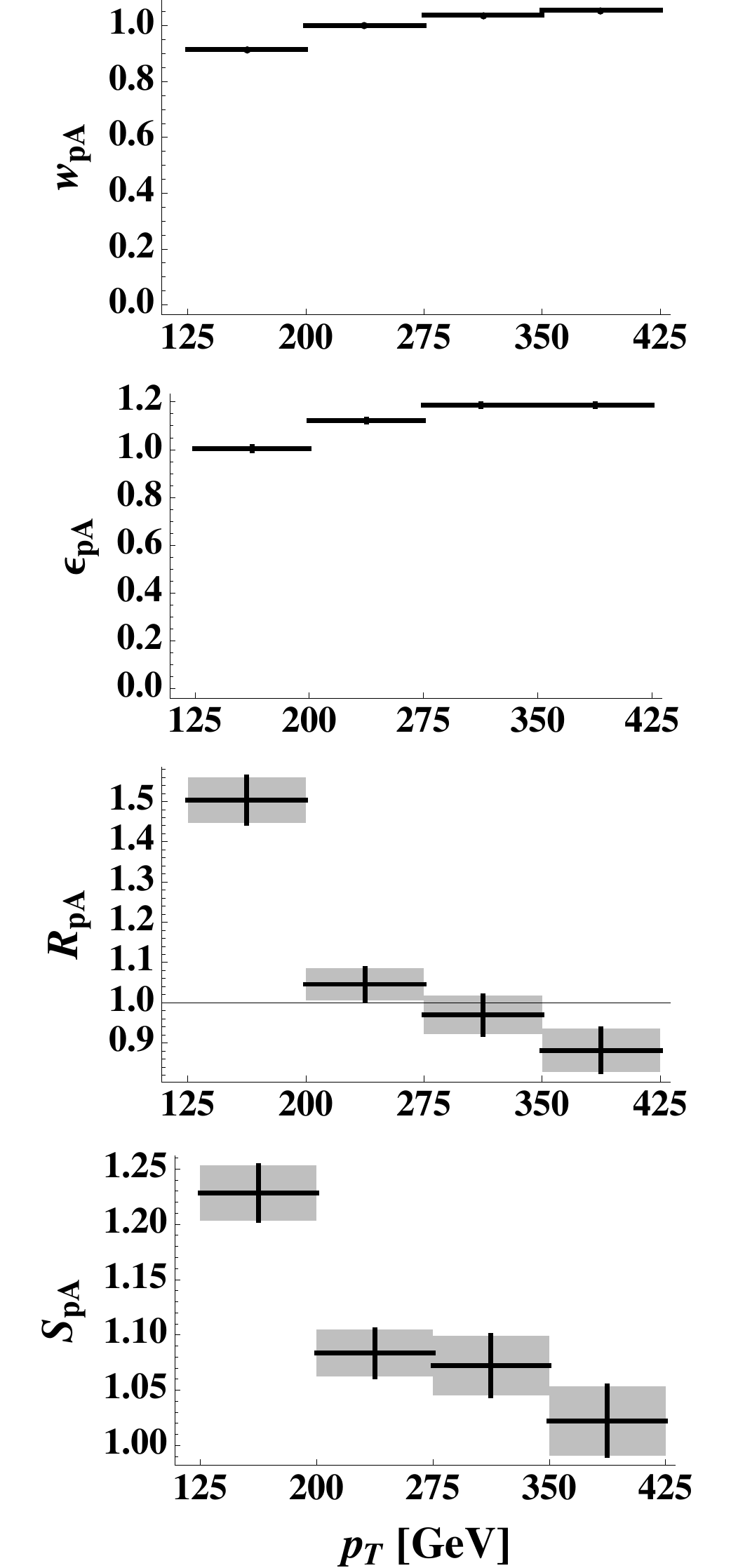}}
\subfloat[tops, pCA vs. p$\kt$]{\label{ComparepCApKT:t}\includegraphics[width=0.24\textwidth]{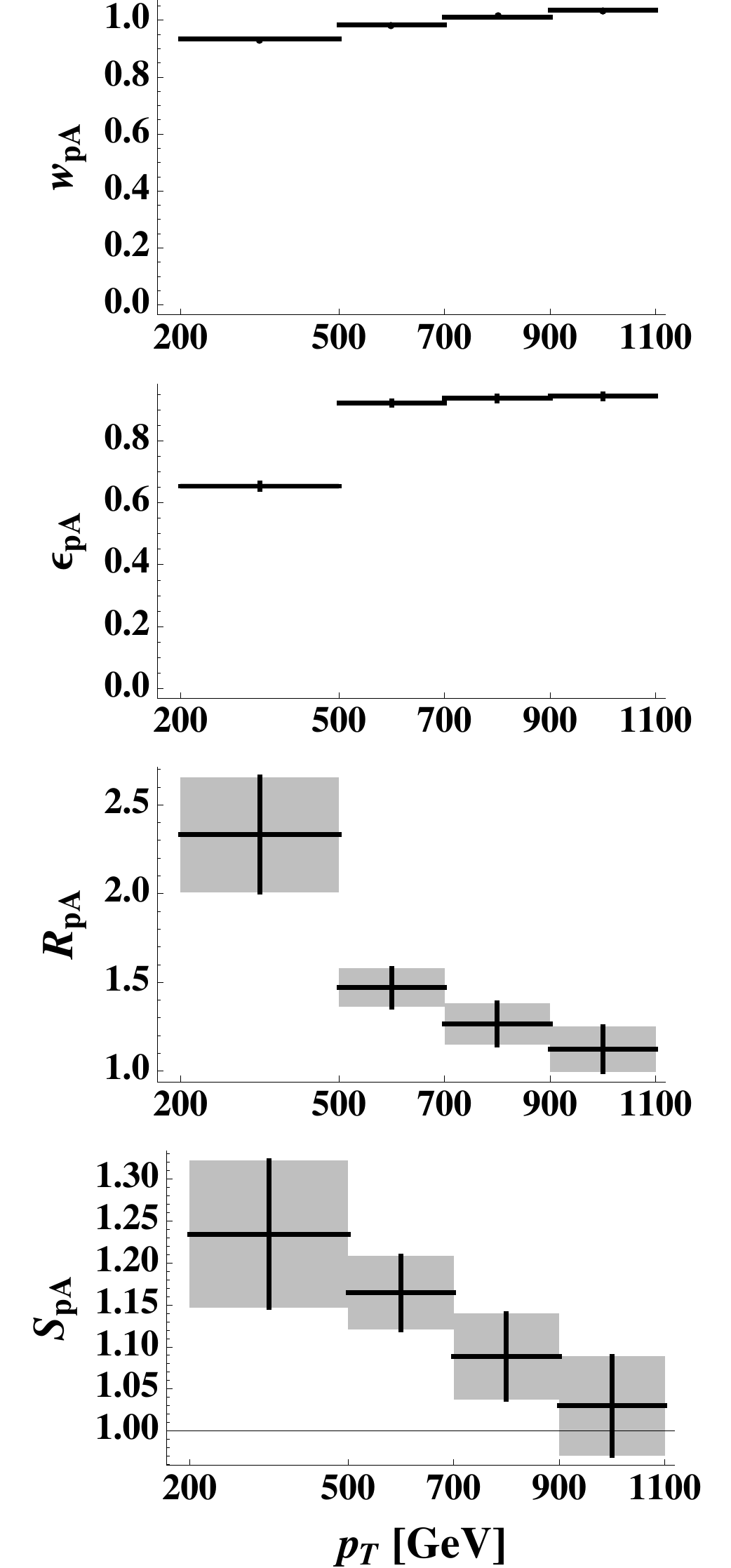}}
\caption{Relative statistical measures comparing CA to $\kt$ jets and pruned CA to pruned $\kt$ jets vs. $p_T$ for $W$'s and tops, using D = 1.0.  Statistical errors are shown.  }%\red{CompareCAKT}}
\label{CompareCAKT}
\end{figure*}

\subsection{Algorithm Comparison}
\label{sec:results:algComparison}

Throughout this paper, we have studied how pruning compares to not pruning for the CA and $\kt$ algorithms.  However, it is also of interest to study how the CA and $\kt$ algorithms compare, with and without pruning.  To do this, we use statistical measures $w_A$, $\epsilon_A$, $R_A$, and $S_A$ analogous to $w_\text{rel}$, $\epsilon$, $R$, and $S$.  For instance,
\be
R_A \equiv \frac{S/B \text{ from the CA algorithm with D = 1.0}}{S/B \text{ from the $\kt$ algorithm with D = 1.0}} .
\ee
We will change the subscript to $pA$ to compare the pruned versions of the algorithms, e.g.,
\be
R_{pA} \equiv \frac{S/B \text{ from pruned CA with D = 1.0}}{S/B \text{ from pruned $\kt$ with D = 1.0}} .
\ee
In Fig.~\ref{CompareCAKT}, we plot the measures comparing CA to $\kt$ and pruned CA to pruned $\kt$ for both the $WW$ and $t\bar{t}$ samples.

These comparisons illustrate many of the effects that we have observed throughout the studies in this paper.  For the unpruned algorithm comparison, CA tends to have a much lower efficiency to identify tops than $\kt$.  As $p_T$ increases, CA performs more poorly relative to $\kt$, with the efficiency decreasing significantly.  This arises because the CA has a decreasing efficiency to identify the $W$ at high $p_T$, when the top quark becomes more localized in the fixed $D$ jet.  Pruning corrects for this, though the performance of CA relative to $\kt$ still decreases at high $p_T$.

The $WW$ sample is instructive because it lets us compare the effectiveness of pruning between CA and $\kt$ across a wide range in $p_T$.  For the unpruned algorithms, the performance of CA relative to $\kt$ is fairly consistent over all $p_T$, reflecting the fact that $W$ identification is simpler than top identification, with accurate mass reconstruction the only requirement.  However, when the jets are pruned, the performance of pruned CA relative to pruned $\kt$ improves in the smallest $p_T$ bin and worsens in the largest $p_T$ bin, as compared to the performance of CA versus $\kt$ for unpruned jets.  This skewing of the statistical measures indicates that pruning is more effective for CA than $\kt$ at small $p_T$, where threshold effects are important, and more effective for $\kt$ than CA at large $p_T$.

\subsection{Detector Effects}
\label{sec:results:smearing}

\begin{figure}[htbp]
\subfloat[tops, CA jets]{\label{CompareTopMassSmeared:CA}\includegraphics[width=0.23\textwidth]{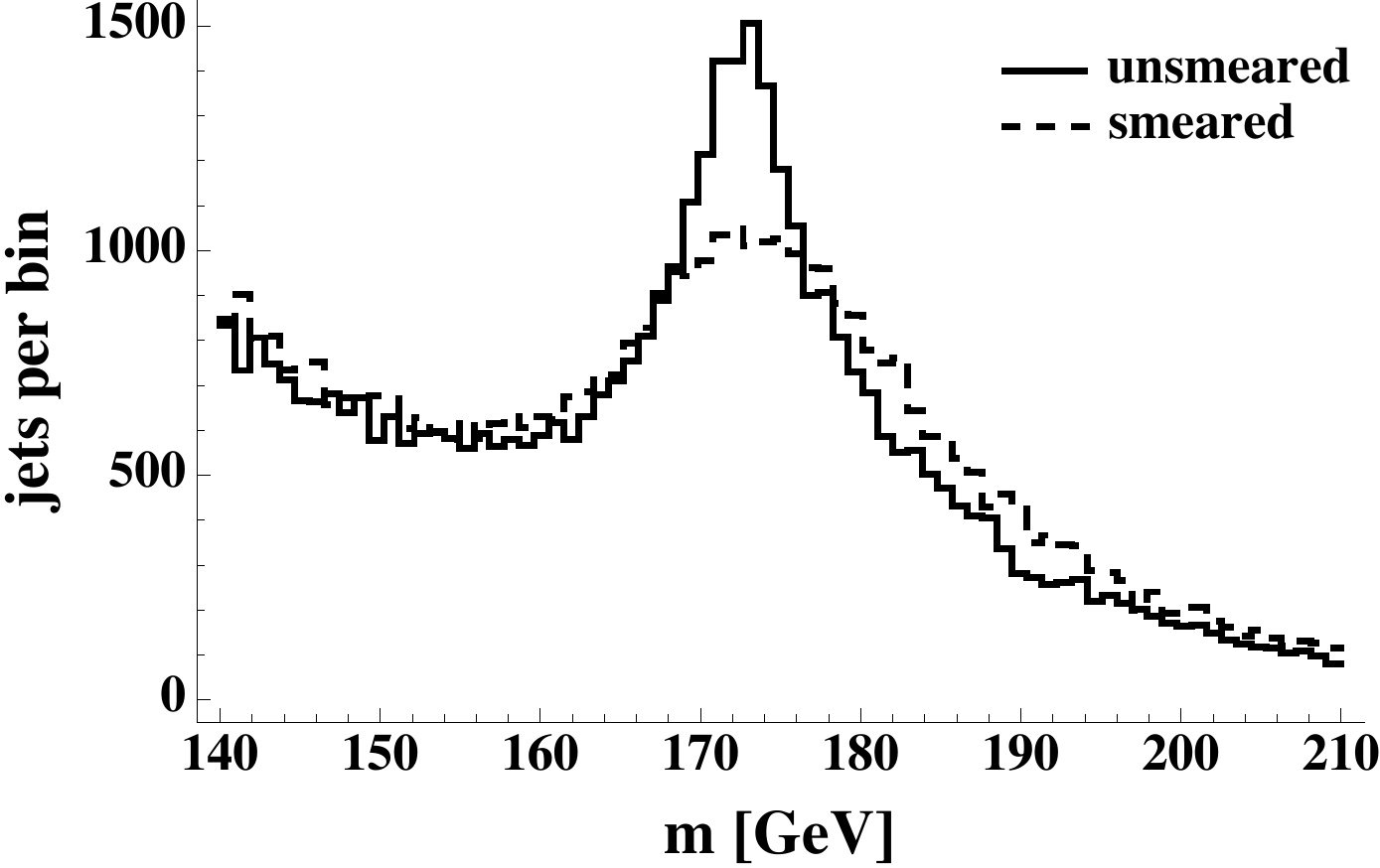}}
\subfloat[tops, $\kt$ jets]{\label{CompareTopMassSmeared:KT}\includegraphics[width=0.23\textwidth]{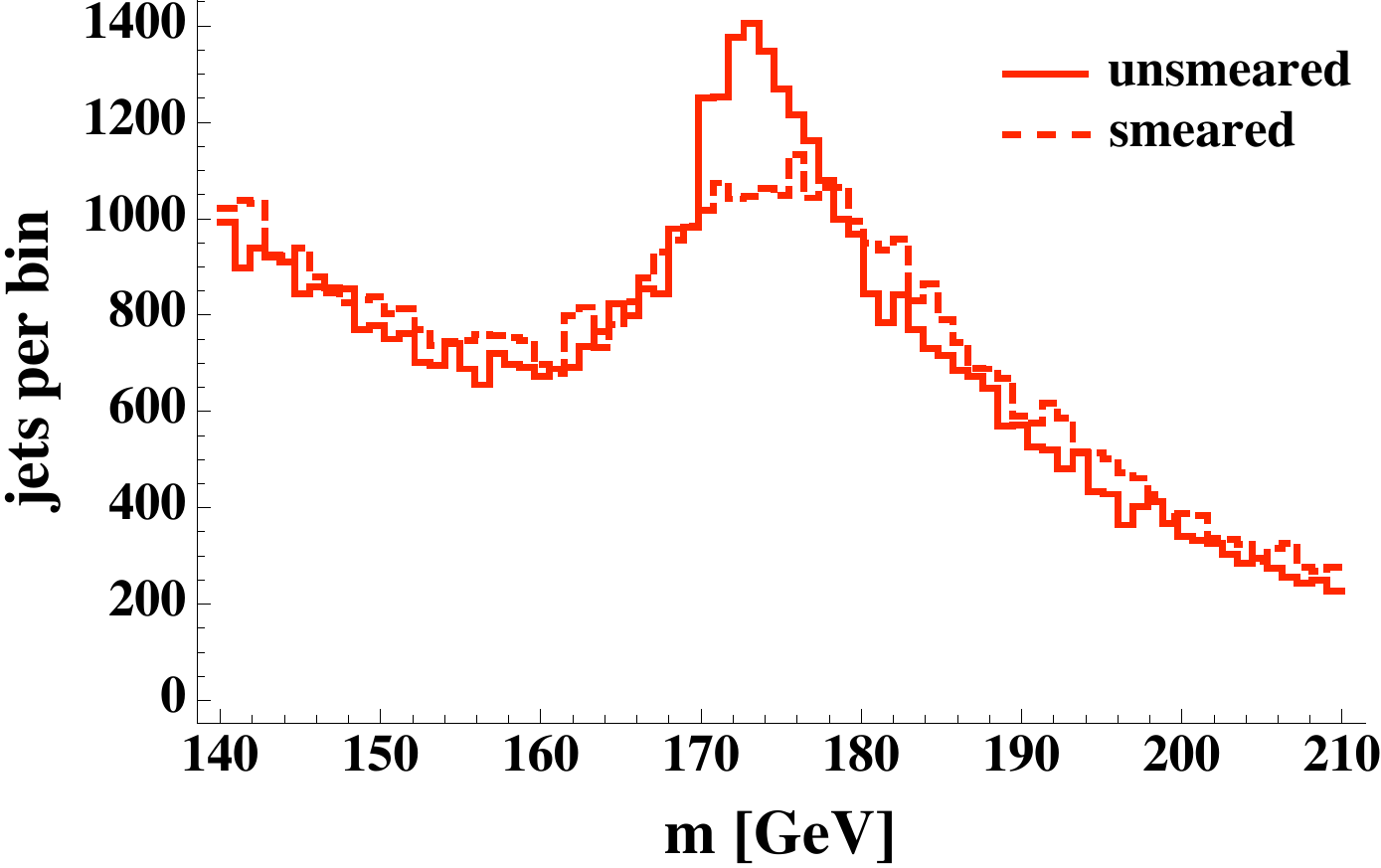}}
\caption{Distribution in jet mass for $t\bar{t}$ events, with (dashed) and without (solid) energy smearing.  The jets have $p_T$ of 200--500 GeV and $D = 1.0$, and there is no pruning.  }%\red{CompareTopMassSmeared}}
\label{CompareTopMassSmeared}
\end{figure}

\begin{figure}[htbp]
\subfloat[tops, pruned CA jets]{\label{CompareTopMassSmearedPruned:CA}\includegraphics[width=0.23\textwidth]{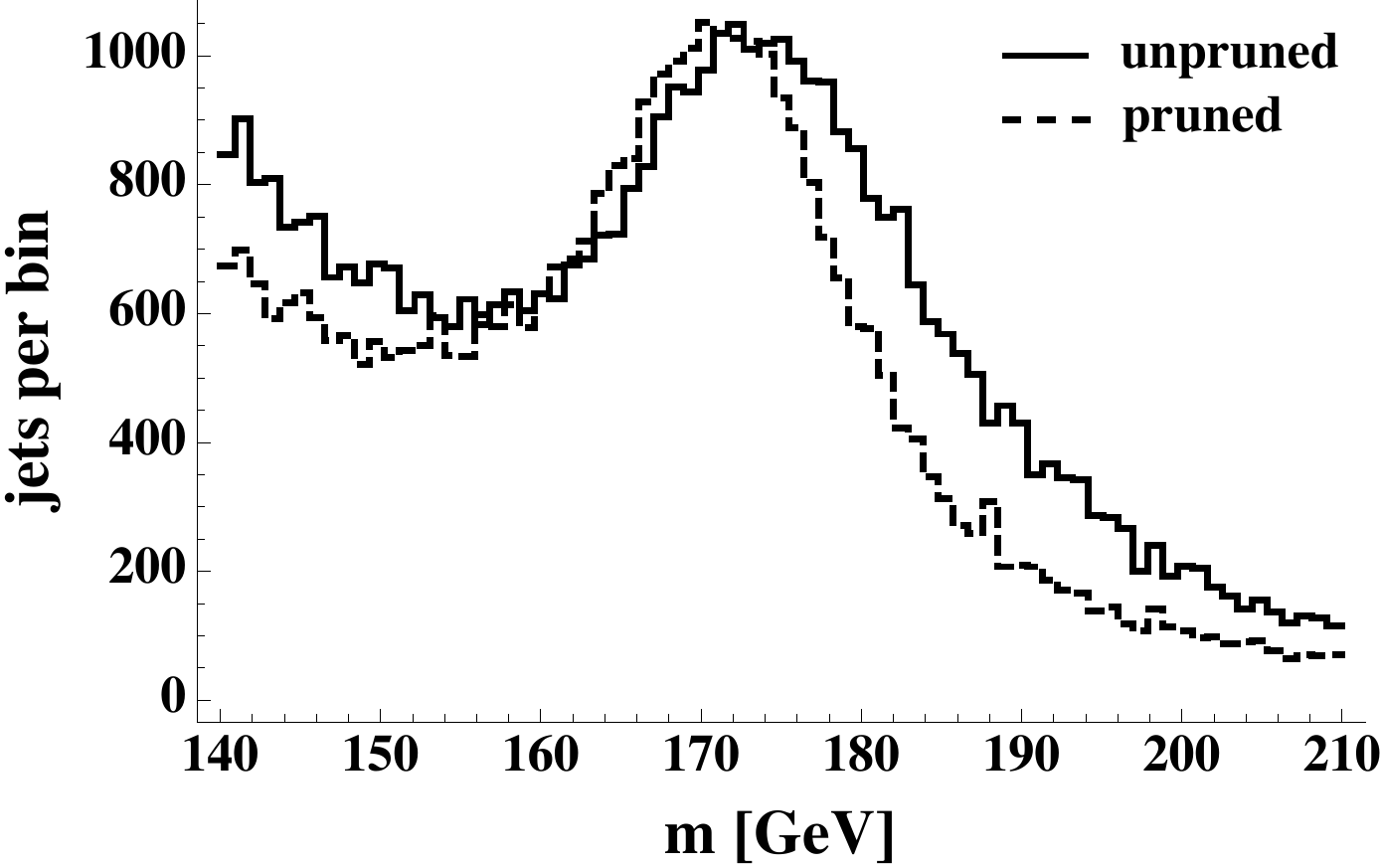}}
\subfloat[tops, pruned $\kt$ jets]{\label{CompareTopMassSmearedPruned:KT}\includegraphics[width=0.23\textwidth]{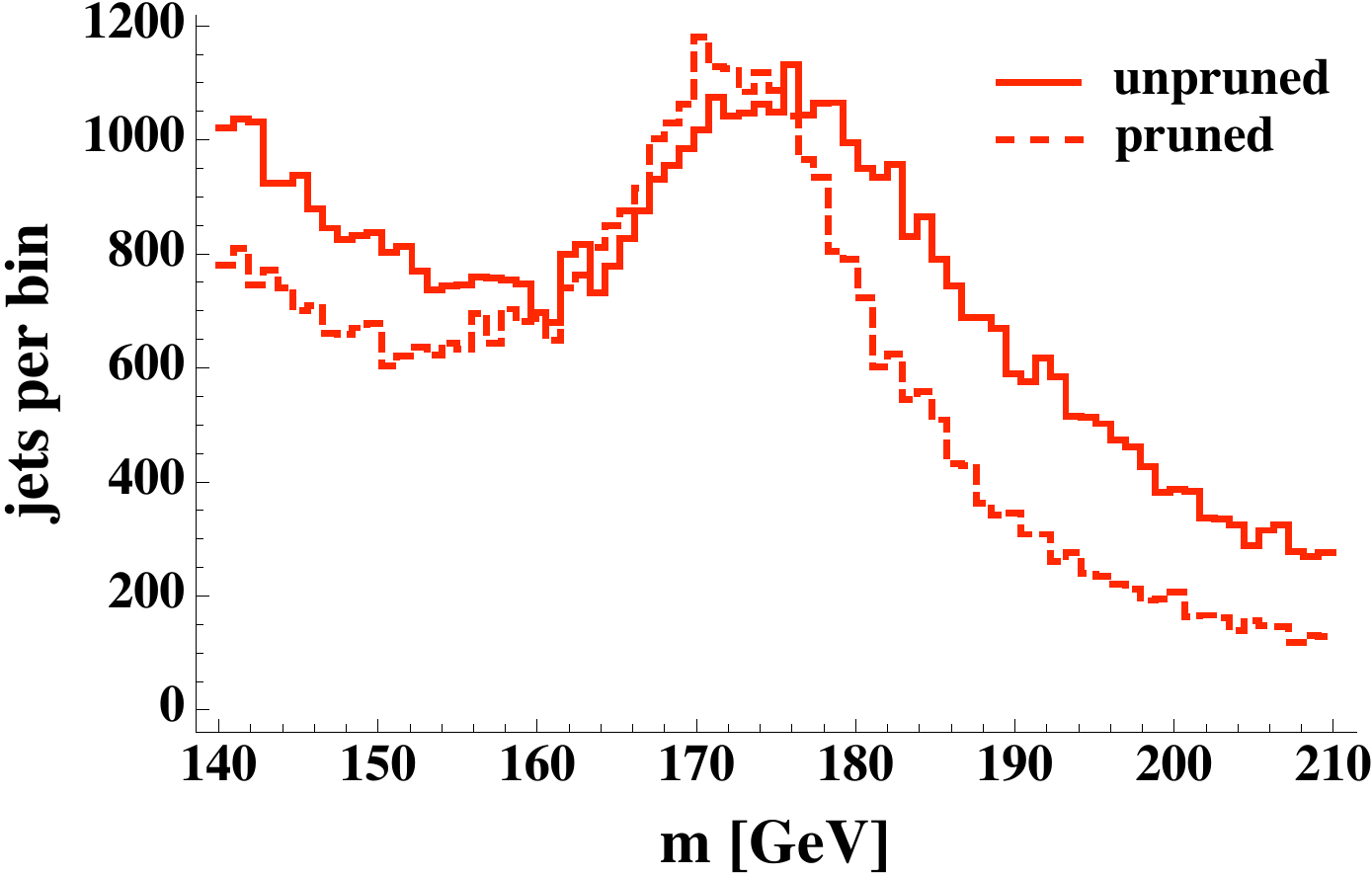}}
\caption{Distribution in jet mass for pruned (dashed) and unpruned (solid) jets, for $t\bar{t}$ events with energy smearing.  The jets have $p_T$ of 200--500 GeV and $D = 1.0$.  }%\red{TopMassSmearedPruned}}
\label{TopMassSmearedPruned}
\end{figure}

So far, no detector simulation has been applied to the simulated events aside from clustering particles into massless calorimeter cells.  We now consider a technique that approximates the impact that detector resolution has on the effectiveness of pruning.  We modify our top and $W$ jet analyses by smearing the energy $E$ of each calorimeter cell with a factor sampled from a Gaussian distribution with mean $E$ and standard deviation $\sigma$ given by
\be
\sigma(E) = \sqrt{a^2 E + b^2 + c^2 E^2} .
\ee
We consider a parameter set motivated by the expected ATLAS hadronic calorimeter resolution \cite{ATLAS:08.1}, $\{a,b,c\} = \{0.65, 0.5, 0.03\}$.  One obvious effect of the detector smearing is degraded mass resolution.  In Fig.~\ref{CompareTopMassSmeared}, we show this effect by plotting the jet mass distribution for the $t\bar{t}$ sample in the first $p_T$ bin.  Even after smearing, however, pruning improves the jet mass resolution.  In Fig.~\ref{TopMassSmearedPruned}, we plot the pruned and unpruned jet mass distribution for the $t\bar{t}$ sample in the first $p_T$ bin.  Note that because the QCD jet mass distribution is smooth, only the overall size of the sample in the mass window changes, so we do not plot these distributions.

\begin{figure*}[htbp]
\subfloat[$W$'s, CA jets]{\label{VarypTSmeared:WCA}\includegraphics[width=0.24\textwidth]{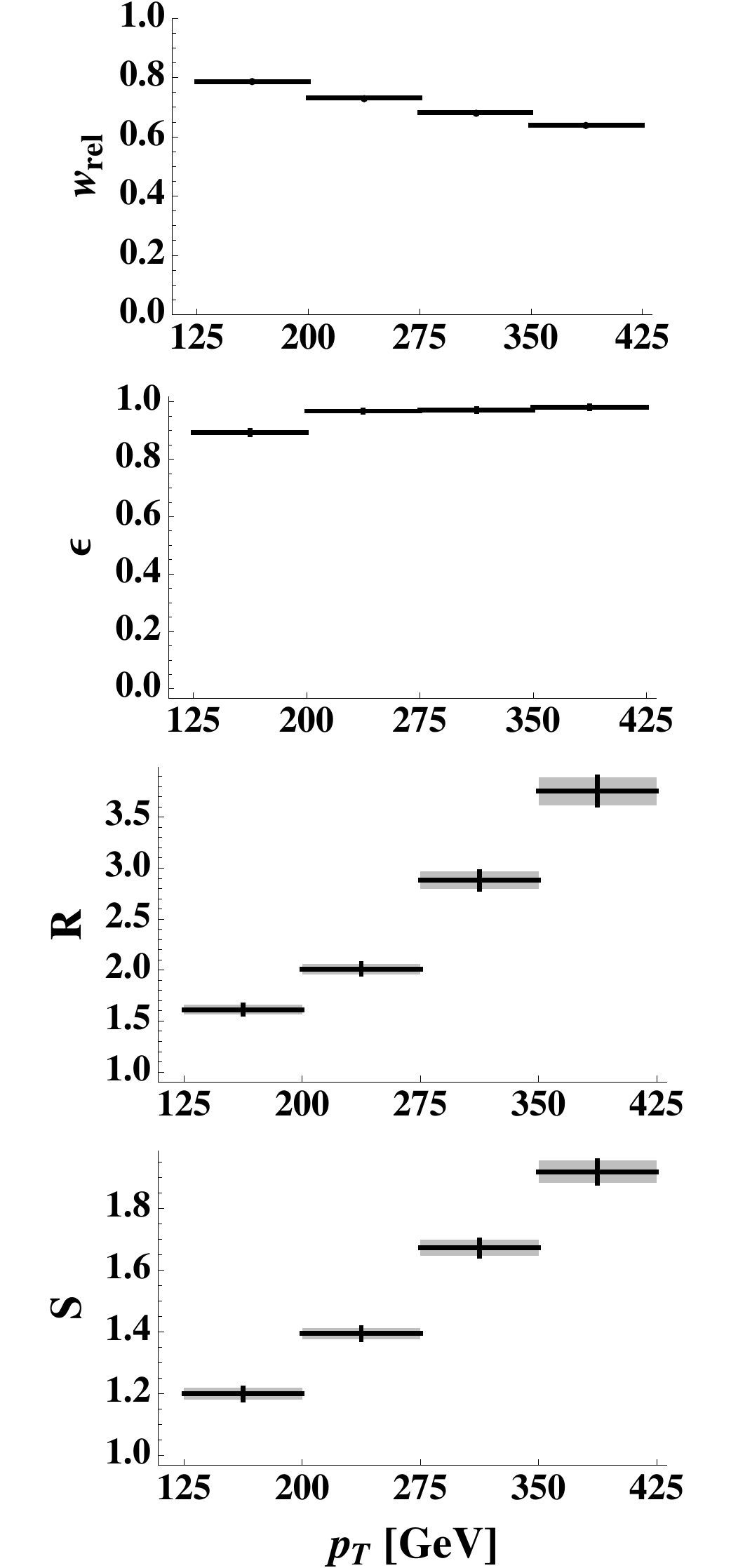}}
\subfloat[tops, CA jets]{\label{VarypTSmeared:tCA}\includegraphics[width=0.24\textwidth]{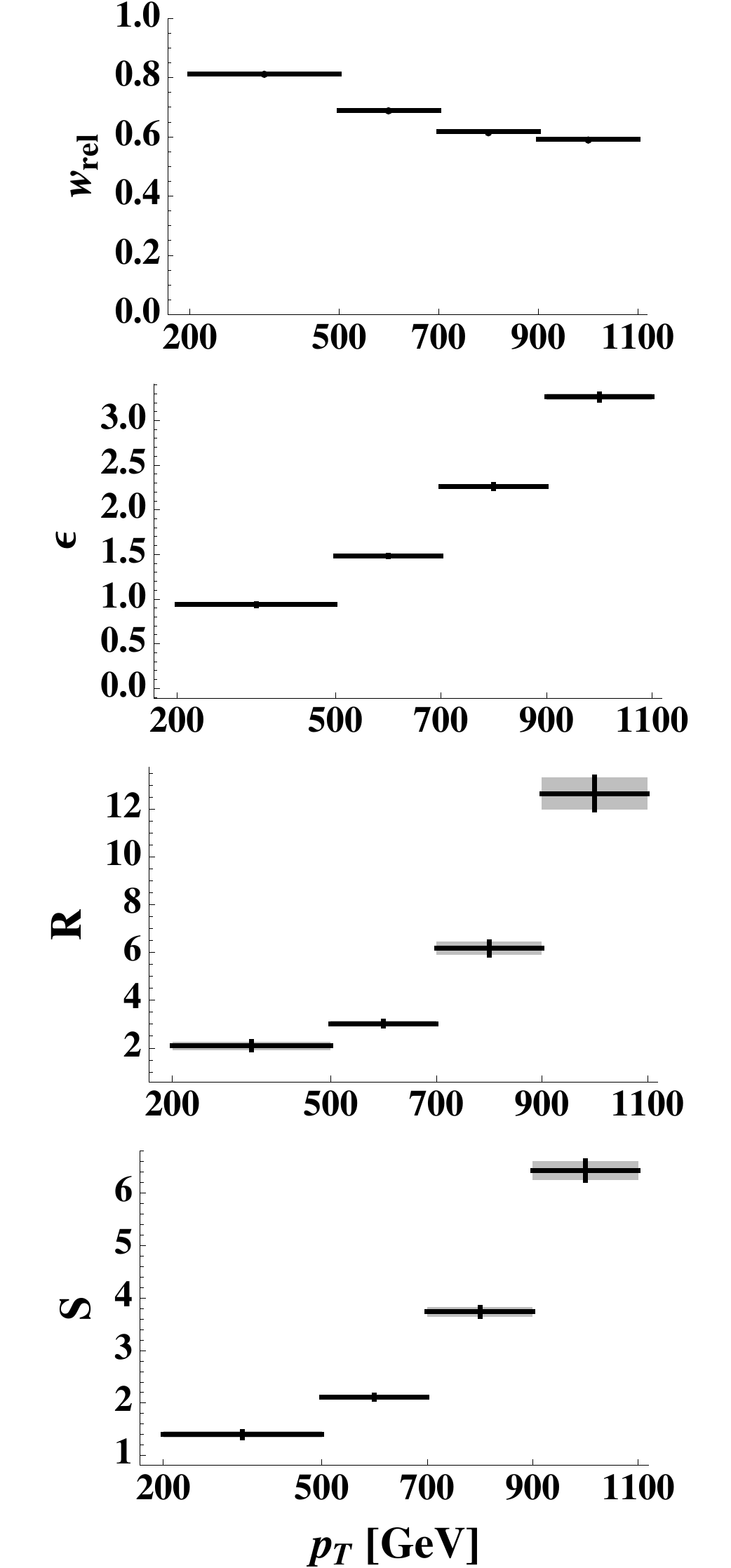}}
\subfloat[$W$'s, $\kt$ jets]{\label{VarypTSmeared:WkT}\includegraphics[width=0.24\textwidth]{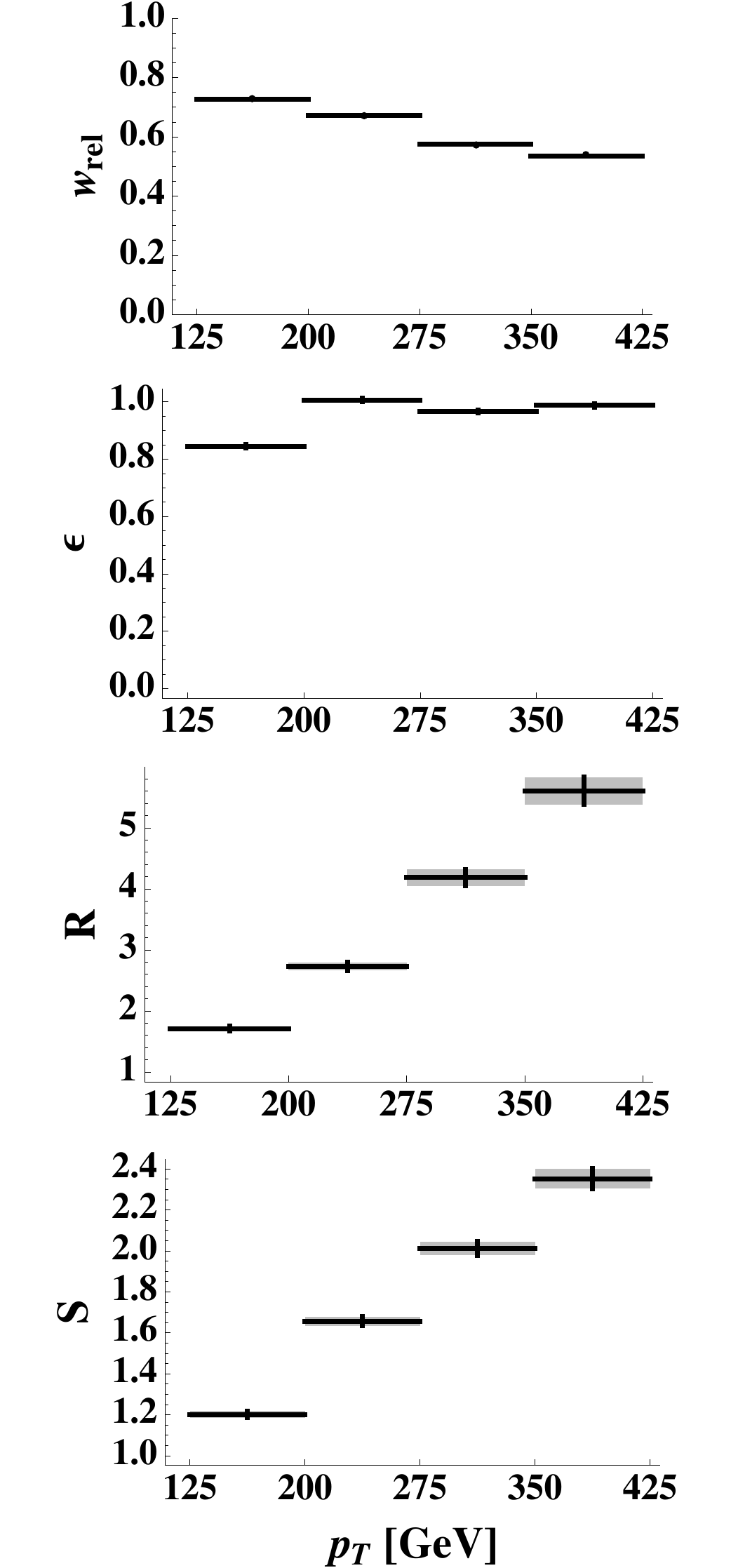}}
\subfloat[tops, $\kt$ jets]{\label{VarypTSmeared:tkT}\includegraphics[width=0.24\textwidth]{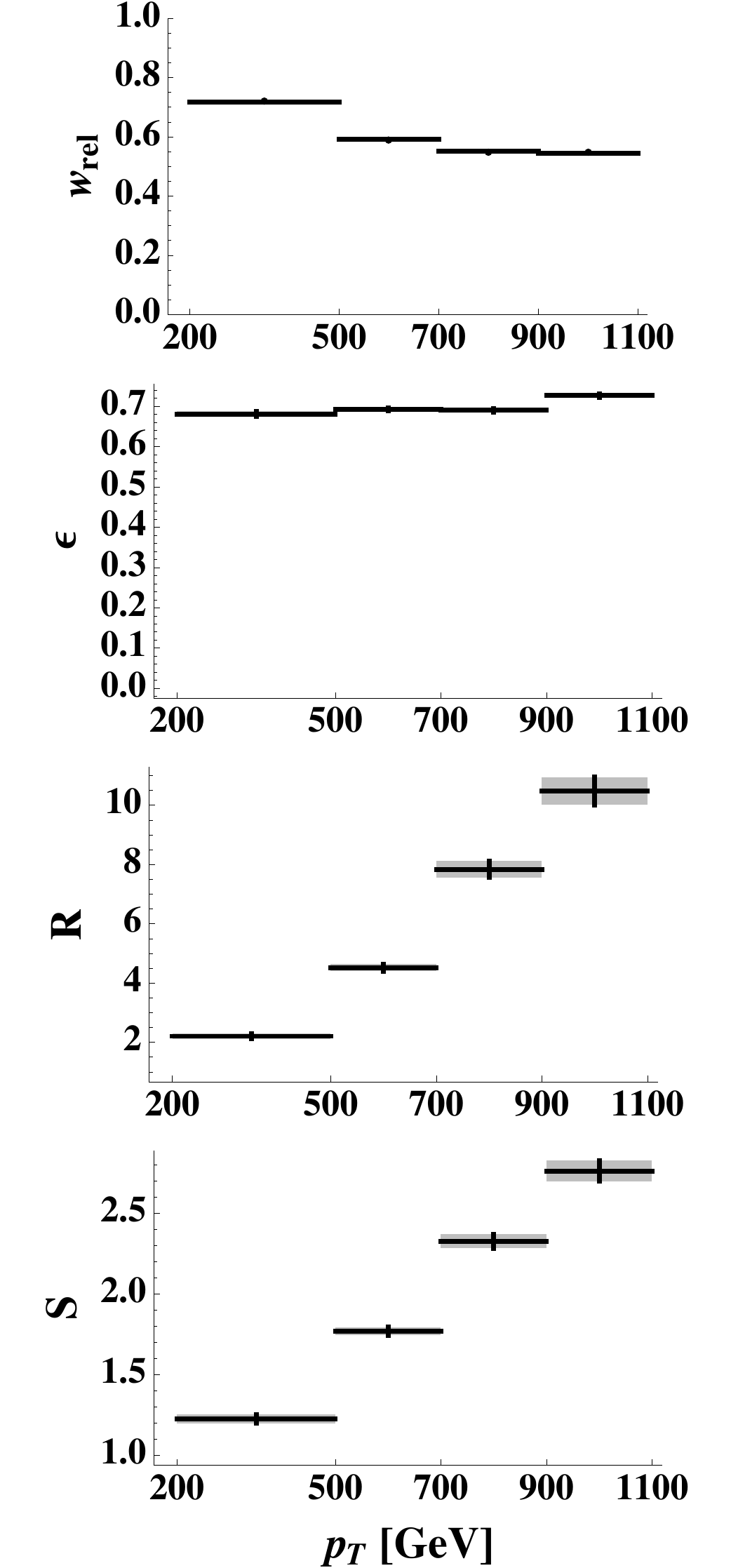}}
\caption{Relative statistical measures $w_\text{rel}$, $\epsilon$, $R$, and $S$ vs. $p_T$ for $W$'s and tops, using CA and $\kt$ jets.  Calorimeter cell energies are smeared as described in the text.  Statistical errors are shown.  }%\red{VarypTSmeared}}
\label{VarypTSmeared}
\end{figure*}

If Fig.~\ref{VarypTSmeared}, we repeat the basic analysis of Sec.~\ref{sec:results:fixedD}, applying the detector smearing described above to events using $D = 1.0$ over all four $p_T$ bins and plotting the measures $w_\text{rel}$, $\epsilon$, $R$, and $S$.  This figure can be compared to Fig.~\ref{VarypT} from the previous analysis, which plots the same measures when no energy smearing is used.  The improvements are very similar to those for unsmeared jets, good evidence that pruning may retain its utility in a more realistic detector simulation or in real data.

%%%%%%%%%% END OF SECTION 8: STUDY RESULTS %%%%%%%%%%

\section{Conclusions and Future Prospects}
\label{sec:conc}

In this work, we have demonstrated that recombination jet algorithms shape the substructure of heavy particles reconstructed in single jets.  We have identified regions in the variables $z$ and $\Delta R$ where individual recombinations are unlikely to represent the kinematics of a reconstructed heavy particle.  Specifically, soft, large-angle recombinations are unlikely to arise from the accurate reconstruction of a heavy particle decay, and are likely to come from QCD jets, uncorrelated radiation, or systematic effects of the jet algorithm.  For the CA algorithm, we have demonstrated that these soft, large-angle recombinations are a key systematic effect that shapes the substructure of the jet, in particular the final recombinations.

We have presented a procedure, calling pruning, that eliminates soft, large-angle recombinations from the substructure of the jet.  Using hadronically decaying top quarks and $W$ bosons as test cases, we have demonstrated that the pruning procedure improves the separation between heavy particles decays and a QCD multijet background.  We have motivated the parameters of the pruning procedure and demonstrated that they roughly optimize the improvements from pruning in our study for both top quarks and $W$ bosons.

Our studies on pruning have demonstrated many positive results of the procedure.  In a heavy particle search, the jet is sensitive to the parameter $D$, and if the value of $D$ is not well matched to the decay of a heavy particle then the ability to identify that particle in single jets is greatly reduced.  Our results indicate that pruning removes much of the jet algorithm's dependence on $D$.  Pruning shows improvements even when $D$ is adjusted to fit the expected decay of the heavy particle.  We have demonstrated that pruning is insensitive to the effects of the underlying event, as the underlying event mainly contributes soft, uncorrelated radiation to a jet.  Additionally, we have shown that the results of pruning are robust to a basic energy-smearing applied to the calorimeter cells used to seed the jet algorithm.  Finally, we have quantified absolute measures of the pruning procedure that can be used to compare to other jet substructure methods.

It should be reiterated that pruning systematizes methods that have been proposed by other authors for specific searches.  Pruning should be applicable to a wide range of searches, and is intended to be a generic jet analysis tool.  We have detailed the ideas behind why pruning works and why it should be used, and presented an in-depth discussion of many of the physics issues arising when studying jet substructure.

\subsection{Future Prospects}

The conclusions in this paper, like those for any analysis technique not demonstrated on real data, must be taken cautiously.  This is especially true for studies like this one on jet substructure, where a majority of the work has been in exploring techniques that may --- or may not --- actually be useful in an experiment.  However, new techniques like jet substructure offer great promise.  All studies thus far indicate that jet substructure, and in general a more innovative approach to jets, will be a useful tool for understanding the physics in events with jets at collider experiments.

The most obvious and immediate application of pruning, and jet substructure tools in general, is in rediscovery of the Standard Model at the LHC.  As the LHC collects data from high-energy collisions, there will be an abundant sample of high-$p_T$ top quarks, and $W$ and $Z$ bosons with fully hadronic decays.  As these channels are observed using standard analyses, jet substructure techniques can be applied and tested.  These channels can also serve as key calibration tools for jet substructure methods applied in the search for new physics.

From the theoretical side, improvements in jet-based analyses can come from a variety of sources.  As calculations in perturbative QCD progress, they can be used to improve predictions for jet-based observables in QCD.  Improved Monte Carlo tools, such as the continued implementation of next-to-leading order matrix elements and better parton showers, will lead to more accurate studies and a better understanding of jet physics.  Additionally, the framework of soft-collinear effective theory (SCET) can improve the understanding of QCD jets  \cite{Bauer:2000yr, Bauer:2000ew, Bauer:2001yt, Bauer:2008dt, Bauer:2008jx}.  As SCET is adapted to describe a wider variety of event topologies and realistic jet algorithms are implemented in the effective theory, it can be used to calculate resummed predictions for jet-based observables and accurately describe processes that are difficult to access with perturbative QCD \cite{Bauer:2006qp, Stewart:2009yx, Cheung:2009sg}.  Jets will likely play a central role in new physics searches at the LHC, and a better understanding of jets and jet substructure can aid in the discovery process.

\subsection*{Acknowledgements}
We would like to thank Matt Strassler, Jacob Miner, and Andrew Larkoski for collaboration in early stages of this work.  We thank Johan Alwall for help with MadGraph/MadEvent, and acknowledge useful discussions with Steve Mrenna, Gavin Salam, Tilman Plehn, Karl Jacobs, Peter Loch, Michael Peskin, and others in the context of the Joint Theoretical-Experimental Terascale Workshops at the
Universities of Washington and Oregon, supported by the U.S. Department of
Energy under Task TeV of Grant No.~DE-FG02-96ER40956. This work was supported in part by the U.S. Department of Energy under Grant No.~DE-FG02-96ER40956. JRW was also supported in part by an LHC Theory Initiative Graduate Fellowship.

\appendix

\section{Computational Details}
\label{sec:appendix}

We give a brief summary of the computational tools employed to do the studies in this paper.  We generate LHC (14 TeV) events using MadGraph/MadEvent v4.4.21 \cite{Alwall:07.1} interfaced with Pythia v6.4 \cite{Sjostrand:06.1}. We employ MLM-style matching, implemented in MadGraph (see, e.g., \cite{Alwall:08.1}), on the backgrounds.  We have checked that our matching parameters are reasonable using the tool MatchChecker \cite{MatchChecker}.  We use the DWT tune \cite{Albrow:06.1} in Pythia to give a ``noisy'' underlying event (UE).  For the hadron-level studies in Secs.~\ref{sec:QCDJets} and \ref{sec:reconHeavy}, we exclude the underlying event by setting the Pythia parameter MSTP(81) to zero, turning off multiple interactions.  The UE comparisons in Sec.~\ref{sec:algEffects} compare samples with this parameter set at 0 or 1.  No detector simulation is performed so we can isolate the ``best case'' effects of our method.  In Sec.~\ref{sec:results:smearing}, we examine the effects of Gaussian smearing on the energies of final state particles from Pythia to get a sense for how much the results may change with a detector.

For the $W$ study, the signal sample is $W^+W^-$ pair production, with exactly one $W$ required to decay leptonically.  The background is a matched sample of a $W$ and one or two light partons (gluons and the four lightest quarks) before showering.  These partons must be in the central region, $|\eta| < 2.5$.  $\eta$ is the pseudorapidity, $\eta \equiv \ln(\cot(\theta_b/2))$, with $\theta_b$ the polar angle with respect to the beam direction ($\eta = y$ for massless particles).  Signal and background samples are divided into four $p_T$ bins: [125, 200], [200, 275], [275, 350], and [350, 425] (all in GeV).  Each bin is defined by a $p_T$ cut that is applied to single jets in the analysis.  These bins confine the $W$ boost to a narrow range and allow us to study the performance of pruning as the jet $p_{T}$ (or $W$ boost) varies.

For each $p_T$ bin $[p_T^\text{min}, p_T^\text{max}]$, both samples are generated with a $p_T$ cut on the leptonic $W$ of $p_T^\text{min} - 25$ GeV.  For the background, we set the matching scales $(Q^\text{ME}_\text{cut}, Q_\text{match})$ to be (10, 15) GeV in all four bins.

For the top quark reconstruction study, the signal sample is $t\bar{t}$ production with fully hadronic decays.  The background is a matched sample of QCD multijet production with two, three, or four light partons, with the same cut on parton centrality as in the $W$ study.  Samples are again divided into four $p_T$ bins: [200, 500], [500, 700], [700, 900], and [900, 1100] (all in GeV).

We generate signal and background samples with a parton-level $h_{T}$ cut for generation efficiency, where $h_{T}$ is the scalar sum of all $p_{T}$ in the event.  For each $p_T$ bin $[p_T^\text{min}, p_T^\text{max}]$, the parton-level $h_T$ cut is $p_T^\text{min} - 25\ \text{GeV} \le h_T/2 \le  p_T^\text{max} + 100\ \text{GeV}$.  For the background, we use matching scales (20, 30) GeV for the smallest $p_T$ bin and (50, 70) GeV in the other three bins.

From the hadron-level output of Pythia, we group final-state particles into ``cells'' based on the segmentation of the ATLAS hadronic calorimeter ($\Delta \eta=0.1$, $\Delta\phi=0.1$ in the central region). We sum the four-momenta of all particles in each cell and rescale the resulting three-momentum to make the cell massless. After a threshold cut on the cell energy of 1 GeV, cells become the inputs to the jet algorithm. Our implementation of recombination algorithms uses FastJet \cite{Cacciari:05.1}, with a pruning plugin we have written \cite{FastPrune}.

Several of the plots in early sections involve mass cuts on jets.  The details of these cuts are provided in Sec.~\ref{sec:MC:Metrics}.

%%%%%%%%%%% END OF APPENDIX %%%%%%%%%%%%%%%%%%%%%%%%%%%

\bibliography{jetcites}

%%%%%%%%%% END OF THE BIBLIOGRAPHY %%%%%%%%%%

\end{document}